\documentclass[12pt]{mycimento}
\usepackage{epsfig}

\def\beq{\begin{equation}}   \def\eeq{\end{equation}}
 
\def\bea{\begin{eqnarray}}   \def\eea{\end{eqnarray}}

\newcommand{\GeV}{\,\mbox{GeV}}
\newcommand{\MeV}{\,\mbox{MeV}}
\newcommand{\matel}[3]{\langle #1|#2|#3\rangle}

\newcommand{\gsim}{\lower.7ex\hbox{$ \;\stackrel{\textstyle>}{\sim}\;$}}
\newcommand{\lsim}{\lower.7ex\hbox{$ \;\stackrel{\textstyle<}{\sim}\;$}}

\def\rarr{ \rightarrow}
\def\epem{ e^+e^- }
\def\c2{CLEO~II.V}
\def\ccb{ c\bar{c} }
\def\d0d0{ D^0\bar{D}^0 }
\def\p0p0{ P^0\bar{P}^0 }
\def\ddb{ D\bar{D} }
\def\qp2{ \Bigl| \frac{q}{p} \Bigr|^2 }
\def\pq2{ \Bigl| \frac{p}{q} \Bigr|^2 }
\def\rarr{ \rightarrow }

\def\jp{ J/\psi }
\def\Jp{ J/\psi }
\def\pspr{ \psi^\prime }
\def\ps2s{  \psi(2S) }
\def\q2{ $q^2$ }
\def\cm2s1{ $\,{\rm cm}^{-2} {\rm s}^{-1}$} 
\begin{document}

\begin{flushright} 
LNF-09(P) 15 August 2003 \\  
UND-HEP-03-BIG\hspace*{.08em}06\\ 
charmreviewv137nc.tex \\ 
\today 
\end{flushright}  
\bigskip  
\boldmath 
\begin{center} 
 {\Large {A Cicerone for the Physics of Charm}}\\ 
 {\large {~}}\\
   ~\\ 
 {\large{S. Bianco and F. L. Fabbri}}\\ 
 \vspace{0.3cm} 
 {\sl Laboratori Nazionali di Frascati dell'INFN} \\ 
  {\sl Frascati (Rome), I-00044,  Italy} \\ 
 \vspace{0.3cm} 
 {\large{D. Benson and I. Bigi}}  \\ 
 \vspace{0.3cm} 
 {\sl Dept. of Physics, University of Notre Dame du Lac} \\ 
 {\sl Notre Dame, IN 46556, U.S.A.}\\ 
 \vspace{0.3cm} 
\end{center} 
 
\begin{abstract} 
After briefly recapitulating the history of the charm 
quantum number we sketch the experimental environments and instruments 
employed to study the behaviour of charm hadrons and then describe 
the theoretical tools for treating charm dynamics. We discuss a 
wide range of inclusive production processes before analyzing the spectroscopy 
of hadrons with hidden and open charm and the weak lifetimes of 
charm mesons and baryons. Then we address leptonic, exclusive semileptonic 
and nonleptonic charm decays. Finally we treat $D^0 - \bar D^0$ 
oscillations and CP (and CPT)  violation before concluding with some comments 
on charm and the quark-gluon plasma. We will make the case that future studies of 
charm dynamics -- in particular of CP violation -- can reveal the presence of New Physics. 
The experimental sensitivity has only recently reached a level where this could reasonably 
happen, yet only as the result of dedicated efforts. 
 
This review is meant to be both a pedagogical 
introduction for the young scholar and a useful reference for the experienced 
researcher. We aim for a self-contained description of the fundamental features while 
providing a guide through the literature for more technical issues. 
\end{abstract} 
 
\tableofcontents 
 
\vspace{1cm} 
\vfill 

\section{Preface} 
"Physicists, colleagues, friends, lend us your ears -- we have come to 
praise charm, not  bury it!"  
We have chosen such a theatrical opening not 
merely to draw   your attention to our review. We feel that charm's 
reputation --   like Caesar's -- has suffered more than its fair share 
from   criticisms by people that are certainly honourable. Of course, 
unlike   in Caesar's case the main charge against charm is not that it 
reaches   for the crown; the charge against charm is one of marginality, 
i.e.   that charm can teach us nothing of true consequence any longer: 
at   best it can serve as a tool facilitating access to something of 
real   interest -- like beauty; at worst it acts as an annoying 
background  -- so goes the saying. 
  
Our contention instead is:   
\begin{itemize}  
\item   
While charm of course had an illustrious past, which should not be  
forgotten and from which we can still learn,   
\item   
it will continue to teach us important lessons on   
{\em Standard Model} {\em (SM)} dynamics, some of which will be   
important for a better understanding of beauty decays, and   
\item   
the best might actually still come concerning manifestations   
of {\em New Physics}. 
 
\end{itemize}  
The case to be made for continuing dedicated studies of charm    
dynamics does {\em not} rest on a single issue or two: there are 
several motivations, and they concern a better understanding of 
various aspects of strong and weak dynamics. 
 
In this article we want to describe the present state-of-the-art in   
experiment and theory for charm studies. We intend it to be a   
self-contained review in that all relevant concepts and tools are   
introduced and the salient features of the data given. Our emphasis will   
be on the essentials rather than technical points.  Yet we will
provide    the truly dedicated reader with a Cicerone through the
literature    where she can find all the details.  We 
sketch charm's place in the $SM$ -- why it was introduced  
and what its  characteristics are -- and the history of its  
discovery. Then we describe the basic features of the 
experimental as well as theoretical tools most relevant in charm 
physics. Subsequent chapters are dedicated to specific 
topics and will be prefaced with more to the point comments on the tools 
required in that context: production, spectroscopy 
and weak lifetimes. 

We shall then address exclusive leptonic, semileptonic and 
nonleptonic transitions, before we cover 
$D^0 - \bar D^0$ oscillations, CP violation and the onset 
of the quark-gluon plasma. This discussion prepares 
the ground for an evaluation of our present understanding; 
on that base we will make a case for future studies of charm physics.

\section{A Bit of History}  
\label{HIST}  
 
\subsection{Charm's Place in the Standard Model}  
\label{PLACE}  

Unlike for strangeness the existence of hadrons with the quantum   
number charm had been predicted for several specific reasons and thus  
with  specific properties as well.  Nevertheless   
their discovery came as a surprise to large parts or even   
most of the community \cite{ROSNER98}. 
\par
Strangeness acted actually as a `midwife' to charm in several respects. 
Extending an earlier proposal by Gell-Mann and Levy, Cabibbo 
\cite{NCABIBBO} made the 
following ansatz in 1963 for the charged current 
\beq 
J_{\mu}^{(+)} [J_{\mu}^{(-)}] = 
{\rm cos}\theta_C \bar d_L\gamma _{\mu}u_L [\bar u_L\gamma _{\mu}d_L]+  
{\rm sin}\theta_C \bar s_L\gamma _{\mu}u_L [\bar u_L\gamma _{\mu}s_L]
\label{CABIBBOCUR}
\eeq 
(written in today's notation), which successfully describes weak decays of 
strange and nonstrange hadrons. Yet commuting $J_{\mu}^{(+)}$ 
with its conjugate  $J_{\mu}^{(-)}$ yields a {\em neutral}  
current that necessarily contains the $\Delta S = \pm 1$ term  
sin$\theta_C$ cos$\theta_C$ 
$(\bar d_L\gamma_{\mu}s_L + \bar s_L\gamma_{\mu}d_L)$. Yet such a  
strangeness changing neutral current (SChNC) 
\index{strangeness changing neutral current} 
is  
phenomenologically unacceptable, since it would produce contributions  
to $\Delta M_K$ and $K_L \to \mu ^+ \mu ^-$ that are too large by  
several orders of magnitude. The match between
leptons and quarks with three leptons -- 
electrons, muons and neutrinos -- and three quarks -- up, down and 
strange -- had been upset already in 1962 by the discovery that there were
two distinct neutrinos. Shortly thereafter the existence of charm quarks 
was postulated to re-establish the match between the two   
known lepton 
families $(\nu _e,e)$ and $(\nu _{\mu},\mu)$   with two quark families 
$(u,d)$ and $(c,s)$ \cite{BJGL,OTHERS}. 
Later it was realized \cite{GIM} that the  
observed huge suppression of strangeness changing neutral  
currents can then be achieved by adopting the form \index{GIM mechanism} 
\bea
J_{\mu}^{(+)} = \bar d_{C,L}\gamma _{\mu}u_L &+& \bar d_{C,L}\gamma _{\mu}c_L 
\nonumber 
\\
d_C = cos\theta_C \, d + sin\theta_C \, s \; &,& \; 
s_C = -sin\theta_C \, d + cos\theta_C \, s
\label{GIMCURRENT}
\eea
for the charged current. The commutator of  $J_{\mu}^{(+)}$ and 
$J_{\mu}^{(-)}$ contains neither a $\Delta S\neq 0$ nor a $\Delta C\neq 0$ piece. 
Even more generally there is no contribution to $\Delta M_K$ in the limit 
$m_c = m_u$; the GIM mechanism yields a suppression 
$\propto (m_c^2 - m_u^2)/M_W^2$. From the value of $\Delta M_K$ one infers  
$m_c \sim 2$ GeV. 

\begin{figure}[t] 
  \centering 
   \includegraphics[height=3.0cm]{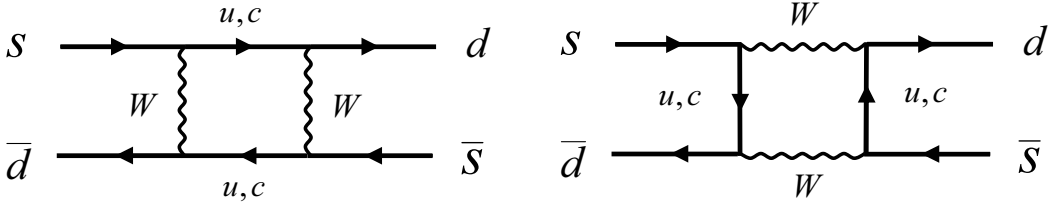} 
 \caption{The box diagram responsible for $K^0 - \bar K^0$ oscillations
 \label{FIG:BOX}}  
\end{figure}

This procedure can be illustrated by the quark box diagram for 
$K^0 - \bar K^0$ oscillations, Fig.(\ref{FIG:BOX}).  It is shown for a 
two-family scenario, since the top quark contribution is insignificant 
for $\Delta m_K$ (though it is essential for $\epsilon _K$). 
 \par
 
 \par
 \begin{figure}[t] 
  \centering 
   \includegraphics[height=3cm]{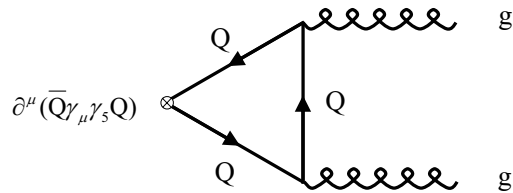} 
 \caption{An example of a triangle diagram contributing to the ABJ anomaly.
 \label{FIG:ANOMALY}}  
\end{figure} 
 
To arrive at a renormalisable theory of the weak interactions one has to 
invoke non-abelian gauge theories \cite{THOOFT}. In those the gauge 
fields couple necessarily to the charged currents and their commutators thus 
making the aforementioned introduction of charm quarks even more 
compelling. Yet one more hurdle had to be passed. For there is 
still one danger spot that could vitiate the renormalizability of the 
Standard Model. The so-called triangle diagram, 
see Fig.(\ref{FIG:ANOMALY}),  has a fermion loop  
to which three external spin-one lines are attached -- all axial vector  
or one axial vector  and two vector: while by itself finite it creates an  
anomaly, the Adler-Bell-Jackiw (ABJ) anomaly\index{ABJ or triangle anomaly}.  
It means that the axial vector current even
for {\em massless} fermions   ceases to be conserved on the loop, i.e.
quantum level  
\footnote{The term `anomaly' is generally applied when a {\em classical}   
symmetry is broken by quantum corrections.}.  
The thus induced nonconservation of the  
axial current even for massless fermions creates infinities in higher  
orders that cannot be removed in the usual way. The only way out  
is to have this anomaly, which does not depend on the mass of the 
internal fermions, cancel among the different fermion loops.  
Within the SM this requires the electric charges of all  
fermions -- quarks and leptons - to add up to zero. With the existence  
of electrons, muons, up, down and strange quarks already established 
and their charges adding up to $-2$,   
this meant that a fourth quark with three colours was needed each 
with charge  
$+\frac {2}{3}$ -- exactly like charm. There is an ironic twist here:  
as described below, the discovery of open charm hadrons was complicated  
and therefore delayed, because the charm threshold is very close to the  
$\tau$ lepton threshold; cancellation of the ABJ anomaly then required  
the existence of a third quark family (which in turn allows for CP  
violation to be implemented in the SM in charged current couplings). 
 
The fact that charm `bans' these  
evils is actually the origin of its name     
\footnote{The name "strangeness" refers  
to the feature -- viewed as odd at the time -- that the production rate of 
these hadrons exceeds their decay rate by many orders of magnitude.}.  
It was the first quark flavour {\em pre}dicted, and even the salient  
features of charm quarks were specified:  
\begin{itemize} 
\item  
They possess the same couplings as $u$ quarks,   
\item  
yet their mass is much heavier, namely about 2 GeV.  
\item  
They form charged and neutral hadrons, of which in the $C=1$ sector  
three mesons and four baryons are stable; i.e., decay only weakly with 
lifetimes of very roughly $10^{-13}$ sec -- an estimate obtained by
scaling from the muon lifetime, as explained below.  
\item  
Charm decay produces direct leptons and preferentially strange hadrons. 
\item 
Charm hadrons are produced in deep inelastic neutrino-nucleon scattering. 
\end{itemize} 
Glashow reiterated these properties in a talk at EMS-74, the 1974 Conference 
on Experimental Meson Spectroscopy and concluded \cite{GLASH74}: 

\noindent 
"What to expect at EMS-76: There are just three possibilities: 
\begin{enumerate}
\item 
Charm is not found, and I eat my hat. 
\item 
Charm is found by hadron spectroscopers, and we celebrate. 
\item 
Charm is found by outlanders, and you eat your hats."
\end{enumerate}
 A crucial element in the acceptance of the $SU(2)_L\times U(1)$ 
theory as the SM for the electroweak forces was the observation 
of flavour-conserving neutral currents by the Gargamelle collab. 
at CERN in 1973. Despite this spectacular success in predicting 
weak neutral currents of normal strength in the 
flavour-conserving sector together with hugely suppressed ones for 
$\Delta S\neq 0$ transitions, the charm hypothesis   
was not readily accepted by the community -- far from it. Even after   
the first sightings of charm hadrons were reported in cosmic ray data   
\cite{NIU}, a wide spread sentiment could be characterized by the 
\index{Niu et al. charm candidate}  
quote:
"Nature is smarter than Shelly [Glashow] ... she can do   
without charm." 
\footnote{It seems, even Glashow did not out rule this
 possibility, see item 1 on his list 
above. 
}
In the preface we have listed three categories of merits that   
charm physics can claim today. Here we want to expand on them, 
before they will be described in detail in subsequent sections.  
\begin{itemize} 
\item  
The production and decays of strange hadrons revealed or at least pointed 
to many features central to the SM, like parity violation, the  
existence of families, the suppression of flavour-changing neutral 
currents and CP violation. Charm physics was likewise essential for the  
development of the SM: its foremost role has been to confirm and 
establish most of those features first suggested by strange physics and 
thus pave the way for the acceptance of the SM. It did so in dramatic 
fashion in the discovery of charmonium, which together with the 
observation of Bjorken scaling in deep inelastic electron-nucleon 
scattering revealed quarks acting as dynamical degrees of freedom rather  
than mere mathematical entities. The demands of charm physics drove 
several lines in the development of accelerators and detectors alike. The  
most notable one is the development of microvertex detectors:  
they found triumphant application in charm as well as in beauty physics 
-- they represent a conditio sine qua non for the observation of CP  
violation in $B \to J/\psi K_S$ -- and in the discovery of top quarks 
through $b$-flavour tagging, to be followed hopefully soon  
by the  discovery of Higgs bosons again through $b$-flavour tagging. Some 
might scoff at such historical merits. We, however, see tremendous value 
in  being aware of the past -- maybe not surprisingly considering where 
two of us live and the other two would love to live 
(we are not referring  
to South Bend here.).  
\item  
The challenge of treating charm physics {\em quantitatively} has lead to  
testing and refining our theoretical tools, in particular 
novel approaches to  
QCD based on heavy quark ideas. This evolutionary process will continue 
to go on. The most vibrant examples are lattice  
QCD and heavy quark expansions described later.  
\item  
Charm can still `come through' as the harbinger or even herald of New  
Physics. It is actually qualified to do so in a unique way, as explained  
in the next section. 
\end{itemize}

\subsection{On the Uniqueness of Charm} 
\label{UNIQUE} 
 
Charm quarks occupy a unique place  among up-type quarks.  
{\em Top} quarks decay before they can 
hadronize \cite{RAPP}, which, by the way, makes searches for CP 
violation there even more challenging.  
On the other end of the mass spectrum there are only two  
weakly decaying light flavour hadrons, namely the neutron and the pion:  
in the former the $d$ quark decays and in the latter the quarks of the  
first family annihilate each other. The charm quark is the only up-type  
quark whose hadronization and subsequent weak decay can be studied. 
Furthermore the charm quark mass $m_c$ provides a new handle on treating
nonperturbative dynamics through an expansion in powers of $1/m_c$. 
 
Decays of the down-type quarks $s$ and $b$ are very promising to reveal  
new physics since their CKM couplings are greatly suppressed, in  
particular for beauty. This is not the case for up-type quarks. 
Yet New Physics beyond the SM could quite conceivably induce 
flavour changing neutral currents that are larger for the 
up-type  than the down-type quarks. In that case charm decays would be best 
suited to  reveal such non-standard dynamics.

\subsection{The Discovery of Charm}  
\label{DISCOVERY}  

\subsubsection{The heroic period}  
 
A candidate event for the decay of a charm hadron was first seen in 
1971 in an
emulsion exposed to cosmic rays \cite{NIU}. It showed a transition 
\index{Niu et al. charm candidate}
$X^{\pm} \to h^{\pm} \pi ^0$ with $h^{\pm}$ denoting a charged hadron
that could be a meson or a baryon. It was recognized that as the 
decaying object $X^{\pm}$ was found in a jet shower, it had to be a
hadron; with an estimated lifetime around few$\times 10^{-14}\; sec$ it 
had to be a weak decay. Assuming $h^{\pm}$ to be a meson, the mass of 
$X^{\pm}$ was about 1.8 GeV. The authors of Ref.\cite{KAWAI} analyzed 
various interpretations for this event and inferred selection rules like
those for charm. It is curious to note that up to the time 
\index{Niu et al. first charm candidate}
of the $J/\psi$ discovery 24 papers published in 
the Japanese journal {\em Prog. Theor. Physics} cited the emulsion event 
versus only 8 in Western journals; a prominent exception was Schwinger in 
an article on neutral currents \cite{SCHWINGER}. The imbalance was even 
more lopsided in experimental papers: while about twenty charm candidates 
had been reported by Japanese groups before 1974, western experimentalists 
were totally silent \cite{NIUNARRATED}.  

It has been suggested that Kobayashi and Maskawa working at Nagoya  
University in the early 70's were encouraged in their work -- namely to  
postulate a third family for implementing CP violation -- by knowing  
about Niu's candidate for charm produced by cosmic rays. Afterwards the 
dams against postulating new quarks broke and a situation arose that can 
be characterized by adapting a well-known quote that  
"... Nature repeats itself twice, ... the second time as a farce". 
\par
It was pointed out already in 1964 \cite{OKUN64} that charm hadrons
could be searched in multilepton events in neutrino production. Indeed 
evidence for their existence was also found by interpreting opposite-sign   
dimuon events in deep inelastic   
neutrino nucleon scattering \cite{Benvenuti:ru} 
as proceeding through 
$\nu N \to \mu ^-  c +... \to \mu ^- D ... \to \mu ^- \mu ^+ ... $.   
\subsubsection{On the eve of a revolution}
\label{EVE}  
%
The October revolution of '74 -- like any true one -- was preceded by a
period where established concepts had to face novel challenges, which
created active fermentation of new  ideas, some of which lead us forward,
while others did not. This period was initiated on the one hand by the
realization that spontaneously broken gauge theories are renormalizable,
and on the other hand by the SLAC-MIT study of deep inelastic
lepton-nucleon scattering. 
The discovery of approximate Bjorken scaling gave rise to the parton  
model to be superseded by QCD; the latter's `asymptotic freedom' 
\index{asymptotic freedom} -- the feature of its coupling 
$\alpha_S(Q^2)$ going to zero (logarithmically) as $Q^2 \to \infty$ -- 
was just beginning to be appreciated. 
\par
 Attention was  
turned to another deep inelastic reaction, namely $e^+e^- \to had$. In 
some quarters there had been the expectation that this reaction would  
be driven merely by the tails of the vector mesons $\rho$, $\omega$  
and $\phi$ leading to a cross section falling off with the  
c.m. energy faster than the $1/E^2_{c.m.}$ dependence of the  
cross section for the `point like' or `scale-free'  
process $e^+e^- \to \mu^+\mu^-$ does. On the other hand it was already
known at that time that within the quark-parton model the transition 
$e^+e^- \to had$ would show the same scale-free behaviour at sufficiently
high energies leading to  the ratio 
$R = \sigma(e^+e^- \to had)/\sigma(e^+e^- \to \mu ^+\mu ^-)$  
being a constant given by the sum of the quark electric charges squared.  
The three known quarks  
$u$, $d$ and $s$ yield $R=2/3$. It was pointed out by  
theorists that having three colours would raise $R$ to a value $2$. 
Yet the data seemed to paint a different picture. Data taken at the ADONE storage 
ring in Frascati yielded $R \sim 3\pm 1$ at $E_{c.m.}=3$ GeV. 
The old Cambridge Electron
Accelerator (CEA) in Massachusetts was converted to an $e^+e^-$ machine in
1972.  Measurements made there showed no signs of $R$ decreasing: 
$R = 4.9 \pm 1.1$ and $6.2 \pm 1.6$ at $E_{c.m.}=4$ and $5$ GeV, 
respectively. Yet these  
findings were not widely accepted as facts due to the low acceptance of 
the detectors. The first measurement of $e^+e^-$ annihilation  
with a large acceptance detector was performed by the MARK I 
collaboration at SLAC's SPEAR storage ring for  
$E_{c.m.} \sim 3 - 5$ GeV. When their initial 
results  were announced at the end of 1973, they caused quite a stir or 
even shock. They established that $R$ was indeed in the range of $2 - 4$   
and not falling with energy. The publicly presented data with their 
sizeable error bars actually seemed to show $R$ rising like $E_{c.m.}^2$ 
meaning  
$\sigma(e^+e^- \to had)$ approaching a constant value 
\cite{LONDONRICHTER}. 
This was taken by  
some, including a very prominent experimentalist, as possible evidence 
for electrons containing a small hadronic core. 
 
The '74 revolution thus shares more features with other revolutions:  
In the end it did not produce the effect that had emerged first; 
furthermore even prominent observers do not own a reliable crystal ball 
for gazing into the future.
Rather  than revealing that electrons are hadrons at heart, it showed
that quarks  are quite similar to leptons at small distances. 
 
The New Physics invoked to induce the rise in $R$ was parameterized 
through  four-fermion operators built from quark and lepton bilinears. 
Some amusing effects were pointed out \cite{BIBJ}: if the new operators 
involved scalar [pseudoscalar] fermion bilinears, one should see  
$\sigma(e^+e^- \to had)$ decrease [increase] {\em with time} from the
turn-on   of the beams. For in that case the cross section would depend
on the   transverse polarization of the incoming leptons, and the latter
would  grow with time due to synchrotron radiation. Later more precise
data did away with these speculations. They showed  
$R$ to change with $E_{c.m.}$ as expected from crossing a production  
threshold. 

Other theoretical developments, however, turned out to be of lasting 
value. In a seminal 1973 paper \cite{GL73} M.K. Gaillard and B. Lee
explored  in detail how charm quarks affect kaon transitions -- 
$K^0 - \bar K^0$ oscillations, $K_L \to \mu^+\mu^-$, 
$K_L \to \gamma \gamma$ etc. -- through quantum corrections. Their 
findings firmed up the bound $m_c \leq 2$ GeV. Together with 
J. Rosner they extended the analysis in a review, most of which was 
written in the summer of 1974, yet published in April 1975 \cite{GLR75} 
with an 
appendix covering the discoveries of the fall of 1974. At the same time
it was suggested \cite{POLITZER} that charm and anticharm quarks form
unusually narrow vector meson bound states due to gluons carrying colour 
and coupling with a strength that decreases for increasing mass scales.

The theoretical tools were thus in place to deal with the surprising 
observations about to be made. 
\subsubsection{The October revolution of '74}   
\label{OCTOBER} 
It is fair to say that the experimental signatures described above   
did not convince the skeptics -- they needed a Damascus experience  
to turn from `Saulus' into `Paulus', from disbelievers into believers. 
\index{October revolution} 
Such an experience was provided by the October revolution of 1974, the 
discovery of the   
$J/\psi$ and $\psi^{\prime}$ viewed as absurdly narrow at the time. 
It provides plenty of yarn for several intriguing story lines 
\cite{ROSNER98}. One is about the complementarity of different 
experiments, one about the value of persistence and of believing in what 
one is doing and there are others more. On the conceptual side these 
events finalized a
fundamental change in the whole outlook of the community onto   
subnuclear  physics that had been initiated a few years earlier, as
sketched above: it revealed  quarks to behave as real
dynamical objects rather than  to represent  merely mathematical
entities. 
\par 
One exotic explanation that the $J/\psi$ represents an  
$\Omega \bar \Omega$ bound state fell by the wayside after the discovery  
of the $\psi^{\prime}$. The two leading explanations for the new  
threshold were charm production and `colour thaw'. Since the early days 
of the quark model there were two types of quarks, namely the 
Gell-Mann-Zweig quarks \index{Gell-Mann-Zweig quarks} with fractional
charges and the   Han-Nambu \cite{HAN} quarks \index{Han-Nambu quarks}
with integer charges. Of those there are actually nine grouped into three  
triplets, of which two contained two neutral and one charged quark and  
the last one two charged and one neutral quark. The Han-Nambu model was 
actually  introduced to solve the spin-statistics problem of baryons  
being S-wave configuration of three quarks. The idea of `colour thaw' is  
to assume that up to a certain energy each of the three triplets acts  
{\em coherently} reproducing results as expected from Gell-Mann-Zweig quarks, 
i.e. $R=2$. Above this energy those `colour' degrees of freedom get  
liberated to act {\em incoherently} as nine quarks producing $R=4$! 
 
Charm gained the upper hand since it could provide a convincing 
explanation for the whole family of narrow resonances as `ortho-' and  
`para-charmonia' in a dramatic demonstration of QCD's asymptotic freedom.  
`Colour thaw' could not match that feat. 
 
Yet the final proof of the charm hypothesis had to be the observation of 
open charm hadrons.   In one of the (fortunately) rare instances of nature 
being   malicious, it had placed the $\tau^+\tau^-$ threshold close to   
the charm threshold. Typical signatures for charm production  --  
increase production of strange hadrons and higher multiplicities in   
the final state -- were counteracted by $\tau ^+\tau ^-$ events,    
the decays of which lead to fewer kaons and lower hadronic   
multiplicities. It took till 1976 till charm hadrons were observed   
in fully reconstructed decays.

\subsubsection{The role of colour} 
\label{COLOUR} 
 
The need for the quantum number `colour' had arisen even before the 
emergence of QCD as  the theory for the strong interactions. On the one 
hand there was  the challenge of reconciling Fermi-Dirac statistics with 
identifying the 
$\Omega^-$ baryon as an $sss$ system in the symmetric $J=3/2$ 
combination: having colour degrees of freedom would allow for the   
wavefunction being odd under exchange for an S-wave configuration. On  
the other hand the aforementioned avoidance of the ABJ anomaly implied  
the existence of three colours for the quarks. 
 
`Colour' is of course central to QCD. Its introduction as  
part of a non-abelian gauge theory is required by the need for a theory  
combining asymptotic freedom in the ultraviolet and confinement in the  
infrared. With three colours $qqq$ combinations can form colour singlets. 
 
It should be noted that studying $e^+e^- \to hadrons$ around the  
charm threshold revealed several other manifestations of colour: 

\noindent  
{\bf (i)} It had been noted before the discovery of the 
$J/\psi$ that three colours  
for quarks are needed to also accommodate the observed value of  
$R=\frac{\sigma (e^+e^- \to had.)}{\sigma (e^+e^- \to \mu ^+\mu ^-)}$
within   quark dynamics. Yet this argument was not viewed as convincing
till data   indeed showed that $R$ below and (well) above the charm
threshold could   be adequately described by two `plateaus' -- i.e.
relatively flat   functions of the c.m. energy -- with their difference
in height   approximately $N_C \sum_i e_i^2 = 4/3$.

\noindent   
{\bf (ii)} The amazingly narrow width of the $J/\psi$ resonance can be ascribed  
naturally to the fact that the decay of this ortho-charmonium state to  
lowest order already requires the $c \bar c$ to annihilate into  
three gluons making the width proportional to $\alpha_S^3$. 
It is amusing to remember that one of the early competitors to the  
$c\bar c$ explanation for the $J/\psi$ was the speculation that the  
colour symmetry is actually broken leading to the existence of  
non-colour singlets in the hadronic spectrum.  

\noindent 
{\bf (iii)} The lifetime of $\tau$ leptons is reproduced correctly by scaling it  
from the muon lifetime  
$\tau _{\tau} \simeq \tau _{\mu} \cdot  
\left( \frac{m_{\mu}}{m_{\tau}}\right)^5 \cdot \frac{1}{2+N_C}$  
with $N_C =3$; $N_C=2$ or $4$ would not do. Likewise for the prediction  
of the leptonic branching ratio  
BR$(\tau \to e \nu \bar \nu) \simeq \frac{1}{2+N_C} = 0.2 $  
for $N_C =3$. This is remarkably close to the experimental number  
$BR(\tau \to e \nu \bar \nu) \simeq 0.1784$ with the difference 
understood as due to the QCD radiative corrections. 
\noindent 
{\bf (iv)}   
Similar estimates were made concerning the lifetime and semileptonic  
branching ratio for charm. Yet the former is a rather iffy statement in  
view of $\tau _c \propto m_c^{-5}$ and the complexity of defining a  
charm {\em quark} mass. The latter, which argues in favour of  
$BR(c \to e \nu s) \sim 1/(2+N_C)$ (again modulo QCD radiative 
corrections) is actually fallacious if taken at face value. These two 
points will be explained in Sect. \ref{WEAKLIFE}. 
%
\section{Experimental Environments and Instruments}  
\label{EXPINST}  
 The birth of the charm paradigm and its experimental confirmation  
fostered a time of development in experimental  
 techniques, which has few parallels in the history of high energy. 
For charm was predicted with a set of properties that facilitate 
their observation. Its mass was 
large by the times' standards, but within reach of existing accelerators. It  
possessed charged current couplings to $d$ and $s$ quarks, 
and therefore should be visible in neutrino beams available then;  
$\epem$ colliders had come into operation. Open charm would decay 
preferentially to final states with strangeness,
 making them taggable by particle ID detectors 
able to discriminate kaons from protons and pions. Hidden charm states 
 would have a large decay rate to lepton pairs providing a clean and signature.   
Charm lifetimes would be small, but within reach experimentally. Charm would
decay semileptonically, thus providing chances of observing the relatively easy
to detect muon. 
 
 In this section we will retrace the historical development, from which we will 
draw lessons on the production environments - focusing on various colliders versus fixed 
target set-ups - and then sketch key detector components. 

 \subsection{On the history of observing charm} 
%
\subsubsection{Hidden charm}
\label{HIDCHA}

The $\jp$ was discovered simultaneously 1974 by  
two experiments, one at the Brookhaven fixed target machine with 
30 GeV protons and the 
other one at SLAC's SPEAR $\epem$ collider, neither of which was 
\index{SPEAR at Stanford}
actually searching for charm. 
Ting's experiment studying $p Be \to \epem + X$, 
after having been rejected at Fermilab and CERN, 
was approved  at BNL to search for the possible existence of a  
{\it heavy photon}, i.e.,  a higher mass 
 recurrence of the $\rho$, $\phi$,  and $\omega$ mesons.  
Richter's group at SPEAR on the other hand was interested in  
the energy dependence of $\epem$ annihilation into hadrons.  
In 1974 Ting's group  
observed a sharp enhancement at $M(\epem) =$ 3.1 GeV. They did 
not announce the result waiting  
some months to confirm it. 
\index{charmonium, discovery of}
Finally they went public together with Richter's SLAC-LBL experiment, 
which observed a sharp resonant peak at the same energy  
in the interactions $\epem \to \mu^+ \mu^-, \epem$. 
\index{ADONE at Frascati}
\index{Brookhaven}
The ADONE $\epem$ collider at Frascati found itself in the
   unfortunate circumstance of having been designed for a maximum
   center-of-mass energy of 3.0 GeV. 
Immediately after the news of the $\jp$ 
   observation was received, currents in ADONE magnets were boosted beyond
   design limits, a scan in the 3.08-3.12 GeV was carried on and the new
   resonance found and confirmed. 
Three papers 
\cite{Aubert:1974js},\cite{Augustin:1974xw}, \cite{Bacci:1974za},
announcing the $\jp$ discovery appeared in early December 1974  
in Physical Review Letters 
\footnote{ 
 The history of the $\jp$ discovery is described in full, 
 including comments of the main actors, in \cite{MAG}.}  
Within ten days of the announcement of the $\jp$'s discovery the 
SLAC-LBL group at SPEAR found another narrow resonance, the 
$\psi ^{\prime}$ at $3.7$ GeV \cite{PSIPRIME}. Soon thereafter other 
actors entered the stage, namely DESY's DORIS 
\index{DORIS at DESY}
storage ring, where the 
DASP collaboration found a resonance just above charm threshold, the 
$\psi ^{\prime \prime}$ at 3.77 GeV \cite{Braunschweig:ac}. 
Over the years a very rich and gratifying experimental 
program was pursued at SPEAR and DORIS by a succession of experiments: 
MARK I - III, Crystal Ball, DASP, PLUTO etc. Their achievements went well 
beyond mapping out charmonium spectroscopy in a detailed way: a host of new 
experimental procedures was established -- actually a whole style of doing physics 
at a heavy flavour `factory' was born that set the standards for the later 
$B$ factories. 

Only charmonium states with $J^{PC}=1^{--}$ can be produced {\em directly} 
in $\epem$ to lowest order in $\alpha$. A novel technique was developed 
allowing the formation of other states as well, namely through low energy 
$\bar pp$ annihilation \index{charmonium formation via $\bar pp$ annihilation}. 
This was pioneered at CERN by experiment R704 
using a $\bar p$ beam on a gas jet target. It led 
to greatly enhanced accuracy in measuring masses and widths of $\chi_{c1,2}$ 
states \cite{Baglin:1986br}. The same 
technique was later used by Fermilab experiment E760 and its successor E835.

The shutdown of SPEAR and the upgrade of DORIS to study $B$ physics
 created a long 
hiatus in this program, before it made a highly welcome comeback with the BES 
program and now with CLEO-c. 

\subsubsection{Open charm}

Hadrons with open charm had to be found before charm could be viewed as 
\index{open charm, discovery of}
\index{charm mesons, discovery of}
the established explanation for the $\jp$. Indirect    
evidence for their existence surfaced in neutrino experiments.  
An event apparently violating the $\Delta Q=-\Delta S$ rule was detected at 
Brookhaven 
\cite{CAZZOLI}, and opposite-sign dimuon  
events were observed as well 
\cite{Benvenuti:ru,BARISH}.  
At CERN 
neutrino-induced $\mu^-e^+V^0$ events were seen \cite{BLIETZ,KROGH} indicating 
that the new resonance was  
correlated with strangeness in weak reactions as required by the  
presence of charm.

An intense hunt for finding charm hadrons at accelerators was begun 
\footnote{ 
 The question whether there are more than four quarks was soon raised 
 \cite{HARARI75}.}; 
the MARK I collaboration
\index{MARK~I observation of charmed mesons}
 found the prey through narrow mass peaks 
in   $K^-\pi^+, K^-\pi^+\pi^+, K^-\pi^+\pi^+\pi^-$  
\cite{GOLDHABER,PERUZZI} for the iso-doublet   $D^0$  and $D^+$, 
i.e. in final states that had been predicted \cite{GLR75}. 
$D$ mesons were soon thereafter detected also 
in neutrino- \cite{BALTAY}, hadron- \cite{DRIJARD} 
and photon-induced \cite{ATIYA} reactions. 

\subsubsection{Measuring charm lifetimes}
\label{MEASLIFE}

\begin{figure}
  \centering 
   \includegraphics[width=12.0cm]{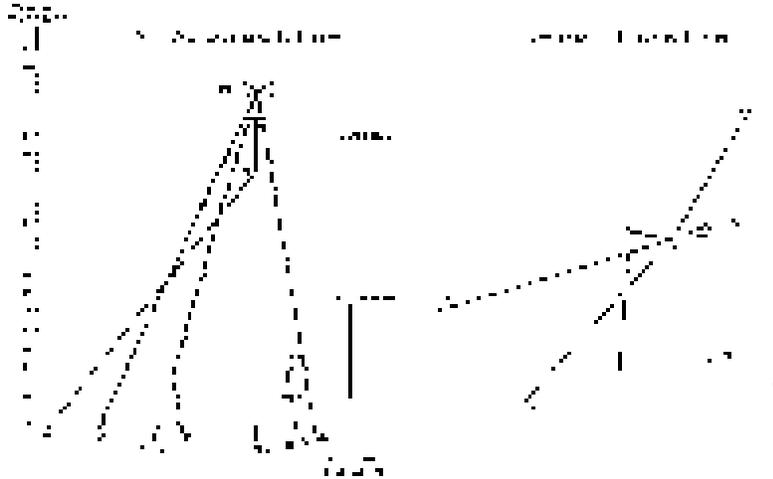} 
 \caption{First charm candidate event in nuclear  emulsions \cite{NIU}. 
 Figure from Ref.~\cite{NIUNARRATED}. 
 \label{FIG:NIU}}   
\end{figure}

Not surprisingly, the first experimental evidence for weakly decaying 
charm hadrons was obtained in an  emulsion experiment 
\index{photographic emulsion} 
\index{lifetimes}
exposed to cosmic rays \cite{NIU}, Fig. \ref{FIG:NIU}. 
For till after the time of the $\jp$  discovery only 
photographic emulsions could provide the spatial resolution needed to 
find particles with lifetimes of about $10^{-13}$ sec.  
Their resolving power of about 1 micron was a very powerful tool 
for tracking charm particles; moreover identification of particles and their
kinematical   
properties could be inferred by measuring ionisation and  
multiple scattering.
 
 \par
 \begin{figure}
  \centering 
   \includegraphics[width=7.0cm]{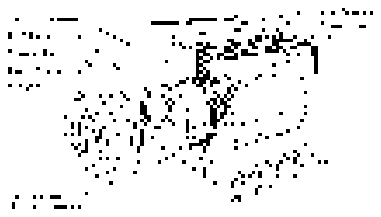} 
   \includegraphics[width=5.0cm]{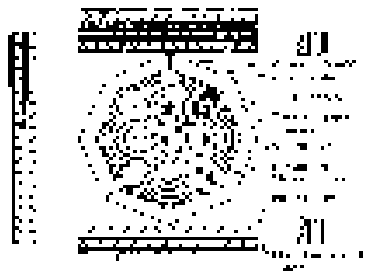} 
 \caption{MARK2 Detector: exploded and beam 
  view (From \cite{Schindler:1980ws}).
 \label{FIG:MARK2b}}   
\end{figure} 
\par

\index{charm in emulsion experiments}
Emulsion experiments had become much more sophisticated since their early 
successes in discovering the pion and the strange particles: in the early 
1950's it had been proposed \cite{MKAPLAN} to combine packs of  
thick metal plates, acting as absorber or target, with  
 thin emulsion layers for tracking. This type of hybrid 
detector was developed 
mainly in Japan and successfully used in cosmic ray studies. 
{\it "One can say that nuclear emulsion  
was the ancestor technique of heavy quark physics"} 
\cite{STROLIN}. 
By 1974 one had already seen lifetime differences between charged 
and neutral charm hadrons in cosmic ray emulsion data \cite{KURAMATA}, 
although that was largely ignored outside Japan. 

Hybrid detectors, where a forward spectrometer complements emulsions, 
were then used to study charm at accelerators. 
Experiments were done at Fermilab from 1975  to 1979 with 
205 GeV \cite{HOSHINO} and 400 GeV \cite{USHIDA} proton beams. 
Those experiments detected the first charm event (and even a charm 
particle pair) at accelerators. By the end of the seventies, the  
numbers of charm detected in emulsions at accelerators exceeded   
the one  from cosmic rays. However statistics  
was still limited to a total of few tens events. 

To overcome this limitation, the traditional visual inspection 
and reconstruction of events in nuclear emulsions  
 was gradually replaced by computer techniques -- from semi-automatic 
scanning machines \cite{NIWA} to fully automatic systems driven by the  
   forward spectrometer tracking information \cite{OKI} . The new  
   technique saved time in both finding and reconstructing 
candidate events without introducing a bias in event selection.  
In 1979, in a few months five thousand events  were analysed in an  
experiment on negative pion beam of 340 GeV at Cern  \cite{FUCHI}, 
and the huge (at the time) number of  
 four charm pairs, five charged and three neutral charm particles  
 were detected.  
 
These improved emulsion techniques were applied in full to   
study charm neutrino production by E531 at Fermilab and 
by CHORUS at CERN.
The E531 collaboration 
\cite{USHIDA88} collected more than 120 charm events; among its most
notable results was the confirmation of the lifetime differences first seen 
in cosmic ray data a few years earlier \cite{USHIDA80}. 

This new technique contributed also to early beauty  searches. 
WA75  at CERN using a 350 GeV pion beam 
was the first to detect beauty hadrons\cite{Albanese:1985wk}
 in a hadron beam.   
In a single event both beauty hadrons  
$B$ and $\bar B$ were detected, and their decays into charmed 
particles observed  
clearly  showing the full sequence of decays from beauty to
 light quark.  WA75 detected about 200 single charm pairs events, 
among them two peculiar 
ones with simultaneous production of two pairs of charm.
 
The CHORUS detector\cite{CHORUS} combined a nuclear emulsion target  
with several electronic devices. By exploiting a fully  
automated scanning system it localized, reconstructed and analysed  
several hundred thousand  
interactions. A sample of  
about 1000 charm events, a ten-fold increase over E531,  
was obtained by CHORUS. This big sample should allow the  
measurement of the,  so far never measured, total charmed-particle  
production inclusive cross-section in antineutrino 
induced event \cite{Delellis03}. 

The scanning speed achievable with fast parallel  
processors increases by about one of magnitude every three years.
 Soon a scanning speed of 20~cm$^2/$s should  
 be possible\cite{Nikano01}. These developments assure a continuing 
presence of emulsion techniques in high energy physics. 

Bubble chambers made important 
contributions as well. Charm decays were seen in the 15~ft bubble chamber at  
 Fermilab 
 \cite{BALLAGH}. 
Very rapid cycle bubble chambers coupled with a forward magnetic  
spectrometer contributed since the early days of charm  
physics at Fermilab 
\cite{BURHOP}  and Cern 
 \cite{ANGELINI}. LEBC was utilized by NA16 and NA27 searching for charm   
  states at CERN , while SLAC operated the SHF (Slac Hybrid facility). 
Yet these devices have remained severely limited in the statistics 
they can generate, due to low repetition rate of 20-40 Hz, the short  
  sensitivity time 200 microseconds, and to the small fiducial volume. 
Thus they are of mainly historical interest now. 

\subsubsection{The silicon revolution}
\label{SILREV}
 Charm quark physics witnessed in a very distinct fashion the very 
 transition from image to logic\cite{GALISON} which is common to several 
 fields of particle physics. Turning point of such transition was the
 replacement of emulsions and bubble chambers with electronic imaging devices. 
\par
 The NA1 experiment at CERN was one of the first
 experiments that introduced silicon and germanium devices 
into the field 
\footnote{The degree to which charm's arrival in the data  produced 
a revolution not merely in our view of fundamental dynamics, 
but also in detector science can be seen from the fact that 
experiments converted their objectives {\em in flight} to 
new quests.  E.g., NA1 at CERN was originally   
designed to study hadronic fragmentation  
(as its FRAMM name recalls).   
}. This was soon followed by one of the major 
 breakthroughs in the detector techniques of the last 20 years: the  
 silicon microstrip high-resolution
\index{silicon detectors}  
 vertex detector \index{silicon microvertex detectors}. 
\par
\begin{figure}
  \centering 
   \includegraphics[width=6.0cm]{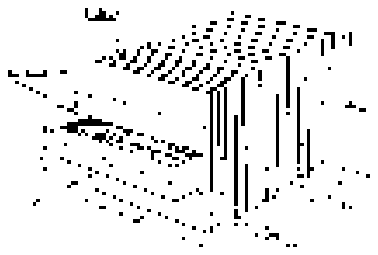} 
   \includegraphics[width=6.0cm]{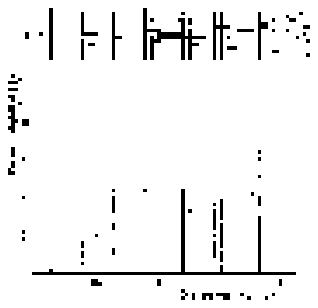}
 \caption{Ge-Si active target of CERN NA1 experiment (left). A $D^+D^-$ event   
 and corresponding pulse height pattern in target (right). From
 Ref.~ \cite{Amendolia84}. 
 \label{FIG:NA1}}   
 \end{figure}

To measure lifetimes, NA1 used  
 a telescope composed of several silicon detectors  
(150-300 microns thick) with beryllium 
  sheet targets in between, installed directly in the  
photon beam Fig.\ref{FIG:NA1}. The telescope acts  
  as an active target: when an interaction occurs, the 
 silicon device detects the energy 
   released by the recoil system (the nuclei or a  
proton) and by particles emerging from  
   the interaction points. The pattern on the  
detected energy in the subsequent detectors  
   identifies production and decay locations along the  
silicon telescope. The recoil  
   fragment or nucleon releases sufficient energy to 
 identify the interaction point even when the emerging  
   particles were neutral. The ideal sequences  of
energy-deposited steps are shown in Fig.\ref{FIG:NA1},
 for photoproduction of both charged  
and neutral charmed mesons pairs compared  
  to a typical event configuration.  
  The first determination of the time evolution curve of a charmed particle  
 was obtained by NA1 collaboration 
  with this innovative device proving a lifetime of
  $9.5^{+3.1}_{-1.9}\,10^{-13}$~s out of a sample 
  of 98 events   \cite{Albini:nb}.
  NA1 published data also on the $\Lambda_c$  lifetime and 
  production asymmetry\cite{Amendolia:1987bx}.
\par
\begin{figure}
  \centering 
   \includegraphics[width=12.0cm]{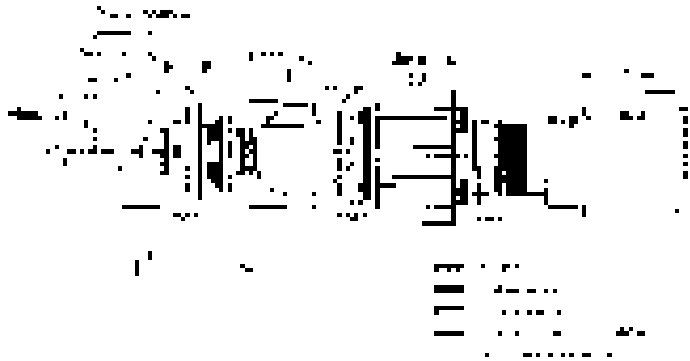} 
 \caption{NA14/2 Spectrometer (from Ref.~\cite{Alvarez90}). 
 \label{FIG:NA14}}   
 \end{figure}

\begin{figure}
  \centering 
     \includegraphics[width=5.0cm,height=2cm]{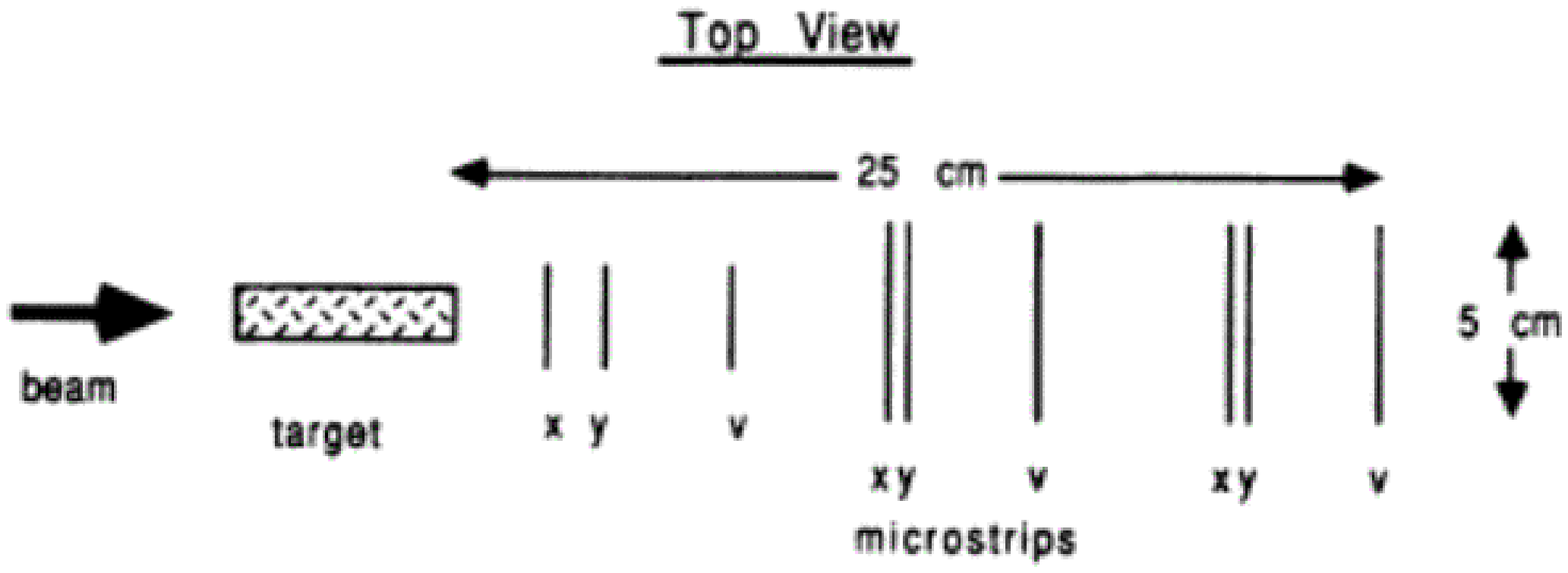} 
   \includegraphics[width=5.0cm]{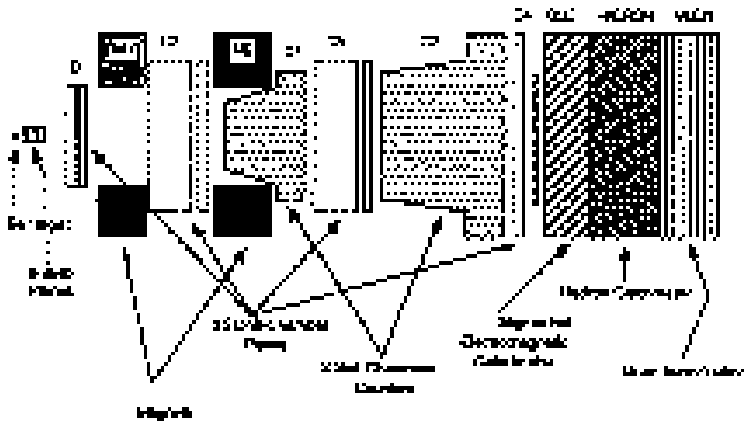}
 \caption{E691 Detector: vertex detector region(left); forward spectrometer
 (right). From ref.\cite{Raab:1987kf}. 
 \label{FIG:E691}}   
\end{figure}

 
Finally  microstrip vertex detectors 
\index{microstrip vertex detectors }
were brought to the 
scene. This new device allowed one to 
 perform tracking of particles trajectories upstream of the forward 
spectrometer magnetic field, and to reconstruct with precision the  
primary (production) and the secondary (decay) 
 vertices of short living particles in the events. 
It moved lifetime determinations to the fully digital state and also 
opened  the field to search and study specific decay channels.
Microstrip vertex detectors  are composed of 
several stations, each formed of three microstrip planes typically 200-300 
micron thick, with strips running at different orientation. Between the target 
(passive Cu or Be bulk or active silicon telescope) where the interaction occurs 
(primary vertex), and the subsequent decay (secondary vertex)  there is an empty 
region were most of the searched decays should happens, whose size must be 
optimised taking into account expected lifetimes and their relativistic boost. A 
second series of microstrip detectors is placed at the secondary vertex location 
and downstream to it. This configuration allows one 
to reconstruct the sequence of 
decay vertices, and to link emitted tracks to those reconstructed in the forward 
spectrometer. The strips typically were 20-100 micron wide, 20-50 micron pitch. 
Spatial resolutions on the plane perpendicular to the beam of the order of 
several microns were obtained. The multiple Coulomb scattering limits to 4-5~mm 
the total thickness allowed.
 The first examples of this kind of apparatus are ACCMOR\cite{Barlag:1987by}
on hadron beams   and 
NA14 (Fig.\ref{FIG:NA14}), E691 (Fig.\ref{FIG:E691}) on photon beams.
\par
By the mid-80's  fixed target experiments  using microstrip  vertex 
detectors had become mature, the technique migrated from CERN to the US, 
 experiments with   thousands of channels were built and took data for 
  more than ten years. The two main experiments at fixed target 
  were operated at FNAL: E691 \cite{Raab:1987kf}
 (later running also as  E769 and E791) and 
 E687\cite{Frabetti:1990au},
 later upgraded to E831-FOCUS.  
  At present the overall largest statistics with more 
  than a million identified charm events has been 
  accumulated by E831-FOCUS, which concluded data taking eight years 
  ago. 
  In the meantime CLEO at Cornell's 
  CESR ring -- for a long time the only $B$ factory in the world -- passed through 
  several upgrades  and developed new methods of analysis.  LEP produced a 
  heavy flavour program at the $Z^0$ that had not been foreseen. Finally  the 
  second generation $B$ factories at KEK and SLAC 
    arrived on the scene at the end of the millennium. They have obtained charm
    samples  
    of similar size to FOCUS and will surpass it considerably in the coming
    years.  
    
 The discovery of charm had been largely a US affair, yet CERN experiments 
 made a dramatic entry in the second act with conceptually new detectors and
 mature  
 measurements. 
    

\par
 Semiconductor detector technology migrated from 
 nuclear to high-energy physics experiments where it  
attained its apogee.  
 It had a truly 
far reaching impact: (i) The resulting technology that allows 
tracing lifetimes of about ${\rm few}\times 10^{-13}$~s for 
charm was `on  the shelves' when beauty 
hadrons were discovered with lifetimes around 1~ps. This was 
a `conditio sine qua non' for the success of the $B$ factories. 
(ii) It is essential for heavy flavour studies at 
hadronic machines. 
(iii) The resulting $B$ flavour tagging was essential - and will 
continue to be so - in finding top production through its decays 
to beauty hadrons. (iv) It will be an indispensable tool in future 
Higgs searches.  
%
\subsection{The past's lessons on the production environment}
\label{PASTLESSONS}

The historical sketch presented above shows that practically 
the high energy physics' whole pantheon of experimental techniques  
has contributed to charm physics. We can draw various lessons 
for the future of heavy flavour physics from the past experiences. 

The cleanest environment is provided by $e^+e^-$ annihilation, where 
threshold machine, $B$ and $Z^0$ factories complement each other. 
Threshold machines like SPEAR and DORIS in the past, BES in the present 
and also CLEO-c in the future allow many unique measurements benefiting 
from low backgrounds and beam-energy constraints. They suffer somewhat 
from the fact that the two charm hadrons are produced basically at rest 
thus denying microvertex detectors their power. 
A $Z^0$ factory, as LEP and 
SLC have been, on the other hand benefits greatly from the hard charm
fragmentation:  
the high momenta of the charm hadrons and their  `hemispheric' 
separation allows to harness the full power of microvertex detectors; similar 
for beauty hadrons. The LEP experiments ALEPH, DELPHI, L3 and OPAL together 
with SLD have made seminal contributions to our knowledge of heavy flavour
physics in  
the areas of spectroscopy, lifetimes, fragmentation functions, production rates 
and forward-back asymmetries.  
The advantage of $B$ factories is their huge statistics with low background
level.  
We are probably not even  near the end of the learning curve of how to make best use 
of it.  

Photoproduction experiments have been a crucially important contributor. The
charm  
production rate is about 1/100 of the total rate with a final
state that is typically of low  
multiplicity. A crucial element for their success was the ability to track the
finite decay  
paths of charm hadrons routinely, which has been acquired due to the dedicated 
R \& D efforts described above. Their forte is thus in the areas of
time-dependent effects  
like lifetimes, $D^0-\bar D^0$ oscillations and CP violation there. 

The largest cross sections for charm production are of course found in
hadroproduction.  
In high energy {\em fixed target} experiments one has to deal with a signal 
to background of a priori about 1/1000 with high multiplicity final states. 
That
this challenge  
could be overcome again speaks highly of the expertise and dedication of the
experiments.  
At hadron  colliders like the TEVATRON the weight of the charm cross section is
higher  
-- about 1/500  of the total cross section 
-- yet so is the
complexity of the final state. CDF, which  
previously 
had surprised the community by its ability to do high-quality beauty physics, is
pioneering  
now the field of charm physics at hadron colliders with its ability to 
trigger on charm  
decays. 
A silicon vertex tracker 
\cite{CDFSVT} reconstructs online the track impact parameters, enriching the
selected data set of charm events, by triggering on decay vertices. First charm
physics results from CDF seem promising\cite{Korn:2003pt}.  
\par
On the novel idea of using cooled antiproton beams impinging on an internal
proton jet target was commented in Sect.\ref{HIDCHA}.  
Such a technique allowed formation studies of charmonium states other than
$1^{--}$,  was pioneered at CERN by experiment R704 and further refined by
Fermilab experiments E760 and 
E835 \cite{Baglin:1986br,Armstrong:1991yk,CAZZOLI}. 
\par
Charm baryon production at fixed target by means of hyperon beams sees SELEX at
Fermilab as
probably the last exponent of a technique which is able to provide
unique information on production mechanisms (Sect.\ref{PROD}), as well as on
charmed baryon properties. 
\par
Studies of charm and beauty production at deep inelastic lepton 
nucleon scattering as done at HERA primarily act as tools for a better 
understanding of the 
nucleon's structure in general and the gluon structure functions in particular. 
\par 
 \begin{figure} 
  \centering 
   \includegraphics[width=5.0cm]{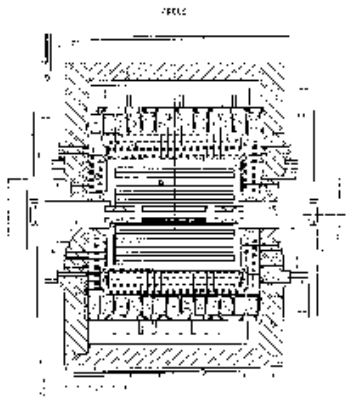} 
 \caption{ARGUS Detector (elevation view from ref.\cite{ARGUS}).
 \label{FIG:ARGUS}}   
\end{figure}
 \begin{figure}
  \centering 
   \includegraphics[width=5.0cm]{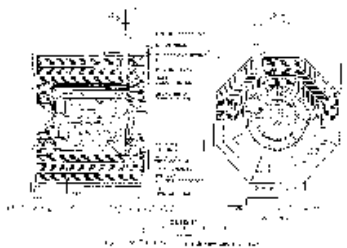} 
 \caption{CLEO-2 Detector (From ref.~\cite{CLEO2}).
 \label{FIG:CLEO2}}   
\end{figure}
%
\subsection{Key detector components}
\label{KEYCOMP}
%
The arrival of charm on  
  the market produced a major revolution, not only in physics,  
but also in detector  
  science.
 The distinct properties predicted for charm decays (mentioned in
 Sect.\ref{EXPINST}: short but finite lifetime, dominance of kaon decays,
 relevant branching ratio for semimuonic decays) gave a definite roadmap for the
 development of new experimental techniques.

   Experiments suddenly converted on flight  
their objectives to the new physics  
  quests, and a big R\&D adventure started to conceive  
 new devices able to reach the  
  needed spatial resolutions. This pushed  the  
migration of the semiconductor detector  
  technique from nuclear to high-energy experiments. 
  First ideas relied on silicon active targets, where jumps in 
 silicon pulse height would be a signal for jumps in charged particle
 multiplicity, i.e., of a charm decay point. Space resolution was limited by the
 thickness of the silicon targets. Key element for the transition to 
  modern charm lifetime
 measurements was the silicon microstrip detector. Such a transition could not
 have been accomplished without the development of DAQ systems able to handle
 the very large dataflow provided by the huge number of channels in microstrip
 detectors. The advantage given by the Lorentz boost at fixed-target experiment
 was immediately realized. It was also realized that, given the statistical
 essence of the lifetime measurement, a very large data sample was needed
  to reach high
 statistical precisions. Porting of silicon microstrips to collider charm
 experiment is a relatively recent history pioneered by CLEO, and fully embraced
 by B-factories by the usage of asymmetric beams in order to avail of some
 Lorentz boost. Silicon pixels were the natural quantum leap from microstrip
 detectors, providing two-dimensional readout, reduced thickness and therefore
 less multiple scattering, lower track occupancy and better space resolution, at
 the cost of a much increased number of readout channels. In the course of R\&D
 for vertex detectors for charm, several good ideas were investigated, such as
 the use of scintillating fibers as micron-resolution tracking
 devices\cite{Ruchti:ar}
  which
 did not last in charm physics but had many applications in HEP, or elsewhere.

 Vertex reconstruction for charm decays is intimately linked to the possibility
 of triggering on it. Charm physics at hadron colliders was born very recently
 with the success obtained by CDF at Tevatron in exploiting a hadronic trigger
 based on the online reconstruction of vertex impact parameters.
 Future experiments such as BTeV at Fermilab (see Sect.\ref{BTEV})
  plan an aggressive
 charm physics program based on a first level trigger selecting events with
 secondary vertices reconstructed in pixel detectors.
 
 Particle identification, namely the rejection of pions and protons against
 kaons,  was immediately recognized as a winner in charm physics. In pioneer
 $\epem$ experiments this was basically limited to an identification based on
 $-dE/dx$ measurement with  gas tracking devices. Thanks to the favourable
 geometry, fixed target experiments could make use of threshold Cerenkov
 counters. Ring-imaging Cerenkov counters only appeared at $\epem$ colliders
 with CLEO, and have been further improved with B-factories. The unique role of
 semielectronic and semimuonic decays in understanding the underlying hadron
 dynamics gave momentum to electron and muon particle identification techniques,
 with collider experiments traditionally more efficient in identifying the
 former, and fixed target experiments favoured by the higher muon momentum is
 deploying muon filters and detectors.
 
  Finally, electromagnetic
   calorimetry was recognized as a necessity in charm physics 
  by CLEO~II,  with
  the operation of a world-class CsI crystal array. Photon 
  and $\pi^0$ detection initially 
  provided  textboook measurements such as measurements \cite{Selen:pt} of
  $BR(D^0\to \pi^0\pi^0)$ decays to study, when 
  compared to the charged pion modes,  isospin amplitudes. Measurements
  \cite{Gronberg:1995qp} of
  $\pi^0$  decays for $D^*_s$ unveiled isospin-violating processes thus opening
   the way to exploring the full L=1 excited mesons spectroscopy with neutrals,
   until the very recent observations by BABAR and CLEO of the enigmatic
   $D^*_{sJ}(2317)$ states discussed in detail in Sect.\ref{LIFE}.
  Such a lesson was deeply metabolized by the physics community and translated
  to B physics and CKM matrix investigations: planned experiments such as BTeV
 do foresee the use of sophisticated em calorimeters for detection of photons
 and $\pi^0$. 

 This section cannot be considered complete without mentioning how the invention 
 of WEB in the 1990's by the HEP community soon was devised as a crucial tool
 for developing online monitoring systems which would actually span borders,
 oceans and frontiers --- the first truly-WEB-based online
  monitoring system was developed for a charm experiment \cite{MENASCE}.
 As a summary of the last two sections we show in Tab.\ref{TAB:EXPTS} and
 Tab.\ref{TAB:FUTURE} 
 features of present and future experiments, reserving a full discussion of
 future initiatives at the end of this paper.
\par
 \par  
\begin{table} 
 \caption{Charm in today's experiments. {\em Sample} column shows 
 number of reconstructed events. $\Delta M$ is the typical mass resolution,
 $\Delta t$ is the typical proper time resolution.
  \label{TAB:EXPTS} 
 } 
 \footnotesize 
 \begin{center} 
 \begin{tabular}{|l|l|l|l|l|l|} \hline 
  & Beam  
          & Sample  
	  & $\Delta M$~MeV  
	  & $\Delta t$~fs 
	  &  $\sigma_{\ccb}/\sigma_T$ \\ 
 \hline  
CLEO  
      & $\epem(\Upsilon(4s))$  & $1.5\, 10^5$~D & 0.3 & 200  &  $\sim$1/5     \\  
BABAR 
      & $\epem(\Upsilon(4s))$  & $1.5\, 10^6$~D & 0.3 & 200  &   $\sim$1/5     \\  
BELLE 
      & $\epem(\Upsilon(4s))$  & $1.5\, 10^6$~D & 0.3 & 200  &    $\sim$1/5    \\  
E791  
      & $\pi$ 500~GeV          & $2.5\, 10^5$~D &  1  & 50   &    1/1000    \\  
SELEX 
      & $\pi,\Sigma, p$ 600~GeV & $1.7\, 10^3\,\Lambda_c^+$ &  1 & 40  & 1/1000 \\ 
FOCUS 
      & $\gamma$ 200~GeV &  $ 1\,10^6$~D &  1      & 40   &  1/100   \\ 
CHORUS 
      & $\nu_\mu$ 27~GeV    & $2\,10^3$~D &  &     &    1/20    \\  
E835  
      & $p\bar p$  $<$8~GeV    & $4\,10^3\,\chi_{c0}$ & 2 &     &    1/70000   
      \\ 
BES  
      & $\epem(\jp)$     & $6\,10^7\,\jp$ & 1 &     &  $\sim$1      \\ 
CDF  
      & $p\bar p$  1~TeV & $ 1.5\, 10^6$~D & 2 &     &    1/500  \\
HERA Expts.  
      & ep 100~GeV   &  & 1 &    &  1/100      \\
LEP  Expts.
      & $\epem(Z^0)$     & $1\, 10^5$~D & 1 &  100   &   1/10     \\
\hline 
 \end{tabular} 
  \vfill 
 \end{center}  
\end{table} 
\begin{table}
\caption{Charm in future experiments. 
 \label{TAB:FUTURE}
 }
 \footnotesize
 \begin{center}
 \begin{tabular}{|l|l|l|l|l|l|l|} \hline
       & Beam  
            & Lumin.         
	    & Cross sect.  
	    & $\int L$    
	    & \# events $\ccb$
	    & S/B   \\
       &       
            & $cm^{-2}s^{-1}$
	    &        
	    & in $10^7$ s 
	    & recon'd/y
	    & \\
 \hline
 BTEV & $p\bar p$~1~TeV 
      & $2\, 10^{32}$    
      & $500\mu$b $\ccb$
      & 2 fb$^{-1}$  
      & $10^8$ 
      & fair \\
 LHCB & $pp$~7~TeV
      & $2\, 10^{32}$    
      & $1000\mu$b $\ccb$
      & 2 fb$^{-1}$  
      & ---    
      & ---       \\
 CLEO-C & $\psi(3770)$
      & $2\, 10^{32}$ 
      & 10~nb  $\ccb$
      & 2 fb$^{-1}$  
      & $2\,10^6$
      & large        \\
 COMPASS & $\pi$Cu FT 
         & $1\, 10^{32}$
         & 10 $\mu$b $\ccb$
	 & 1 fb$^{-1}$
	 & $5\,10^6$
	 & fair \\
 BABAR    & $\epem(\Upsilon(4s))$
          & $3\, 10^{33}$	
	  & 1.2~nb $b \bar b$
	  & 30 fb$^{-1}$ 
	  & $4\,10^6$
	  & large \\
 BELLE    & $\epem(\Upsilon(4s))$
          & $3\, 10^{33}$	
	  & 1.2~nb $b \bar b$
	  & 30 fb$^{-1}$ 
	  & $4\,10^6$
	  & large \\	  
\hline
 \end{tabular}
  \vfill
 \end{center}
\end{table}
%

\section{Theoretical Technologies}  
\label{THTOOLS}  

The relationship between the world of {\em hadrons} experiments have to 
deal with and the world of quarks (and gluons) in which our basic 
theories  are formulated is a highly complex one. Quarks undergo   
various processes of {\em hadronization}, 
\index{hadronization} namely how they exchange   
energy with their environment, how they end up asymptotically in hadrons  
and specifically in what kinds of hadrons, etc.

Almost all theoretical concepts and tools used in high  
energy physics are relevant for treating charm physics in particular,  
albeit often with quite specific features. From the outset it had been
realized -- or at least   conjectured --  that hadronization's impact on
charm transitions    would become more treatable than for ordinary
hadrons due to the    large charm mass:   
\begin{itemize}  
\item   
{\em Producing} charm from a charmless initial state requires an   
energy concentration that places in into the realm of short distance   
dynamics, which can be described perturbatively with sufficient   
accuracy. It is understood here that one considers production well   
above threshold since complexities associated with the opening  
of individual channels can 
invalidate a short-distance treatment, as discussed in 
Sect.\ref{QHDUALITY}. At such high
energies it is expected that (inclusive) hadronic rates can be  
approximated with rates  
evaluated at the  quark-gluon level, i.e. that   
{\em quark-hadron duality} should hold with sufficient accuracy.  
This topic will be addressed in Sect.\ref{QHDUALITY}.   
\item   
To identify charm production experimentally one typically has to   
deal with charm fragmentation, i.e. the fact that charm quarks give   
off energy before they hadronize. For asymptotically   
heavy quarks such fragmentation functions are expected to turn into   
delta functions \cite{BJ}. For charm quarks they should already be `hard'  
with  the final charm hadron retaining a large fraction of the   
primary charm quark energy. A simple quantum mechanical   
ansatz yields a single parameter description that describes data   
quite well \cite{PET}.     
\item   
The lifetimes of weakly decaying charm hadrons were expected to exhibit   
an approximate `partonic' pattern with lifetime ratios   
close to unity, in marked contrast to strange hadrons.  We will sketch  
the reasons for such expectations and explain their shortcomings.  
\end{itemize}  
Very significant progress has happened in formalizing these ideas   
into definite frameworks that allow further refinements.   
\begin{itemize}  
\item    
Corrections of higher orders in $\alpha_S$ have been computed for   
cross sections, structure and fragmentation functions.   
\item   
Different parameterizations have been explored for the latter.   
\item   
Heavy quark expansions have been developed to describe, among   
other things, weak decays of charm hadrons.   
\item   
Considerable efforts have been made to treat charm hadrons on   
the lattice. 
 
\end{itemize} 
The goal in sketching these tools and some of their intrinsic   
limitations is to give the reader   
a better appreciation of the results to be presented later rather than   
complete descriptions. Those can be found in dedicated   
reviews we are going to list at the appropriate places. 
 
\subsection{The stalwarts: quark (and bag) models} 
\label{QUARKMODELS}  
 
Quark models (actually different classes of them, nonrelativistic as 
well as  
relativistic ones) have been developed well before the emergence of  
charm. They cannot capture all aspects of the quantum world. Their  
relationship with QCD is actually somewhat tenuous, unlike for the second
generation technologies described below.   Quark model quantities like
quark masses, potential parameters etc.   cannot be related reliably to
SM quantities defined in a quantum field theory. Varying these model 
quantities or even comparing predictions from different quark models
does  not necessarily yield a reliable yardstick for the theoretical 
uncertainties, and no {\em systematic} improvement on the error is 
possible. 
 
Nevertheless considerable mutual benefits arise when quark models are  
applied to charm physics. Often quark models are the tool of last resort,  
when tools of choice, like lattice QCD, cannot be applied (yet). They  
can certainly educate our intuition and help our interpretation of the  
results obtained from more refined methods. Lastly they can be invoked  
to at least estimate certain matrix elements arising in heavy quark  
expansions, QCD sum rules etc. 
 
Quark models on the other hand are trained  
and improved by the challenges and  
insights offered by charm physics. Charmonia constitute the most 
suitable systems for a description based on inter-quark potential. Open  
charm mesons consisting of a heavy and a light quark represent a more
direct analogy to the hydrogen atom than light-flavour hadrons. Charm
baryons, in particular those with  
$C=2$, offer novel probes for quark dynamics: the two charm quarks move 
in close orbits around each other like binary stars surrounded by a light  
quark farther out. 
 
Bag models -- in particular their protagonist, the MIT bag model 
\cite{MITBAG} -- were 
very much en vogue in the 1970's and 1980's. The underlying idea  
actually impresses by its simplicity. One implements the intuitive picture 
of quarks being free at short distances while permanently confined at 
long  distances in the following way: one describes a hadron at rest as a 
cavity of fixed shape (typically a spherical one), yet a priori 
undetermined size; the quark fields are assumed to be free inside the  
cavity or "bag", while to vanish outside; this is achieved by imposing  
certain boundary conditions on the quark fields on the interface  
between the inside and outside of the bag. The resulting wavefunctions  
are expressed in terms of spherical Bessel functions; they are  
relativistic and can be used to evaluate matrix elements. Again 
open charm mesons lend themselves quite readily to a description by a 
spherical cavity.  Bag models have gained a second lease on life in 
nuclear  physics under the names of "cloudy bag models" or  
"chiral bag models"; clouds of pions and kaons are added to the bag to 
implement chiral invariance.

\subsubsection{Quarkonium potential}
\label{QUARKPOT}

Since QCD dynamics at small distances can be treated perturbatively,
one expects the interactions between very heavy flavour quarks to be 
well approximated by a Coulombic potential. This expectation can
actually be proven using NRQCD to be sketched below; the resulting 
description is an excellent one for top quarks \cite{HOANGMANOHAR} 
due to their enormous mass and their decay width 
$\Gamma _t > \Lambda _{QCD}$ \cite{RAPP} providing an infrared 
cutoff.  

The situation is much more involved for charm quarks. Unlike for $t$  
quarks, $\bar cc$ bound states have to exist. The fact that $m_c$ 
exceeds ordinary hadronic scales suggests that a potential description 
might yield a decent approximation for $\bar cc$ dynamics as a sequence 
of resonances with a narrow width, since they possess only Zweig rule 
(see Sect.\ref{ZWEIG}) violating decays, and with
mass splittings small compared to their  masses in qualitative analogy
with positronium, hence the moniker charmonium. That analogy can be
pursued even further: there are s-wave vector and
pseudoscalar resonances named  ortho- and para-charmonium, respectively,
with the former being even narrower than the latter. For while 
paracharmonia can decay into two gluons, orthocharmonia annihilation has 
to yield at least three gluons: 
$\Gamma ([\bar cc]_{J=1}) \propto 
\alpha_S^3(m_c)|\psi(0)|^2$ vs. $\Gamma ([\bar cc]_{J=0}) \propto 
\alpha_S^2(m_c)|\psi(0)|^2$; $\psi(0)$ denotes the $\bar cc$ wavefunction
at zero spatial separation, which can be calculated for a given 
$\bar cc$-potential. 

For the latter one knows that it is Coulombic at small distances and
confining at large ones. The simplest implementation of this scenario 
is given the ansatz   
\beq
 V(r)=\frac{A}{r}+B\,r+V_0.
\eeq
One finds the energy eigenvalues and wavefunctions by solving the 
resulting Schr\"odinger equation as a function of the three parameters 
$A,B,V_0$, which are then fitted to the data.

\subsection{Charm Production and fragmentation}  
\label{CPRODTH}  

Producing charm hadrons from a charmless initial state requires an   
energy deposition of at least $2M_D$ into a small domain   
(or at least $M_D$ in neutrino production).   
Such production processes are thus controlled by short distance   
dynamics -- unless one asks for the production of individual   
species of charm hadrons, considers only a very limited    
kinematical range or special cases like leading particle   
effects. It has to be understood also that charm production  
close to its threshold cannot be described by short distance dynamics  
since relative momenta between the $c$ and $\bar c$ are low and the 
opening of individual channels  can dominate the rate. For a perturbative 
treatment one has to stay at least a certain amount of 
energy above threshold, so that the relevant momenta are sufficiently  
large and a sufficient number of exclusive channels contribute;  
some  averaging  or `smearing' over energy might still be 
required. This  
minimal amount  of energy above threshold is determined by nonperturbative 
dynamics. Therefore we refer to it generically as $\Lambda _{NPD}$;  
sometimes we will invoke a more specifically defined energy scale like   
$\bar \Lambda$ denoting the asymptotic mass difference    
for heavy flavour  
hadrons and quarks --   
$\bar \Lambda \equiv {\rm lim}_{m_Q \to \infty}   
\left( M(H_Q) - m_Q \right)$. On general grounds one guestimates values  
like $0.5 - 1$ GeV for them  
\footnote{Strictly speaking they should not be identified with  
$\Lambda _{QCD}$ entering in the argument of the   
running strong coupling --   
$\alpha _S (Q^2)= 4\pi /\left(\beta _0  
{\rm log}\frac{Q^2}{\Lambda _{QCD}^2}\right)$ -- although they are all  
related   
to each other.}. 
 
We do not have the theoretical tools to describe reliably charm  
production close to threshold -- a region characterized by resonances  
and other prominent structures. Yet well above threshold violations of 
duality will be of no real   significance; the practical limitations are 
then due to uncertainties in the value of $m_c$ and the input 
structure functions. It is important to keep in mind that $m_c$ has  
to be defined not merely as a parameter in a quark model, but  
in a field theoretical sense. Among other things that means that it  
will be a scale-dependent quantity like the QCD coupling $\alpha_S$.

Furthermore one cannot automatically use the same  value for $m_c$ as 
extracted from heavy flavour decays, since the  impact of nonperturbative 
dynamics will differ in the two scenarios.  The charm quark mass that 
enters in production and in decay processes  is of course related.   
The tools to identify this relationship are available;  
however it has not been determined explicitly yet. Similar comments  
apply to the masses of strange and beauty quarks.

After charm quarks have been produced well above threshold, they move   
relativistically and as such are the source of gluon radiation:   
$ c \to c + gluons$. 
Such reactions can be treated perturbatively for which  
well-defined  prescriptions exist based on shower models. This radiation  
degrades  the charm quark energy till its momentum has been lowered to   
the GeV scale, when nonperturbative dynamics becomes crucial,   
since the charm quark will hadronize now:   
\beq   
Q \to H_Q (=[Q \bar q]) + q  
\eeq   
On very general grounds 
\cite{BJ} one expects the fragmentation function   
for asymptotically heavy quarks to peak at $z\simeq 1$, where   
$z\equiv p_{H_Q}/p_{Q}$ denotes the ratio between the momentum of the
emerging hadron and of the primary (heavy) quark with a width   
$\Lambda_{NPD} /m_Q$ on dimensional grounds. This conjecture   
has been turned into an explicit ansatz by approximating the   
amplitude $T(Q \to H_Q + q)$ with the energy denominator   
$(\Delta E)^{-1}$,  where 
$\Delta E = E_{H_Q} + E_q - E_Q =  \sqrt{z^2P^2 + m_Q^2} + 
\sqrt{(1-z)^2P^2 + m_q^2} - \sqrt{P^2 + m_Q^2}$ 
$\propto 1 - \frac{1}{z} + \frac{\epsilon}{1-z}$
 with   
$\epsilon = m_q^2/m_Q^2$. Hence one arrives at   
the following expression for the fragmentation function 
\index{Peterson fragmentation}:   
\beq   
D(z) \propto \frac{z(1-z)^2}{[(1-z)^2 + \epsilon z]^2} \; ,   
\label{FRAG}  
\eeq  
which is strongly peaked at $z=1$.

Naively one expects $z\leq 1$ even though the function in   
Eq.(\ref{FRAG}) yields $D(z) \neq 0$ also for $z>1$. However   
this might not hold necessarily; i.e., the heavy flavour hadron   
$H_Q$ might pick up some extra energy from the "environment". In   
particular in hadronic collisions charm and beauty production is   
central; the energy of the $Q\bar Q$ subsystem is quite small   
compared to the overall energy of the collision. A very small   
`leakage' from the huge amount of energy in the environment into   
$Q\bar Q$ and finally $H_Q$ system can increase the latter's   
energy -- as well as $p_{\perp}$ -- very significantly. Since those   
primary distributions are steeply falling such energy leakage   
would `fake' a larger charm (or beauty) production cross section   
than is actually the case \cite{FEINDT}. 
 
\subsection{Effective field theories (EFTh)} 
\label{EFT} 
 
Nature exhibits processes evolving at a vast array of different scales.  
To describe them, we typically need an explicit theory only for the  
dynamics at `nearby' scales; this is called the {\em effective} theory.  
The impact from a more fundamental underlying theory at {\em smaller}   
distance scales is mainly indirect: the fundamental dynamics create and  
shape certain quantities that appear in the {\em effective} theory as
free   input parameters. 
 
This general concept is realized in quantum field theories as well. 
For illustrative purposes let us consider a theory with two sets of 
fields $\Phi_i$ and $\phi_j$ with masses $M_{\Phi_i}$ and 
$m_{\phi_j}$, respectively, where min$\{ M_{\Phi_i}\} \gg$ 
max$\{ m_{\phi_j}\}$. Let us also assume that the theory is
asymptotically free, i.e. that at ultraviolet  
scales $\Lambda_{UV} \gg$ max $\{ M_{\Phi_i}\}$ the theory describing the 
interactions of the `heavy' and `light' fields $\Phi_i$ and 
$\phi_j$ can be treated perturbatively. At lower scales $\mu$ s.t. 
min$\{ M_{\Phi_i}\} > \mu >$ max$\{ m_{\phi_j}\}$ only the light 
fields $\phi_j$ remain fully `dynamical', i.e. can be excited as 
on-shell fields. The dynamics occurring around such scales $\mu$ 
can then be described by an effective Lagrangian ${\cal L}_{eff}$ 
containing operators built from the light fields only. Yet the heavy
fields are not irrelevant: they can contribute to observables as 
off-shell fields and through quantum corrections. Such effects enter 
through the c number coefficients, with which the light field operators 
appear in ${\cal L}_{eff}$:  
\beq  
{\cal L}(\Lambda_{UV} \gg M_{\Phi}) = 
\sum _i c_i{\cal O}_i (\Phi , \phi ) 
\Rightarrow {\cal L}(M_{\Phi} \gg \mu > m_{\phi}) \simeq   
\sum _i \bar c_i (\Phi ) \overline{{\cal  O}}_i (\phi ) 
\eeq 
One typically obtains a larger set of operators involving a smaller  
set of {\em dynamical} fields. 
 
This factoring is usually referred to as `integrating out' the  
heavy fields. We will present two examples explicitly below, namely  
the effective weak Lagrangian in Sect.\ref{EFFWEAK} and the QCD  
Lagrangian for static heavy quarks in Sect.\ref{HQET}. The latter  
example will also illustrate that effective Lagrangian are  
typically non-renormalizable; this does not pose a problem, though,  
since they are introduced to tackle low- rather than high-energy  
dynamics. 
 %
\subsection{$1/N_C$ expansions} 
\label{1OVERN} 
 
As described in Sect.\ref{COLOUR} there are several reasons why the  
number of colours $N_C$ has to be three. Yet in the limit of  
$N_C \to \infty$ QCD's nonperturbative dynamics becomes  
tractable \cite{THOOFTNC} with the emerging results highly welcome 
\index{'t Hooft model}:  
to leading order in $1/N_C$ only planar diagrams contribute  
to hadronic scattering amplitudes, and the  
asymptotic states are mesons and baryons; i.e., confinement can be proven  
then; also the Zweig rule (also called the OZI rule) holds. 

Such expansions are employed as follows to estimate at least the size 
of nonperturbative contributions: one treats short distance dynamics 
perturbatively with  
$N_C=3$ kept fixed to derive the effective Lagrangian at lower and 
lower scales, see Sect.\ref{EFFWEAK}.  Once it has been evolved down 
to scales, where one wants to evaluate hadronic matrix elements, which
are shaped by long distance dynamics, one expands those in powers of
$1/N_C$:  
\beq  
\matel{f}{{\cal L}_{eff}}{i} \propto b_0 +\frac{b_1}{N_C} +  
{\cal O}(1/N_C^2)  
\eeq 
How this is done, will be exemplified in 
Sect.\ref{TWOBODY}.  
In almost all applications only the leading term $b_0$ is retained,  
since the next-to-leading term $b_1$ is in general beyond theoretical  
control. In that sense one indeed invokes the $N_C \to \infty$ limit. 
 
While $1/N_C$ expansions offer us novel perspectives onto nonperturbative  
dynamics, they do {\em not} enable us to decrease the uncertainties 
{\em systematically}, since we have little theoretical control over the 
nonleading term $b_1$, let alone even higher order contributions.

\subsection{Heavy quark symmetry (HQS)}  
\label{HQS}  

The nonrelativistic dynamics of a spin 1/2 particle with charge $g$ is   
described by the Pauli Hamiltonian \index{Pauli Hamiltonian}:   
\beq   
{\cal H}_{\rm Pauli} =   
- g A_0 + \frac{(i \vec \partial - g \vec A)^2}{2m}  +   
\frac{g\vec \sigma \cdot \vec B}{2m}   
\label{PAULIHAM}
\eeq   
where $A_0$ and $\vec A$ denote the scalar and vector potential,   
respectively, and $\vec B$ the magnetic field. In the heavy mass   
limit only the first term survives:   
\beq   
{\cal H}_{\rm Pauli} \to - g A_0 \; \; \; {\rm as} \; \; \;   
m\to \infty \; ;   
\eeq   
i.e., an infinitely heavy `electron'  is static: it does not   
propagate, it interacts only via the Coulomb potential and its spin   
dynamics have become decoupled. Likewise for an infinitely heavy quark   
its mass drops out from its dynamics (though not its kinematics of   
course); it is the source of a static colour Coulomb field independent   
of the heavy quark spin. This is the statement of heavy quark symmetry   
of QCD in a nutshell. 
 
There are several immediate consequences for the spectrum of heavy-light   
systems, namely mesons = $[\bar Qq]$ or baryons = $[Qq_1q_2]$:   
\begin{itemize}  
\item   
In the limit $m_Q \to \infty$   
the spin of the heavy quark $Q$ decouples, and the spectroscopy of   
heavy flavour hadrons can be described in terms of the spin and orbital   
degrees of freedom of the {\em light} quarks alone.  
\item  
Therefore to leading order one has no hyperfine splitting:  
\beq  
M_D \simeq M_{D^*} \; , \; M_B \simeq M_{B^*} 
\eeq  
\item  
In the baryons $\Lambda_Q = [Qud]$ and $\Xi_Q = [Qsu/d]$ the  
light diquark system forms a scalar; to leading  
order in $1/m_Q$ {\em baryons} accordingly  
constitute a simpler dynamical system than {\em mesons},  
where the light degrees of freedom carry spin one-half. Among other 
things  this feature reduces the number  
of {\em independent} form factors describing semileptonic decays of 
heavy flavour baryons. We will return to this point in  Sect.\ref{SLXCL}.   
\item  
Some hadronic properties are   
independent of the mass of the heavy quark flavour.  For example,  
in a transition $Q_1 \bar q\to Q_2 \bar q +"W/\gamma /Z^0"$ between 
{\em two heavy} quarks $Q_{1,2}$ the formfactor, which reflects the  
response of the cloud of light degrees of freedom, has to be  
\begin{itemize} 
\item  
normalized to unity asymptotically for zero-recoil -- i.e. when  
there is no momentum transfer;  
\item  
in general dependent on the velocity $v=p/m_Q$ only.  
\end{itemize}  
\item  
There are simple scaling laws about the approach to asymptotia:  
\bea  
M_{B^*} - M_B &\simeq& \frac{m_c}{m_b}\left(M_{D^*}-M_D\right) \\ 
M_B - M_D &\simeq& m_b - m_c \simeq M_{\Lambda_b} - M_{\Lambda_c} 
 \label{SIMPLESCALING}
\eea 
\end{itemize}   
The question how quickly the heavy quark case is approached can be  
addressed through $1/m_Q$ expansions sketched below.  
A priori it is not clear to which degree the statements   
listed above apply to the   
actual charm hadrons with their marginally heavy mass.

Beyond its intrinsic interest of probing QCD in a novel environment   
there is also another motivation for studying the spectroscopy of the   
excitations of charm mesons, namely to enhance our understanding   
of semileptonic $B$ meson decays and how to extract the CKM   
parameter $V(cb)$ there. Rigorous sum rules can be derived from   
QCD that relate basic heavy quark parameters relevant to $B$ decays   
-- like quark masses, hadronic expectation values, the slope of the   
Isgur-Wise functions -- to the observable transition rates for   
$B \to \ell \nu D^{JLP}$, where the produced charm meson   
$D^{JLP}$ carries fixed spin $J$, orbital $L$ and parity  
$P$ quantum numbers. We will discuss some explicit examples later on.

\subsection{Heavy quark expansions (HQE)}  
\label{OPE}  

With HQS representing an asymptotic scenario, one can ask whether   
one can evaluate pre-asymptotic effects. The example of the Pauli   
Hamiltonian already shows that the heavy quark mass   
constitutes an expansion parameter for describing its dynamics   
in general and its   
nonperturbative aspects in particular. There are two variants  
for the implementation of such expansions, namely for (a) describing  
the dynamics of heavy quarks purely within QCD and  
(b) the weak decays of hadrons containing heavy quarks when electroweak  
forces are included.

\subsubsection{QCD for heavy quarks} 
\label{HQET} 
 
One starts by decomposing the QCD Lagrangian at scales larger than  
$m_Q$ into a 
part that contains  only light degrees of freedom and one that contains 
the heavy quarks:  
\bea  
{\cal L}_{QCD}(\mu \gg m_Q) &=& {\cal L}_{light}(\mu \gg m_Q)  
+ {\cal L}_{heavy}(\mu \gg m_Q)\\ 
{\cal L}_{light}(\mu \gg m_Q) &=& - \frac{1}{4} G_{\mu \nu}G_{\mu \nu}  
+ \sum_q\bar q i(\not D-m_q)q \\ 
{\cal L}_{heavy}(\mu \gg m_Q) &=& \sum_Q\bar Q i(\not D-m_Q)Q 
\eea 
with $D$ denoting the covariant derivative. At scales below $m_Q$ -- yet  
still above normal hadronic scales -- ${\cal L}_{light}$ remains
basically  the same since its degrees of freedom are still fully
dynamical, whereas  
${\cal L}_{heavy}$ undergoes a fundamental change since the heavy quark  
cease to be dynamical degrees of freedom:  
\beq  
{\cal L}^{heavy} = \sum _Q  
\left[ \bar Q (i \not D-m_Q)Q +  
\frac{c_G}{2m_Q}\bar Q \frac{i}{2} \sigma \cdot G Q +   
\sum _{q,\Gamma}\frac{d_{Qq}^{(\Gamma )}}{m_Q^2}  
\bar Q \Gamma Q \bar q \Gamma q \right]  
+{\cal O}\left( 1/m_Q^3 \right)  
\label{LHEAVYEXP}  
\eeq  
where $c_G$ and $d_{Qq}^{(\Gamma )}$ are coefficient  
functions: the $\Gamma$ denote the possible  
Lorentz covariant fermion bilinears and  
$\sigma \cdot G = \sigma _{\mu \nu}  
G_{\mu \nu}$ with the gluonic field strength tensor  
$G_{\mu \nu} = g t^a G^a_{\mu \nu}$. Thus a 
dimension five operator arises -- usually referred  
to as {\em chromomagnetic} operator -- and  
various dimension six four-fermion operators.  
When expressing the heavy quark fields through their static  
nonrelativistic fields  
also the so-called kinetic energy operator of  
dimension five  $O_{kin} = \bar Q (i\vec D) ^2 Q$ 
enters.     
Since it is not Lorentz invariant, it cannot appear  
in the Lagrangian. 
 
This effective Lagrangian is not renormalizable since it contains
operators of dimension higher than four. This is no drawback, though,
when treating hadronic spectroscopy. 
 
\subsubsection{The Operator Product Expansion (OPE) and weak decays of
heavy flavour hadrons} 
\label{WEAKDK} 

A powerful theoretical tool of wide applicability is provided by the 
operator product expansion a la Wilson \cite{WILSON}. One can apply it 
profitably when {\em inclusive} transitions involving hadrons are 
driven by short-distance dynamics characterized by a high momentum 
or energy scale $\sqrt{Q^2}$. `Classical' examples are provided 
by deep-inelastic lepton-nucleon scattering with {\em space-like} 
$Q^2$ and by $\sigma (e^+e^- \to had.)$ for {\em time-like} 
$Q^2=s$. Starting in 1984 \cite{SV1} another application has been 
developed for decays of a heavy flavour hadron $H_Q$, where the 
width for a sufficiently inclusive final state can be expressed as 
follows \cite{HQT}:  
$$   
\Gamma (H_Q \to f) = \frac{G_F^2m_Q^5(\mu )}{192 \pi ^3}  
|V_{CKM}|^2  \left[ c_3(m_f;  \mu)   
\matel{H_Q}{\bar QQ}{H_Q}|_{(\mu )} 
 + \right. 
$$
\beq  
\left.   
c_5(m_f; \mu) \frac{\mu _G^2(H_Q,\mu )}{m_Q^2}   
 + \sum _i c_{6,i}(m_f;  \mu)   
 \frac{  
\matel{H_Q}{(\bar Q\Gamma _iq)(\bar q\Gamma _iQ)}  
{H_Q}|_{(\mu )}}{m_Q^3} + {\cal O}( 1/m_Q^4 )   
\right]    
\label{MASTER}  
\eeq   
with 
$\mu _G^2(H_Q) \equiv \matel{H_Q}{\bar Q\frac{i}{2}\sigma \cdot GQ}{H_Q}$;
$\Gamma _i$ denote the various Lorentz structures for the quark bilinears
and 
$V_{CKM}$ the appropriate combination of CKM parameters. 
Eq.(\ref{MASTER}) exhibits the following important features:  
\begin{itemize} 
\item 
The expansion involves 
\begin{itemize}
\item 
c-number coefficients $c_{...}(m_f; \mu)$ given by short-distance 
dynamics; they depend on the final state as characterized by quark masses
$m_f$;
\item 
expectation values of {\em local} operators controlled by long-distance 
physics; 
\item 
inverse powers of the heavy quark $m_Q$ scaling with the known dimension 
of the operator they accompany. 
\end{itemize} 
\item 
A central element of Wilson's prescription \index{Wilson's prescription}
is to provide a  self consistent
separation of short-distance  and long-distance dynamics implied 
above. This is
achieved by introducing an {\em auxiliary} energy scale $\mu$ demarking 
the border: short-distance $< \mu ^{-1} <$ long-distance. 
The {\em heavy} degrees  of freedom -- i.e. with masses exceeding
$\mu$ -- are `integrated out'  meaning that their contributions are
lumped into the {\em coefficients} $c_i$, which are thus shaped by 
short-distance dynamics. Only degrees of freedom with masses
{\em below} $\mu$ -- the `light' fields -- appear  as fully dynamical
fields in the  local {\em operators}. The one exception from this
rule  are the heavy quark fields $Q$; the operators have to be bilinear
in 
$\bar Q$ and $Q$, since the initial state -- the decaying hadron 
$H_Q$ -- carries heavy flavour.  
\item 
As a matter of principle
observables have to be independent   
of the auxiliary scale $\mu$ since nature cannot be   
sensitive to how we arrange our   
computational tasks. The $\mu$ dependence of the coefficients $c_i$ 
has therefore to cancel against that of the expectation values due to 
the operators. In practice, though, the value of $\mu$ is not arbitrary, 
but has to be chosen judiciously for those very tasks: on one hand one
would like to  choose $\mu$ as high as possible to obtain a reliable
{\em  perturbative}  expression in powers of $\alpha _S(\mu )$; on the
other  hand one   likes to have it as low as possible to evaluate the 
nonperturbative {\em expectation values} in powers of $\mu /m_Q$:  
\beq 
\Lambda_{QCD} \ll \mu \ll m_Q 
\label{SCYLLA} 
\eeq
For simplicity we will not state the dependence on $\mu$
explicitly.  

\item 
The expectation values of these local operators are shaped by
long-distance dynamics. The nonperturbative effects on the decay width --   
a {\em dynamical} quantity -- can thus be expressed through expectation   
values and quark masses. 
Such {\em static} quantities are treated more easily; their values can be
inferred from symmetry  arguments, other observables, QCD sum rules,
lattice studies or quark models.   
   
\item  
The same cast of local
operators  
$(\bar Q ... Q)$ appears whether one deals with nonleptonic,  
semileptonic or radiative decays of mesons or baryons containing one, two  
or even three heavy quarks or antiquarks. The weight of the  
different operators depends on the specifics of the transition though.  
\item     
{\em No} ${\cal O}(1/m_Q)$ contribution can arise in the OPE   
since there is no independent dimension four operator in QCD
\footnote{The operator $\bar Q i \not D Q$ can be reduced   
to the leading operator $\bar QQ$ through the equation   
of motion.} \cite{QCDTHEOR}.   
A $1/m_Q$   
contribution can arise only due to a massive duality violation; 
this concept will be discussed in Sect.\ref{QHDUALITY}. 
Even then it cannot lead to a {\em systematic} excess or deficit 
in the predicted rate; for a duality violating contribution has to
{\em oscillate} around the `true' result as a function of $m_Q$ 
\cite{VADE}.  Thus
one should set a rather high threshold  before accepting the need for
such a contribution.   
The absence of such corrections gives rise to the hope that a $1/m_c$
expansion can provide a meaningful description.   
 
\item    
The free quark model   
or spectator   
expression emerges asymptotically   
(for $m_Q \to \infty$) from the $\bar QQ$ operator   
since $\matel{H_Q}{\bar QQ}{H_Q}    
= 1 + {\cal O}(1/m_Q^2)$, see Eq.(\ref{QBARQ}). 
 
\end{itemize}

\subsubsection{Heavy Quark Parameters (HQP): Quark masses  
and expectation values}  
\label{HQP}  
 
An internally consistent definition of the heavy quark mass is crucial  
for $1/m_Q$ expansions conceptually as well as numerically. While this 
remark is obvious in hindsight, the theoretical implications were at  
first not fully appreciated. 
 
In QED one naturally adopts the {\em pole} mass for the electron, which 
is defined as the position of the pole in the electron Green function  
(actually the beginning of the cut, to be more precise). It is gauge 
invariant, and can be measured, since it represents the mass of an 
isolated electron. For quarks the situation is qualitatively different 
due to confinement (except for top quarks since they decay before they  
can hadronize \cite{RAPP}). Yet computational convenience suggested 
to use the pole mass for quarks as well: while not measurable per se, it  
is still gauge invariant and infrared stable {\em order by order in  
perturbation theory}. It thus constitutes a useful theoretical construct 
-- as long as one addresses purely perturbative effects. Yet the pole 
mass is {\em not} infrared stable in {\em full} QCD -- it exhibits an  
{\em irreducible theoretical} uncertainty called a renormalon 
ambiguity \cite{POLEPAPERS} \index{renormalons}:  
$\frac{\delta m_Q^{pole}}{m_Q} \sim  
{\cal O}\left( \frac{\bar \Lambda}{m_Q}\right)$.   
Its origin can be understood intuitively by considering the  
energy stored in the chromoelectric field  
$\vec E_{Coul}$ in a sphere of radius  
$R\gg 1/m_Q$ around a static colour source of mass $m_Q$ \cite{HQT}:  
\beq  
\delta {\cal E}_{Coul}(R) \propto \int _{1/m_Q \leq |\vec x|\leq R}  
d^3x \vec E_{Coul}^2 \propto {\rm const.} -  
\frac{\alpha_S(R)}{\pi}\frac{1}{R}   
\eeq 
The definition of the pole mass amounts to setting $R \to \infty$; i.e.,  
in evaluating the pole mass one undertakes to integrate the energy 
density associated with the colour source over {\em all space} assuming 
it to have a Coulombic form as inferred from perturbation theory. Yet in 
the full theory the colour interaction becomes strong at distances  
approaching $R_0 \sim 1/\Lambda _{QCD}$, and the colour field can no 
longer be approximated by a $1/R$ field. Thus the long distance or 
infrared region around and beyond $R_0$ cannot be included in a  
meaningful way, and its contribution has to be viewed as an intrinsic 
uncertainty in the pole mass, which is then estimated as stated above. 
 Using the pole mass in 
the width $\Gamma \propto m_Q^5$ would generate an uncertainty 
$\sim\; 5\bar \Lambda /m_Q$ and thus dominate (at least parameterically) the  
leading nonperturbative contributions of order $1/m_Q^2$ one works so 
hard to incorporate.

Instead one has to employ a running mass $m_Q(\mu)$ defined at a scale  
$\mu$ that shields it against the strong infrared dynamics. There are two  
kinds of well defined running masses one can rely on, namely the  
`$\overline{MS}$' mass $\overline m_Q(m_Q)$ 
\footnote{$\overline{MS}$ stands for `modified minimal subtraction
scheme'.} 
 and the `kinetic' mass  
$m^{kin}_Q(\mu)$. The former represents a quantity of computational 
convenience -- in particular when calculating perturbative contributions  
in dimensional regularization -- rather than one that can be measured  
directly. For $\mu \geq m_Q$ it basically coincides with the 
running mass in the Lagrangian and is best normalized at $\mu \sim m_Q$.
However it diverges logarithmically for  
$ \mu \to 0$ . It is quite  
appropriate for describing heavy flavour {\em production} like in  
$Z^0 \to Q \bar Q$, but not for treating $H_Q$ {\em decays}, since there 
the dynamics are characterized by scales {\em below} $m_Q$. 
 
The kinetic mass \index{kinetic quark mass} 
(so-called since it enters in the kinetic energy of the 
heavy quark) on the other hand is regular in the infrared regime with  
\cite{KINETICPAPER,POLEPAPERS,FIVE}
\beq  
\frac{dm_Q^{kin}(\mu)}{d\mu} = - \frac{16}{9}\frac{\alpha_S}{\pi}  
- \frac{4}{3}\frac{\alpha_S}{\pi}\frac{\mu}{m_Q} +  
{\cal O}(\alpha_S^2)  
\eeq 
and is well suited for treating decay processes. It can be shown that  
for $b$ quarks $\mu \sim 1$ GeV is an appropriate scale for these 
purposes whereas $\mu \sim m_b$ leads to higher order  
perturbative corrections that are artificially large \cite{FIVE}. For 
charm quarks on the other hand this distinction disappears since $m_c$  
exceeds the 1 GeV scale by a moderate amount only. 
 
There are four classes of observables from which one can infer the value 
of $m_c$, listed in descending order of the achieved theoretical
reliability: (i) the spectroscopy of hadrons with hidden or open charm;
(ii) the {\em shape} of spectra in semileptonic $B$ decays driven by  
$b\to c$; (iii) charm production in deep inelastic lepton nucleon
scattering or  
$e^+e^-$ annihilation;  (iv) the weak decays of charm hadrons.    
 
Another approach to the value of $m_c$ is provided by relating the 
difference $m_b - m_c$ to the spin averaged masses of charm and beauty 
mesons:  
\beq  
m_b - m_c = \langle M_B\rangle - \langle M_D\rangle +  
\frac{\mu_{\pi}^2}{2} \left( \frac{1}{m_c} - \frac{1}{m_b} \right)  
+ {\cal O}(1/m^2_{c,b}) 
\label{MBMINUSMC} 
\eeq 
with $\langle M_{B[D]}\rangle \equiv M_{B[D]}/4 + 3M^*_{B[D]}/4$ and  
\beq  
\mu_{\pi}^2 \equiv \matel{H_Q}{\bar Q(i\vec D)^2Q}{H_Q}/2M_{H_Q} \; ; 
\label{MUPISQ} 
\eeq 
$\vec D$ denotes the covariant derivative and $i\vec D$ thus the  
(generalized) momentum; $\mu_{\pi}^2/2m_Q$ therefore 
represents the average  
kinetic energy of the quark $Q$ inside the hadron $H_Q$. This relation is  
free of the renormalon ambiguity mentioned above. On the down side it  
represents an expansion in $1/m_c$, which is of uncertain  
numerical reliability. Furthermore in order $1/m^3_{c,b}$ {\em non}local  
operators appear. Later we will give numerical values for $m_c$. 
 
The expectation value of the leading operator $\bar QQ$   
can be related to the flavour quantum number of the hadron   
$H_Q$ and operators of dimension five and higher:   
\beq   
\matel{H_Q}{\bar QQ}{H_Q}/2M_{H_Q} = 1 -   
\frac{1}{2} \frac{\mu _{\pi}^2}{m_Q^2} +   
\frac{1}{2} \frac{\mu _{G}^2}{m_Q^2} + {\cal O}(1/m_Q^3)    
\label{QBARQ}  
\eeq 
 
The chromomagnetic moment   
$\matel{H_Q}{\bar Q\frac{i}{2}\sigma \cdot GQ}{H_Q}$ is known with  
sufficient  accuracy for the present purposes from the hyperfine  
splittings in the masses of vector and pseudoscalar mesons   
$V_Q$ and $P_Q$, respectively:  
\beq   
\mu _G^2(H_Q, 1\; \GeV) \equiv   
\frac{\matel{H_Q}{\bar Q \frac{i}{2}\sigma \cdot GQ}{H_Q}}  
{2M_{H_Q}} \simeq \frac{3}{4}(M(V_Q)^2 - M(P_Q)^2)  
\label{MUGSQ} 
\eeq 
The size of the charm chromomagnetic moment is similar to what   
is found for beauty hadrons   
\beq   
\mu _G^2(D, 1\; \GeV) \simeq 0.41\; (\GeV)^2 \; \;   
vs. \; \; \mu _G^2(B, 1\; \GeV) \simeq 0.37\; (\GeV)^2   
\eeq    
and thus in line what one expects for   
a heavy quark   
\footnote{One should keep in mind though that for reasons   
we do not understand the hyperfine splittings are quite universal:   
$M^2_{\rho} -M^2_{\pi} \sim 0.43 \; (\GeV)^2$,   
$M^2_{K^*} - M^2_K \sim 0.41 \; (\GeV)^2$.}. 
  
For the $\Lambda _Q \equiv [Qdu]$ and   
$\Xi_Q \equiv [Qsq]$ baryons we have   
\beq   
\mu _G^2(\Lambda_Q, 1\; \GeV) \simeq 0 \simeq   
\mu _G^2(\Xi_Q, 1\; \GeV) \; ,   
\eeq   
since the light diquark system $q_1q_2$ in   
[$Qq_1q_2$] carries spin 0 in $\Lambda _c$ and $\Xi_c$.   
Yet in the $\Omega_Q \equiv [Qss]$ baryon the $ss$ diquark carries   
spin one and we have   
\beq  
\mu _G^2(\Omega_c, 1\; \GeV) \simeq \frac{2}{3}   
\left( M^2(\Omega _c^{(3/2)}) - M^2(\Omega_c) \;   
\right). 
\eeq

The kinetic energy expectation values are less precisely known beyond   
the inequality \cite{SVSRPAPER,HQT}  
\beq   
\mu_{\pi}^2 (H_Q) \geq \mu_G^2(H_Q)   
\eeq   
derived in QCD. To the degree   
that charm quarks fill the role of heavy quarks one expects   
very similar values as for $B$ mesons; i.e.    
\beq   
\mu _{\pi}^2(D, 1\; \GeV) \sim 0.45\pm 0.10 \; (\GeV)^2 \; .   
\eeq

The largest uncertainties enter in the expectation values for the   
dimension-six four-fermion operators in order $1/m_Q^3$.   
In general there are two classes of expectation values, namely   
for $SU(3)_C$ singlet and octet quark bilinears   
$\matel{H_Q}{(\bar Q_L\gamma _{\mu}q_L)(\bar q_L\gamma_{\mu}Q)}{H_Q}$   
and   
$\matel{H_Q}{(\bar Q_L\gamma _{\mu}\lambda_i q_L)  
(\bar q_L\gamma_{\mu}\lambda_iQ)}{H_Q}$,  respectively:   
\bea   
\matel{H_Q}  
{(\bar Q_L\gamma _{\mu}q_L)(\bar q_L\gamma_{\mu}Q)}{H_Q}  
&=& \frac{1}{4} f^2_{H_Q}M^2_{H_Q} B_{H_Q}    \\  
\matel{H_Q}{(\bar Q_L\gamma _{\mu}\lambda_iq_L)  
(\bar q_L\gamma_{\mu}\lambda_i Q)}{H_Q}  &=& 
f^2_{H_Q}M^2_{H_Q}\epsilon _{H_Q}    
\eea 
A natural way to estimate the mesonic expectation values is to assume   
{\em factorization} \index{factorization} or {\em vacuum saturation} 
\index{vacuum saturation} at a
low scale of around  1 GeV; i.e. $B_{H_Q} =1$ and $\epsilon _{H_Q}=0$.
Such an approximation should be sufficient considering we cannot, as
already mentioned, count on more than semi-quantitative predictions 
about charm decays:   
\bea   
\matel{P_Q}{(\bar c_L\gamma _{\mu}q_L)}{0}\cdot   
\matel{0}{(\bar q_L\gamma_{\mu}c)}{P_Q} &=& 
\frac{1}{4} f^2_{P_Q}M^2_{P_Q}    \\  
\matel{P_Q}{(\bar Q_L\gamma _{\mu}\lambda_i q_L)}{0}   
\matel{0}{(\bar q_L\gamma_{\mu}\lambda_iQ)}{P_Q} &=&  0   
\eea  
with the last equation reflecting invariance under colour $SU(3)_C$.   
One should note that factorizable  contributions at a low scale $\sim$
1 GeV will be partially  nonfactorizable at the high scale $m_Q$. These
expectation values are  then controlled by the decay constants.

For {\em baryons} there is no   
concept of factorization for estimating or at least calibrating the 
expectation values of four-fermion operators, and
we have to rely  on quark model results. 

Since the moments $\mu_{\pi}^2$ and $\mu_G^2$ represent long-distance  
contributions of order $1/m_Q^2$, one can use their values to estimate 
the scale characterising nonperturbative dynamics as 
\beq 
\Lambda _{NPD} \sim \sqrt{\mu_{\pi}^2} \sim 700 \; {\rm MeV}
\label{NPDSCALE} 
\eeq
Later we will see that this scale agrees with what one infers from 
$\bar \Lambda \equiv {\rm lim}_{m_Q \to \infty}   
\left( M(H_Q) - m_Q \right)$. 
 
These considerations lead to a first resume:  
\begin{itemize}
\item 
There is little  
`plausible deniability' if a  
description based on HQE  fails for $B$ decays: 
since $m_b$ is a multiple of $\Lambda _{NPD}$ given in 
Eq.(\ref{NPDSCALE}), Furthermore for the scale   
$\mu$ separating short and long distance dynamics in the OPE   
one can adopt $\mu \sim 1$ GeV, which satisfies the   
computational `Scylla and Charybdis' requirements stated in 
Eq.(\ref{SCYLLA}) for $m_b$. 
\item  
For charm decays, on the other hand, the situation is much more   
iffy on both counts: with the expansion 
parameter $\sim \Lambda _{NPD}/m_c$ being only
moderately  smaller than unity, higher order nonperturbative corrections 
decrease only slowly, if at all. Furthermore the computational 
requirement of Eq.(\ref{SCYLLA}) is
hardly  satisfied.     
 The one saving  
grace might be provided by the  absence of an ${\cal O}(1/m_c)$  
contribution noted above.  Finally one expects limitations to  
quark-hadron  duality to be characterized by a factor  
$e^{-\Lambda_{NPD}/m_c}$ with   
$\Lambda_{NPD}$ reflecting the onset of nonperturbative dynamics.   
It is obviously of essential importance then if this scale is indeed   
about 700  MeV or 1 GeV (or even higher), which would be bad news.    
\item 
For these reasons one cannot count on more than a semi-quantitative   
description and going beyond ${\cal O}(1/m_c^3)$ would   
then seem pretentious. More generally, 
it is not clear to which degree charm quarks act
dynamically as heavy quarks with respect to QCD. It is unlikely    that
the answer to the question "Is charm heavy ?" will be a    universal `yes'
or universal `no'. The answer will probably depend    on the type of
transition one is considering. Yet this uncertainty should not be seen as 
necessarily evil. For charm transitions allow us to probe the onset of 
the regime where duality provides a useful concept. 
 
\end{itemize} 
 
We will adopt as working hypothesis that charm quarks are sufficiently   
heavy so that HQE can provide a semi quantitative description. We   
treat it as a learning exercise in the sense that we fully expect the   
HQE description to fail in some cases. We will apply it to fully   
inclusive observables like weak lifetimes and integrated semileptonic   
widths of mesons and baryons. Counting on HQE to describe energy  
{\em distributions} in inclusive semileptonic decays is presumably   
not realistic since the averaging or `smearing' over the lepton energies   
etc. required in particular near the end points is such that it amounts   
to a large fraction of the kinematical range anyway. Furthermore there 
is   no justification for treating strange quarks in the final state of  
semileptonic decays as heavy.  
%
\subsection{HQET}  
\label{HQET2}  
%
Heavy Quark Effective Theory (HQET) is another implementation   
of HQS, where one calculates pre-asymptotic contributions as   
an expansion in $1/m_Q$ \cite{HQET}. 
While the core applications of HQET used to be 
hadronic spectroscopy and the evaluation of form factors for exclusive 
semileptonic decays of heavy flavour hadrons, the name HQET has been
applied to    
more and more types of observables like lifetimes     
with varying degrees of justification. Yet we will address here only 
how HQET deals with spectroscopy and hadronic form factors. 

The heavy flavour part of the QCD Lagrangian is expressed with the help 
of non-relativistic spinor fields $\Phi_Q(x)$ \cite{VARENNAURI}: 
$$ 
{\cal L}_{HQET} = \sum _Q \left\{ -m_Q\Phi_Q^{\dagger}\Phi_Q 
+ \Phi_Q^{\dagger}iD_0\Phi_Q - 
\frac{1}{2m_Q} \Phi_Q^{\dagger} (i\vec \sigma \cdot \vec D)^2\Phi_Q - 
\right. 
$$ 
\beq 
\left. \frac{1}{8m_Q^2}\Phi_Q^{\dagger}\left[ 
-\vec D \cdot \vec E + \vec \sigma \cdot 
(\vec E \times \vec \pi - \vec \pi \cdot \vec E 
\right]\Phi_Q
\right\} + {\cal O}(1/m_Q^3) 
\label{HQETLAG}
\eeq
where 
\beq 
\vec \pi \equiv i \vec D = \vec p - \vec A\; , \; 
(\vec \sigma \cdot \vec \pi)^2 = \vec \pi ^2 + \vec \sigma \cdot \vec B 
\eeq
with $\vec D$, $\vec A$, $\vec B$ and $\vec E$ denoting the 
covariant derivative, the gluon vector
potential,  the colour magnetic and electric fields, respectively. 

On the other hand forces outside QCD -- namely the electroweak ones -- 
are given for the full relativistic fields $Q$. To obtain the relation 
between the $\Phi_Q$ and $Q$ fields one first factors off the time 
dependence associated with $m_Q$, which makes up the lion share of 
$Q$'s energy: $Q(x) = e^{-im_Qt}\hat Q(x)$. This can be written 
covariantly in terms of the four-velocity $v_{\mu}$: 
\beq 
Q(x) = e^{-im_Q x\cdot v} \hat Q(x) 
\eeq
Yet a consistent separation of the `large' and `small' components of the 
Dirac spinor $\hat Q$ cannot be achieved by simply using 
$h(x) = \frac{1+\gamma_0}{2}\hat Q(x)$. A Foldy-Wouthuysen 
transformation \cite{BJDR1} has to be applied yielding 
\cite{PIRJOL,VARENNAURI}: 
\beq 
\Phi_Q(x) = \left( 1+ \frac{\vec \sigma \cdot \vec \pi}{8m_Q^2} 
+ ... \right) \frac{1+\gamma_0}{2} Q(x) 
\eeq

There is another complication -- both conceptual and technical -- in 
the way HQET is usually defined, namely without introducing an auxiliary 
scale $\mu > 0$ to separate self-consistently heavy and light degrees of 
freedom as discussed in Sect.\ref{WEAKDK}. With proper care this
problem can be  cured, though \cite{VARENNAURI}.  

HQET has actually become an important tool for inferring lessons on the
dynamics  of heavy flavour hadrons from lattice QCD results. 

\subsubsection{Basics of the spectroscopy}
\label{BASICSPECT}

Like for any hadron the mass of a heavy flavour hadron $H_Q$ is given 
by the properly normalized expectation value of the QCD Hamiltonian; 
the only difference, which actually amounts to a considerable
simplification, is that the latter can be expanded in powers of 
$1/m_Q$: 
$$ 
M_{H_Q} = \matel{H_Q}{{\cal H}}{H_Q} 
$$
\beq 
{\cal H} =  m_Q + {\cal H}_Q + {\cal H}_{\rm light} \; , \; 
{\cal H}_Q= {\cal H}_0 + \frac{1}{m_Q} {\cal H}_1 + 
\frac{1}{m_Q^2} {\cal H}_2 + ... 
\eeq
where ${\cal H}_{\rm light}$ contains the dynamics for the light degrees 
of freedom and 
\beq 
{\cal H}_0 = - A_0 \; , \; 
{\cal H}_1 = \frac{1}{2}\left( \vec \pi^2 + \vec \sigma \cdot \vec B
\right) \; , \;  
{\cal H}_2 = \frac{1}{8} \left[ - \vec D \cdot \vec E + 
\vec \sigma \cdot (\vec E \times \vec \pi - \vec \pi \times \vec E )
\right] 
\eeq
with the first and second term in ${\cal H}_2$ being the Darwin and 
LS term, respectively, familiar from atomic physics. Hence 
\beq 
M_{H_Q} = m_Q + \bar \Lambda + 
\frac{(\mu_{\pi}^2 - \mu_G^2)_{H_Q}}{2m_Q} + ... 
\label{HADMASSEXP} 
\eeq
with $\mu_{\pi}^2$ and $\mu_{G}^2$ defined in 
Eqs.(\ref{MUPISQ},\ref{MUGSQ}). Eq.(\ref{HADMASSEXP}) has an obvious 
intuitive interpretation: the mass of $H_Q$ is given by the 
heavy quark mass $m_Q$, the `binding energy' $\bar \Lambda$, 
the average kinetic energy $\mu_{\pi}^2$ of the heavy quark $Q$ inside 
$H_Q$ and its chromomagnetic moment $\mu_G^2$. Since the latter term,
which is spin dependent, vanishes in the limit $m_Q \to \infty$, 
the spin of a heavy flavour hadron can be labeled by the total 
spin $J$ as well as the spin of the light degrees of freedom $j$. 
S-wave pseudoscalar and vector mesons thus form a pair of $j=1/2$ 
ground states
that in the heavy quark limit are degenerate. Baryons 
$\Lambda_Q$ and $\Xi_Q$ carry $j=0$ and thus represent actually a simpler 
state than the mesons. 

With these expressions one can derive and extend to higher orders the 
expression for $m_b - m_c$ already stated in Eq.(\ref{MBMINUSMC}): 
\beq  
m_b - m_c = \langle M_B\rangle - \langle M_D\rangle +  
\frac{\mu_{\pi}^2}{2} \left( \frac{1}{m_c} - \frac{1}{m_b} \right)  
+ \frac{\rho_D^3 - \bar \rho ^3}{4}
\left( \frac{1}{m^2_c} - \frac{1}{m^2_b}
\right)  + {\cal O}(1/m^3_{c,b}) 
\label{MBMINUSMC2} 
\eeq 
where $\rho_D^3$ denotes the Darwin term and $\bar \rho ^3$ the sum of 
two positive {\em non}local correlators.

\subsubsection{Semileptonic form factors}
\label{SLFF}

Consider a hadron $H_Q$, where the heavy quark $Q$ is surrounded by 
-- in a terminology coined by Nathan Isgur -- 
the `brown muck' of the light degrees of freedom in analogy to the
situation in an atom. When $Q$ decays weakly into a lighter, yet still
heavy quark $Q^{\prime}$ plus a $l\nu$ pair with 
invariant mass $\sqrt{q^2}$, the
surrounding cloud  of light degrees of freedom will not feel this change
in its center instantaneously -- the hadronization process requires time
to  adjust. Heavy quark symmetry has two main consequences here, one
concerning the normalization of the hadronic formfactors and one their 
$q^2$ dependence. 
\begin{itemize}
\item 
In the infinite mass limit 
$m_Q > m_{Q^{\prime}} \to \infty$ the rate for an
exclusive semileptonic transition $H_Q \to H_{Q^{\prime}} \ell \nu$   
at {\em zero recoil} for the final state hadron $H_{Q^{\prime}}$ will 
depend neither on $m_{Q^{\prime}}$ nor on the heavy quark spin 
as can be inferred from the Pauli Hamiltonian given in 
Eq.(\ref{PAULIHAM}) \cite{SV1}. I.e., the form factor for 
$H_Q \to H_{Q^{\prime}}$ at zero recoil is asymptotically (ignoring 
also perturbative gluon corrections) unity
in  `heavy-to-heavy' transitions for pseudoscalar and vector hadrons 
$H_{Q^{(\prime})}$. 
\item 
The $q^2$ of the lepton pair can be expressed through the four-momenta 
$p$ and $p^{\prime}$ of $H_Q$ and $H_{Q^{\prime}}$, respectively, and 
their four-velocities $v$ and $v^{\prime}$: 
\beq 
q^2 = 2M^2_{H_Q} \left(1 - \frac{p\cdot p^{\prime}}{M^2_{H_Q}}\right) = 
2M^2_{H_Q} (1 - v \cdot v^{\prime})
\eeq
For both $H_Q$ and $H_{Q^{\prime}}$ being pseudoscalars one can write 
$$ 
\matel{H_{Q^{\prime}}(v^{\prime})}{\bar Q^{\prime}\gamma_{\mu}Q}
{H_Q(v)} = 
$$
\beq
\left(
\frac{M(H_Q)+M(H_{Q^{\prime}})}{2\sqrt{M(H_Q)M(H_{Q^{\prime}})}}
(p + p^{\prime})_{\mu} - 
\frac{M(H_Q)-M(H_{Q^{\prime}})}{2\sqrt{M(H_Q)M(H_{Q^{\prime}})}}
(p - p^{\prime})_{\mu}
\right) \xi (v\cdot v^{\prime}) \; . 
\eeq
I.e., there is a single independent form factor 
$\xi (v\cdot v^{\prime})$, which is `universal' in the double sense that
it is independent of the heavy quark masses and that it also controls 
the $q^2$ dependence, when $H_{Q^{\prime}}$ is a vector meson; it is 
usually referred to as the `Isgur-Wise' function 
\index{Isgur-Wise function}.

\end{itemize}
At finite values of $m_{Q^{(\prime})}$ there 
are corrections to both these features. 

Such a scenario is realized with reasonable accuracy for 
$B \to D^{(*)} \ell \nu$ channels. On the other hand the charm decays 
$c\to s$ as well as $c\to d$ are   
of the type `(moderately) heavy to light'. Even then heavy quark 
symmetry allows to relate the form factors in, say, 
$D \to \ell \nu \pi$, $D \to \ell \nu \rho$ etc. to those for 
$B \to \ell \nu \pi$, $B \to \ell \nu \rho$ etc. at the same values of 
$v\cdot v^{\prime}$. Yet this relation is not overly useful 
quantitatively due to the potentially large $1/m_c$ corrections.

\subsection{NRQCD}  
\label{NRQCD}  
 
Heavy quark bound states like $[c\bar c]$, $B_c = [b\bar c]$,  
$[b\bar b]$ etc. are nonrelativistic bound states in the heavy-quark 
limit. Pre-asymptotic corrections to this limit can conveniently be  
calculated employing another effective theory, namely  
nonrelativistic QCD (NRQCD). The local operators that appear in NRQCD  
are formally very similar as in HQET, Eqs(\ref{LHEAVYEXP}.
Yet at the same time there is a basic difference in the dynamical stage  
for $[Q\bar Q]$ and $[Q\bar q]$ systems: the light antiquark $\bar q$  
in the latter has to be treated fully relativistically.   
Formally the same operators can thus appear in different orders in the
two   schemes. Technically it is easily understood how this comes about:
since  
$E_Q/m_Q$ and $p_Q^2/m_Q^2$ are of the same order in a nonrelativistic  
expansion, the primary expansion parameter in NRQCD is the heavy quark  
velocity $v_Q = p_Q/m_Q$ rather than $1/m_Q$. Among other things this  
implies that the average heavy quark kinetic energy  
$\matel{H_Q}{\bar Q|\vec D|^2Q}{H_Q}/2m_Q$, which enters as a leading   
order pre-asymptotic {\em correction} in HQET appears already as part of 
the asymptotic contribution in NRQCD. Similar to the situation with
lattice QCD, see below, there are   alternative formulations of NRQCD. 

One of the main applications of NRQCD is describing the production of 
quarkonia in different reactions. The basic picture is the following: 
one invokes {\em short}-distance dynamics to produce a $\bar QQ$ pair 
without restriction on the latter's spin, angular and colour quantum
numbers from two initial partons $i$ and $j$. This $\bar QQ$ pair is then
assumed to evolve into the final  state quarkonium $H$ -- a process
involving long-distance dynamics. The  analysis is thus based on three
main elements 
\cite{ORIGINAL,LEIBO}: 
\begin{itemize}
\item 
One makes a factorization ansatz for the (differential) cross section for
producing a quarkonium $h$ from partons $i$ and $j$: 
\beq 
d\sigma = \sum _n d \hat \sigma (ij\to \bar QQ(n) +X)
\langle O^H(n)\rangle  
\label{QQFACT}
\eeq
\item 
The quantities $d \hat \sigma (ij\to \bar QQ(n) +X)$ are calculated 
perturbatively and convoluted with parton distribution functions, when 
necessary. 
\item 
The long-distance matrix elements $\langle O^H(n)\rangle$, which encode
the hadronization of $\bar QQ(n)$ into $H$ are assumed to be universal, 
i.e. irrespective of the subprocess $ij\to \bar QQ(n) +X$. On fairly
general grounds one can infer how they scale with the heavy quark
velocity $v$. One should note that both colour singlet and octet 
configurations are included. It thus goes well beyond models assuming 
dominance by colour singlet configurations (or colour octet ones for that
matter). One can extract these matrix elements most cleanly from
quarkonia production at LEP and apply them to Tevatron or HERA data. 
While data from LEP have very limited statistics, the predictions for
rates at Tevatron and HERA depend sensitively on the parton distribution
functions adopted. Some quantitative information on them can also be
inferred from quark models and lattice QCD. 

\end{itemize}
The basic philosophy is similar to what was described above for the OPE 
treatment of charm decays, and the factorization ansatz of 
Eq.(\ref{QQFACT}) is quite reasonable. However it has not (yet) been
proven in a rigorous fashion. One might also be concerned about treating 
the matrix elements $\langle O^H(n)\rangle$ as universal quantities 
\footnote{This latter concern could be overcome by including some 
nonperturbative corrections in the subprocess $ij\to QQ(n) +X$.}. 

Looking beyond these general caveats one expects this formulism to 
apply to sufficiently heavy quarkonia, like the $\Upsilon$. Whether 
it can be applied already for charmonia is another question of course, 
for which we do not know the answer a priori. As it is with applying HQE
to charm decays we should use NRQCD as a tool for {\em learning} about
nonperturbative dynamics and incorporating such lessons rather than
ruling out models.

\subsection{Lattice QCD}  
\label{LATTICEQCD}  
 
Monte Carlo simulations of   
QCD on the lattice or lattice QCD for short provides a very different  
framework to deal with QCD's complementary features of asymtotic  
freedom in the ultraviolet and infrared slavery. The four-dimensional  
space-time continuum is replaced by a discrete lattice with  
spacing $a$ between lattice sites. This is 
(usually)  viewed not as representing physical reality, but providing  
the mathematical means to deal with long-distance dynamics through an  
expansion in the {\em inverse} coupling. Distances $\sim a$ and smaller  
obviously cannot be treated in this way. This can be expressed by  
saying that the finite spacing introduces  
an ultraviolet cut-off $\sim \pi/a$ for the lattice version of QCD.  
The short distance dynamics is treated by perturbative QCD and 
considerable care has to be applied in matching the two theories at  
a distance scales $\sim a$. One uses the technique of  
effective field theory sketched in Sect.\ref{EFT} to incorporate 
short-distance  dynamics cut off by the finite lattice spacing; the 
discretization  effects are described through an expansion in powers of 
$a$:  
\beq  
{\cal L}_{eff} = {\cal L}_{QCD} + a {\cal L}_{1}+ a^2{\cal L}_{2} + ...  
\label{SYM} 
\eeq 
With the ${\cal L}_{i}$ containing operators of dimension higher  
than four, they are nonrenormalizable; this poses no problem since they  
are constructed to describe long-distance dynamics. 
 
There are actually two measures for the quality of the lattice for  
our purposes: (i) To get as close as possible to the continuum case one 
would like to have  
$a$ as small as possible. (ii) At the same time one wants to have a 
sufficient number of lattice sites in each dimension to be not overly 
sensitive to finite size  effects. I.e., effectively one has put the  
particles inside a box to study the response to the forces they  
experience; yet one does not want having them bounce off the walls of the 
box too frequently since that is an artifact of the framework. 
 
Obviously there are practical 
limitations in the available computing power to achieve these  desirable 
goals. Yet as a condition sine qua non for treating light  degrees of 
freedom one requires $am_q$, 
$a\Lambda_{NPD} \ll 1$ for the expansion of Eq.(\ref{SYM}) to have 
practical value. 
 
There are actually a number of different implementations of lattice  
QCD; they differ mainly in three areas 
\cite{LELLOUCHAMST}:  
\begin{itemize} 
\item  
Different expressions for the action defined on the lattice will merge 
into the same QCD action in the continuum limit. The lattice action  
can be optimized or `improved' for example by eliminating  
${\cal L}_1$; for in that case the continuum case $a\to 0$ is approached  
like $a^2$, i.e. much faster.  
\item  
Putting fermions on the lattice creates problems between the  
`Scylla' of `fermion doubling' and the `Charybdis' of vitiating  
chiral invariance. For very general theorems tell us that in four  
dimensions chiral invariance is either violated for $a\neq 0$ or 
maintained at the price of getting too many fermions.  
\item  
For heavy quarks one needs actually $am_Q \ll 1$. However  
with presently available computing power we  
can achieve merely $am_b \sim 1-2$ and $am_c$ about a third of it. It 
seems unlikely that in the near future one can achieve $am_b \ll 1$.  
Several strategies have been suggested to overcome this limitation,  
namely relying on the static approximation, lattice NRQCD, matching up 
with HQET and/or extrapolating from $m_c$ up to $m_b$. 
 
This is actually another example, where charm hadrons and their decays  
can provide us with an important bridge on the road towards a deeper  
understanding of the dynamics of beauty hadrons. 
\item  
Including light quarks as fully dynamical degrees of freedom that can be  
produced and annihilated slows down lattice computations tremendously.  
This Gordian knot has been treated mostly in the tradition of Alexander  
the Great, i.e. by `cutting' or ignoring it. This is called the  
quenched approximation. The first partially unquenched studies have been  
presented recently, where two different flavours of light quarks have  
been fully included in the Monte Carlo simulations. 
 
\end{itemize} 

From the start the primary goal of lattice QCD has been to provide a 
framework for dealing {\em quantitatively} with nonperturbative 
dynamics in all its aspects and in ways that are genuinely based on 
the first principles of QCD and where the uncertainties can be reduced 
in a {\em systematic} way. Indeed no other method has surfaced which 
can lay claim to a similarly `universal' validity. On the other hand 
there are other theoretical technologies that provide a `first principles' 
treatment of 
nonperturbative dynamics, albeit in a more restricted domain; examples  
are chiral invariance and HQE. Those most definitely benefit from input 
lattice QCD produces, as described above. Yet lattice QCD benefits also 
from them, which serve not only as a cross check, but can also provide 
valuable insights for interpreting findings by lattice QCD. 

There are some observables where there is no plausible deniability if 
lattice QCD failed to reproduce them. Matrix elements involving at most a
single hadron each in the initial and final state are in that category. 
The best developed case history 
is provided  by the decay constants. Studies show an 
enhancement by  8\% in the values for $f_{D_s}$ and $f_D$ when going from 
quenched to  partially unquenched (with $N_f =2$) while not affecting the 
ratio  $f_{D_s}/f_D$ \cite{RYAN}:  
\bea  
f_D(N_f=0) = 203\pm 14\, {\rm MeV}, \; &\Longrightarrow& \;  
f_D(N_f=2) = 226\pm 15\, {\rm MeV} \\ 
f_{D_s}(N_f=0) = 230\pm14\, {\rm MeV}\; &\Longrightarrow& \;   
f_{D_s}(N_f=2) = 250\pm 30\, {\rm MeV} \\  
(f_{D_s}/f_{D})(N_f=0) = 1.12 \pm 0.02 \; &\Longrightarrow& \;  
(f_{D_s}/f_{D})(N_f=2) = 1.12 \pm 0.04  \label{FDTHEO}
\eea 
which can be compared with the experimental findings from   
$D \to \mu \nu$ as explained later  
\beq   
f_D|_{exp} \sim 200 \div  300 \; {\rm MeV} \; .   
\eeq 

Very recently a short paper appeared \cite{LATTICEPREC} with the 
ambitious title "High-Precision Lattice QCD Confronts Experiment" stating
that  "... realistic simulations are possible now, with all three 
flavors of light quark" due to a breakthrough in the treatment of light
quarks. The authors point out that the treatment in particular of heavy
quark physics will benefit greatly. 

A note of caution seems appropriate (and it is also sounded by the
authors of Ref.\cite{LATTICEPREC}. Before a difference in a measured 
and predicted rate -- with the latter based solely on lattice QCD -- can
be taken as conclusive evidence for the intervention of New Physics,
lattice QCD has to be subjected to a whole battery of tests through 
different types of observables. Charm physics -- and this is one 
of the recurring themes of this review -- provides ample opportunity 
for such a comprehensive program as described later. 
As an extra bonus, one can, at least in principle,    
approach the charm scales in both direction, namely from below by   
using finer and finer lattices and from above by extrapolating   
from the limit of {\em static} quarks, for which $b$ quarks provide   
a good approximation. 
 
\subsection{Special tools} 
 
In the preceding Subsections we have described theoretical technologies  
that are most relevant for dealing with heavy flavour hadrons, yet at the 
same time apply to many other areas as well. 
Now we sketch some tools with a more limited range of application or  
more special nature, 
namely  the short distance renormalization of the weak Lagrangian, QCD 
sum rules, dispersion relations  and the concept of final state 
interactions.

\subsubsection{Effective weak Lagrangian} 
\label{EFFWEAK} 
 
The weak Lagrangian responsible for Cabibbo allowed nonleptonic charm 
decays is given by  a single term at scales just below $M_W$:  
\beq  
{\cal L}_W^{\Delta C=1} (\mu < M_W) = (4G_F\sqrt{2}) V(cs)V^*(ud)  
(\bar s_L\gamma_{\nu}c_L)(\bar u_L\gamma_{\nu}d_L) + h.c.  
\label{BARELAG1} 
\eeq 
Radiative QCD corrections lead to a renormalization at scale $m_c$,  
often referred to as {\em ultraviolet} renormalization 
\index{ultraviolet renormalization}, since its scales 
are larger and thus more ultraviolet than $m_c$.  
One-loop contributions generate an operator different from  
$(\bar s_L\gamma_{\nu}c_L)(\bar u_L\gamma_{\nu}d_L)$, namely  
$(\bar s_L\gamma_{\nu}\frac{\lambda_i}{2}c_L) 
(\bar u_L\gamma_{\nu}\frac{\lambda_i}{2}d_L)$ with the  
$\lambda_i$ denoting the $SU(3)$ matrices. The renormalization is  
therefore additive and not multiplicative  
\index{additive renormalization}, i.e.  
${\cal L}_W^{\Delta C=1} (\mu = m_c) \not \propto  
{\cal L}_W^{\Delta C=1} (\mu < M_W)$. Considering all operators that 
under QCD renormalization can mix with the original transition 
operator(s) one can determine which are the multiplicatively  
renormalized operators and with which coefficients they appear  
in the effective Lagrangian by diagonalizing the matrix with the one-loop  
corrections of these operators. 
 
However there is a more direct way to 
understand why QCD corrections double the number of transition operators 
and which operators are multiplicatively renormalized. Already on the  
one-loop level one has two types of couplings in colour space,  
namely $1\otimes 1$ and $\lambda_i \otimes \lambda _i$ with  
$\lambda_i$ denoting the eight Gell-Mann matrices. Higher loop 
contributions do not change this pattern since  
$\lambda_i \lambda_j\otimes \lambda _i\lambda_j$ can again be expressed  
through a linear combination of $1\otimes 1$ and  
$\lambda_i \otimes \lambda _i$. This holds no matter what the Lorentz  
structure of the coupling is. For couplings involving left-handed quark  
currents only we have  
the simplification that the product of two such currents remains a  
product of two left-handed currents under Fierz transformations 
\index{Fierz transformations} 
\cite{PESKIN}. This allows us to write a current 
product of the  form $\lambda_i \otimes \lambda _i$ as 
linear combination of $1\otimes 1$ and $[1 \otimes 1]_{Fierz}$, where  
the second term is the Fierz transformed product.  
Consider now the interaction described by Eq.(\ref{BARELAG1}) 
$ c_L \to s_L \bar d_L u_L $. 
Its operator is purely isovector. Yet under $V$ spin  
\index{$V$ spin},   
which groups $(u,s)$ into a doublet  
with $c$ and $d$ being singlets the final state is a combination of  
$V=0$ and $V=1$. Fermi-Dirac statistics tells us that if $u$ and $s$ are  
in the antisymmetric $V=0$ [ symmetric $V=1$] configuration, they have to 
be  in the symmetric [antisymmetric] $SU(3)_C$ $6$ [$\bar 3$] 
representation. I.e., the two multiplicatively renormalized operators  
have to be Fierz even and odd:  
\beq  
O_{\pm}^{\Delta C=1} = \frac{1}{2}  
[ (\bar s_L\gamma_{\nu}c_L)(\bar u_L\gamma_{\nu}d_L)  
\pm (\bar s_L\gamma_{\nu}d_L)(\bar u_L\gamma_{\nu}c_L) ] 
\label{OEVENODD}  
\eeq  
Therefore   
$$  
\frac{{\cal L}_W^{\Delta C=1} (\mu = m_c)} 
{(4G_F\sqrt{2}) V(cs)V^*(ud)} =   
c_+O_{+}^{\Delta C=1} + c_- O_{-}^{\Delta C=1}  =  
$$   
\beq  
[c_1(\bar s_L\gamma_{\nu}c_L)(\bar u_L\gamma_{\nu}d_L)  
+c_2(\bar u_L\gamma_{\nu}c_L)(\bar s_L\gamma_{\nu}d_L) ] 
\eeq  
The coefficients $c_{1,2}$ can be expressed as  
follows at leading log level: 
\beq  
c_1|_{LL} = \frac{1}{2}(c_+ + c_-) \; , \; \;  
c_2|_{LL} = \frac{1}{2}(c_+ - c_-)  
\eeq 
\beq 
c_{\pm} \equiv \left[  
\frac{\alpha_S(M_W^2)}{\alpha_S(m_c^2)}\right]^{\gamma_{\pm}} \; , \; \;  
\gamma_+=\frac{6}{33-2N_f} = - \frac{1}{2}\gamma_-  
\label{CPLUSMINUS} 
\eeq 
$N_f$ denotes the active flavours. Next-to-leading log corrections  
are sizeable at the charm scale. They cannot be expressed in a compact  
analytical way; numerically one finds when including these contributions: 
\beq  
c_1|_{LL+NLL} \simeq 1.32 \; , \; \; c_2|_{LL+NLL} \simeq - 0.58 
\eeq 
Noting that without QCD radiative corrections one has  
$c_1 =1$ and $c_2=0$, QCD renormalization constitutes a quite  
sizeable effect. This is not surprising since the leading log result  
represents an expansion in powers of $\alpha_S$ log$M_W^2$ rather than  
just $\alpha_S$. 
 
With hadronic matrix elements evaluated at ordinary hadronic  
scales $\Lambda_{NPD}$ rather than the heavy quark mass, one has to  
consider also renormalization from $m_c$ down to $\Lambda_{NPD}$.  
This is often called {\em hybrid} renormalization 
\index{hybrid renormalization} since its scales  
are in the infrared relative to $m_c$ and in the ultraviolet  
relative to $\Lambda_{NPD}$. Yet since $m_c$ -- unlike $m_b$ --  
exceeds $\Lambda_{NPD}$ by a moderate amount only, one does not  
expect hybrid renormalization to play a major role in most cases  
for charm. One notable exception is the $D^+-D^0$ lifetime ratio,  
which will be discussed later. 
 
There are analogous effects on the Cabibbo once and twice suppressed  
levels. The analogues of the Fierz even and odd operators of  
Eq.(\ref{OEVENODD}) are multiplicatively renormalized with the  
coefficients $c_{\pm}$ as in Eq.(\ref{CPLUSMINUS}). In addition  
Penguin operators emerge on the Cabibbo disfavoured level. These  
renormalization effects with $c_- \simeq 1.9 > c_+\simeq 0.74$  
lead to the enhancement of the $\Delta I=0$ [$\Delta I=1/2$] over  
the $\Delta I=1$ [$\Delta I=3/2$] transition operators for  
once [doubly] Cabibbo suppressed modes. These issues will be addressed  
further in Sect.\ref{NONLEPT}.

\subsubsection{Sum Rules} 
\label{SUMRULES} 
 
Sum rules are an ubiquitous tool in many branches of physics where sums  
or integrals over observables -- rates, moments of rates etc. --  
are related to a normalization condition  reflecting unitarity etc. or  
a quantity that can be calculated in the underlying theory. They form  
an important ingredient in our treatment of deep inelastic  
lepton-nucleon scattering for example, where moments of structure  
functions are related to terms in an OPE. Examples are the  
Adler and the Gross-Llewellyn-Smith sum rules 
\cite{KAKU,PESKIN}. 
 
Another celebrated case  
are the SVZ QCD sum rules named after Shifman, Vainshtein and Zakharov
\cite{SVZSR}, which allow to express {\em low energy} hadronic  
quantities through basic QCD parameters. The starting point is again  
provided by an OPE in terms of local operators.   
Nonperturbative dynamics are parameterized through vacuum expectation 
values -- or {\em condensates} \index{condensates} -- $\matel{0}{\bar
qq}{0}$, 
$\matel{0}{G^2}{0}$ etc.,  since those have to vanish in perturbation 
theory. Those  condensates are treated as free parameters the values of 
which are  fitted from some observables. One typically  
matches up a quantity calculated on the quark-gluon level through a  
dispersion relation -- see the next Subsection -- with an ansatz for the 
hadronic observables;  the stability of the match under variations 
of input values  provides an intrinsic gauge for the theoretical control 
in this case. Introducing nonperturbative dynamics through condensates  
represents an approximation of less than universal validity: such an 
ansatz cannot be counted on to reproduce observables exhibiting rapid 
variations in, say, energy like narrow resonances and their phase 
shifts. In such situations one can hope at best for being able to treat  
`smeared'  hadronic observables, i.e. ones that have been averaged over  
some energy interval.  
Manifold experience shows that one has to allow for an 
irreducible  theoretical uncertainty of 
 about 20-30 \% 
 due to unknown  
contributions from higher operators in the OPE, excited states in  
the dispersion relations and due to the ansatz with condensates. Contrary  
to the situation with lattice QCD, one cannot hope for a systematic  
reduction in the theoretical uncertainty. 
 
A very similar approach under 
the name of "lightcone sum rules"  \cite{LCSR} has been developed for
describing the  formfactors in exclusive  semileptonic decays of heavy
flavour hadrons to  be mentioned later. 
 
There is a second and third class of sum rules that are relevant here,  
namely the so-called S(mall)V(elocity) \cite{SVSRPAPER} and 
the spin sum rules \cite{SPINSR} 
that have   been formulated for semileptonic decays of
heavy flavour hadrons.   They are based on {\em systematic} expansions in
$1/m_Q$ and --   for $b\to c$ transitions -- in the velocity of the final
state quark.   Accordingly they do not exhibit this brickwall of about 20
- 30\%   in theoretical uncertainty, but can be improved successively. 
 
\subsubsection{Dispersion relations} 
\label{DISPREL} 
Dispersion relations are encountered in many branches of physics and in 
quite different contexts. The common element is that certain fundamental 
features of general validity can be imposed by requiring that physical 
quantities have to be analytical functions of their variables, when they 
are allowed to be complex. One then invokes Cauchy's theorem on 
path integrals in the complex plane to relate the real and imaginary 
part of these quantities to each other. 
 
For example in classical electrodynamics causality implies field 
amplitudes to be analytic. This holds in particular for the 
dielectric constant $\epsilon$ when it is frequency dependent: 
$\vec D(\vec x,\omega) = \epsilon (\omega) \vec E(\vec x,\omega)$. 
Causality implies the Kramers-Kronig relation \cite{JACKSON} 
\beq 
{\rm Im}\epsilon (\omega)/\epsilon _0 = 
- \frac{2\omega}{\pi} P \int _0^{\infty} d\omega^{\prime} 
\frac{{\rm Re}\, \epsilon(\omega^{\prime})/\epsilon_0 - 1} 
{\omega^{\prime 2} - \omega^2} \; , 
\eeq 
where $P$ denotes the principal part computation of this 
integral. 
In S matrix theory one postulates unitarity, Lorentz and 
crossing symmetry and analyticity. Dispersion relations relate the 
scattering amplitudes for different reactions through integrals. 
 
Likewise one can relate the values of a two-point function 
$\Pi (q^2)$ in a quantum field theory at different complex values 
of $q^2$ to each other through an integral representation; $q$ denotes a 
four-momentum. 
In  particular one can evaluate $\Pi(q^2)$ for large {\em Euclidean} 
values $-q^2 = q_0^2 + |\vec q|^2$ with the help of an OPE and then 
relate the coefficients $I_n^{OPE}$ of local operators $O_n$ to 
observables  like $\sigma (e^+e^- \to {\rm had.})$ and their moments in 
the  physical, i.e. Minkowskian domain through an integral 
over the discontinuity along the real axis; the integral 
over the asymptotic arcs vanishes \cite{PESKIN}: 
\beq 
I_n^{OPE} \simeq \frac{1}{\pi} \int _0^{\infty} ds 
\frac{s}{(s+q^2)^{n+1}} \cdot \sigma (s) 
\eeq 
Such a procedure is based on there being only physical singularities 
-- poles and cuts -- on the real axis of $q^2$: then one can 
first calculate  two-point functions for large Euclidean values of $q^2$ 
and secondly one will not pick up extra unphysical contributions from 
poles etc. This is the basis of the derivation of the 
celebrated QCD sum rules by the ITEP group \cite{SVZSR}. 
 
Such dispersion relations are used to calculate transition rates in the 
HQE and to derive new classes of sum rules \cite{SVSRPAPER}. 
 
\subsubsection{Final State Interactions (FSI) and Watson's theorem}  
\label{FSI}  
 
The mass of charm hadrons places them into an environment populated by 
many non-charm resonances, hadronic thresholds etc. making FSI  
quite virulent. This provides for a particularly challenging dynamical  
environment. 
\index{Final State Interactions (FSI)} 
 \index{Watson's theorem}
Let us consider the decay of a meson. The primary weak force 
transmogrifies the initially present valence quark and antiquark into  
two quarks and antiquarks. Yet those will not rearrange themselves  
immediately into two mesons that emerge as asymptotic states. Typically  
quarks and antiquarks will be exchanged, they can change their flavour  
identity thus giving rise to final states that are absent  
otherwise, and even 
additional 
$q\bar q$ pairs can be excited. Precisely  since the forces driving these 
processes  are strong, those secondary  interactions cannot be ignored or 
treated to first (or any finite)  order only. They can induce even 
spectacular resonance enhancements  (or depletions for that matter). This 
is sometimes described by saying  that the initially produced two 
quark-antiquark clusters can and  typically will {\em rescatter} into 
different kinds of two-meson or  even multi-meson final states. 
\index{rescattering}
 
Fortunately there is a modicum of theoretical  
guidance for dealing with this quagmire as sketched by the following 
remarks.   
\begin{itemize} 
\item  
While FSI can change the nature of the final state 
dramatically,  they mainly re-arrange the rate between different channels 
{\em without}   create overall rate. I.e., typically they do not increase 
or decrease  the total nonleptonic or semileptonic widths.  
\item  
However 
FSI can affect even fully inclusive transitions. As we will discuss  
later the nearby presence of a hadronic resonance of appropriate  quantum 
numbers can enhance or suppress significantly the width of a  charm 
hadron -- an effect that would constitute a violation of  quark-hadron 
duality.   
\item  
With the strong interactions conserving isospin and G 
parity, possible rescatterings are constrained by these quantum numbers.  
\item  
The most treatable case after total rates is provided by two-body  
final states, where we include hadronic resonances in the latter, due to  
their `trivial' kinematics. A small number of quark-level diagrams can  
drive a large number of hadronic transitions.  
Consider for example $D^0 \to K^-\pi^+$ where two different  
four-quark operators contribute changing isospin by $1/2$ and $3/2$:  
\beq  
T(D^0 \to K^-\pi^+) = e^{i\alpha_{1/2}} T_{1/2} +  
e^{i\alpha_{3/2}} T_{3/2} 
\eeq 
A priori one expects -- correctly, as it turns out -- that the FSI  
generate a nontrivial relative phase between the two  
different isospin amplitudes $T_{1/2,3/2}$  
-- $\alpha_{1/2}\neq \alpha_{3/2}$ --  and affect also their size.  
As we will discuss later in detail, such relative  
strong phases are a conditio sine 
qua non for {\em direct} CP asymmetries to arise.  
A well-known theorem is frequently quoted in this context, namely  
Watson's theorem. Below we will describe it mainly to make explicit  
the underlying assumptions and corresponding limitations.  
\item  
Novel theoretical frameworks have been put forward recently to  
treat nonleptonic two-body decays of $B$ mesons \cite{BENEKE,SANDA}.  
However there is no a priori justification for applying such treatments  
to $D$ decays.  
\item  
Three-body final states can be and are subjected to Dalitz plot  
analyses. Unfortunately theory can provide very little  
guidance beyond that. 
 
\end{itemize} 
 
In describing {\em Watson's theorem} we follow  
the discussion in Ref.\cite{CPBOOK}. For reasons that will become  
clear we consider $K \to n\pi$. 
 
A $\Delta S[C,...] \neq 0$ process has to be initiated by weak forces  
which can be treated perturbatively. Yet the final state is shaped  
largely by strong dynamics mostly beyond the reach of a  
perturbative description. Nevertheless one can make some reliable  
theoretical statements based on symmetry considerations -- sometimes. 
 
With the strong interactions conserving G-parity a state of an even 
number of pions cannot evolve {\em strongly} into a state with an odd  
number. Therefore  
\beq  
K \stackrel{H_{weak}}\longrightarrow 2\pi \stackrel{H_{strong}}  
{\not \longrightarrow} 3\pi 
\eeq 
On the other hand, the two pions emerging from the weak decay are  
not asymptotic states yet; due to the strong forces they will undergo  
rescattering before they lose sight of each other. Deriving the  
properties of these strong FSI from first principles is beyond our  
present computational capabilities. However, we can relate some  
of their properties to other observables. 
 
Let us assume the weak interactions to be invariant under  
time reversal:  
\beq  
TH_WT^{-1} = H_W 
\eeq  
We will show now that even then the amplitude for $K^0 \to 2\pi$ is 
complex; the strong FSI generate a phase, which   
actually  coincides with the S wave $\pi\pi$ phase shift $\delta_I$ taken 
at  energy $M_K$ \cite{WATSON}. That is, the amplitude is real, except
for the fact   that the two pions interact before becoming asymptotic
states. 
 
At first we allow the phase for the $K^0 \to 2\pi$ amplitude to  
be arbitrary:   
\beq  
\matel{(2\pi)_I^{out}}{H_W}{K^0} = |A_I|e^{i\phi_I}   
\eeq 
where the label $I$ denotes the isopin of the $2\pi$ state.    
With $T$ being an antiunitary operator and using  
$T^{\dagger}T= \Im^{\dagger}\Im$ with $\Im$ denoting  
the complex conjugation 
operator we have  
\beq 
\matel{(\pi\pi)_I; out}{H_W}{K} =  
\matel{(\pi\pi)_I; out}{T^{\dagger}TH_WT^{-1}T}{K}^* =  
\matel{(\pi\pi)_I; in}{H_W}{K}^* \; ,  
\eeq 
since for a single state -- the kaon in this case -- there is no  
distinction between an $in$ and $out$ state. After inserting  
a complete set of $out$ states  
\beq  
\matel{(\pi\pi)_I; out}{H_W}{K} =  
\sum _n \langle (\pi\pi)_I; in|n; out \rangle \matel{n; out}{H_W}{K}^*  
\; ,  
\eeq  
where the $S$ matrix element $\langle (\pi\pi)_I; in|n; out \rangle$  
contains the delta function describing conservation of energy and  
momentum, we can analyze the possible final states. The only hadronic  
states allowed kinematically are $2\pi$ and $3\pi$ combinations. With G 
parity enforcing  
\beq  
\langle (\pi\pi)_I; in |3\pi; out \rangle = 0 
\eeq 
only the $2\pi$ $out$ state can contribute in the sum:  
\beq  
\matel{(\pi\pi)_I; out}{H_W}{K} =  
\langle(\pi\pi)_I; {\rm in}|(\pi\pi)_I; {\rm out}\rangle  
\matel{(\pi\pi)_I; out}{H_W}{K}^*  
\eeq 
This is usually referred to as the condition of {\em elastic}  
unitarity \index{elastic unitarity}. With the S matrix for 
$(\pi\pi)_I \to (\pi\pi)_I$   given by  
\beq  
S_{\rm elastic} =  
\langle(\pi\pi)_I; {\rm out}|(\pi\pi)_I; {\rm in}\rangle =  
e^{-2i\delta _I}  
\eeq 
we have  
\beq  
\matel{(\pi\pi)_I; {\rm out}}{H_W}{K^0} =  
|\matel{(\pi\pi)_I; {\rm out}}{H_W}{K^0}|e^{i\delta_I} \; ;  
\eeq 
i.e., as long as $H_W$ conserves $T$, the decay amplitude remains  
real after having the strong phase shift factored out. This is Watson's  
theorem in a nutshell. 
 
FSI also affect the decays  of {\em heavy} flavour hadrons, yet we 
{\em cannot}   apply Watson's theorem blindly even for  
$T$ conserving $H_W$. In particular it would be absurd to assume  
{\em elastic} unitarity to apply in two-body or even quasi-two-body  
beauty decays: strong FSI are bound to generate additional hadrons in the  
final state. The decays of {\em charm} hadrons provide a borderline  
case: while the FSI can change the identity of the emerging particles  
and can produce additional hadrons, their impact is    
moderated since the available phase space is less than abundant.  
This is consistent with the observation that (quasi-)two-body  
modes constitute the bulk of nonleptonic $D$ decays, although we have  
not learnt yet how to assign precise numbers to this statement, see  
Sect.\ref{TWOBODY}.  
Introducing the concept of {\em absorption} \index{absorption}   
-- $T_f \to \eta_f T_f$ with $|\eta _f| < 1$  --  
provides a useful  {\em phenomenological} approximation for parameterising 
such inelasticities.  

\subsubsection{Zweig's rule}
\label{ZWEIG}

The Zweig rule goes back to the earliest days of the quark model 
\cite{ZW64}. It can be expressed as follows: In scattering or decay 
processes driven by the {\em strong} interactions those quark 
diagrams dominate where all valence quarks and antiquarks from the 
initial state are still present in the final state; i.e., 
initially present quarks and antiquarks do not annihilate. 

The motivation for this selection rule came from the observation 
that the $\phi$ meson interpreted as an $s\bar s$ bound state decays 
mainly into a $K\bar K$ pair rather than a kinematically favoured pion 
pair. The rule was later somewhat extended by stating that all 
{\em disconnected} quark diagrams are suppressed. 

Obviously such a rule holds only approximately. It was the 
discovery of the extremely narrow $J/\psi$ resonance that turned the Zweig
rule from respectable folklore into a dynamical notion
based on colour symmetry and QCD's asymptotic freedom. For it was
realized that an ortho[para]-quarkonium state has to 
annihilation into (at least) three [two] gluons to decay and that their 
couplings become smaller for increasing quarkonium masses: 
\beq 
\Gamma [\bar QQ]_{\rm ortho} \propto \alpha_S^3(m_Q)  <  
\Gamma [\bar QQ]_{\rm para} \propto \alpha_S^2(m_Q)
\eeq
Thus one can estimate how much `Zweig forbidden' transitions are
suppressed, and how it depends on the specifics of the decaying state.

\subsection{On quark-hadron duality} 
\label{QHDUALITY}
 
Quark-hadron duality  
-- or duality for short -- is one of the central  
concepts in contemporary particle physics.  
It is invoked to connect quantities evaluated on the  
quark-gluon level to the (observable) world of hadrons.  
It is used all the 
time as it has been since the early days of the quark model and of  
QCD,  more 
often than not without explicit reference to it. A striking example of the 
confidence the HEP community has in  the asymptotic validity of duality 
was provided by the discussion  of the width $\Gamma (Z^0 \to H_b 
H_b^{\prime}X)$.  There was about a 2\% difference in the predicted and 
measured  decay width, which lead to lively debates on its significance  
vis-a-vis the {\em experimental} error. No concern was expressed about  
the fact that the $Z^0$ width was calculated on the quark-gluon  
level, yet measured for hadrons. Likewise the strong coupling  
$\alpha_S(M_Z)$ is routinely extracted from the  
perturbatively computed hadronic $Z^0$ width with a stated  
theoretical uncertainty of $\pm$ 0.003 
which translates into a  
theoretical error in $\Gamma _{had}(Z^0)$ of about 0.1\%. 
 
There  are, however, several different versions and implementations of 
the concept of duality. The problem with invoking duality  
{\em implicitly} is 
that it is  very often unclear which version is used. In $B$ physics -- 
in  particular when determining $|V(cb)|$ and $|V(ub)|$ -- the  
measurements have become so precise that theory can no longer  
hide behind experimental errors. To estimate theoretical  
uncertainties in a meaningful way one has to give clear meaning  
to the concept of duality; only then can one analyze its  
limitations. In response to the demands of heavy flavour physics a 
considerable literature  has been created on duality over the last few 
years, which we want to summarize. We will sketch the  
underlying principles; technical details can be found in the references 
we list. 
 
Duality for processes involving time-like momenta was first addressed 
theoretically in the late  '70's in references \cite{PQW76} and 
\cite{GRECO}. We sketch  here the argument of Ref.\cite{PQW76},  
since it contains  several of the relevant elements in a nutshell. The 
cross section  for $e^+e^- \to hadrons$ can be expressed  
through an operator product expansion (OPE) of two hadronic  
currents.  One might be tempted to think that by invoking  
QCD's asymptotic freedom one can compute  
$\sigma (e^+e^- \to {\rm hadrons})$ for large c.m. energies  
$\sqrt{s} \gg \Lambda _{QCD}$ in terms of quarks (and gluons) since  
it is shaped by short distance dynamics. However production  
thresholds like for charm induce singularities that vitiate  
such a straightforward computation. This complication can be  
handled in the following way:  One 
evaluates the OPE in the (deep)  Euclidean region thus avoiding proximity 
to singularities induced by  hadronic thresholds; then one analytically 
continues it into the Minkowskian  domain through a dispersion relation. 
There is a price to be paid:   in general one 
cannot obtain the cross section as a  point-for-point function of $s$,  
only averaged -- or `smeared' -- over an energy interval,  
which can be written symbolically as follows:   
\beq    
\langle \sigma (e^+e^- \to hadrons) \rangle \simeq  
\int _{s_0}^{s_0+\Delta s} ds\sigma (e^+e^- \to hadrons)  
\eeq  
This feature is immediately obvious: for the smooth $s$ dependence  
that the OPE necessarily yields in Euclidean space  
has to be compared to the measured 
cross section $e^+ e^- \to$ hadrons as a function of $s$, which has 
pronounced structures, in particular close to thresholds for  
$c \bar{c}$-production. 
 
This simple illustration already points to the salient elements  
and features of duality and its limitations 
\cite{SHIFMANDUAL,VADE}:  
\begin{itemize} 
\item  
An OPE description for the observable under study is required in  
terms of quark {\em and} gluon degrees of freedom.  
\footnote{The name {\em parton}-hadron duality is actually more  
appropriate in the sense that gluon effects have to be included for  
duality to hold.}  
\item  
The extrapolation from the Euclidean to the Minkowskian  
domain implies some loss of information: in general one can calculate  
only hadronic observables that are averaged over energy.  
\begin{equation} 
\langle \sigma^{hadronic} \rangle _w \simeq 
          \langle \sigma^{partonic} \rangle _w 
\label{ANSATZ}  
\end{equation} 
where $\langle ... \rangle _w$ denotes the smearing which is 
an average using a smooth weight function $w(s)$; it generalizes  
the simplistic use of a fixed energy interval: 
\begin{equation} 
\langle ... \rangle _w = \int ds \, ... \, w(s) 
\end{equation} 
 
\item  
Some contributions that are quite insignificant in the Euclidean  
regime and therefore cannot be captured through the OPE can become  
relevant after the analytical continuation to the Minkowskian domain,  
as explained below.  
For that reason we have used the approximate rather than the equality  
sign in Eq.(\ref{ANSATZ}). 
 
\item  
One can make few {\em universal} statements on the numerical validity of  
duality. How much and what kind of smearing is required depends on the  
specifics of the reaction under study.  
\end{itemize} 
The last item needs expanding right away. The degree to which  
$\langle \sigma^{partonic} \rangle _w$ can be trusted as a 
theoretical description of the observable  
$\langle \sigma^{hadronic}\rangle _w $ depends on the weight  
function $w$, in particular its width. It can be  
broad compared to the structures that may appear 
in the hadronic spectral function, or it could be quite narrow,  
as an extreme case even $w(s) \sim \delta(s-s_0)$. It has become  
popular to refer to the first and second scenarios as  
{\em global} and {\em local} duality, respectively. Other authors  
use different names, and one can argue that this nomenclature  
is actually misleading. 
 
When one treats distributions rather than fully integrated widths,  
another complication arises. Consider for example inclusive semileptonic  
transitions $H_Q \to \ell \nu X_q$. The lepton spectrum is expressed  
through an expansion in powers of $1/m_Q(1-x_l)$ rather than $1/m_Q$  
where $x_l = 2E_l/m_Q$. It obviously is singular for $x_l \to 1$ and  
thus breaks down in the endpoint region. One can still make  
statements on partially integrated spectra; yet for semileptonic 
{\em charm}   
decays the situation becomes somewhat marginal since $\mu/m_c$ is not  
a small number to start with. 
 
A fundamental distinction concerning   
duality is often drawn between semileptonic and nonleptonic widths.  
Since the former necessarily involves smearing  
with a smooth weight function due to the  
integration over neutrino momenta, it is often argued that predictions  
for the former are fundamentally more trustworthy than for the latter.  
However, such a categorical distinction is  
overstated and artificial. Of much  
more relevance is the differentiation between distributions and  
fully integrated rates sketched above. 
 
No real progress beyond the more qualitative arguments of  
Refs. \cite{PQW76} and \cite{GRECO} occurred for many years. For as long  
as one has very limited control over nonperturbative effects,  
there is little meaningful that can be said about duality violations.  
Yet this has changed for heavy flavour physics with the development  
of heavy quark expansions. 
 
The possibility of duality violations clearly represents a theoretical 
uncertainty.  However it is not helpful to lump all such uncertainties into  
a single `black box'. For proper evaluation and analysis it is useful  
to distinguish between three sources of theoretical errors:  
\begin{enumerate} 
\item  
unknown terms of higher order in $\alpha_S$;  
\item  
unknown terms of higher order in $1/m_Q$; 
\item  
uncertainties in the input parameters $\alpha_S$, $m_Q$  
and the expectation values.   
\end{enumerate} 
Duality violations constitute uncertainties {\em over and above}  
these; i.e. they represent contributions not accounted for due to  
\begin{itemize} 
\item  
truncating these expansions at finite order and  
\item  
limitations in the algorithm employed. 
\end{itemize} 
These two effects are not unrelated. The first one means that the  
OPE in practice is insensitive to contributions of the type  
$e^{-m_Q /\mu}$ with $\mu$ denoting some hadronic scale; the second  
one reflects the fact that under a analytic continuation the  
term $e^{-m_Q /\mu}$ turns into an oscillating  
rather than suppressed term sin$(m_Q /\mu)$. 
 
Of course we do not have (yet) a full theory for duality and its  
violations. Yet we know that without an OPE the question of  
duality is ill-posed.  
Furthermore in the last few years we have moved beyond the  
stage, where we could merely point to folklore. This progress has  
come about because theorists have -- driven by the availability 
of data of higher and higher quality -- developed a better understanding  
of the physical origins of duality violations and of the mathematical  
portals through which they enter the formalism. 
 
Again charm studies can teach us lessons on duality that are neatly  
complementary to those from light quark studies on one hand and beauty  
physics on the other:  
\begin{itemize} 
\item  
It has been estimated that duality violating contributions to  
$\Gamma_{SL}(B)$ fall safely below 1/2 \% and thus are expected to  
remain in the `noise' of other theoretical uncertainties, i.e.  
to not exceed unknown higher order (in $\alpha_S$ as well as  
$1/m_Q$) contributions \cite{VADE}. 
 
The expansion parameter $\mu/m_c$ in  
charm decays on the other hand provides at  
best only a moderate suppression for higher order contributions,    
and at the same time limitations to duality will become more relevant  
and noticeable. This means that while we cannot have confidence in  
quantitative predictions, we can learn valuable lessons from a  
careful analysis of the data.  
\item  
A duality violating contribution $e^{-m_b /\mu}$ will remain in the  
theoretical `noise' level. Yet the charm analogue $e^{-m_c /\mu}$  
might become visible, again meaning that a careful study of charm 
dynamics  can teach us lessons on the transition from short- to 
long-distance dynamics that could not be obtained in beauty decays. 
 
\end{itemize}

\subsection{Resume on the theoretical tools}  
\label{RESUME}  
 
The fact that the charm mass exceeds ordinary hadronic scales  
$\Lambda_{NPD}$ provides a  
new expansion parameter -- $\Lambda_{NPD}/m_c < 1$ -- and thus a  
very useful handle on treating nonperturbative dynamics. Yet the excess is 
only moderate.  Therefore -- unlike the situation for beauty -- 
nonperturbative  effects  can still be sizeable or even large, and it 
constitutes a  theoretical challenge to  bring them under control.   
However we view the glass as (at least) half full rather than  
(at most) half empty. Exactly because nonperturbative  
effects are sizeable, one can learn important lessons on nonperturbative   
dynamics in a novel, yet still controlled environment by analysing   
charm interactions in a detailed way. 
 
Encouraging evidence that this is not an idle hope --   
that we are developing a better understanding of nonperturbative  
dynamics at the charm scale -- is 
provided by the realization that, as described later in detail, HQS 
provides an  approximate understanding of charm spectroscopy and HQE 
reproduce  correctly -- in part even correctly {\em pre}dicted --  
the observed pattern of charm lifetimes. NRQCD is yielding complementary  
new insights, and there is the expectation that lattice QCD will provide  
us not only with valuable guidance in charm physics, but even with  
reliable quantitative answers.

\subsection{On Future Lessons} 
\label{FUTURELESS} 
 
Our intent is not to write a historical review or present a mere status  
report. We want to emphasize the importance of future charm studies  
based on a triple motivation: 
\begin{itemize} 
\item  
As sketched in this section there is a vast array of theoretical  
technologies that are truly based on QCD, yet require some additional  
assumptions. They apply to beauty physics with considerably enhanced  
validity and thus can be {\em tested} there. Yet we view the fact that  
nonperturbative effects are larger in charm than in beauty physics as  
a virtue rather than a vice, at least for the discriminating observer:  
charm physics constitutes a rich lab to {\em probe} (rather than test)  
these methods, to provide new insights into the transition from the  
nonperturbative to the perturbative domain. We have to be prepared  
that these methods will occasionally fail; yet we shall be able to obtain  
valuable lessons even from such failures.  
\item  
A more detailed knowledge and understanding of charm physics than 
presently available is also essential for a better understanding of  
beauty physics and for a fuller exploitation of the discovery  
potential for New Physics there. This starts with the trivial  
observation that knowing charm branching ratios and decay sequences  
are important for interpreting beauty decays. Secondly, as indicated 
above, the theoretical technologies employed in beauty decays can be  
cross-referenced in charm decays. Lastly, a detailed understanding of  
charm spectroscopy is important in properly interpreting certain  
$B\to \ell \nu X_c$ transitions and the information they can yield  
concerning the underlying QCD treatment. This last more subtle point  
will be explained later.  
\item  
High sensitivity studies of $D^0 - \bar D^0$ oscillations, CP violation 
and rare decays provide a novel window onto conceivable New Physics --  
actually of {\em non-}standard extensions of the SM as indicated in the  
previous subsection. 
\end{itemize}  

\section{Production dynamics}  
\label{PROD}  
Understanding the production processes for hadrons containing charm
quarks is of obvious practical importance if one wants to obtain a 
well-defined sample of those hadrons for studying their decays. 
Yet new conceptual insights into QCD can be gained as well. 
In the following we address these two issues for different 
production reactions. In doing so one has to treat separately the cases 
of hidden and open charm hadrons, whose scenarios are quite different 
theoretically as well as experimentally. 
Rather than give an exhaustive
discussion we aim  at describing a few telling examples. Recent reviews can be
found in
\cite{Gottschalk:2002pf,Gladilin02,Migliozzi02,Bertolin:2003is,Kraemer03}, and
predictions in \cite{Frixione:1997ma}. 
 \par 
\begin{figure}[t] 
  \epsfig{file=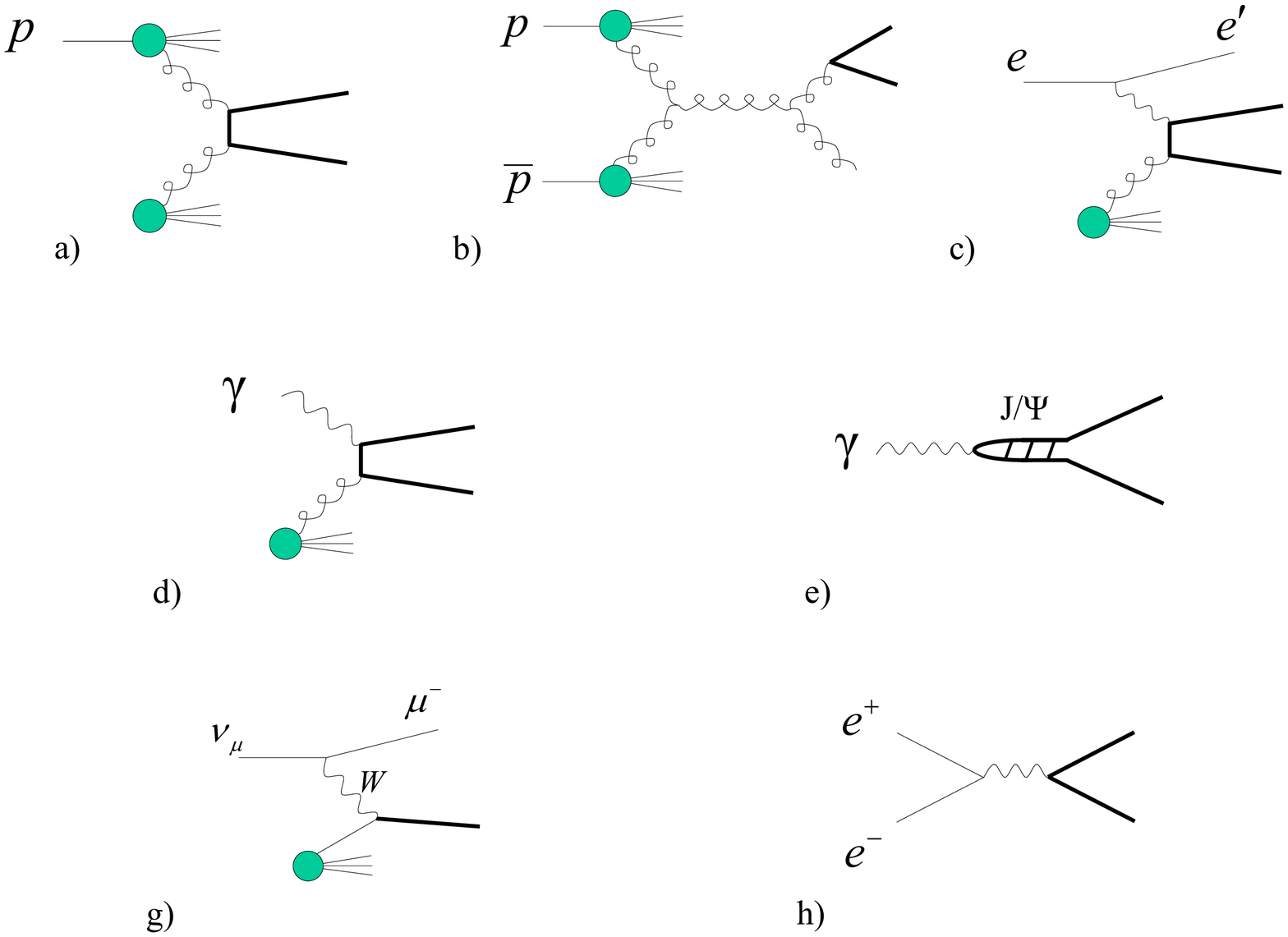,width=12cm,height=6.5cm}  
 \caption{ 
 Diagrams for charm production: Hadroproduction (a,b); electron-proton
 production (c);
 photoproduction point like photon (d), resolved photon (e); neutrino (f);
 $\epem$ (g).
  \label{FIG:PRODMECHS} 
 } 
\end{figure} 

 As described before in our
historical sketch of Sect.\ref{HIST},
 the use of a variety of intense particle beams
on a wide variety of nuclear fixed targets dates back to the beginning of
the charm adventure, and it constitutes a mature technique for
investigating charm production. On the other hand, heavy flavour physics
at hadron colliders,  after pioneering work at the ISR, has undergone a
renaissance at CDF. There are multiple motivations for studying
hadroproduction of  open and hidden  
charm states:  
\begin{itemize} 
 \item  
 The production of heavy flavour hadrons presents new tests of our  
 quantitative  
 understanding of QCD. Their worth is enhanced by the fact that there  
 are similar ingredients in the theoretical treatment of charm and beauty  
 production.  
 \item  
 It serves as a sensitive and efficient probe for determining gluon  
 distributions inside nucleons. 
\item 
Understanding the production mechanisms helps us in fully 
harnessing the statistical muscle of hadroproduction for 
studies of weak {\em decays} of charm hadrons. 
 \item  
 Analyzing charm production inside heavy nuclei provides us with  
 insights into how QCD's dynamics act under exotic or even  
 extreme conditions. Furthermore it can signal the onset of the  
 quark-gluon plasma as discussed later. 
\end{itemize} 
We have chosen to organize the vast material in the following way: 
first we will describe hidden charm production in the different settings, 
then we will turn to open charm produced through $\epem$ annihilation, at fixed
target experiments, hadronic colliders and deep inelastic lepton-nucleon 
scattering and conclude with charm production inside heavy nuclei. 

\subsection{Charmonium production}
\label{ONIUMPROD}
\begin{figure}
  \begin{center}
   \epsfig{file=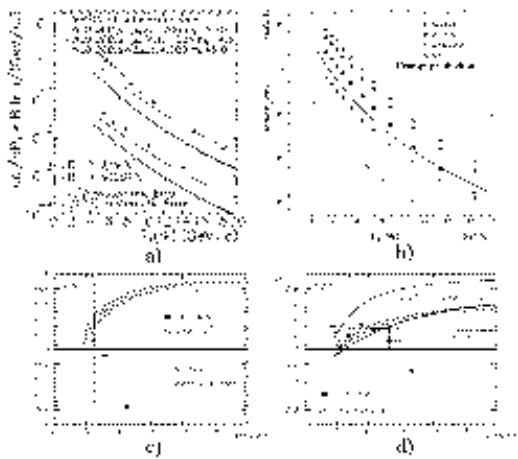,width=10cm}  
    \caption{ 
       CDF results on cross section for $\jp$ (a), $\psi^\prime$ (b), and
       polarization (c,d). Data from \cite{Abe:1997jz,Abe:1997yz}, theoretical
       predictions from \cite{BENEKEPOL,BRAATENPOL}.
       \label{FIG:CDFJPSI} 
       } 
   \end{center}
\end{figure}

\begin{figure}[t] 
  \begin{center}
   \epsfig{file=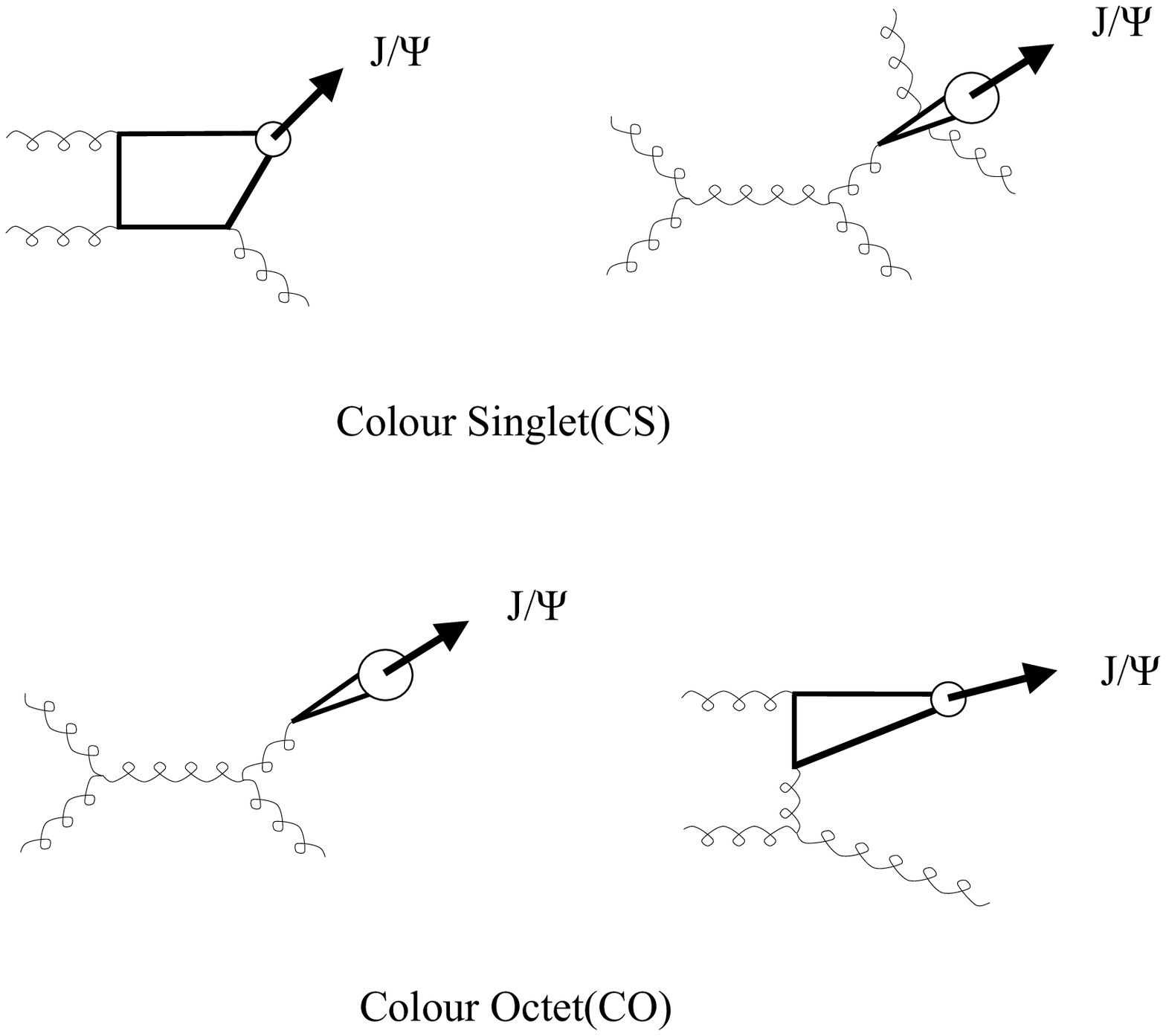,width=5cm,height=5cm}  
   \epsfig{file=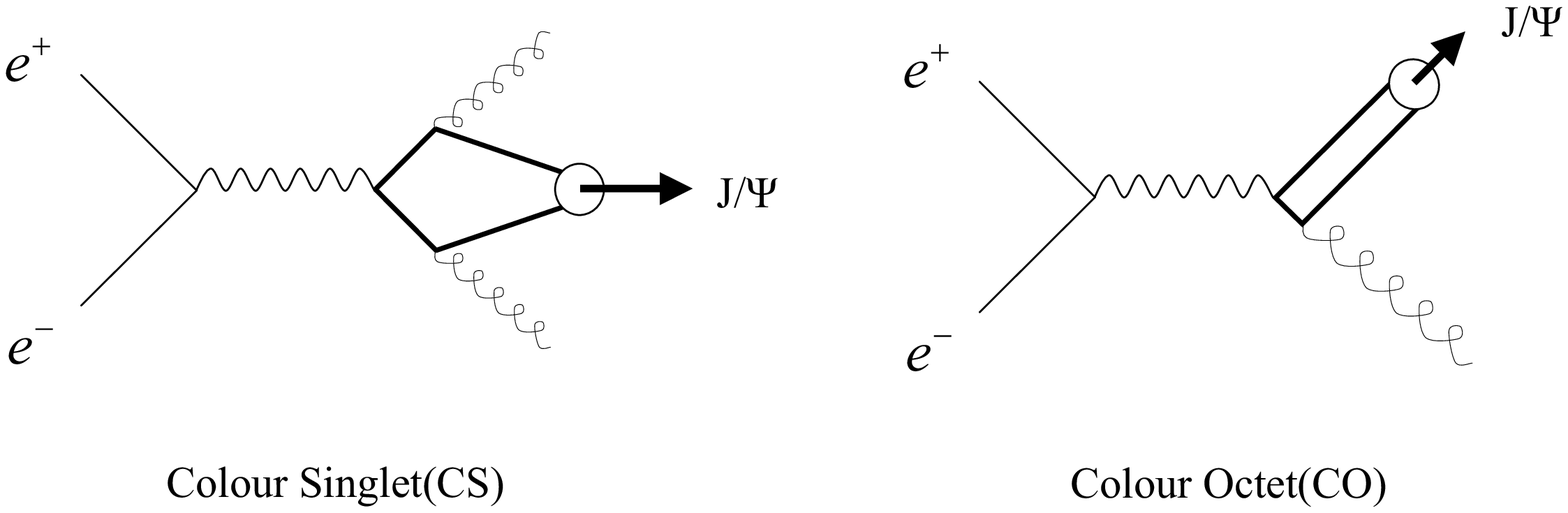,width=6cm,height=4cm}  
   \end{center}
 \caption{ 
 Colour-singlet and colour-octet diagrams for hadroproduction (left) and
 $\epem$ production (right) of charmonium; photons and gluons are 
 denoted by wave and curly lines, respectively.   \label{FIG:CSCO} 
 } 
\end{figure}

A priori there are three 
experimentally distinct scenarios for the production of 
prompt $J/\psi$: the {\em secondary} 
production via a para-charmonium state $\chi_c$ cascading
down $e^+e^- \to \chi_c +X \to J/\psi + \gamma +X $ or {\em primary} production of
$J/\psi$ together with the excitation of two charm hadrons -- like 
$e^+ e^- \to J/\psi + D \bar D^{\prime} +X$ --, which is a Zweig allowed 
process,
or without such additional charm states, which is not. In 1995 the 
CDF collaboration \cite{Abe:1997jz,Abe:1997yz} discovered that 
$B$ meson decays are not the major source of $J/\psi$ production 
in hadronic collisions \index{$\jp$ production at CDF}: 
many $\jp$ are prompt rather than the decay products of an object 
with a lifetime of around 1~ps. The production of these 
`direct' charmonia was found to be enhanced by a factor of about 
fifty 
(Fig.~\ref{FIG:CDFJPSI}
with respect to predictions of the theoretical model of that time, 
the colour-singlet model,
\index{colour singlet model} Fig.\ref{FIG:CSCO}. 
 In this model it is assumed that charmonium states can get excited 
only via their $\ccb$ component making the production of para-charmonium 
-- $\chi_c$ -- to 
dominate over that for ortho-charmonium -- $\Jp$. There is no reason beyond 
simplicity, why the $\Jp$ cannot be produced via a $\ccb$ octet component. 
The most radical of such colour octet models
\index{colour octet model} is often called the 
`colour evaporation model', where the octet sheds its colour with unit 
probability via soft gluons. 
\par 
These models can be embedded in NRQCD, see  
Sect.~\ref{NRQCD}, which  was
developed partly in  
response to the challenge posed by $\Jp$ production at the TEVATRON. By including
charmonium production off  colour {\em octet} 
$\bar cc$ configurations, where colour is shed via soft gluons, 
NRQCD is able to reproduce these data; 
the colour octet component Fig.\ref{FIG:CSCO} thus represents by far the
dominant source  
of prompt charmonia at TEVATRON energies -- in clear contrast to the 
situation at lower energies. 
 \par 
\par
Our understanding can be further tested by measuring the polarization 
$\alpha$ of the $\jp$ and $\pspr$ defined as 
 $(d\Gamma / d\cos\theta) \propto 1 +\alpha \cos^2\theta $
 in the angular distribution of decay leptons pairs from charmonium; 
 $\alpha=1[-1]$ corresponds to pure transverse [longitudinal] polarization. 
 Both charmonia states are predicted \cite{BENEKEPOL,BRAATENPOL}  to 
 be increasingly transversely polarized with growing $p_{\perp}$ since one expects 
 the transverse polarization of almost `on-shell' gluons to be transferred  to the 
 $\bar cc$ bound state produced from them. However this effect is certainly not 
 apparent in the data, see Fig.\ref{FIG:CDFJPSI} c), d)
 \cite{Cester02,Skwarnicki03}.  More data has become available from CDF 
 out of RUN~II data taking period presently in progress \cite{CDFRUNII}.
Again, this might not be a
fatal flaw in NRQCD; it might just mean that contributions of higher
order in $\alpha_S$ and in $v$ are still sizeable for charmonia and
affect polarization more than cross sections, which are more robust
against higher order contributions. 
\par
Real photoproduction experiments provide also results on hidden charm states,
whose diffractive production proceeds via VMD coupling of the beam photon to
$J^{PC}=1^{--}$ mesons such as $\jp$. Such studies are generally limited to
dimuon final states, since dielectron decay modes are hindered by the
presence of electron-positron pairs copiously produced by Bethe-Heitler
mechanisms.
 Perhaps more importantly, the $\epem$ decay mode is made difficult
by the presence of a very long tail in the dielectron 
invariant mass spectrum, due to bremmsstrahlung, which needs
 to be corrected for.
Very recently, the first observation of $\psi(3770)$ was preliminarily reported by
FOCUS \cite{Gottschalk:2002mr}.
\par 
Experiments E760 and E835 at the
Fermilab antiproton source have performed precision measurements of all
charmonium states, also  measuring the $\chi_{c1}$ for
the first time. Charmonium states are produced by the collisions of antiprotons
on a hydrogen jet target, thus providing interaction whose geometry is
effectively the one typical of fixed target experiments.  
Experiment E835 charmonium results are discussed in Sect.\ref{LIFE}.

\subsection{Charm at LEP (mainly)}

With no hadron being present in the initial state, 
$\epem$ annihilation  
represents the simplest scenario. 
At the same time charm  was the first
quantum number high enough in mass that one can invoke  
perturbative QCD to describe
$e^+e^- \to hadrons$ {\em below} as well as 
{\em above} its threshold; i.e.,
the ratio $R$ had reached a constant value below  threshold and a higher
one above it. 

One of the most intriguing aspects theoretically has been discussed in 
Sect.\ref{QHDUALITY}, namely how quark-hadron duality and the approach 
to it is realized in nature. The conclusion is that starting at about 
1 GeV above threshold perturbative QCD can be employed for predicting 
the total cross section for $e^+e^- \to {\rm hadrons}$. 
The experimental situation just above charm
threshold had been somewhat unsettled with some
data sets showing an unusually large cross section. Measurements
\cite{Bai:2001ct} done
by the BES collaboration  in Beijing have clarified the situation; 
the value of $R$ does not seem to be excessively large. 
Future studies from CLEO-c  should settle it completely. 
The transition region around 4 GeV will presumably  remain beyond 
theoretical control, since so many thresholds for exclusive final states 
open up: $e^+e^- \to D \bar D$, $D^*\bar D + D \bar D^*$, $D^*\bar D^*$, 
$D^+_s  D^-_s$, ..., $\Lambda_c \bar \Lambda_c$ etc.. One can
attempt to describe  this highly complex landscape  
through models involving a coupled-channel approach 
\index{coupled channel approach} \cite{CCHPAP,GOTTFRIED}; yet 
the predictions  based on such models are not reliable, since they are
quite unstable under variations of the model parameters. 
Nevertheless important measurements can be performed there: in particular 
the {\em absolute} branching ratios for the different charm hadrons 
can be measured in a model independent way as explained in Sect.\ref{ABSBR} 
of this review. 

BELLE has shown highly surprising data on double $c\bar c$ production: 
it finds the $J/\psi$ to be accompanied more often than not by an additional 
$\bar cc$ pair \cite{Abe:2002rb}:
\beq 
\frac{\sigma (e^+e^- \to J/\psi \bar cc)}
{\sigma (e^+e^- \to J/\psi X)} = 0.59 ^{+0.15}_{-0.13} \pm 0.12
\label{BELLEJCC} 
\eeq 
There is no good idea from theory (yet) how such a large ratio could be
accommodated. 

Measurements by HRS, MARK II and
TASSO at the PETRA and PEP storage rings at DESY and 
SLAC, respectively,  had provided  the first reliable information on the fragmentation
functions  
\index{fragmentation functions}
of charm 
and beauty quarks described in Sect. \ref{DISPROD}. Yet they have been
superseded by  
ARGUS and CLEO measurements with their
much higher statistics and by LEP and SLD data, 
the latter for beauty as well as charm quarks; see 
\cite{Hagiwara:fs,Biebel:es} for concise reviews. 
CLEO finds a deviation from  a Peterson {\em et. al.}-type fragmentation
function  
for $D_s^+, D_s^{*+}$ \cite{Briere:2000np}. 
It is actually 
more difficult to measure the charm than the beauty fragmentation fragmentation at LEP, 
since in the case of charm one has to rely on exclusively reconstructed 
charm hadrons to obtain sufficient purity of the sample, which reduces
considerably  
the statistics \cite{ROUDEAUPRIV}. Furthermore an interpretation of the data is 
less straightforward, among other reasons due to the secondary production of charm 
via gluon splitting \index{gluon splitting} $g \to c \bar c$.  Nevertheless there exists a 
strong twofold motivation for determining the charm fragmentation function as accurately as 
possible: (i) it is an important ingredient in predicting charm production cross sections and 
distributions to be measured at the TEVATRON; (ii) comparing it to the beauty fragmentation 
function will shed further light on the nonperturbative dynamics driving it.  

The level of production 
can be computed (with $m_c=1.5\pm 0.3$ GeV), and when compared to recent data
\cite{Biebel:es} it is found a 
couple of standard deviations below.  
\par 
LEP experiments provide information on charm production through studies of
$\gamma\gamma \to \ccb$, where the initial state is realized by
initial state radiation off both beams 
\index{two photon physics}. Recent results are discussed
in Ref.\cite{Gladilin02}. $D^{*\pm}$ production has been measured by LEP
experiments at $\sqrt{s}=183-209$ GeV and found in agreement with NLO QCD
predictions. L3 also measured $\sigma(\gamma\gamma\to \ccb X)$ as a
function of  the $\gamma\gamma$ invariant mass finding reasonable agreement
with NLO QCD for $m_c=1.2$ GeV; for $m_c=1.5$ GeV one predicts a 50\% lower 
cross section. 
\par
   Another  observable   that  can be computed in NLO QCD is
   the aforementioned gluon splitting probability $g_{\ccb}$ of $\ccb$    for 
 $(\epem \rarr q\bar{q}g, g\rarr Q\bar{Q})$. 
 An OPAL result yields 
   $g_{\ccb}=(3.20\pm0.21\pm0.38)\times 10^{-2}$ \cite{Abbiendi:1999sx}, 
 which is higher than theoretical estimates as well as the L3 measurement 
 \cite{Acciarri:1999pw} $g_{\ccb}=(2.45\pm0.29\pm0.53)\times 10^{-2}$. 
 Likewise ALEPH and DELPHI find higher than predicted values for the 
 corresponding quantity $g_{b\bar{b}}$. 
\par
 An area of great importance to the validation of the electroweak sector of the 
 Standard Model is the determination of the forward-backward asymmetries for 
 charm and beauty jets 
$A_{FB}^{c\bar c},A_{FB}^{b\bar b}$, and the ratios of charm and beauty quark
partial widths $R_b,R_c$ 
$R_c = 
\Gamma(Z^0\to c\bar c)/\Gamma(Z^0\to  hadrons)$, 
see \cite{Roudeau:1997zn,Roudeau:2001ay}
 for an extensive review.
The charm FB asymmetries had been measured before the LEP era 
\cite{Nakano:bg},
\cite{Okamoto:1991ry},
\cite{Gerhards:yb},  but only the huge data sets gathered at LEP
allowed meaningful searches for manifestations of New Physics. 
The situation has changed considerably over the years, as it can be realized
browsing the LEP Electroweak Working group pages \cite{LEPEWWG}.
 Measurements of $R_b,R_c$ in 1995 
\cite{Renton:1995jj}
differed from SM predictions by $+3.7\sigma,-2.56\sigma$, while
they appeared totally consistent in 2002 
\cite{Grunewald:2002wg}
$+1.01\sigma,-0.15\sigma$. On the other hand,  
$A_{FB}^{c\bar c},A_{FB}^{b\bar b}$ that in 1995
were completely consistent with the SM within their relatively large errors, in
2002 represent a pull of  $-0.84\sigma,-2.62\sigma$ respectively, with
$A_{FB}^{b\bar b}$ being the second largest contribution of pulls in the fit to
the SM parameters, after the intriguing NuTeV result on $\sin^2\theta_W$ 
\cite{NUTEV}. 
\subsection{Photoproduction}
\label{FOTOPROD}  
The real photon has two components, namely a hadronic one {\em a la}
 vector meson 
dominance \index{vector meson dominance}, 
and one coupling directly to quarks via their electric charge. Due to 
the small weight of the former, photon beams provide a cleaner environment than 
hadron beams since mostly there is no hadronic jet from beam 
fragmentation. One still has to contend with a large background of 
light hadrons. Yet the charm-to-total cross-section ratio of about 1/100 
 is 
considerably higher than the 1/1000 for hadroproduction. 
Also the theoretical treatment of 
photoproduction is easier than of hadroproduction since only one hadron 
participates in the collision.  

The recent success in 
gathering sizeable samples of events with both charm and anti-charm hadrons 
allows novel probes of perturbative QCD and has already lead to new 
insights on QCD dynamics. 

Scattering electrons off protons at small values of $Q^2$ provides a 
fluid transitions to {\it real} photoproduction experiments with 
$Q^2=0$. 
In real photoproduction 
 high-energy, high-intensity beams impinge on nuclear 
targets and can produce charm states. The tree-level production mechanism proceeds 
via the fusion\cite{Jones:1977wx} of beam photon off a gluon emitted by the
nucleus (Fig.~\ref{FIG:PRODMECHS}). 
\par
The charm cross-section is sensitive to the charm quark 
mass, and it has been thoroughly measured from threshold up to HERA energies, 
and compared to QCD predictions \cite{Frixione:1997ma}. A value of 1.5 GeV for the 
{\em pole} charm quark mass is favoured by the data 
with large errors due to the choice of other 
theory parameters. 
\par
Real photoproduction is studied via the very large data samples collected by
fixed target experiments \cite{Gottschalk:2002pf}. The reconstruction of
both $D$ and $\bar D$ in the same event allows one to study $D\bar D$
correlations that can be predicted in principle by QCD. Important
variables are $\Delta \phi$, the angle between the particle and the antiparticle
in the plane transverse to the beam, and the transverse $P_T^2$ momentum squared
of the pair. 
At leading order, with $\ccb$ quarks produced back to back, one expects $\Delta
\phi=\pi$ and $P_T=0$. Recent results from FOCUS 
\cite{Link:2003uj}
 show that data
disagree with predictions even when taking into account NLO
contributions.  There is also a small but highly
significant excess of data at $\Delta \phi=0$ suggesting that a small  
fraction of $D\bar D$ pairs are produced collinearly
rather than back-to-back. 
These studies are a valuable tool to tune charm production computational
algorithms, such as PYTHIA \cite{Sjostrand:2000wi}.

\par
Particle-antiparticle asymmetry studies have been carried out in the past by
photoproduction experiments NA14/2\cite{Alvarez:1992yb}, E691
\cite{Anjos:1989bz} and E687\cite{Frabetti:1996vi}.  E691 and E687 measure a 
significant asymmetry for $D^+, D^0$ and $D^{*+}$, and one compatible with
zero within large experimental errors for $D_s^+, \Lambda_c^+$. The
asymmetry measured is ten times the one predicted by perturbative QCD.
Mechanisms based on heavy-quark recombination have been proposed
\cite{Braaten:2001uu}. High
statistics results from FOCUS are expected soon.
\par
Finally, FOCUS \cite{Riccardi02} showed recently a null result on the production
of double charm baryons reported by hyperon beam experiment SELEX 
\cite{MATTSON02} \cite{RUSS02}.
The
issue is addressed in Sects.\ref{LIFE} and \ref{C2BAR}. 
\subsection{Fixed target hadroproduction}
\label{FTHADPROD}

Experiments have been performed with a host of extracted meson as well as  
baryon beams and also internal beams on gas targets. 
\par
The leading particle effect \index{leading particle effect} is the most 
interesting phenomenological feature in charm production studies with extracted 
beams. This is the enhancement of the production of particles compared to the 
production of antiparticles, and it is due to the presence of quarks present 
both in the produced particle, and in the target nucleon or in the beam 
particle. The enhancement is represented (usually in a differential fashion in 
$x_F$ and $P_T^2$) via the asymmetry variable 
\beq
 A\equiv \frac{N_{particle}-N_{antiparticle}/R}{N_{particle}+N_{antiparticle}/R}
\eeq
where $R\equiv \bar \epsilon / \epsilon$ is the ratio of acceptances for 
particles and antiparticles.
\par
As a matter of principle perturbative QCD can yield only a very small asymmetry in charm 
production; in contrast asymmetries as large as 50\% have often been reported in the data.
  In the following we shall limit ourselves to outline the most striking features 
of charm production at extracted beams, referring the interested reader to 
recent reviews \cite{Ratti:1996gi},
\cite{Appel:2000iy},
\cite{Gottschalk:2002pf}. 
\par
Tree-level hadroproduction proceeds via diagrams shown in
Fig.~\ref{FIG:PRODMECHS}(a,b). At fixed target, the availability of several kinds
of 
beam allows one to study the leading particle effect. Recent results come from
E791 ($\pi^-$ beam) and SELEX ($\pi^-$, proton and $\Sigma^-$ beams) .
Asymmetry data for $\Lambda_c^+$ produced by $\pi^-$ beams in E791 and SELEX do
agree, while the asymmetry for proton and $\Sigma^-$ beams in SELEX is much more
pronounced, a clear manifestation of leading particle effect, since baryon beams
will produce preferentially $\Lambda_c^+$ baryons, not antibaryons.
\par
SELEX recently reported observation of four different $C=2$ baryons
\cite{MATTSON02} \cite{RUSS02}.  If confirmed, it would have profound implications 
for our understanding of charm production. We will discuss this issue in
Sect.\ref{C2BAR}. 
\subsection{Hadroproduction at colliders} 
\label{COLLPROD}

The study of charm physics at hadronic colliders was pioneered at the CERN ISR
(see Sect.\ref{PASTLESSONS}). 
Experiments done there showed 
evidence for much larger charm cross sections than expected, in
particular in the forward region of up to 1.4 mb.
It was finally understood that such high values were due to efficiency and
acceptance corrections used to get cross sections out of low-acceptance
mass-peak observations \cite{D'Almagne:1987gd},\cite{Garvey:rp}.

The most lasting legacy is maybe the concept of 
{\em intrinsic charm}\index{intrinsic charm} suggested 
a long tome ago  by Brodsky and collaborators \cite{INTRINSIC} 
to account for the larger charm production in the forward 
or projectile fragmentation region. It says that protons (and other 
hadrons) have a $\bar cc$ component that unlike in the conventional 
picture is {\em not} concentrated in the `sea' at very small values of 
fractional momenta $x$: 
$|p\rangle \propto |uud\rangle + |uud\bar cc\rangle + `sea' $. 
This has been referred to as a higher Fock state in the proton wave
function. There has been and still is an ongoing debate over the validity 
of this intriguing picture and the danger of double-counting. It seems now 
that in the framework of the heavy quark 
expansions one can assign the concept of intrinsic charm an 
unambiguous meaning \cite{IMPRECATED}. 
\par
Little new work theoretically as well
as experimentally has been done on charm production at hadron colliders 
until recently. This lull seems to be coming to an end now. 
Most of our community was quite surprised when in 1991  CDF \cite{Sansoni:1991vj} 
demonstrated the ability of studying 
beauty quark  physics in a $p_T$ regime totally unsuited for the mainstream W
physics the detector was conceived for. CDF's capabilities have been 
further boosted by the implementation of a
detached vertex trigger described in Sect.\ref{PASTLESSONS}, which provides
online  
selection of events based on
reconstructed decay vertices by the microstrip detector. The vertex trigger 
 might allow reconstruction of charm correlations.  

So far we have results on  $D^*$ production \cite{Korn:2003pt,Gladilin02}. In 2000
CDF published the 
only available measurements of open charm production cross sections in $p\bar
p$ collisions at $\sqrt{s}=1.8 TeV$ for the process $D^{*+}\to D^0 \pi^+ \to
(K^-\mu^+X) \pi^+$ for in the rapidity and transverse momentum intervals
$|\eta(D^{*+})|<1.0, p_T(D^{*+}>10 GeV)$. The integrated cross section found
$\sigma=347\pm 65\pm 58 nb$ exceeds calculations based on both NLO and FONLL.
Such a result was confirmed in 2003 by the cross section measurement \cite{Acosta:2003ax} for exclusive processes $D^0\to K^-\pi^+$, 
$D^{*+}\to D^0\pi^+$, $D_s^+\to \phi\pi^+$, obtained on a dataset  selected by the new detached vertex trigger. Cross section found exceeds by 100\% central value of theory predictions\cite{Cacciari:2003zu}, although theory and experiment are compatible when considering theoretical uncertainties.
\subsection{Deep inelastic lepton-nucleon scattering} 
\label{DISPROD} 
%
As already mentioned in Sect.\ref{HIST}, the first experimental signal for charm
production  
outside cosmic rays came from dimuon events in deep inelastic neutrino 
nucleon scattering \cite{Benvenuti:ru}
\beq 
\nu_\mu N \to \mu ^- H_c +X \to \mu^- \mu^+ + \tilde X
\eeq
Also some early charm spectroscopy had been done in neutrino induced 
events. Today's main lessons from charm production by neutrinos are 
the following: 
\begin{itemize}
\item 
It provides important information on $|V(cs)|$ \& $|V(cd)|$, as 
described in Sect.\ref{VCS}. 
\item 
It allows extraction of the structure functions for $d$ and $s$ quarks 
and antiquarks from open charm and for the gluons from $J/\psi$ 
production. 
\item 
In $\nu X \to l^- \Lambda_c^+ X$ one can measure the form factor of
$\Lambda_c$ baryons in the  space-like region \cite{Bigi:2001xb}. 
\item 
Since $\Lambda_c$ is expected to be produced with a high degree of 
longitudinal polarization, one could search for a T odd correlation 
$C_{\rm T\; odd} \equiv \langle \vec \sigma _{\Lambda_c}\cdot 
(\vec p_{\Lambda} \times \vec p_l) \rangle $
in semileptonic $\Lambda_c$ decays
$\nu N \to \Lambda_c^+ X \to (l^+ \nu \Lambda)_{\Lambda_c} + X$
\item 
To determine the fundamental electroweak parameters at lower energies 
in different kinematical domains one has to understand -- or at least 
model reliably -- charm production, since it varies with energy and 
is different in charged vs. neutral current reactions. A very simple
ansatz is often used  here, namely the `slow rescaling' model, where one
replaces the usual scaling variable $x$ by 
\beq 
x \to \xi = x\left( 1 + \frac{"m_c"^2}{Q^2}  
\right)\left( 1 - \frac{x^2M_N}{Q^2}  
\right) 
\label{SLOWRESC}
\eeq
It should be noted that $"m_c"$ here is merely a quark model 
parameter. Measuring its value with high accuracy from $\nu$ data does 
{\em not} mean we know the charm quark mass till we can derive 
Eq.(\ref{SLOWRESC}) from QCD. 
\end{itemize}
Charm production occurs in high energy neutrino interactions at the few percent 
level and to lowest order is described by the  diagram  in
Fig.\ref{FIG:PRODMECHS} f), with strong dependence to the strange quark sea,  
since charm production off $d$ quarks is
Cabibbo-suppressed.
 This sensitivity is
further enhanced in the case of antineutrino scattering, where only sea $\bar d$
and $\bar s$ quarks contribute with the latter dominating. 
\par
A wealth of  results keeps coming from charm neutrino experiments 
using emulsion and electronic techniques, namely NOMAD and CHORUS at CERN 
and NuTeV at FNAL \cite{Migliozzi02}:
\begin{itemize}
 \item 
 Emulsion experiments have been able to measure the inclusive charm
 production cross-section $\sigma(\nu_\mu N \to c \mu^- X)/\sigma(\nu_\mu N \to 
  \mu^- X)$ which is of order five percent, while electronic experiments
 measured the inclusive  
 $D$ production rate $\sigma(\nu_\mu N \to D^0 \mu^- X)/\sigma(\nu_\mu N \to 
 \mu^- X)$ about two percent. A recent analysis \cite{DeLellis:2002mf} combines
 both electronic and emulsion experiments results.
 \item 
 Insights are gained on the hadronization of charm
 quark described through fragmentation functions, see Sect.\ref{CPRODTH}.  
 It should be noted that 
 the fragmentation process is expected to be universal, i.e.,
 independent of the {\em hard} scattering process under study; i.e.,  
 charm quarks emerging from, say, 
 $\epem$ collisions dress into charmed hadrons in the same fashion as those
 produced in lepton-nucleon scattering.
  Cross-section data for neutrino
 production are parameterized via the usual Peterson form 
 $D(z)\propto [z(1-z^{-1}-\epsilon_P/(1-z))]^{-1}$, and the customary
 kinematical variables $p_T^2$, $f_h$ (the mean multiplicity of charmed hadron
 $h$) and $z$ (the fraction of the quark longitudinal momentum carried by the
 charmed hadron). The 
 \index{fragmentation functions}
 fragmentation function $D(z)$ is peaked at $z=0.8$ which
 means that the hadronization process is hard, and relatively energetic.
 Neutrino experiments measure the $\epsilon_P$ parameter and compare it to
 $\epem$ data, generally finding good agreement for $D^*$ production.
\end{itemize}
 Charm production in neutrino physics is thus an alive field, where  great interest
 exists for 
 the  huge improvements which are expected at a Neutrino Factory 
 \cite{Bigi:2001xb}, \cite{Mangano:2001mj}.
\par
A new realm of analyzing heavy flavour production -- of charm and beauty, 
open and hidden -- has opened up in high energy $\sqrt{s}=300-318\,{\rm
GeV}$ electron-proton  
collisions  
studied at HERA by the H1, Zeus and HERA-C collaborations. Production of 
charm hadrons and charmonia can occur off gluons, $\bar cc$ pairs in the 
sea at small values of $x$ and off an intrinsic charm 
\index{intrinsic charm} component at medium and large values of $x$ 
\cite{INTRINSIC}. The stage is thus more complex
than in $e^+e^-$ annihilation -- yet that should be viewed as a virtue, 
since data allow us access to these parton distribution functions. 
\par
Not only the proton {\em target} adds complexity to the phenomenology, 
but also the electron {\it projectile}, which effectively acts either as in 
deep inelastic lepton-nucleon scattering, 
\index{deep inelastic scattering}
 (Fig.\ref{FIG:PRODMECHS} c), or in 
photoproduction (Fig.\ref{FIG:PRODMECHS} d), 
depending on the $Q^2$ region considered.
In the photoproduction regime $(Q^2\sim 0)$, then, the photon can produce charm 
via a direct point-like coupling to partons in the proton\index{direct photons}
(Fig.\ref{FIG:PRODMECHS} d), or it can effectively act as a hadron a la vector 
meson dominance\index{resolved photons}, Fig.\ref{FIG:PRODMECHS} e).
In the latter case, any intrinsic charm components in the photon or
proton particles may give origin to charm excitation processes, such as 
$cg\to cg$. The variable $x_\gamma^{obs}$ is normally used to discriminate direct from
resolved photon processes. 
\par
The experimental panorama is discussed in recent reviews 
\cite{Gladilin02,Meyer02,Bertolin:2003is}.
 Photoproduction cross-sections for
$D^*$ and $D_s$ 
generally exceed the next to leading order (NLO) QCD predictions, as well as the
fixed order plus next to leading logarithm (FONLL) calculations
\cite{Cacciari:2001td}.
The photoproduction cross section is also measured as a function of
$x_\gamma^{obs}$. This allows to show that a relevant contribution from charm
excitation processes needs to be taken into account by theory.
In DIS electroproduction regime $(Q^2> 1$~GeV$^2)$ $D^*$ cross sections are
compared to predictions and found in fair agreement, although somehow
undershooting data. 
The NRQCD prediction for $J/\psi$ production yields a rising 
cross section for $z \equiv E_{J/\psi}/E_{\gamma} \to 1$, i.e. the 
kinematic boundary -- in conflict with observation. 
\par
Another observable predicted by NRQCD predictions is the ratio of
diffractive photoproduction rates  of $J/\psi$ vs. $\psi(2S)$. New data from H1
are found to be consistent with NRQCD predictions \cite{Ma:2002gc}.
\par
Resolved photon processes are expected to dominate the low-$z$ inelastic region,
while direct photon processes should dominate the region up to about 
$z\sim 0.9$, 
with diffractive photoproduction taking over at $z\sim 1$. Recent H1 and ZEUS
results are reviewed and compared \cite{Bertolin:2003is} to colour singlet (CS) 
and colour singlet + colour octet (CS+CO) 
predictions. As explained in Sect.\ref{NRQCD}, the CO component 
enters naturally in NRQCD model, and is fitted to the large $\jp$ cross section
measured by CDF in 1995. HERA data are consistent with CS+CO contributions,
although data do not rise as a function of $z$ as rapidly as CS+CO predictions
do. On the other hand, in electroproduction regime $Q^2>2 GeV^2$, the inelastic
$\jp$ 
cross section measured   by H1 clearly favours CS predictions.
\par
Yet it would be 
premature to condemn NRQCD for this apparent discrepancy; for in its 
present level of sophistication it is not applicable in this 
kinematical domain. Future refinements of NRQCD should enable us to extend
its applicability there. 

\subsection{Hadroproduction inside heavy nuclei} 
\label{HVYNUCL} 
The fabric of QCD is such that it can create an extremely rich  
dynamical landscape. To explore it fully one has to go beyond  
observing reactions involving single hadrons. When heavy nuclei collide  
with hadrons or other heavy nuclei the interactions between individual  
hadrons take place against the background of nuclear matter; this can 
lead to highly intriguing phenomena, of which we sketch two examples,  
namely the {\em lowering of the $D$ meson mass} and  
{\em colour screening 
induced  by the quark-gluon plasma}
\index{colour screening}. 
 
Most of the mass of pions and kaons, which are Goldstone bosons,   
is due to  
how approximate chiral symmetry \index{chiral symmetry} 
 is realized in QCD. Spontaneous  chiral 
symmetry breaking leads to the emergence of non-vanishing  quark and 
gluon condensates. Chiral invariance is partially restored  in the medium 
of nuclear matter. It is expected that the masses pions  and kaons 
exhibit inside nuclei get changed relative to their  vacuum values, and 
that there is even a split between the masses  of charge conjugate pairs 
with the nuclear medium providing an  effective CPT breaking; it has been 
predicted that the masses of  
$\pi^+$ and $\pi^-$ [$K^+$ and $K^-$] get shifted by about  
25 [100] MeV. Experimental evidence for such effects has been 
inferred from the observation of pionic atoms and the study  
of the onset of $K^+$ and $K^-$ production in heavy-ion collisions. 
 
The situation is qualitatively different -- and richer -- in the charm 
sector  since there the mass is due mostly to the $c$ quark mass,  
and different scales enter the dynamics for the interactions with  
the nuclear medium. For the $J/\psi$ and $\eta_c$ only a small mass 
reduction of around 5 - 10 MeV is predicted, since charmonium 
masses are affected by mostly gluon condensates. $D$ mesons on  
the other hand offer the unique opportunity to study the restoration  
of chiral invariance in a system with a single light valence  
quark \cite{GSIPROP}. A lowering of both $D^{\pm}$ masses is  
predicted with a relative shift of $\sim 50$ MeV in  
$M(D^+)$ vs. $M(D^-)$. One of the items in the GSI HESR proposal  
is to study these effects in detail in  
$\bar p Au$ collisions. Very intriguing effects are expected in the  
charm threshold region: at normal nuclear density the $D\bar D$  
thresholds falls below the $\psi ^{\prime}$ resonance; at twice  
nuclear density this threshold moves below even the  
$\chi_{c2}$! 
 
The fact that hidden charm states are significantly less  
extended than open charm states has been invoked as a signature  
for the quark-gluon plasma\index{quark-gluon plasma}, 
where the correlation between colour  
sources and sinks is broken up over small distances. If in  
heavy ion collision a phase transition to the quark-gluon  
plasma is achieved, one expects a reduction in $J/\psi$  
production. 
The data are intriguing in this respect, yet not conclusive. 
 
Charmonium production is investigated in 
 relativistic heavy ion collisions\cite{Saito02}, 
 where the NA50 experiment \cite{Gonin:wn,Santos:2003qa} 
 using  1996 data (158~GeV per nucleon Pb beams on Pb target) provided 
 circumstantial evidence for charmonium suppression, which may be explained 
 by the onset of a quark-gluon plasma regime.  
They  measure $\jp$ 
 production relative to Drell-Yan pair production. After accounting for 
 conventional  nuclear absorption, their data  show 
 evidence for a  suddenly lower production, due to the attracting 
 force  between the $\ccb$ quarks being screened by gluons, and fewer $\ccb$ 
 pairs  hadronizing into $\jp$. 
 
  To conclude this section, we discuss a 
  fascinating as much as hypothetical possibility uniquely provided by the  
study 
  of charm particles in close contact to nuclear media, i.e., the formation of 
  {\it supernuclei}. In complete analogy to what has been studied in great detail 
 for 
  several decades in $\Lambda$-hypernuclei 
\cite{Alberico:2001jb},
a charm quark produced at 
  rest, or brought to rest, could interact with the nuclear matter, replace a 
  light quark, and form a $\Lambda_C$  baryon inside the nucleus. 
  The $\Lambda_C$ would then decay. This is an appealing process because the
  $\Lambda_C$  
  does not need to obey the Pauli exclusion principle, and can occupy nuclear 
  levels forbidden to the nucleons. The lifetime is also expected to differ 
  from that for   
  free $\Lambda_C$, and it would be possible to study both mesonic and
  nonmesonic decays. The only attempts carried out so fare have been in
  emulsions \cite{Lyukov:nn}. Supernuclei studies are foreseen at GSI with the
  \index{supernuclei}
  PANDA experiment (see Sect.\ref{GSI}).

\section{Spectroscopy and Lifetimes}  
\label{LIFE}  
%
The minimal information to describe a particle are its  
mass and its spin  
\footnote{This can be expressed for the mathematically minded  
reader by saying that elementary particles are defined by  
irreducible representations of the Poincare group; for those are  
labeled by the eigenvalues of two Casimir operators, which happen to be  
the mass and spin (or helicity for massless particles).}. Under the term  
`mass' we can include also the width as the imaginary part of the mass.  
The width or lifetime of a particle actually characterizes its underlying  
dynamics in a way that the (real) mass cannot, namely whether they are  
strong (even if reduced), electromagnetic or weak, and in the latter case
whether they are  CKM suppressed or not. Beyond these general remarks the
situation is different for  hidden and open charm hadrons.
 
{\em Hidden} charm
states $\bar cc$ are  characterized by a Compton wave length  
$\sim 2/m_c \sim 1/3\; fm$, i.e. their extension  is somewhat smaller than
for  light-flavour hadrons. They can decay electromagnetically and even
strongly, the latter however with a very reduced width since it is order
$\alpha_S^3(m_c)$ (for $J/\psi$). 
Powerful algorithms have been and are being developed to obtain very accurate 
predictions on charmonium spectroscopy from lattice QCD thus turning their experimental 
study into precision tests of QCD proper .

As already explained in Sect.\ref{HQS} HQS tells us that for  
{\em open} heavy flavour hadrons  
the two S wave configurations $P_Q$ and $V_Q$ become mass degenerate  
for $m_Q \to \infty$,  
while their mass exceeds $m_Q$ by the scale  
$\sim \Lambda_{NPD}$  
\footnote{It is usually denoted by $\bar \Lambda$.}  
as do the P wave configuration. I.e.:  
\beq  
m_c + \Lambda_{NPD} \sim M_D \simeq M_{D^*} \; , \;  
M_{D^{**}} \sim M_D + \Lambda_{NPD}  
\eeq  
The degree to which the hyperfine splitting $M_{D^*} - M_D$ is  
small compared to $\Lambda_{NPD}$ is one measure for whether  
charm is a heavy flavour. It is, though not by a large
factor:
\beq
M_{D^*} - M_D \sim \; 140 \; {\rm MeV} <
M_{D^{**}} - \langle M_D \rangle \sim \; 480 \; {\rm MeV}
\eeq
with $\langle M_D \rangle =
\frac{1}{4} M_D + \frac{3}{4}M_{D^*}$ denoting the spin averaged
meson mass. 
Also the simple
scaling law of Eq.(\ref{SIMPLESCALING}) is well satisfied:
\beq
M_B - M_D \simeq 3.41 \; \GeV \; vs. \;
M_{\Lambda_b} - M_{\Lambda_c} \simeq 3.34 \; \GeV
\eeq
There are further reasons to study the  
mass spectroscopy of charm resonances:
\begin{itemize}
\item
For a better understanding  
of the transition $B \to \ell \nu D^*$ that figures prominently in  
determinations of $V(cb)$ -- and of $B \to \ell \nu X_c$ in general --  
one needs information on the mass and width of $D^{**}$ and other higher
resonances.
\item
More specifically, the SV sum rules \cite{SVSRPAPER} relate the basic
HQP to the production of certain charm states in semileptonic $B$
meson decays. E.g. \cite{HQT}:
\bea
\frac{1}{2} &=& - 2 \sum _n \left|\tau _{1/2}^{(n)} \right|^2
+ \sum _m \left|\tau _{3/2}^{(m)} \right|^2  
\nonumber
\\
\bar \Lambda (\mu) &=& 2 \left( \sum _n \epsilon_n
\left|\tau _{1/2}^{(n)} \right|^2 + 2\sum _m \epsilon_m
\left|\tau _{3/2}^{(m)} \right|^2
\right)
\nonumber
\\
\frac{\mu_{\pi}^2 (\mu )}{3} &=&  \sum _n \epsilon_n^2
\left|\tau _{1/2}^{(n)} \right|^2 + 2\sum _m \epsilon_m^2
\left|\tau _{3/2}^{(m)} \right|^2
\nonumber
 \\
\frac{\mu_{G}^2 (\mu )}{3} &=&  -2 \sum _n \epsilon_n^2
\left|\tau _{1/2}^{(n)} \right|^2 + 2\sum _m \epsilon_m^2
\left|\tau _{3/2}^{(m)} \right|^2 \; ;
\label{SRSLBDEC}
\eea
here $\epsilon_k$ denotes the excitation energy of the final state
$D^{k}$ beyond the ground states $D$ and $D^*$ 
($\epsilon_k = M_{D^{k}} - M_D$) while
$\tau _{1/2}^{(n)}$ and $\tau _{3/2}^{(m)}$ denote the transition
amplitudes for producing a state, where the light degrees of freedom
carry angular momentum $j_q=1/2$ or $3/2$, respectively \cite{WISGUR}.
Obviously, the masses
of these charm resonances matter, as does their interpretation in terms
of the quantum numbers $1/2$ or $3/2$.
\item
The mass splittings of baryonic charm resonances provide  
important cross checks for the evaluation of expectation values of  
four-quark operators that are highly relevant for predicting charm baryon  
lifetimes as discussed below.
\end{itemize}

Beyond classification there are other reasons for measuring total  
widths as precisely as possible. One needs them as an {\em engineering}  
input to translate branching ratios into partial widths. This is needed,  
for example, to infer the value of CKM parameters from semileptonic  
decays. On the {\em phenomenological} level a precise analysis of the  
$D^0$ lifetime is a prerequisite for studying $D^0 - \bar D^0$
oscillations. Finally on the {\em theoretical} side the  
lifetime ratios for the different charm hadrons provide the best, since  
most inclusive observables to probe hadrodynamics at the charm scale.

From the {\it raison d'etre for charm quarks}, namely to suppress
strangeness   changing neutral currents to the observed levels, one
infers  
$m_c \leq 2$ GeV. The lifetime of charm {\em quarks} can be estimated by  
relating it to the muon lifetime and the number of colours and lepton  
flavours  
$\tau _c \sim \tau _{\mu} \cdot \left( \frac{m_{\mu}}{m_c}\right)^5  
\cdot \frac{1}{N_C + 2} \sim ({\rm few}\,  10^{-13}s) \cdot  
\left(\frac{1.5\; {\rm GeV}}{m_c}\right) ^5$ 
with an obviously high sensitivity to the value of $m_c$.

These very simple estimates have turned out to be remarkably on target.
Yet before we describe it, a few comments might be in order on the 
charm quark mass.

\subsection{On the charm quark mass}  
\label{CHARMMASS}
 
{\em Within} a given quark {\em model} a quark mass has a clear
meaning as a fixed parameter; however it depends on the specifics of the
dynamical treatment adopted there, and therefore differs from model  
to model. Yet even more importantly the connection between such quark  
model {\em parameters} and fundamental quantities appearing in, say, the  
Lagrangian of the SM is rather tenuous. For example one can model single  
and double charm production in deep inelastic $\nu$-nucleon scattering
by charged and neutral currents with a parton model ansatz, where  
$m_c$ plays of course a central role. Fitting data can yield a highly  
constrained value for $m_c$. Yet such a `precise' value cannot be taken
at  face value to describe charm hadroproduction, let alone charmonium
physics or charm decays. For that purpose one needs a proper  
field theoretical definition of the charm quark mass, which takes into
account that the dynamical environments for these reactions differ in
their perturbative as well as nonperturbative aspects. The resulting
quantity has to be a `running', i.e. scale dependent mass, where one  
has to specify its normalization scale; these issues have been discussed  
in Sect.\ref{HQP}.
 
The two areas where quark masses have been discussed  
with considerable care are  
charmonium spectroscopy and the weak decays of heavy flavour hadrons.  
\begin{enumerate}
\item  
The first analysis  was based on charmonium sum rules that  
approximate nonperturbative dynamics through including quark and gluon  
condensates in the OPE \cite{SVVZ}. One finds for the $\overline{MS}$  
mass  
\beq  
\overline m_c(m_c) = 1.25 \pm 0.10 \; {\rm GeV}
\eeq
More recent analyses find fully consistent values:
\beq
\overline m_c(m_c) = \left\{
\begin{array}{ll} 1.19\pm 0.11\; {\rm GeV} & {\rm Ref.\cite{EIDE}}\\
1.30 \pm 0.03 \; {\rm GeV} & {\rm Ref.\cite{STEIN}}
\end{array}
\right.
\eeq
Lattice studies yield in the quenched approximation \cite{ROLF}  
\beq  
\overline m_c(m_c) = 1.301 \pm 0.034 \pm 0.13_{quench}\; {\rm GeV} \; . 
\eeq
  
\item  
The expansion for $m_b - m_c$ given in Eq.(\ref{MBMINUSMC}) yields  
\beq  
m_b - m_c = 3.50 \; {\rm GeV} + 40 \; {\rm MeV}  
\left( \frac{\mu_{\pi}^2 - 0.5\; {\rm GeV^2}}{0.1\; {\rm GeV^2}}\right)  
\pm 20 \; {\rm MeV}  
\label{MBMMCDATA}
\eeq   
Using the value for the $b$ quark mass that has been extracted  
from $e^+ e^- \to b \bar b$ near threshold by several groups  
\cite{CKMPROC}
\beq  
m_b^{kin} (1\; {\rm GeV}) = 4.57 \pm 0.08 \; {\rm GeV} \; \; 
\hat = \; \; \overline m_b(m_b) = 4.21 \pm 0.08 \; {\rm GeV}
\eeq  
which is in nice agreement with what one infers from a moment analysis  
of semileptonic $B$ decays \cite{BATTAGLIAETAL},
and Eq.(\ref{MBMMCDATA}) one arrives at  
\beq  
\overline m_c (m_c) = 1.13 \pm 0.1 \; {\rm GeV}      \; .  
\eeq  
This value is completely consistent with what one obtains directly  
from the aforementioned moment analysis, namely  
\beq  
\overline m_c (m_c) = 1.14 \pm 0.1 \; {\rm GeV}  
\eeq  
despite the caveats stated in Sect.\ref{HQP} about the reliability of
this expansion.  
\item  
As will become clear from our discussion below, one cannot infer  
(yet) a reliable value for $m_c$ from the charm lifetimes.  
\end{enumerate}
To summarize: a quite consistent picture has emerged, which supports
treating charm as a heavy flavour.
%
\subsection{Spectroscopy in the hidden charm sector}  
\label{SPECONIA}
%
Charm entered reality in a most dramatic fashion through the
discovery of hidden charm mesons and their striking properties, 
and our knowledge about them increased at breathtaking speed for some 
time due to very favourable experimental features.
\par
\begin{figure}
  \centering
   \includegraphics[width=10.0cm,height=5.0cm]{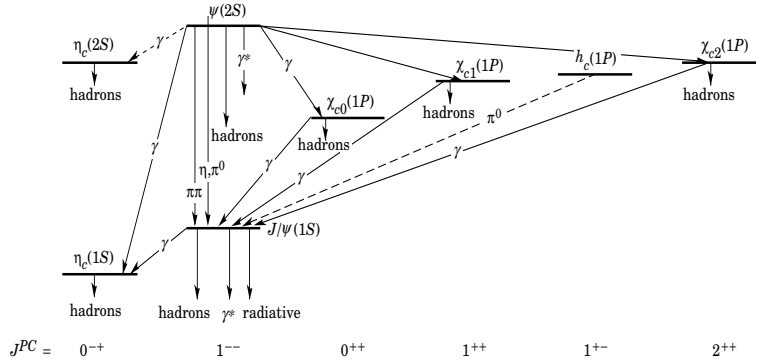}
 \caption{\it
 Chart of charmonium states \cite{Hagiwara:fs}.
    \label{FIG:ONIA} }
\end{figure}

 Most of the spectroscopy results 
have come from $\epem$ storage rings, where 
$J^{PC}=1^{--}$ states can be formed directly to lowest order. The three 
prominent states $\Jp(3100)$, $\psi^{\prime}(3700)$ and 
$\psi^{\prime\prime}(3770)$ have been well established for a long time 
as the $^3S_1$, $2^3S_1$ and $^3D_1$ states, respectively, with the 
last one being broad since above $D\bar D$ production threshold. The 
nonvector states such $^3P_J$ (also referred to as $\chi_{cJ}$) 
and $^1S_0$ can be reached by $E1$ and $M1$ transitions from them 
and thus be observed in two--step processes like 
$\epem \rarr \psi^\prime \rarr (\ccb)_{\chi_{cJ}}+\gamma$, 
see Fig.~\ref{FIG:ONIA}. 
This area of research pioneered by SPEAR and DORIS has experienced a
welcome renaissance due to the operation of the Beijing Spectrometer
(BES); in 2002 the BES collaboration has completed a four-month
run which yielded 14 million $\ps2s$, to be added to the 4
million events previously collected. 

A qualitatively new access to charmonium dynamics has been provided 
by low energy $p\bar p$ annihilation, since all $J^{PC}$ quantum numbers
then become accessible, in particular also $^1P_1$ and $^1D_2$ and 
$^3D_2$ states. 
The idea (pioneered by R704 at the ISR and carried forward by 
E760, E835 at FNAL) is to study the {\em formation} 
of charmonia states in the annihilation of
 antiprotons   on a jet hydrogen target. 
  E835 showed \cite{Cester02,Negrini02}  preliminary  measurements of masses,
 widths and branching ratios of the three $\chi_{cJ}$ states with an
 unprecedented level of precision.  
 \par
 Finally a third actor has appeared: the $B$ factories CLEO, BABAR and 
 BELLE have such large statistics that one can study charmonia in 
 $B\to [\bar cc] X$.  This has been demonstrated quite dramatically by BELLE 
 finding 5 $\sigma$ and 3.5 $\sigma$ signals for $\eta_c(1S)$ and $\eta_c(2S)$, 
 respectively \cite{Choi02,Abe:2002rb}. The  $\eta_c(2S)$ can boast of quite a saga 
 \cite{Martin03}. 
 Previous simultaneous observations of $\eta_c$ and 
 $\eta_c(2S)$ date back to conflicting measurements in the 1980's
 (DASP, Serpukhov, MARK~II and Crystal Ball).
  While the $\eta_c$ has become well established, the
 $\eta_c(2S)$
  was not confirmed by either DELPHI or E835 in extensive searches
  ($30\,{\rm pb}^{-1}$ in the range $3666$ to $3575\,\MeV$). 2000
  E835 searched with higher statistics for the
  $\eta_c(2S)$, with negative results. The
  $\eta_c(2S)$ was instead spotted in 2002 by BELLE at  
  $3622 \pm 12 \, \MeV$ in B decays, and $3654 \pm 6 \, \
  MeV$ in the recoil spectrum of $\jp c\bar c$ events. 
 Similarly frustrating is
 the search for the singlet P-state called $h_c$.
 Claimed by R704 at the ISR in 1986 and seen 
 by E760 in 1993, the $h_c$ has not, as yet, been confirmed by E760's
successor E835 in its 2001 data set. 
 
One expects \cite{EICHTEN02} four charmonium states {\em below} $D\bar D$ 
threshold (and thus narrow), whose existence has not been established, namely 
$\eta_c^{\prime}(2^1S_0)$, $h_c(1^1P_1)$, $\eta_{c2}(1^1D_2)$ and 
$\psi_2(1^3D_2)$; they can be identified in $B$ decays.

The potential model ansatz pioneered by 
the Cornell group \cite{EICHTEN} was successful in describing the 
charmonium spectroscopy of  Fig.~\ref{FIG:ONIA}).
The factorization of  nonperturbative and perturbative effects into a wave function 
and $\alpha_S$ corrections, respectively, as mentioned 
in Sect.\ref{QUARKPOT}  can be seen from the theoretical expression for
the hadronic width of the $\jp$:
$\Gamma (J/\psi \to {\rm hadrons})=\frac{80(\pi^2-9)}{81 \pi}\alpha_s(M_{J/\psi})^3
 (1+4.9\frac{\alpha_s}{\pi})|\Psi(0)|^2$. 
The ratio of this to the leptonic width can be used to extract a
 value of $\alpha_s$: 
 $R_{\mu\mu}=\frac{J/\psi \to {\rm hadrons}}{J/\psi \to
 \mu^+\mu^-}=\frac{5(\pi^2-9)}{81
 \pi}\frac{\alpha_s(M_{J/\psi})^3}{\alpha_{em}}(1+10.3\frac{\alpha_s}{\pi})$. 
The experimental value $R_{\mu\mu}\approx 14.9$ leads to a reasonable
 result: $\alpha_s(M_{\jp})\approx 0.2$.  Relativistic effects of
 order $v^2/c^2$ can also be included. Yet this
 method of extracting $\alpha_s$ is not as theoretically sound as
 others.  The first order
 radiative correction is as large as the lowest order correction calling
 into question the validity of the perturbative expansion.  Furthermore,
 the expressions for the various widths ultimately depend on the expression
 that is chosen to describe the quark-antiquark potential, which  
 is based upon phenomenological aspects of QCD rather than rigorously
 derived from it.  By taking ratios, in which the
 dependence on the wavefunctions vanish, this source of uncertainty can be
 reduced.  It still remains unclear how valid the factorization assumption
 is for the charmonium system in which $m_c$ is only moderately larger than
 typical hadronic scales.

Radiative transitions between charmomium states can similarly be 
described.  The radius of the bound state is
 typically much smaller than the wavelength of the emitted radiation so a 
 multipole expansion is expected to converge quite rapidly.  Electric
 dipole (E1) transitions are responsible for $\Delta S=0,\Delta L=1$
 processes.  The rate for transitions between S- and P- wave states is:
\beq
 \Gamma_{\gamma}(S \leftrightarrow P) =
 \frac{4}{9}\left(\frac{2J_f+1}{2J_i+1}\right)Q^2 \alpha |E_{if}|^2
 E_{\gamma}^3.
\label{E3GAMMA}
 \eeq
 \noindent
 Here $J_{f[i]}$ denotes the total angular momentum of the final[initial]
 state, $Q=2/3$ is the charge of the charmed quark, $E_{\gamma}$ is the
photon energy and $E_{if}$ is the matrix element of the transition dipole
 operator: $ E_{if}=\int^{ \infty}_0 r^2 \Psi_i(r)\, r\, \Psi_f(r)$. Since this
matrix element is more sensitive to the exact {\em shape} of the wavefunction 
unlike $|\Psi(0)|^2$ that appeared previously, considerable differences 
emerge among theoretical predictions.  Even so, there is reasonable agreement
 with experiment \cite{GOTTFRIED}. 
 Magnetic dipole (M1) transitions are responsible for $\Delta S=1,\Delta
 L=0$ processes and are suppressed by $E_{\gamma}/m_c$ with respect to the E1
 transitions.  The transition rate between spin 0 and 1 S- wave states
 is given by the following expression:
\beq
 \Gamma_{\gamma}(^3S_1 \leftrightarrow ^1S_0) =
 \frac{16}{3}\left(2J_f+1\right)\left(\frac{Q^2}{2m_c}\right) \alpha
 |M_{if}|^2 E_{\gamma} ^3.
 \eeq
 Here the magnetic dipole moment is the expectation value of the zeroth
 order spherical Bessel function:
$|M_{if}|=\int^{ \infty}_0 r^2 \Psi_i(r)\, j_0(\frac{1}{2}E_{\gamma} r)\,
 \Psi_f(r)$. 
Since these matrix elements depend quite sensitively on details 
of the wave functions, it is not surprising that the 
 agreement between theory and experiment for M1 transitions is rather
 poor. 
 
 The lattice community is able now to treat charmonium physics 
 with three flavours of dynamical quarks; from the spin-averaged 
 $1P-1S$ and $2S-1S$ splittings  one infers for the strong 
 coupling $\alpha_S^{\overline{MS}}(M_Z) = 0.119 \pm 0.004$ 
 \cite{AIDA}. 

 Hadronic transitions like $\psi ' \to \psi \pi \pi$ are also treated using
 a multipole expansion to describe the gluonic radiation.  An added
 complication is the hadronization of the emitted gluonic radiation.  By
 introducing a chiral Lagrangian to describe the effective low energy
 behaviour of the hadronic state, a semi-quantitative analysis can be
 carried out for these transitions.

\par
The transition $J\psi \to \gamma X$ driven by $J\psi \to \gamma gg$ provides 
a gluonic origin for the final state $X$. Accordingly states with a particular affinity 
to gluons should figure prominently in $X$. Narrow states would show up 
as mass peaks in the $\gamma$ recoil spectrum; no prominent signal has been 
found yet. One can search for them also in exclusive final states, as discussed 
in Sect.\ref{NONLEPT}.  
\par
An update overview of the experimental panorama can be found in
\cite{Skwarnicki03}. Breaking news in summer 2003 was the preliminary result by
BELLE on the observation of a $\jp \pi^+ \pi^-$ state in decay $B^+\to K^+ (jp
\pi^+ \pi^-)$. BELLE finds a clear ( 8.6$\sigma$ ) signal at 3871.8 $\pm$ 0.7
$\pm$ 0.4 MeV, which is suggestive of
\index{$DD^*$ molecule}
 a $DD^*$ molecule.
%
\subsection{Spectroscopy in the $C\neq 0$ sector}  
\label{SPECC12}

Adding charm as the fourth quark leads to a very rich spectroscopy. There
are six $C=1$ pseudoscalar states (plus one $\bar cc$ state already
discussed) in addition to the familiar $SU(3)$ meson nonet, namely
$D^{\pm}$, $D_s^{\pm}$ and $D^0/\bar D^0$; likewise for the vector
mesons with $D^{*\pm}$, $D_s^{*\pm}$ and $D^{*0}/\bar D^{*0}$. For
baryons even more facets emerge, as described later.

These states can be fitted into $SU(4)$ multiplets. Yet $SU(4)$ breaking driven by
$m_c > \Lambda_{NPD} \gg m_s$ is much larger than
$SU(3)$ breaking
\footnote{Another way to put it is to say that charm mesons --
unlike pions and kaons -- cannot be viewed as Goldstone bosons.}. Heavy
quark symmetry provides a much more useful classification scheme.
As explained in Sect.(\ref{HQS}) for heavy flavour hadrons $H_Q$ the spin
of $Q$ -- ${\bf S_Q}$ -- decouples  from the light quark degrees of
freedom, and ${\bf j_q} \equiv {\bf s_q} +  {\bf L}$  and ${\bf S_Q}$
become {\em separately} conserved quantum numbers.
A meson [baryon] can then be characterised by
the spin of the light antiquark [diquark] and the orbital angular  
momentum.
\par
While  the masses of the groundstates $D/D^*$ and $D_s/D_s^*$ have been known
with $1\,\MeV$ precision for a decade now, the experimental information
available on other states is still
unsatisfactory. In this section we shall discuss open problems and very recent 
surprises.
%
\subsubsection{$D^*$ width}
\label{MesonSpectrDWidth}
Measuring $\Gamma(D^{*+})$ represents an experimental challenge:  
such widths are predicted in the range of tens or hundreds  keV, and must be
experimentally deconvoluted to the experimental resolution of detectors.
CLEO has presented the first measurement
\cite{Anastassov:2001cw,Ahmed:2001xc} 
based on  the sequence $D^{*+}\to \pi^+ D^0, D^0\to K^-\pi^+$ from a
$9\; fb^{-1}$ sample of $\epem$  data collected with
the CLEO II.V detector. $\Gamma(D^{*+})$ is controlled by the strong
coupling constant, since the
electromagnetic contribution  $D^{*+}\to \gamma D^+$ produces  
a very small branching ratio, and can thus be neglected.

Based on an complex analysis of 11,000 reconstructed $D^*$ decays 
CLEO finds $\Gamma(D^{*+})=96 \pm 4 \pm 22 \, {\rm keV}$.  
Hence they infer
for the $D^*D\pi$ coupling:
$g_{D^*D\pi} = 10 \pm 3.5$.
Light-cone sum rules have been employed to obtain the prediction \cite{KHODJA1}
$g_{D^* D\pi} = 17.9 \pm 0.3 \pm 1.9$, 
where the value $13.5$ is viewed as a rather firm upper bound \cite{KHODJA2}.
\par
 \begin{figure}
  \centering
   \includegraphics[width=3.0cm]{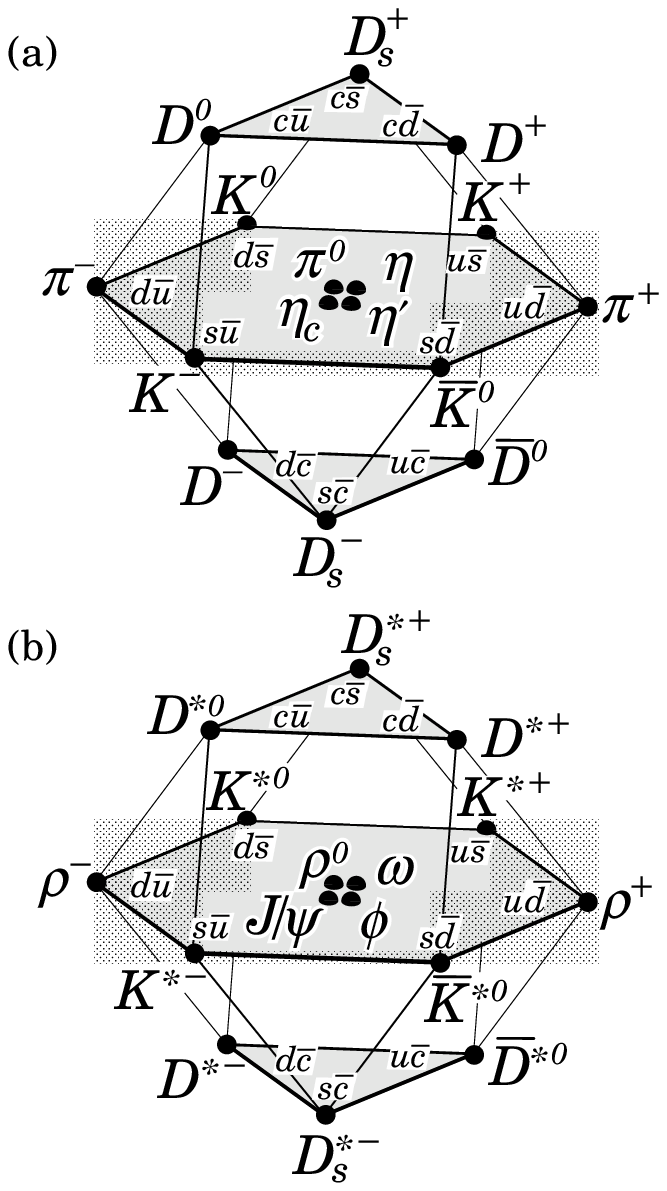}
   \includegraphics[width=3.0cm]{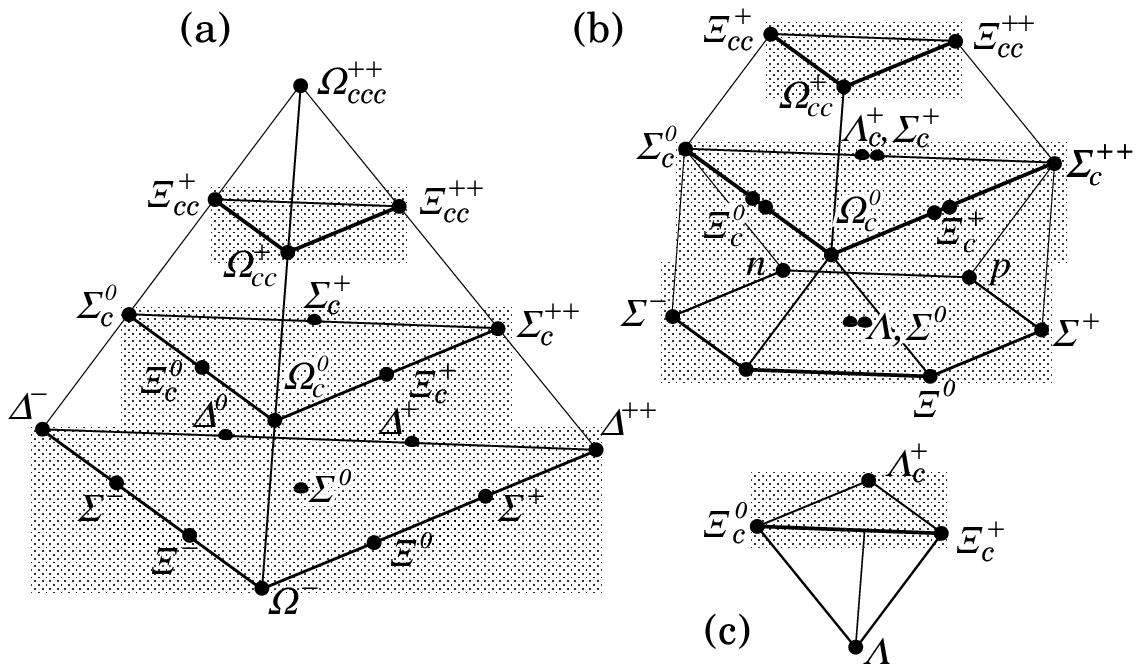}
   \includegraphics[width=3.0cm]{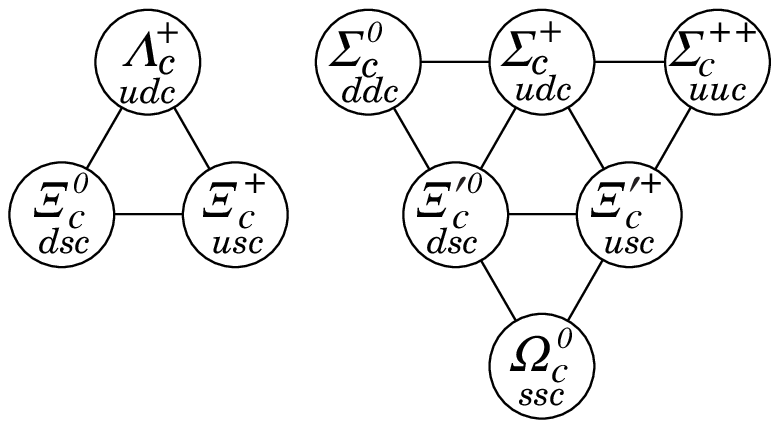}  
 \caption{\it
  Chart of ground state charmed meson and baryon
  multiplets \cite{Hagiwara:fs}.
    \label{FIG:MULTI} }
\end{figure}
The $B$ factories with their large charm samples should be able to
check CLEO's results, once detector simulation will reach the level of
accuracy necessary to tame the severe systematic uncertainty.

\subsubsection{Charm mesons - $L=1$ excited states}
\label{MesonSpectrL1States}
\begin{figure}
\centering
\includegraphics[width=10.0cm,height=5.0cm]{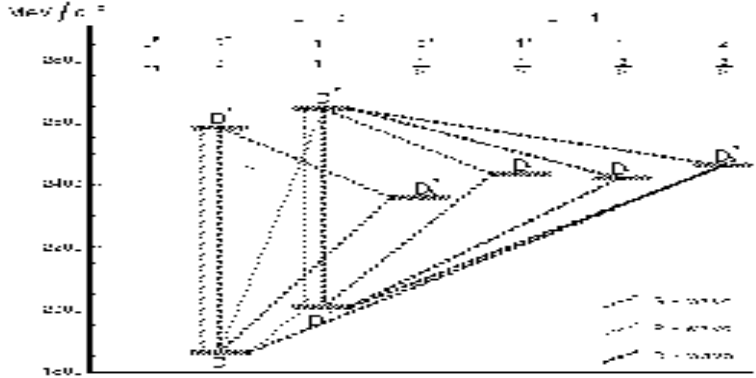}
 \caption{\it
  Masses and transitions predicted for the excited D meson states (pre-Spring
  2003).
    \label{FIG:DSPEC} }
\end{figure}

For each of the $c\bar u$, $c\bar d$ and $c \bar s$ systems four
P-wave and two $n=2$ radial excitations have been studied.
There are four $L=1$ states, namely two with $j_q=1/2$ and total spin
$J=0,1$ and two with $j_q=3/2$ and $J=1,2$. These four states
are named respectively $D_0^*$, $D_1(j_q=1/2)$, $D_1(j_q=3/2)$ and
$D_2^*$ (Fig.\ref{FIG:DSPEC}). 
Parity and
angular momentum conservation force the $(j_q=1/2)$ states to
decay to the ground states via S-wave transitions (broad width),
while  $(j_q=3/2)$
states decay via D-wave (narrow width). To be more specific, for the
$1/2$ one predicts widths of $\sim 100$~MeV and for the
$3/2$ of about $\sim 10$ MeV with the exception of the
$D_{s1}(j_q=3/2)(2536)$ which is kinematically forced to a $\sim 1$~MeV  width. 
\par
All six $L=1, j=3/2$  {\em narrow} states are well established,  
 with precisions on masses at the
 1 MeV level and on widths at the few MeV level. This is due to the fact that
 excited D states are abundantly produced both at FT experiments, in
$\epem$ continuum production, in B decays and at the
 $Z^0$ \cite{Roudeau:1997zn}.
 Common analysis techniques are the selection of a clean
 sample of D meson candidates typically via a candidate-driven algorithm
 \index{candidate-driven algorithm} (see Sect.\ref{HISLIFE}), a cut on the $D\pi$ mass to 
reject $D^*$  
 compatible combinations, and the pairing of the D candidate with one (two) soft
 pion (pions) in   the primary vertex to form the $D^{**}$ candidate.  
 A selection on the helicity angle is effective in selecting L=1 states
 from background, as well as in selecting different states in the same
 final channel.
\par
At $\epem$ one also invokes the
 mass constraint of  the fixed center of mass energy and kinematical helicity 
 cuts which exploit the constraint of the parent
 B  mass. Dalitz plot and partial wave analyses have also been presented
 \cite{Belle02}. 
\par
A review of the data from different experiments can be found in 
 \cite{Bartelt:rq}. A common
 feature is a prominent peak in the 2460 MeV  region, escorted by
 satellite 
 peaks at one $\pi^0$ mass below, due to feeddown from decays
 where the $\pi^0$ escapes detection, such as the decay chain $D_2^{*+} \to D^0
 \pi^+$ with the $D^0$ wrongly assigned to $D^{*0}\to D^0\pi^0$ decay.
\par  
 Generally speaking, narrow L=1 signals at FT have higher statistics but sit
 on a more prominent background, due to the larger combinatorics coming from
 larger primary multiplicities. On the other hand 
 photoproduction signals with their lower primary multiplicity  have less
 combinatorics  than hadroproduction.  

The status of the {\em broad} L=1  states is much less clear, and    
the assignments of the quantum numbers are largely based on theory
expectations for their masses and widths. 
In 1998 CLEO \cite{Anderson99}  showed evidence for
 the $D_1(j_q=1/2)$ broad state. The authors
of Ref.\cite{ITOISHIDA} propose an alternative interpretation of this state as
the 
axial chiral partner.  Such an alternative interpretation is actually supported
by the SV and spin sum rules describing semileptonic $B$ decays, see
Eq.(\ref{SRSLBDEC}): they strongly suggest that the $D^{**}$ states around
$2.4 - 2.6 {\rm GeV}$ have to come mainly from a $3/2$ state -- or they do 
not represent a P wave configuration \cite{URIDURHAM}.

Tab.\ref{TAB:DST} gives a summary  of our knowledge of excited $D$ mesons
as it appeared in early 2003. At that time it seemed all one needed was
to fill in a few gaps. 

\begin{table}[t]
 \caption{ Winter '02/'03 status of (L=1, n=1) and (L=0, n=2)  
 $c\bar q$ and $c\bar s$ mesons (MeV).
Statistical and
  systematical errors added in quadrature.
   Experimental results not included in PDG \cite{Hagiwara:fs} are from BELLE
   \cite{Abe:2003zm},   CLEO   \cite{Anderson99}, 
   DELPHI  \cite{Bloch02},  
   FOCUS \cite{Fabri:2000tr} \cite{Kutschke:2000sm}.
    Theory predictions from \cite{THDST}. 
\label{TAB:DST}
 }
 \footnotesize
 \begin{center}
 \begin{tabular}{|l|c|c|c|c|c|c|} \hline
 $j_q$     & $1/2$    & $1/2$  & $3/2$ & $3/2$   & $1/2$ & $1/2$  \\
 $J^P$     & $0^+$    & $1^+$  & $1^+$ & $2^+$   & $0^-$ & $1^-$  \\
 $L,n$     & $1,1$    & $1,1$  & $1,1$ & $1,1$   & $0,2$ & $0,2$  \\
\hline
 & $D_0^*$ & $D_1$     & $D_1(2420)$   & $D_2^*(2460)$ &$D^\prime$&$D^{*\prime}$
 \\ 
  Decay Mode                &
    $D\pi$                  &
    $D^*\pi$                &
    $D^*\pi$                 &  
    $D\pi,D^*\pi$             &
                             &
  $D^*\pi\pi$  \\      
\hline
 \multicolumn{7}{|c|}{Mass (MeV)}    \\
   PDG $0$            &          
                                &
                                &  
    2422 $\pm$  2               &
    2459 $\pm$  2               &  
                                &
                                 \\ 
 PDG  $\pm$                         &          
                                &
                                &  
    2427 $\pm$  5               &
    2459 $\pm$  4               &  
                                &
    2637 $\pm$  7    \\   
 FOCUS   $0$                & 
        $\sim$ 2420                     & 
                              &  
                            &
       2463 $\pm 2$        &  
                           &
	                       \\  
 FOCUS   $\pm$                & 
       $\sim$ 2420                      & 
                              &  
                            &
       2468 $\pm 2$         &  
                           &
	                       \\  			       
 BELLE   $0$                  & 
        2308 $\pm$  36                     & 
	2427 $\pm$ 36 &  
        2421 $\pm$ 2     &
	2461 $\pm$ 4        &  
	                      &
	                          \\    
 DELPHI  $\pm$                 &
                              & 
	2470 $\pm$ 58 &  
             &
                &  
                              &
                               \\  
 CLEO  $0$                 &
                              & 
	2461 $\pm$ 51 &  
             &
               &  
                              &
                               \\    			          
  Theory                    & 
          2400               & 
	  2490               & 
	  2440               &  
	  2500               &  
	  2580               & 
	  2640                 \\  
 \multicolumn{7}{|c|}{Width (MeV)}    \\	  
 PDG  $0$      &          
                             & 
	                     &
	  19 $\pm$ 4        &  
          23 $\pm$ 5     &
	                    &
	                  \\  
 PDG $\pm$                        &          
                             & 
	                     &
	  28 $\pm$ 8        &  
          25 $\pm$ 7     &
	                    &
	     $<$ 15           \\  
	     
 FOCUS   $0$                & 
       $\sim$ 185                      & 
                              &  
                            &
       30 $\pm$ 4       &  
                           &
	                       \\  
 FOCUS   $\pm$                & 
       $\sim$ 185                      & 
                              &  
                            &
       29 $\pm$  4       &  
                           &
	                       \\  			       
 BELLE   $0$                  & 
            276 $\pm$ 66                   & 
	 384 $\pm$ 114&  
          24 $\pm$ 5   &
	  46 $\pm$ 8      &  
	                      &
	                          \\   
 DELPHI  $\pm$                 &
                              & 
	160 $\pm$ 77 &  
             &
               &  
                              &
                               \\ 
 CLEO  $0$                 &
                              & 
	290 $\pm$ 100 &  
             &
              &  
                              &
                               \\   			       
 Theory                 & 
       $>$170               & 
       $>$250              &
       20-40             & 
       20-40             &
                            &
      40-200            \\
\hline
 &$D_{s0}^*$         & 
 $D_{s1}$             & 
 $D_{s1}(2536)$         & 
 $D_{sJ}^*(2573)$           &
  $D_s^\prime$&
   $D_s^{*\prime}$ \\ 
 Decay Mode                &
                                &
			        &
	  $D^*K$           &
	   $DK$            &
	                   &
			    \\      
\hline
 \multicolumn{7}{|c|}{Mass (MeV)}    \\
  PDG $\pm$                 &
                          &  
			        &
            2535.3 $\pm$   0.6       &
	   2572.4 $\pm$ 1.5           &
	                     & \\    
 FOCUS   $\pm$                & 
                            & 
                              &  
            2535.1 $\pm$ 0.3                &
            2567.3 $\pm$ 1.4    &  
                           &
	                       \\ 
  Theory                   &
             2480               &
	      2570         &
	       2530           &
	        2590          &
		 2670          & 
		 2730               \\   			        		 
 \multicolumn{7}{|c|}{Width (MeV)}    \\
 PDG $\pm$.                 &
                           &
	                   &
	    $<$2.3 \@ 90 \% cl       &
	     15 $\pm$ 5        &
	                     &
			        \\  
 FOCUS   $\pm$                & 
                             & 
                              &  
               1.6 $\pm$ 1.0             &
             28 $\pm$ 5   &  
                           &
	                       \\ 				
Theory                  &
                            &
			            &
	 $<1$              &
       $10-20$             &
	                   &        
                           \\
 \hline
 \end{tabular}
  \vfill
 \end{center}  
\end{table}

A major confirmation for HQS would be the definite
observation of the missing $(L=1, j_q=1/2, J^P=0^+)$ $D_0^{*}$ and
 $(L=1, j_q=1/2, J^P=1^+)$ $D_1$ broad states.
  More recently, FOCUS\cite{Fabri:2000tr} and BELLE\cite{Belle02} showed new
 preliminary results and evidence for 
 $D_0^*$. Errors on both masses and widths are still very large and we
 expect better measurements from the B-factories due to their larger data sets.

The overall information on L=1 $c\bar s$ states is
unsatisfactory anyway. The $D_{s1}(j_q=3/2)$  has been seen in $D^*K^0$ final
state, and not in $D^+K^0$ or in $D^0K^+$. The
 $D_{s12}$ (called  $D_{sJ}(2573)$ by PDG) has been seen in  $D^0K^+$  and
 recently in $D^+K^0$\cite{Kutschke:2000sm}.
Furthermore no candidate for the $D_s(1/2)$ doublet had been seen yet.
Since their masses were firmly expected to be
about 80 MeV larger than for the corresponding nonstrange states, they
would have enough phase space for the decays into $D^{(*)}K$ final states leading
to large widths.
\par
 An open question remains the first evidence\cite{Abreu:1998vk} seen by DELPHI 
  of a charm
 radial excitation $D^{*\prime+}$ in the $D^{*+}\pi^-\pi^+$ final state (called  
$D*(2640)^\pm$ by PDG); it has  not been confirmed by any experiment 
   (OPAL\cite{Abbiendi:2001qp},
   CLEO\cite{Rodriguez:1998ng},
  ZEUS\cite{Sefkow:2000eu}),
 and
 questioned by theory  predictions \cite{Melikhov:1998jm}.
\subsubsection{Charm mesons - New $L=1$ $D_s$ states}  
\label{MesonSpectrDs}
Analyses presented by BABAR 
\cite{BABARDS**} and CLEO
\cite{CLEODS**} in the spring of 2003 are challenging
the whole picture.
\begin{enumerate}
\item
BABAR reported finding a narrow resonance $D^*_{sJ}(2317)$ with  
$D^*_{sJ}(2317) \to D_s^+\pi ^0$ in 90$fb^{-1}$ of data. With the observed width 
consistent with the experimental resolution, the {\em intrinsic} width
has to be below 10 MeV. This discovery has been
confirmed by CLEO.
\item
CLEO with 13$fb^{-1}$ has observed another similarly narrow state at a mass 2.46 GeV, 
for which BABAR had found evidence before: 
$D^*_{sJ}(2463) \to D_s^{*+}\pi ^0$. 
\item
 BELLE\cite{Abe:2003jk,Abe:2003vu} out of $86.9 fb^{-1}$ dataset finds evidence for $D^*_{sJ}(2463)\to D_s^+\gamma$ decay, measures relative branching ratio to $D_s^*\pi^0$, and determines $J^P=1^+$ assignment for $D^*_{sJ}(2463)$. 
\end{enumerate} 
It seems natural to interpret $D^*_{sJ}(2317)$ and $D^*_{sJ}(2463)$ as $0^+$ and
$1^+$ states, respectively. The decay distributions are consistent with such 
assignments, yet do not
establish them. They together with the mass values would explain the narrow
widths:
for $D^*_{s1^+}(2463) \to D K$ is forbidden by parity, 
$D^*_{s0^+}(2317) \to D K$ and $D^*_{s1^+}(2463) \to D K^*$ by kinematics 
and $D^*_{s0^+}(2317) \to D_s^+ \pi^0$ and
$D^*_{s1^+}(2463) \to D_s^{*+} \pi^0$ are isospin violating transition and thus
suppressed. Also $D^*_{s0^+}(2317) \to D_s^+ \gamma$ is forbidden.

There are three puzzling aspects to these states:
\begin{itemize}
\item
Why have no other decay modes been seen ? In particular CLEO places a low upper
bound
\beq
BR(D^*_{s0^+}(2317) \to D_s^{*+} \gamma ) < 0.078  \; \; 90\% \; C.L.
\eeq
Why is it not more prominent, when $D^*_{s0^+}(2317) \to D_s \pi^0$ is isospin
violating ?
\item
Why are their masses so much below predictions ? One should note that a deficit
of $\sim 160$ and $\sim 100$ MeV is quite significant on the scale of
$M(D^*_{sJ}) - M(D)$.  Why is the mass splitting to the previously found narrow states 
$D_{s1}(2536)$
and $D_{sJ}(2573)$ so much larger than anticipated ?
\item
A related mystery is the following: where are the corresponding 
{\em non}-strange
charm resonances ? They should be lighter, not heavier than $D^*_{s0^+}(2317)$
and $D^*_{s1^+}(2463)$.  
\end{itemize}
Potential model results have been re-analyzed, lattice QCD is being consulted
\cite{BALI} and more exotic scenarios like a $DK$ molecule below threshold
\cite{BARNES} and
other four-quark interpretations have been put forward. 
The latter are hard pressed to explain the narrow width in the first place, 
since the transition would 
not have to be isospin violating; yet in addition CDF sees no evidence for 
$D^*_{sJ} \to D_s \pi^{\pm}$ \cite{SHAPIRO},
although the preliminary CDF data of well known L=1 states (such as the $D_2^*$)
 reported so far show rather modest signal-to-noise ratios. 
\par
 \begin{figure}[t]
  \centering
   \includegraphics[width=3.0cm]{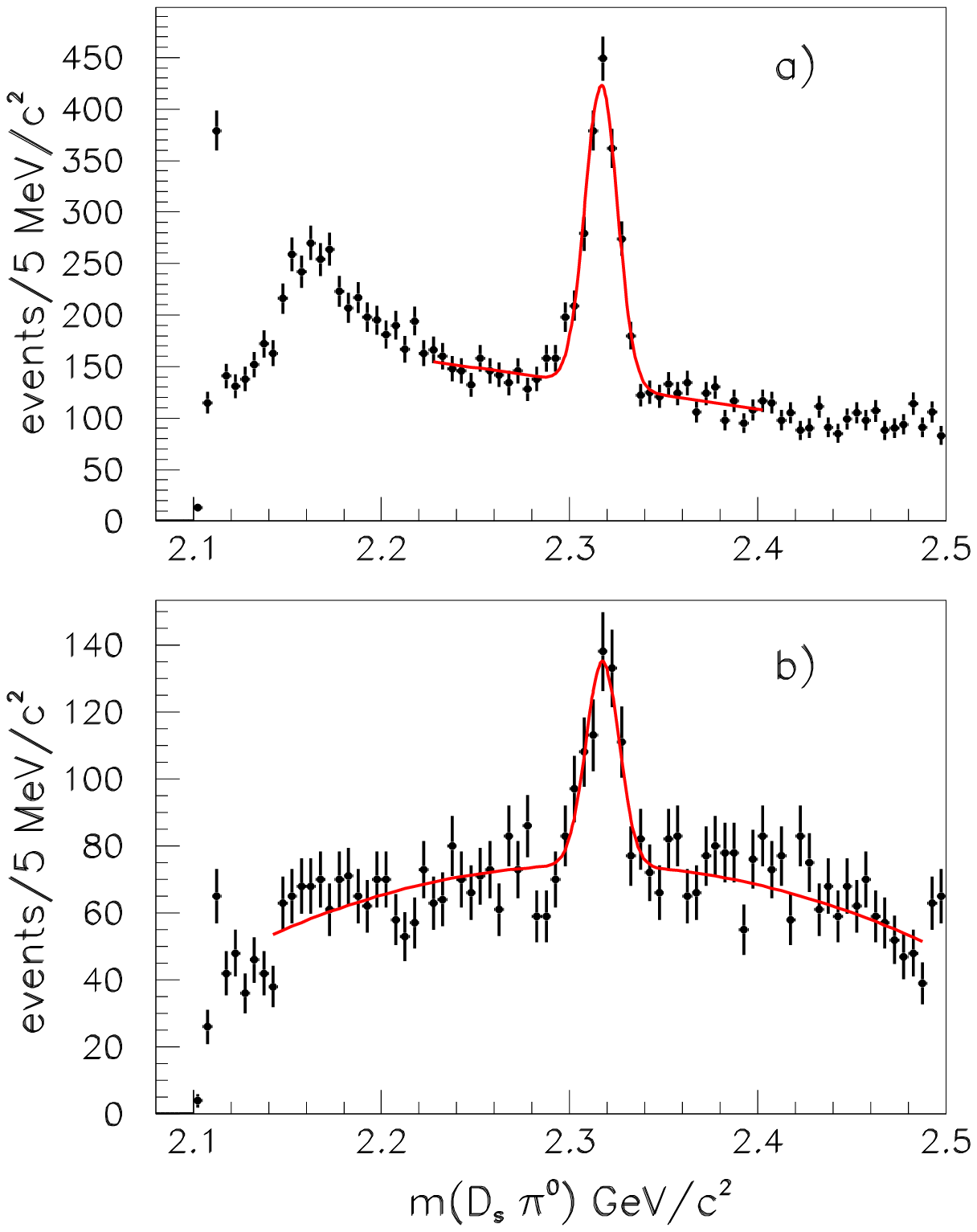}
   \includegraphics[width=3.0cm]{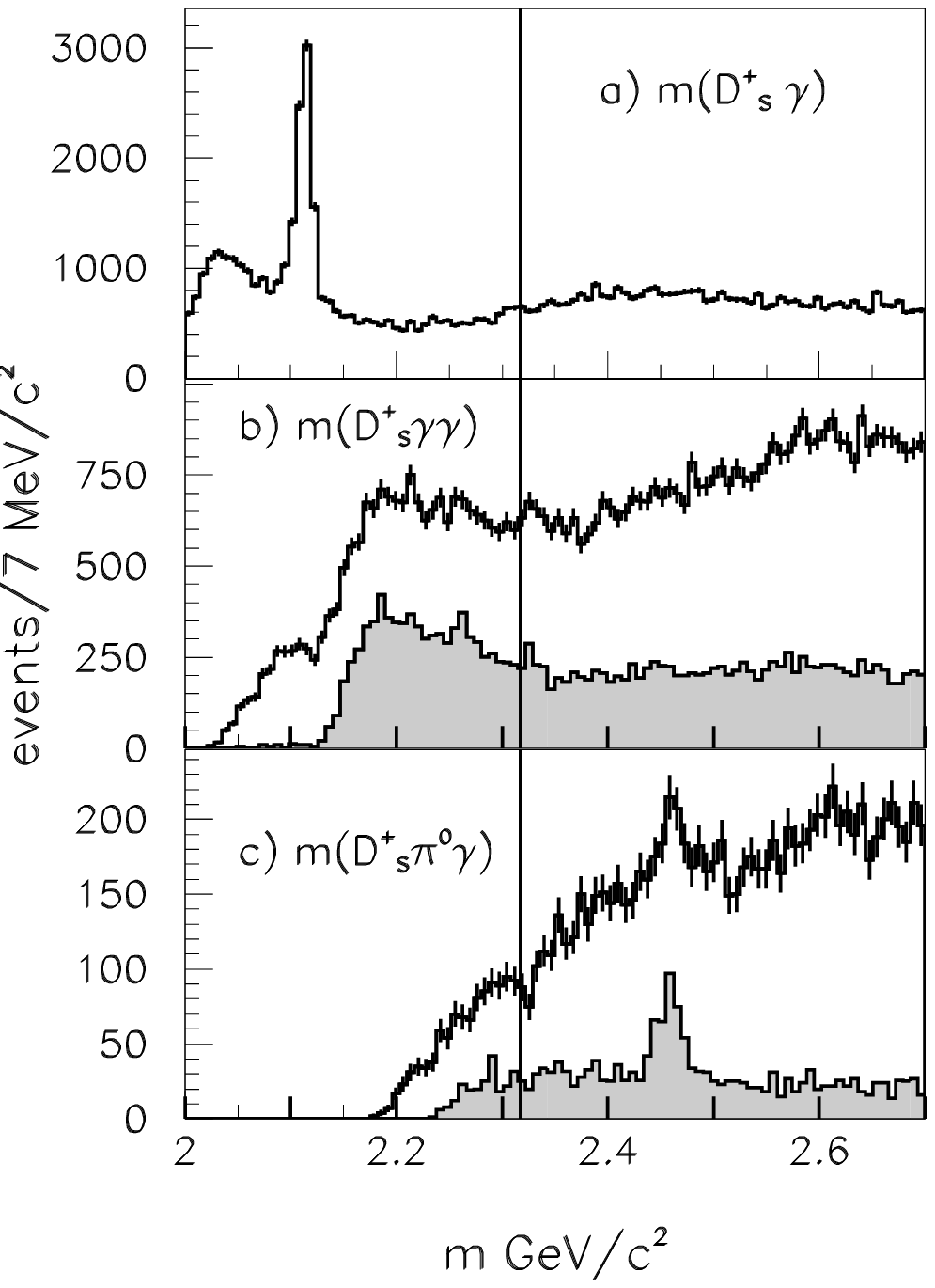}
   \includegraphics[width=3.0cm]{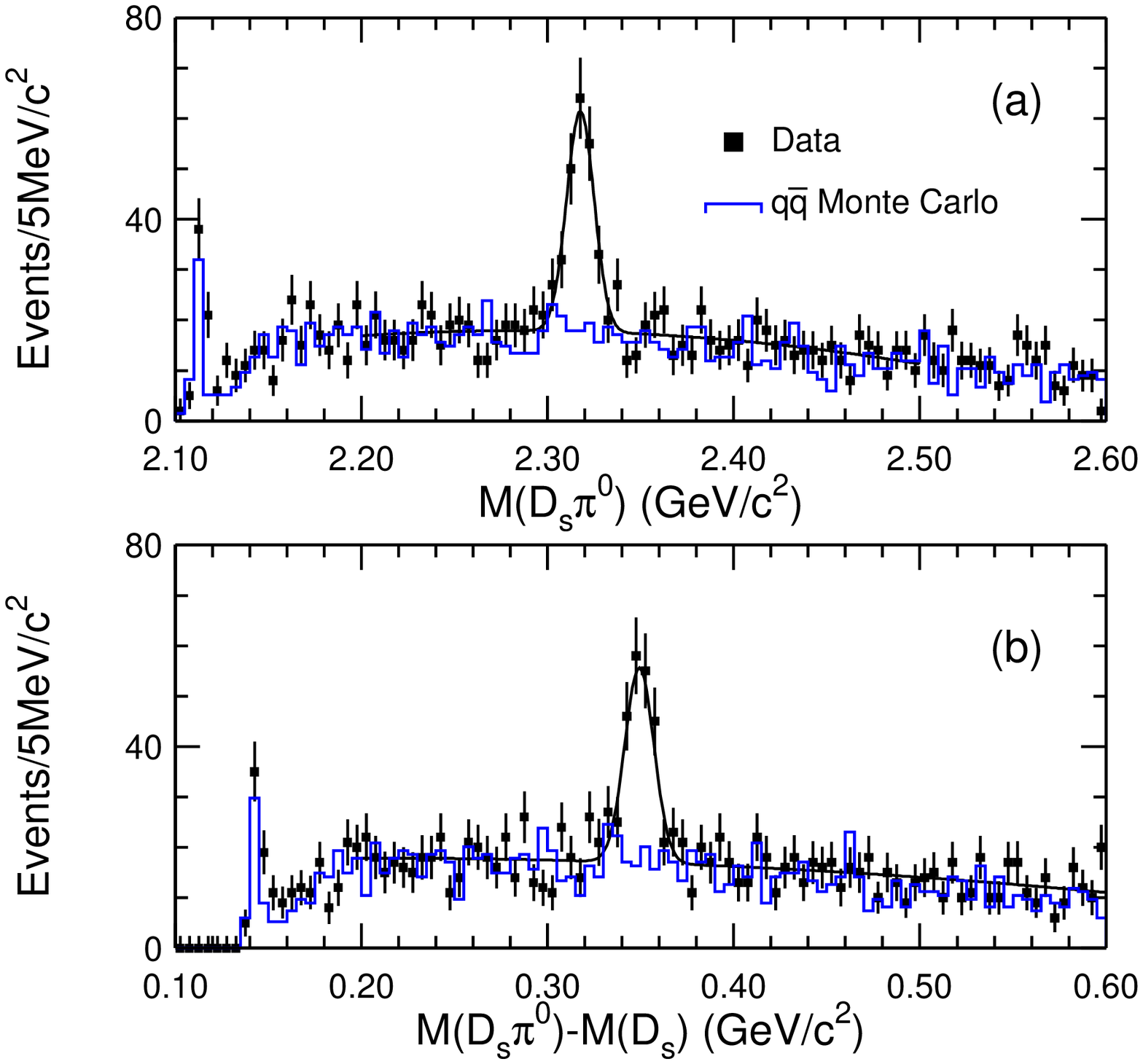}
   \includegraphics[width=3.0cm]{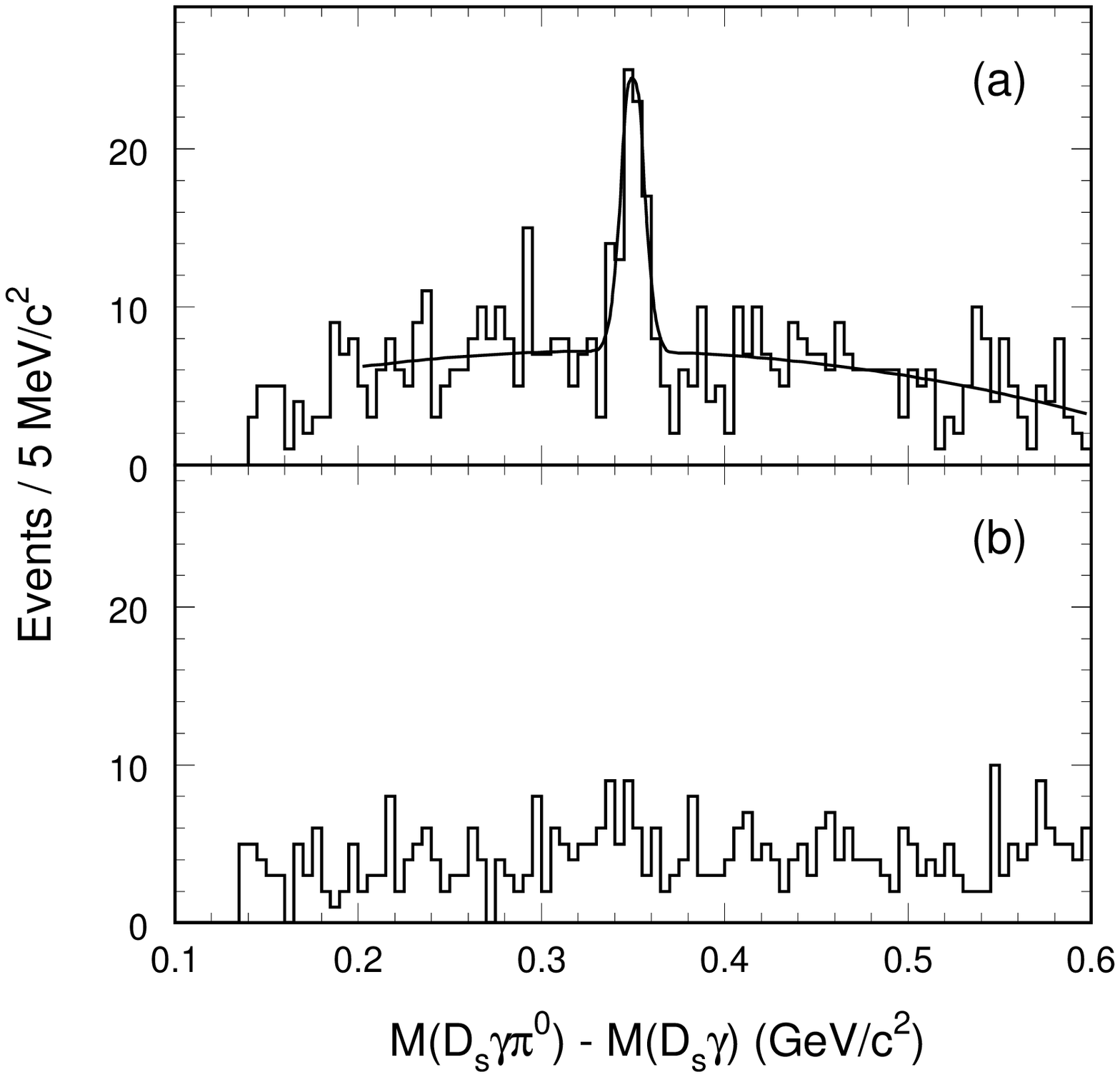}
 \caption{\it New $D^*_{s0^+}(2317)$ and $D^*_{s1^+}(2463)$ states observed by
 BABAR (a,b)  \cite{BABARDS**} and CLEO (c,d) \cite{CLEODS**}.
    \label{FIG:D0DIAG} }
\end{figure}
Maybe the most intriguing explanation, since it would represent
a new paradigm for the implementation of chiral invariance, is the
suggestion made in Ref.\cite{BARDEEN} to combine heavy quark symmetry and
chiral invariance \index{chiral invariance} in a novel way: the latter is realized through parity
doublets pairing $(D_s,D_s^*)$ with $(D^*_{s0^+},D^*_{s1^+})$. Chiral
dynamics induces a mass splitting $\Delta M$ between the heavy quark doublets:
invoking a Goldberger-Treiman relation \cite{GT}
\index{Goldberger-Treiman relation} the authors of Ref.\cite{BARDEEN}
estimate
\beq
\Delta M = M(D^*_{s0^+}) - M(D_s) \simeq M(D^*_{s1^+}) - M(D^*_s) \sim
m_N/3
\eeq
in agreement with the experimental findings. They also find that
the radiative width is reduced:
$BR(D^*_{s0^+}(2317) \to D_s^{*+} \gamma ) \sim 0.08$, i.e. right about CLEO's
upper
limit. The same chiral symmetry and $\Delta M$ apply to $B$ mesons and double
charm
baryons $[ccq]$.

Last -- but certainly not least -- the same chiral splitting should arise
for nonstrange charm mesons leading to
\bea
M(D^{\pm}(0^+) &\simeq& 2217 \; {\rm MeV} \; ,
\; M(D^{\pm}(1^+) \simeq 2358 \; {\rm MeV} \\
M(D^{0}(0^+) &\simeq& 2212 \; {\rm MeV} \; ,
\; M(D^{0}(1^+) \simeq 2355 \; {\rm MeV}
\eea
Of course, these predictions could be modified significantly by order
$\Lambda_{NPD}/m_c$ corrections.
These states can decay strongly into $D\pi$ and $D^*\pi$ and thus would be
broad. This makes it difficult to distinguish the decays 
of these resonances from phase space distributions of $D\pi$ and $D^*\pi$.
It is very important to search for them; finding them would
neccessitate a new interpretation for the previously listed $D_0^*$ and $D_1$. 
$B$ factories analysing the final states of $B$ decays  have an 
advantage in such analyses. 

These remarks indicate that a fundamental re-evaluation of strange as well as
non-strange charm hadrons might be in store. This would have an important
impact not only on charm spectroscopy, but also on the
aforementioned sum rules describing semileptonic $B$ decays,
Eq.(\ref{SRSLBDEC}), and on the latter's interpretation.

\subsubsection{$C=1$ baryons}
\label{C1BAR}

In the framework of SU(4) (Fig.\ref{FIG:MULTI})   
we expect nine ground state $cqq$ $J^P=1/2^+$ baryons (all of them detected after the 
1999  \c2
observation    of $\Xi^\prime_c$) and six $cqq$ $J^P=3/2^+$
baryons (with only the $\Omega_c^{*0}$ remaining  undetected, 
after the  observation of $\Sigma_c^{*+}$ by \c2 in 2002 \cite{Yelton:xv}); 
for a concise and for a more detailed review
 see \cite{Hagiwara:fs} and  \cite{Yelton:2002ru}, respectively.
The   $\Omega_c^{*0}$ is  expected to decay via the experimentally challenging
channel $\Omega_c^{*0} \rarr \gamma    \Omega_c^0$ )
\footnote{We adopt for excited baryon states the
 nomenclature  in \cite{Appel:1993hn}. Thus, members of 3/2 multiplets are given a
 $(*)$, the subscript is the orbital light diquark momentum ${\bf L}$, and
    $(\prime)$     indicates symmetric quark wavefunctions $c\{q_1q_2\}$
with  respect to interchange of light quarks, opposed to 
antisymmetric wavefunctions $c[q_1q_2]$.}.  
\par  
Only  a few of the orbitally excited P-wave $(L=1)$
  baryons have been     observed. The first doublet $\Lambda_{c1}(2593)$
  $(1/2^-)$     and $\Lambda^*_{c1}(2625)$ $(3/2^-)$ was observed several
    years ago by \c2, ARGUS, and E687.
    In 1999 \c2
    presented evidence\cite{Alexander:1999ud} 
    for the charmed-strange baryon analogous
    to $\Lambda^*_{c1}(2625)$, called $\Xi^*_{c1}(2815)$, in
    its decay to $\Xi_c \pi \pi$ via an intermediate $\Xi_c^*$ state.
    The presence of an intermediate $\Xi_c^*$ instead of $\Xi_c^\prime$ supports 
  an $3/2^-$ assignment, while HQS explicitly forbids
  a direct transition to the $\Xi_c \pi $   ground state due to 
  angular momentum and parity conservation. 
  More recently, CLEO has reported on the observation of both a broad and a narrow
  state in the $\Lambda_c^+\pi^+\pi^-$ channel, identified as the $\Sigma_{c1}$
  and the $\Lambda_{c0} \, 1/2^-$ \cite{Artuso01}.
\par
Mass splittings within isospin multiplets are caused by QCD
corrections due to $m_d > m_u$ and by electromagnetic effects. Interest in this
subject has attracted new interest recently: while for all well-measured
isodoublets one increases the baryon mass by replacing a
 $u$--quark with a $d$--quark, the opposite happens in the case of the
 poorly measured  $\Sigma_c^{++}(cuu) -\Sigma_c^{0}(cdd)$ isosplit
\cite{Rosner:1998zc,Varga:1998wp}. Also, as discussed below, the SELEX
candidates for the two isodoublet $C=2$ baryons $\Xi_{cc}$ show larger
than expected mass splitting. 
FOCUS' new number $M(\Sigma_c^{++}) - M(\Sigma_c^{0}) = 
-0.03 \pm 0.28 \pm 0.11 \; {\rm MeV}$ \cite{Link:2000qs} 
indicates a trend towards a smaller
splitting, although still statistically  consistent
with both  E791 one's of $0.38 \pm 0.40 \pm 0.15\; {\rm MeV}$ as well as 
PDG02's $0.35  \pm 0.18\; {\rm MeV}$. 

There is another use of baryonic spectroscopy in a somewhat unexpected 
quarter: as explained in Sect. \ref{WEAKDK} and discussed more specifically below, 
the weak decay widths of charm baryons can be expressed through the expectation 
values of local operators. The numerically leading contributions are due to 
four-quark operators. At present we can compute those only with the help of 
wave functions obtained in quark models. Yet those wave functions allow us also 
to calculate baryon mass splittings. One can then relate the needed baryonic 
expectation values to static observables like $M(\Sigma _c) - M(\Lambda _c)$ or 
$M(\Sigma _c^*) - M(\Sigma _c)$ 
\cite{ROSNERURI}. Uraltsev, Phys. Lett. B 376 (1996) 303
\par
 Finally, measurements of the $\Sigma_c, \Sigma_c^{++}$ natural 
 widths were presented by CLEO\cite{Artuso:2001us}
 and FOCUS\cite{Link:2001ee}. They both agree on a few MeV width, but the level
 of precision is not enough yet  to discriminate among theoretical models.

\subsubsection{$C\geq 2$ baryons}
\label{C2BAR}

Combining the large charm production rates in hadronic collisions with  
state-of-the-art microvertex detectors that allow to {\em trigger} on  
charm decays opens the window to a more exotic class of hadrons,  
namely baryons containing two (or even three) charm quarks.  
There is an  
$SU(3)$ triplet of such  
states: $\Xi_{cc}^{++} = [ccu]$, $\Xi_{cc}^{+} = [ccd]$ and  
$\Omega_{cc}^{+} = [ccs]$ (plus the superheavy  
$\Omega _{ccc}=[ccc]$).
 
Like for $C=1$ baryons, one can employ quark models of various stripes to  
predict their masses. However there are some qualitative differences:  
as stated before, in the $\Lambda_c$ and $\Xi_c$ bound states a light  
spin-zero diquark surrounds the heavy $c$ quark, whose spin is decoupled  
to leading order in $1/m_c$ due to QCD's HQS. In $\Xi_{cc}$ and  
$\Omega_{cc}$ on the other hand the light degrees of freedom carry  
spin 1/2 implying degeneracy among several ground states to leading order  
in $1/m_c$. It has been suggested to model $C=2$ baryons as a heavy-light  
system consisting of a $cc$-diquark and a light quark. Accordingly there 
will be two kinds of mass spectra, namely due to excitations  of the light 
quark and of the $cc$ `core'. 
Based on such an ansatz the following  
predictions on the masses were made more than ten years ago
\cite{RICHARD}:  
\beq  
M(\Xi _{cc}) \simeq 3.61 \; {\rm GeV},\;  
M(\Omega _{cc}) \simeq 3.70 \; {\rm GeV},\;  
M(\Omega _{ccc}) \simeq 4.80 \; {\rm GeV},\;
\label{CCPRED}
\eeq
These numbers still reflect today's theoretical expectations  
\cite{KISELEVMASS,RICHARDTODAY}. Ref.\cite{KISELEVMASS} lists various  
models; their predictions are in the range 3.48 - 3.74 GeV for  
$M(\Xi_{cc})$ and 3.59 - 3.89 GeV for $M(\Omega _{cc})$.

In 2002, the SELEX Collaboration claimed the observation  
\cite{MATTSON02} of the $\Xi^+_{cc}$ (ccd)  
through its decay mode to $\Lambda_c^+K^-\pi^+$. In this experiment
charmed baryons are produced by a 600 GeV charged hyperon beam
(Tab.\ref{TAB:EXPTS}). 
Charged tracks are detected and reconstructed by a silicon vertex detector,
coupled  
to a forward magnetic spectrometer. After analyses cuts, a
sample of 1630  fully reconstructed $\Lambda_c \to p K^-\pi^+$ events is
selected.
Double charm baryon candidates are searched for in events which assign a
$\Lambda_c$
to a $K^-\pi^+$ secondary vertex.
SELEX found a 15.9 events signal over an
expected background of $6.1 \pm 0.5$ events, that they translate to a
$6.3\sigma$  
significance, at a mass of $3519\pm 1\MeV$, and a width of $3 \MeV$
compatible to the  
experimental resolution. SELEX has also shown  
\cite{RUSS02} preliminary results on an excess of 9 events (over an
expected background of 1), at  
$3460 \MeV$, which they translate to a $7.9 \sigma$ significance in  
the $\Lambda_c^+ K^- \pi^+\pi^+$ final state. This was interpreted as
evidence for
the  isodoublet partner, the $\Xi^{++}_{cc}$. 
Alternate statistical approaches have been proposed 
\cite{Rolke:2000ij}
 which treat
 differently the background fluctuations, and  provide
a signal significance for the SELEX candidates of about 3-4~$\sigma$.
The very serious trouble with this interpretation is the 
apparent $\sim 60$ MeV mass splitting
between the isospin  
partners $\Xi_{cc}^+$ and $\Xi_{cc}^{++}$: it causes a major headache for  
theorists, in particular when one keeps in mind that the proper
yardstick for comparing this mass splitting to is {\em not} the total
mass, but the much smaller binding energy of a few hundred MeV.
Furthermore as discussed above there is no evidence for such an exotic
effect in single charm baryons. 
\par
In May 2003 a very intriguing new twist has been added to the story: based on 
further studies, in particular of the angular distributions of the decay 
products, SELEX \cite{Cooper03}
now concludes that there are actually {\em four} $C=2$ baryons,
namely two $J=1/2^+$ states with $L=0$, namely $\Xi_{cc}^+(3443)$ and 
$\Xi_{cc}^{++}(3460)$ decaying isotropically into $\Lambda_c^+K^-\pi^+$
and $\Lambda_c^+K^-\pi^+\pi^+$, respectively, and a heavier pair 
$\Xi_{cc}^+(3520)$ and $\Xi_{cc}^{++}(3541)$ presumably with 
$J=1/2^-$ and $L=1$ decaying nonisotropically into the same final states. 
The problem with the isospin mass splittings mentioned above has 
been alleviated now, since the two isodoublets have mass splittings 
of 17 and 21 MeV, although this is still anomalously large. The heavier 
doublet could be understood as an excitation of the $cc$ core; preliminary 
estimates yield for this excitation energy a range of about 70 to 200 MeV and 
for the isospin mass splitting about 6 MeV
\cite{BARDEENPRIV}.

The SELEX evidence has not been confirmed by the photoproduction experiment
FOCUS\cite{Focusccq}.  In FOCUS, following analysis techniques similar to SELEX,
a
sample of 19,444 $\Lambda_c \to p K^-\pi^+$ events is selected, with neither
$\Xi_{cc}^+$ nor $\Xi_{cc}^{++}$ candidates. FOCUS concludes
(Tab.~\ref{TAB:FOCSEL}) that this implies a production difference between double
charm baryons and $\Lambda_c$ baryons of $> 42$ for the $\Xi_{cc}^+$, and $>111$
for the $\Xi_{cc}^{++}$.

\par
 
\begin{table}  
 \caption{Comparison of SELEX and FOCUS results on double-charm baryons (from
 \cite{Focusccq}). Limits are at the 90\% cl.
  \label{TAB:FOCSEL}
 }
 \footnotesize
 \begin{center}
 \begin{tabular}{|l|r|r|r|r|} \hline
                 & \multicolumn{2}{r|}{$\Xi_{cc}^{+}\to\Lambda_c^+ K^- \pi^+$}
	&  \multicolumn{2}{r|}{$\Xi_{cc}^{++}\to\Lambda_c^+ K^- \pi^+\pi^+$}
	 \\
 \hline  
                 & FOCUS            &    SELEX  &    FOCUS       &    SELEX   \\
$\Xi_{cc}$ Candidate events   &  
    $<$2.21 \%  &
    15.8    &  
    $<$2.21 \% &  
    8         \\
Reconstructed $\Lambda_c$ &
     19~500&
              1~650&
      19~500         &
         1~650      \\
$\Xi_{cc}/\Lambda_c$ Relative Efficiency &
    5\%     &
    10\%   &
    13\%   &
    5\%     \\
Relative Yield   $\Xi_{cc}/\Lambda_c$ &
   $<$0.23 \%  &
   9.6 \%       &
   $<$0.09\%     &
    9.7\% \\
Relative Production $\Xi_{cc}/\Lambda_c$  &    
  \multicolumn{2}{r|}{ SELEX/FOCUS $>$42}     &
  \multicolumn{2}{r|}{ SELEX/FOCUS $>$111}                  \\                
\hline
 \end{tabular}
  \vfill
 \end{center}  
\end{table}
\par

SELEX' interpretation of its data creates very considerable headaches for theory. 
The observed isospin splitting is larger than expected, yet the main problem concerns 
the reported lifetimes. None of the four states exhibit a finite lifetime, and the apparent 
upper bound is about 33 fs. One should note that the upper doublet can decay 
electromagnetically through $M2$ transitions, for which one can estimate a lifetime of about 1 
fs 
\cite{BARDEENPRIV}. Therefore if they indeed represent $C=2$  baryons, 
their weak lifetimes have to be indeed on the femtosecond level. 
The poses a very 
serious challenge to theory, as will be
discussed in Sect. \ref{CGEQ2BARY}. 
\par
Not unlike BELLE's observation of $J/\psi \ccb$  events \cite{Abe:2002rb} 
SELEX' findings point to unexpectedly large  $\ccb \ccb$ production. Yet  
they present another puzzle as well, namely why two charm quarks each produced 
in a hard collision presumably incoherently end up in the same $C=2$ baryon. 
One 
would expect that much more often  than not they hadronize separately leading to 
 sizeable $DD$ and $\bar D \bar D$ production. 

\subsubsection{Production of charm resonances}
\label{HCPROD}

A very naive estimate for the relative production of different charm states
is based on `spin counting', i.e. assigning equal probability to the
production of each spin component of a given charm hadron. Consider
the simplest case of vector ($V$) vs. pseudoscalar ($P$) production
like $D^*$ vs. $D$. Spin counting suggests a ratio
$r= V/(V+P) = 3/4$.  
There is
no justification (beyond its simplicity) for such an ansatz and it fails
already for continuum $D_s*$ vs. $D_s$ production in $e^+e^-$ annihilation,
where CLEO finds \cite{CLEOSPINCOUNT} $r=0.44\pm0.04$ which is less than
even equal production
for $D_s^*$ and $D_s$. In
semileptonic $B$ decays on the other hand one finds
$\Gamma (B \to \ell \nu D^*) \sim 2\Gamma (B \to \ell \nu D^)$.

With the increase in data sets, and the simultaneous refinement in the level of  
sophistication in understanding the sources of systematics, measurements
of relative production yields of L=1 states in $B$ decays have become
available.

There is a double motivation for understanding charm production in
$B$ decays -- in particular semileptonic $B$ decays -- that goes
well beyond testing hadronization models per se.
\begin{itemize}
\item
The CKM parameter $|V(cb)|$ is best extracted from
$B \to \ell \nu X_c $, $B \to \ell \nu D^*$ and $B \to \ell \nu D$. To
fully understand detection efficiencies, feed-downs etc. one has to
know the quantum numbers of the charm final states produced.
\item
The sum rules stated in Eq.(\ref{SRSLBDEC}) relate the basic  
heavy quark parameters to moments of the production rates for
$j_q=1/2$ and $j_q =3/2$ charm resonances. Those heavy quark parameters form
an essential input for the theoretical treatment of semileptonic $B$ decays,
beauty lifetimes etc. and also provide a valuable quantitative test ground
for lattice QCD.

\end{itemize}

It has been known for a long time that charm production in $B$ decays is
characterized by the dominance of broad over narrow states.  
Since the former are
usually identified with $j_q=1/2$ and the latter with $j_q=3/2$ this is again in
clear
contrast to spin counting. 
 
The BELLE paper \cite{Belle02}
   discusses the issue. Their measurements {\it show that narrow
   resonances compose
   33$\pm$4\% of the $B\to (D\pi)\pi$ decays, and 66$\pm$7\% of $B\to (D^*\pi)\pi$
   decays.}
   This trend is consistent with the excess of broad states component in
   semileptonic B decays $B\to D^{**} \ell \nu$ at LEP \cite{Buskulic97}

One should keep in mind though that these assignments $1/2$ vs. $3/2$ are
typically inferred from theory rather than the data. It was already mentioned
that the sum rules of Eq.(\ref{SRSLBDEC}) cast serious doubts on some of these
assignments.

\par
  Only a few theoretical papers have addressed the issue of relative
   production rates of L=1 states, from B decays. Neubert\cite{neubert97}
   predicts   the ratio J=2 / J=1(j=3/2)
       $R=B(B\to D_2^{*0}\pi^-)/B(B\to D_1^0(j=3/2) \pi^-) \sim 0.35$
   while BELLE measures \cite{Belle02} R=0.89$\pm$0.14; both numbers actually  
   contradict spin counting.  

 The Orsay group\cite{Leyaouanc} has developed a model for 
 describing charm production
in exclusive semileptonic $B$ decays.   
 The model predicts dominance
   of narrow states    in B semileptonic decays. It can claim reasonable
reliability, since it implements the SV and spin sum rules referred to
in Eq.(\ref{SRSLBDEC}).
\par
 As for L=1 D mesons not produced from B decays, there are no  theoretical
 predictions. The only experimental evidences here for broad states come from
 FOCUS \cite{Fabri:2000tr}, where the $D_0^*$ is observed
  relative production yields with respect to the $D^*_2$ of about 3:1.
\subsection{Weak lifetimes and semileptonic  
branching ratios of $C=1$ hadrons}  
\label{WEAKLIFE}
 
The decay rate of a charm {\em quark} provides only an order of magnitude  
estimate for the lifetimes of the weakly decaying charm {\em hadrons};  
their individual lifetimes could differ quite substantially.
 
To illustrate this point, let us look at strange quarks and hadrons.  
Very naively one would expect for a strange quark of mass 150 MeV a  
(Cabibbo suppressed) lifetime of roughly $10^{-6}\; s$. Not surprisingly,  
such a guestimate is considerably off the mark. Furthermore strange  
hadrons exhibit huge lifetime differences which are fed by
two  sources:  
\beq  
\frac{\tau (K_L)}{\tau (K_S)} \sim 600 \; , \; 
\frac{\tau (K^+)}{\tau (K_S)} \sim {\cal O}(100) \sim  
\frac{\tau (K^+)}{\tau (\Lambda )}  
\label{RATIO}
\eeq  
The first ratio is understood as the combination of (approximate)  
CP invariance in kaon decays, which forbids $K_L$ to decay into  
two pions, with the accidental fact that the $K_L$ mass is
barely above the three-pion threshold. Such an effect cannot induce a
significant lifetime difference among charm hadrons.  
For the second and third ratio in
Eq.(\ref{RATIO}) a name has been coined -- the  
$\Delta I = 1/2$ rule -- yet no conclusive dynamical explanation given.  
Since all weakly decaying strange baryons benefit from  
$\Delta I=1/2$ transitions, no large lifetime differences among them  
arise. The impact of the $\Delta I=1/2$ rule is seen directly  
in the ratio of the two major modes of $\Lambda$ decays:  
$\Gamma (\Lambda \to n \pi^0)/\Gamma (\Lambda \to p \pi^-) \simeq 1/2$.
 
Even before charm lifetimes were measured, it had been anticipated that  
as a `first' for hadronic flavours    
the lifetime of a few $\cdot 10^{-13}$ sec predicted for $c$  
{\em quarks}   provides a meaningful benchmark for the lifetimes of weakly
decaying  charm {\em hadrons} and 
the lifetime ratios for the latter
would differ relatively little from   unity, certainly much less than for
strange hadrons.  

The semileptonic branching ratios provide a complementary perspective  
onto nonperturbative hadrodynamics. The semileptonic widths of charm  
{\em mesons} are basically universal meaning that the ratios of their  
semileptonic branching ratios coincide with the ratios of their
lifetimes. Yet the semileptonic widths of charm {\em baryons} are  
expected to vary substantially meaning that the ratios of their  
semileptonic branching ratios yield information over and above  
what one can learn from the ratios of their lifetimes.
 
\subsubsection{Brief History, and Current Status of Lifetime Measurements}  
\label{HISLIFE}
 Unstable particles decay  following an exponential law $P(t)=\exp(-t/\tau)$,
  whose constant slope $\tau$ is defined as the mean decay time in the
  particle's rest frame, i.e., the lifetime.
 Charm lifetimes were expected in the range of $10^{-12}-10^{-13}$.
  In the lab frame, a particle gets time-dilated by a
  factor $\gamma=E/m$, and a measurement of the decay length $\ell$ provides
  determination of the proper decay time   $\ell=\gamma\beta c t$ for each  
decay
  event.
  The slope of the exponential distribution of decay lengths $\ell$ is the
  particle's lifetime.
  The decay is a probabilistic effect following an exponential distribution,
  therefore one needs adequate statistics, and precise measurement of decay
  length and particle momentum.
 \par 
 For an experiment at fixed target, where charm particles are produced with a
  typical average momentum of $50-60\,{\rm  GeV}$, the expected charm lifetime
  translates to decay lengths of order one centimeter. At  symmetrical  
colliders,  
where
  the charm particle is produced at rest in the lab frame, the decay length is
  very short, and one cannot determine it by directly observing the separation
  between primary (production) vertex and secondary (decay) vertex, but needs  
to
  use an impact parameter technique. For  a 
  detailed review on experimental techniques see \cite{PRO,cheung99b}.

The space resolution at fixed target in the transverse plan is often of a few  
microns,
  which translates on a resolution in reconstructing the decay vertex which does
  depend on the charm particle momentum, and it is typically of order
  10 microns in x,y and 300 microns in z.  When coupled to a forward magnetic
  spectrometer with good momentum resolution, the resolution on the decay  
time is of order 30-50~fs.  
  At collider experiments the decay length is very small, due to the lack of
  substancial Lorentz boost. High-resolution drift chambers are used for  
vertex
  detection, or, recently, microstrip arrays deployed in $4\pi$ geometry. 
  The primary vertex is 
detected as the beam spot in the  
interaction
  region. Tracks of secondaries are reconstructed as helices in  
$(\rho,\varphi)$
  plane. Distribution of proper decay times is not an exponential as in the
  fixed target case, but a gaussian with an right-hand exponential tail, due  
to
  the relatively poorer space resolution. Proper time resolution, thanks to
  better momentum resolution, of  state of the art $\epem$ collider  
experiments is now comparable to fixed target experiments. 
 \par
 The output signals from the vertex detector are used in track- and
 vertex-finding algorithms. The principal approach for  vertex finding
 is the candidate-driven algorithm. \index{candidate-driven algorithm}
 It consists of determining a set of tracks which have been  
particle-identified
 and that are compatible with a charm decay topology.
  Tracks are  
requested
 to be compatible as coming from a common decay vertex.
 The reconstructed charm particle momentum is
 then projected  to the
 primary vertex, thus determining the primary-secondary separation
 $\ell$ and error $\sigma_\ell$. 
At $\epem$ (Ref.\cite{Abe:2001ed},\cite{Pompili:2002jy}),
 candidate tracks are fit to a
 common vertex, the reconstructed momentum is projected to the
 interaction region to obtain the decay length.
\par
A selection cut which requires the
 presence of a parent $D^*$ for $D^0$ lifetimes ({\it $D^*$-tagging})
 may be used, albeit at the cost of a reduction in statistics. To
 cope with the reduced
 reconstruction efficiency at small decay lengths, one uses the
 reduced proper time
 $t^\prime \equiv (\ell - N \sigma_\ell)/\beta \gamma  c$
 where $N$ is the primary-secondary detachment cut applied.
 Using $t^\prime$ instead of $t$ corresponds to starting the clock
 for each event at a fixed detachment significance, and thus the
 distribution of $t^\prime$ recovers exponential behaviour.
 Additional bonus is given by the correction function
 determined by montecarlo simulation which, when espressed in terms of 
 $t^\prime$, provides very small corrections thus reducing the contribution to
 the systematic error \cite{PEDRINIPRIV}.
 \par
 Main sources of systematics at fixed-target are absorption
 of both secondary tracks and charm in target, knowledge of D
 momentum, backgrounds, and montecarlo event sample size. For $\epem$
 colliders,
 the determination of the decay vertex, beam spot, knowledge of D
 momentum, time-mass correlations, large $t$ outlier events,
 decay length bias, backgrounds, and montecarlo event sample
 size.
\par
 New results  are shown in Tab.\ref{tab:life},  with updated world
 averages with respect to PDG02. Lifetimes ratios significant for
 comparison to theoretical predictions are listed in
 Tab.\ref{TABLECHARM}.
 While the accuracy on the lifetimes of long-lived mesons is now
 at the level of the percent, essentially systematics-dominated, the
 measurement of very short-lived
 charm states, such as the
 $\Omega_C$, still poses relevant challenges. In this case the
 superior decay times resolution of fixed-target experiments (of order
 30~fs), although comparable to the lifetime itself, does allow
 lifetime determination at the level of 15\%.
\begin{table}
 \caption{Summary of world averages from \cite{Hagiwara:fs}, new results, and
 updated world averages. Statistical and systematical errors are summed in
 quadrature. 
  \label{tab:life}
 }
 \footnotesize
 \begin{center}
 \begin{tabular}{|c|l|l|c|c|c|} \hline
 & Experiment & Lifetime (fs) & Events & Channels & Techn. \\
\hline
\hline
 {\bf  $D^+$} &
 New Average & $1045\pm 8$                    &      &   &       \\
             &
 FOCUS \cite{Link:2002bx}
      & $1039.4 \pm 8 $ & 110k & $K2\pi$ & $\gamma N$   \\
             &
 BELLE \cite{belletanaka01}prel & $ 1037 \pm 13 $ & 8k & $K2\pi$ & $\epem$   \\
             &
 PDG02       & $1051 \pm 13$    &      &    &  \\
 \hline  
{\bf  $D^0$} &
 New Average & $410.6\pm1.3$                    &      &    &  \\
             &
 FOCUS \cite{Link:2002bx}
      & $409.6 \pm 1.5$  & 210k  & $K\pi,K3\pi$ &  $\gamma N$ \\
             &
 BELLE\cite{YABSLEY03,Abe:2003ys} prel & $412.6 \pm 1.1 (st)$   & 448k  &  $K\pi$ &  $e^+e^-$  
\\
             &
 PDG02       & $411.7 \pm 2.7$       &      &    &  \\  
\hline
 {\bf  $D^+_s$} &
 New Average & $494\pm 5$                    &      &    &  \\
             &
 FOCUS \cite{cheung99} prel  & $506 \pm 8(st)$       & 6k & $\phi\pi$ &
 $\gamma N$  \\
             &
 BELLE \cite{belletanaka01} prel  & $485 \pm 9 $ & 6k & $\phi\pi$ & $\epem$ 
 \\ 
             &
 PDG02       & $ 490 \pm 9$                    &      &    &       \\  
\hline
 {\bf  $\Lambda_c^+$} &
  PDG02       & $ 200 \pm 6$                   &      &    &       \\
\hline
 {\bf  $\Xi_c^+$} &
 PDG02       & $ 442 \pm 26$                   &      &    &       \\
\hline
 {\bf  $\Xi_c^0$} &
 New Average  & $108 \pm 15$                   &      &    &       \\
              &
 FOCUS  \cite{Link:2002xu}      & $118\pm 14$        & $110\pm17$  &
 $\Xi^-\pi^+,\Omega^-K^+$ &  
$\gamma N$ \\
              &
 PDG02        & $ 98  ^{+23}_{-15}$             &      &    &      \\
\hline
 {\bf  $\Omega_c^0$} &
 New Average  & $76 \pm 11$                    &      &    &       \\
              &
 FOCUS \cite{Link:2003nq}   & $79\pm 15 $                & 64 $\pm$ 14  & 
 $\Omega^-\pi^+,\Xi^-K^-\,2\pi^+$ &
 $\gamma N$                        \\
              &
 PDG02       & $ 64  \pm 20$                   &      &    &       \\
 \hline  
\end{tabular}
  \vfill
 \end{center}  
\end{table}  

\subsubsection{Early phenomenology}  
 
\begin{figure}
        \centering
        \epsfysize=2.0cm
      \epsffile{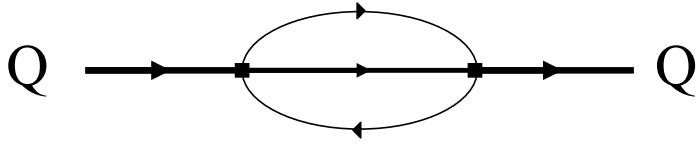}
 \caption{The diagram describing the weak decay of the charm quark.  This
spectator process contributes to the width of all charmed hadrons.
    \label{FIG:SPECT} }
\end{figure}

All charm hadrons share one contribution, namely the weak decay of the  
charm quark, which to leading order is not modified by its hadronic  
environment, see Fig.~\ref{FIG:SPECT}. It is often called the spectator
process, since  
the other partons present in the hadron (antiquarks, quarks and gluons)
remain passive bystanders  
\footnote{The reader should be warned that some authors use the term  
`spectator contribution' for processes where the  
`spectators' become active.}. This reaction contributes to all  
charm hadrons equally scaling like  
\beq  
\Gamma _{Spect} \propto G_F^2 m_c^5 
\label{SPECTWIDTH} 
\eeq  
Originally it had been expected that this term dominates the lifetime  
already for charm hadrons implying a small difference between    
$\tau (D^0)$ and $\tau (D^+)$. It caused quite a stir in the  
community when the lifetime measurements showed the $D^+$ to be  
longer lived than the $D^0$ by a considerable factor 
\cite{NUSS,STIR}. It enhanced the  
drama that the first data `overshot the target', i.e. yielded  
$\tau (D^+)/\tau (D^0) \sim 5$ before `retreating' to a ratio of  
$\sim 2.5$.  
This surprise caused considerable activities on the  
theory side to accommodate the data and make new predictions for
the other lifetime ratios.

\begin{figure}
        \centering
        \epsfysize=2.0cm
      \epsffile{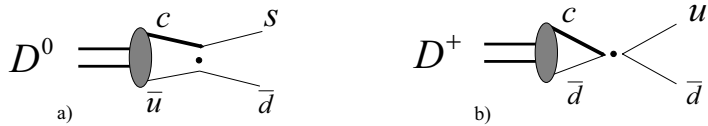}
 
\caption{a) The Cabibbo-favoured weak exchange contribution to the $D^0$ width.
b) The Cabibbo-suppressed weak annihilation contribution to the  $D^+$ width.
    \label{FIG:DDecays} }
\end{figure}

\begin{figure}
        \centering
        \epsfysize=2.5cm
    \epsffile{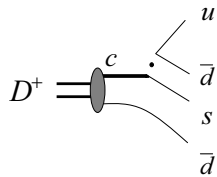}
 \caption{The presence of two identical quarks in the final state of the $D^+$
decay leads to Pauli Interference.
    \label{FIG:PauliInt} }
\end{figure}

Two mechanisms were quickly put forward as to induce  
$\tau (D^+)/\tau (D^0) > 1$:  
\begin{enumerate}
\item  
On the Cabibbo-favoured level $W$-exchange can contribute to $D^0$,  
as shown in Fig.~\ref{FIG:DDecays}~a), but not to $D^+$ transitions. The
latter are
affected only on the Cabibbo suppressed level by $W$ exchange in the  
s-channel, Fig.~\ref{FIG:DDecays}~b).  
\item  
In the reaction $D^+ =[c \bar d] \to s \bar d u \bar d$ one has two  
identical quark flavours in the final state, see Fig.~\ref{FIG:PauliInt}.
Interference
effects thus have to be included, which turn out to be destructive for
nontrivial reasons. This effect is  referred to as Pauli Interference
(PI) \index{Pauli Interference}.
 
\end{enumerate}  
The first mechanism had been known all along, yet its contributions had
been discarded for a reason. For it suffers from two suppression factors,  
namely helicity suppression and wavefunction suppression:  
(i) With spin-one couplings conserving chirality, a pseudoscalar meson  
cannot decay into a massless fermion-antifermion pair. Thus the  
{\em amplitude}  
for $W$-exchange is proportional to the mass of the heaviest quark in the  
{\em final} state: $T(D^0 \to s \bar d) \propto m_s/M_D$ 
\index{helicity suppression}. This is a  
repetition of the well known tale why  
$\Gamma (\pi^+ \to \mu ^+ \nu) \gg \Gamma (\pi^+ \to e^+ \nu)$ holds. 
(ii) Due to the almost zero range of the weak force the $c$ and $\bar u$  
quark wavefunctions have to overlap to exchange a $W$ boson. The decay
constant
$f_D$  provides a measure for this overlap:  
$f_D \simeq \sqrt{12}|\psi _{c\bar u}(0)|/\sqrt{M_D}$. Accordingly  
$T(D^0 \to s \bar d) \propto f_D/M_D$.  

Putting everything together one obtains  
$\Gamma _{WX}(D^0) \propto G_F^2|f_D|^2 m_s^2m_D$   
rather than $\Gamma _{Spect} \propto G_F^2 m_c^5 $.   
Since $m_s, f_D \ll m_c$ such a W-exchange contribution is very small.
Yet after the data revealed a large lifetime ratio, various mechanisms  
were suggested that might vitiate helicity suppression and overcome  
wavefunction suppression. Most of them were quite ad-hoc. One that
appeared natural was to invoke gluon emission from the initial light  
(anti)quark line leading to  
\beq  
\Gamma _{WX}(D^0 \to s \bar d g) \propto  
\frac{\alpha_S}{\pi}\left( \frac{f_D}{\langle E_{\bar u}\rangle}
\right)^2 G_F^2 m_c^5
\label{SONI}
\eeq  
with $\langle E_{\bar u}\rangle$ denoting the average energy of  
$\bar u$ inside the $D^0$. Despite the double penalty one pays here --  
it is a term $\sim {\cal O}(\alpha_S/\pi)$ controlled by three-body  
rather than two-body phase space -- such a contribution could be sizeable
for  
$\langle E_{\bar u}\rangle \sim f_D$.


A distinction had been made between $W$ {\em exchange} and  
{\em weak annihilation} with the weak boson being exchanged in the  
$t$ and $s$ channels, respectively. Yet since QCD renormalization  
mixes those two diagrams, it is more meaningful to lump both of them  
together under the term weak annihilation (WA) as far as  
{\em meson} decays are concerned; $W$ exchange in {\em baryon} decays,  
which is not helicity suppressed, is denoted by $W$ scattering (WS) 
\index{weak annihilation, $W$ scattering}.
 
While the occurrence of PI in exclusive $D^+$ decays had been noted,  
its relevance for inclusive $D^+$ rates had been discarded for various
reasons. Yet in a seminal paper \cite{NUSS} it was put forward as the  
source of the observed lifetime ratio.
 
Both mechanisms can raise $\tau(D^+)/\tau(D^0)$ considerably above  
unity, yet in different ways and with different consequences:  
(i) While PI reduces $\Gamma (D^+)$ without touching $\Gamma (D^0)$,  
WA was introduced to enhance $\Gamma (D^0)$ while hardly affecting  
$\Gamma (D^+)$. As a consequence, since $\Gamma_{SL}(D^+) \simeq
\Gamma_{SL}(D^0)$ due to isospin invariance, PI will enhance
BR$_{SL}(D^+)$ while not changing   BR$_{SL}(D^0)$, whereas WA will
decrease BR$_{SL}(D^0)$ while hardly   affecting BR$_{SL}(D^+)$.  
(ii) For PI one expects $\tau (D^0) \sim \tau (D_s)$, whereas WA should
induce   a difference in $\tau (D_s)$ vs. $\tau (D^0)$ roughly similar
to $\tau (D^+)$ vs. $\tau (D^0)$.  
(iii) One would expect to see different footprints of what is driving
the   lifetime differences in the weight of different {\em exclusive}
modes   like $D_s \to \pi's$ vs. $D_s \to \phi + \pi's$ vs.  
$D_s \to K \bar K + \pi's$.

Weak baryon decays provide a rich laboratory for these effects:  WS is
not helicity suppressed in baryon decays, and one can count
  on it making significant contributions here. Furthermore PI can be
constructive as well as destructive, and its weight is
  more stable under QCD radiative corrections than is the case for meson
  decays. On fairly general grounds a hierarchy is predicted 
\cite{GUBERINA}:
 \beq
 \tau (\Omega_c) < \tau (\Xi_c^0) < \tau (\Lambda_c) <
 \tau (\Xi_c^+)
\label{BARHIER}
 \eeq

It should be noted that all these analyses invoked -- usually  
implicitly -- the assumption that a valence quark description provides  
a good approximation for computing such transition rates. For if these  
hadrons contained large `sea' components, they would all share the same  
basic reactions, albeit in somewhat different mixtures; this would  
greatly dilute the weight of processes specific to a given hadron and
thus even out lifetime differences.
 
 These phenomenological treatments laid important groundwork, in  
particular in exploring various possibilities. Yet there were some  
serious shortcomings as well: in particular there was no agreement  
on the size of WA contributions in $D$ decays, i.e. to which degree  
the helicity suppression could be overcome; even their scaling with  
$m_c$ was controversial. One should also be quite surprised -- actually
mystified -- by  Eq.(\ref{SONI}): for one would expect an inclusive  
width to be described by short-distance dynamics, 
which Eq.(\ref{SONI})  
manifestly is not due to the low scale $\langle E_{\bar u}\rangle$  
appearing in the denominator. Lastly, as we will explain below,  
those treatments overlooked one fact of considerable significance  
concerning semileptonic branching ratios.

\subsubsection{The HQE description}
\label{HQEDESCR}
\begin{figure}[t]
        \centering
        \epsfysize=2.0cm
      \epsffile{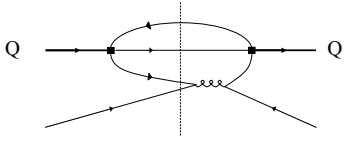}
 \caption{The interference of the spectator and WA diagrams that must be
included for fully inclusive transitions.
    \label{FIG:SPECTWA} }
\end{figure}

The HQE implemented through the OPE can overcome the shortcomings  
inherent in the phenomenological models.  
Yet before  
it could be fully developed, the just mentioned apparent paradox  
posed by the expression for  
$\Gamma _{WX}(D^0 \to s \bar d g)$, Eq.(\ref{SONI}) had to be resolved.  
This was achieved in Ref.\cite{MIRAGE}.  For a truly  
inclusive transition to order $\alpha_S$ one had to include also  
the interference between the spectator and WA diagrams, where the  
latter contains an off-shell gluon going into a $q\bar q$ pair,  
see Fig.(~\ref{FIG:SPECTWA}).  
It was shown that when one sums over all contributions  
through order $\alpha_S$, the small energy denominators  
$1/\langle E_{\bar u}\rangle^2$ and $1/\langle E_{\bar u}\rangle$  
disappear due to cancellations among the different diagrams, as it  
has to on general grounds. This is just another `toy model' example  
that while fully inclusive rates are short-distance dominated,  
partially inclusive ones -- let alone exclusive ones -- are not.

The basic method of the HQE has been described in Sect.(\ref{OPE}); it
yields:  
\bea
  \Gamma (H_Q\rarr f) & = & \frac{G_F^2 m_Q^5}{192\, \pi^3} |KM|^2    
\Bigl[ A_0 + \frac{A_2}{m^2_Q} + \frac{A_3}{m^3_Q} + {\cal O}(1/m_Q^4)
\Bigr]
\label{eq:bi96}  
\eea
The quantities $A_n$ contain the phase space factors as appropriate for  
the final state, the  
QCD radiative corrections and the short-distance coefficients appearing  
in the OPE and  
$\matel{H_c}{{\cal O}_n}{H_c}$,  
the hadronic expectation values of local operators ${\cal O}_n$  
of dimension $n+3$. They scale like  
$\mu ^n$, where $\mu$ denotes an ordinary hadronic scale  
close to and hopefully below 1 GeV.    
Each term has a transparent physical meaning;  
let us stress those features that are particularly relevant for
lifetimes and semileptonic  branching ratios.  
\begin{itemize}     
\item  
In all cases one has to use the same value of the  
charm quark mass $m_Q$ properly defined in a field theoretical sense.
Furthermore this value is in principle  not a free parameter to be fitted
to the data on lifetimes, but should be  inferred from other
observables. Choosing different values for $m_Q$ when describing, say,
meson and baryon  lifetimes can serve merely as a {\em temporary crutch}
to parameterize an  observed difference between mesons and baryons one does
not understand at all.
 
 When using quark models to evaluate these  
expectation values, charm quark masses will also enter through the quark
wave functions. Yet those are like `constituent' masses, i.e. parameters  
specific to the model used, {\em not} quantities in the QCD Lagrangian;  
therefore they can be adjusted for the task at hand.
 
\item  
The leading term $A_0$ represents the spectator diagram contribution
common to all hadrons, see Fig.(~\ref{FIG:SPECT}).  
\item  
The {\em leading} nonperturbative corrections    
arise at order $1/m_Q^2$. $A_2$ reflects the motion of the heavy quark  
inside the hadron and  
its spin interaction with the light quark degrees of
freedom, see Fig.(~\ref{FIG:GLUON}). This latter effect had not been anticipated
in the
phenomenological descriptions of the 1980's.

\begin{figure}
        \centering
        \epsfysize=2.0cm
      \epsffile{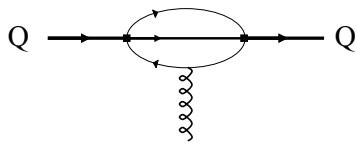}
 \caption{The spin interaction of the heavy quark with the light degrees of
freedom, a leading nonperturbative effect.
    \label{FIG:GLUON} }
\end{figure}
 
These terms in general differentiate  
between baryons on one side and mesons on the other, yet have  
practically the same impact on all mesons. However the contribution
proportional to $\mu_G^2$ due to the aforementioned spin interaction 
{\em unequivocally enhance}   the nonleptonic width over the semileptonic
one and thus reduce   the  semileptonic branching ratios of $D$ mesons.  
This does not happen in $\Lambda_c$ decays; for  
$\mu_G^2(\Lambda_c)\simeq 0$, since the light diquark system carries  
spin zero there.

\item  
Pauli Interference (PI) \cite{NUSS},  
Weak Annihilation (WA) for mesons and W-scattering (WS) for
baryons arise unambiguously and naturally in order  
$1/m_Q^3$ with WA being helicity  
suppressed and/or nonfactorisable\cite{MIRAGE}. They mainly
drive the differences in the lifetimes of the various hadrons of a given
heavy flavour.
\item  
For the {\em lifetime ratios} effects of order $1/m_Q^3$ rather than
$1/m_Q^2$ are numerically dominant. There are two reasons for that,  
one straightforward and one more subtle: (i) As illustrated by
Figs.(~\ref{FIG:FOURFER}~a-d)
the ${\cal O}(1/m_Q^3)$ contributions from four-fermion operators  
involve integrating over only two partons in the final state rather  
than the three as for the decay contribution, Fig.(~\ref{FIG:SPECT}~a).
$A_3$ is  
therefore enhanced relative to $A_{0,2}$ by a phase space factor;  
however the latter amounts effectively to considerably less than the  
often quoted $16\pi^2$. (ii) As  explained in Sect. \ref{OPE}   
there are no ${\cal O}(1/m_Q)$ contributions. One can  
actually see that there are various sources for such contributions,  
but that they cancel exactly between initial and final state  
radiation. Such cancellations still arise for ${\cal O}(1/m_Q^2)$, which
are thus reduced relative to their `natural' scale. In  
${\cal O}(1/m_Q^3)$ etc., however, they have become ineffective.  
The latter are thus of normal size, whereas the $1/m_Q^2$ contributions
are `anomalously' small. Accordingly there is no reason to suspect  
$1/m_Q^4$ terms to be larger than $1/m_Q^3$ ones.
\item  
The HQE is better equipped to predict the {\em ratios} of lifetimes,
rather than lifetimes themselves. For the (leading term in the) full  
width scales with $m_Q^5$, whereas the numerically leading contributions  
generating lifetime differences are due to dimension-six operators  
and scale with $m_Q^2$, i.e. are of order $1/m_Q^3$. Uncertainties in the  
value of $m_Q$ thus affect lifetime ratios much less. 
\end{itemize}

\subsubsection{Theoretical interpretation of the lifetime ratios}  
\label{THINTRATIOS}
 
As already stated the three weakly decaying mesons $D^+$, $D_s^+$  
and $D^0$ receive identical contributions from the leading term $A_0$  
in Eq.(\ref{eq:bi96})  
\footnote{One should note that for inclusive transitions the distinction
of {\em internal} vs. {\em external} spectator  diagrams makes little
sense, since in fully inclusive processes -- in contrast to exclusive  
channels -- one does not specify which  
quarks end up in the same hadron.}.  
This is largely true also for $A_2$ although  
less obvious \cite{DSPAPER}. Yet to order $1/m_Q^3$ their lifetimes  
get differentiated: on the Cabibbo favoured level PI contributes to  
$D^+$ and WA to $D^0$ and $D^+_s$ decays. Yet a careful HQE analysis  
reveals that the WA contributions are helicity suppressed and/or  
suppressed due to being nonfactorizable etc. \cite{MIRAGE}. Thus one  
expects approximate equality between the $D^0$ and $D^+_s$ lifetimes:  
$\Gamma (D^0) \simeq \Gamma_{\rm spect}(D) \simeq \Gamma (D_s^+)$. 
In the $D^+$ width on the other hand PI occurs due to interference  
between two $\bar d$ quark fields, one from the wavefunction, while the  
second one emerges from the decay. A priori there is no reason  
for this effect to be small. More specifically one finds  
\beq  
\Gamma (D^+) \simeq \Gamma_{\rm spect}(D) + \Delta \Gamma _{PI}(D^+)  
\eeq
$$
\Delta \Gamma _{PI}(D^+) \simeq \Gamma_0\cdot 24\pi^2
\left( f_D^2/m_c^2\right) \kappa ^{-4} \cdot  
$$
\beq  
\cdot \left[  
(c_+^2 - c_-^2)\kappa^{9/2} + \frac{1}{N_C}(c_+^2 + c_-^2)  
- \frac{1}{9}(\kappa^{9/2} - 1)(c_+^2 - c_-^2)
\right]  
\eeq
where  
\beq  
\kappa = [\alpha_S(\mu_{had}^2)/\alpha_S(m_c^2)]^{1/b}\; ,  
b = 11 -\frac{2}{3}N_f
\eeq
reflects hybrid renormalization mentioned in Sect.(\ref{EFFWEAK}).  
A few comments are  
in order:    
without QCD corrections one has $c_- = c_+=1=\kappa$ and thus  
$\Delta \Gamma _{PI}(D^+) > 0$, i.e.  
{\em constructive} interference meaning $\tau (D^+) < \tau
(D^0)$; including UV renormalization flips the {\em sign} of  
$\Delta \Gamma _{PI}(D^+)$ and hybrid renormalization makes this effect  
quite robust.  

One arrives then at  
\beq  
\frac{\tau (D^+)}{\tau (D^0)} \simeq 1 + (f_D/200\; \MeV)^2  
\simeq 2.4
\eeq  
for $f_D \simeq 240$ MeV, see Eq.(\ref{FDTHEO}); i.e., PI is capable of
reproducing  the observed lifetime ratio by itself even without WA. Of
course this has to be taken as a semi quantitative statement  only, since
we cannot claim (yet) precise knowledge of $f_D$,  the size of the {\em
non}factorizable contributions in the  expectation value for the
four-fermion operator or of the ${\cal O}(1/m_c^4)$ contributions.
 
Next one has to compare $\tau (D^+_s)$ and $\tau (D^0)$.
The statement underlying
$\Gamma (D^0) \simeq \Gamma_{\rm spect}(D) \simeq
 \Gamma(D_s^+)$ is
actually that  
$\frac{\tau
(D_s^+)}{\tau (D^0)} \ll \frac{\tau (D^+)}{\tau (D^0)}  
\simeq 2.4$.   
It is nontrivial since one would not expect it to hold if WA were the  
main effect generating the lifetime differences among charm mesons.
 
A first milestone was reached with the E687 measurement:  
\beq  
\frac{\tau(D^+_s)}{\tau (D^0)}= 1.12 \pm 0.04 \; .  
\eeq
It provided significant, though not conclusive evidence that  
$\tau (D^+_s)$ exceeds $\tau (D^0)$. Even more importantly it  
clearly confirmed the prediction of WA being suppressed.

\begin{figure}
        \centering
        \epsfysize=5.0cm
      \epsffile{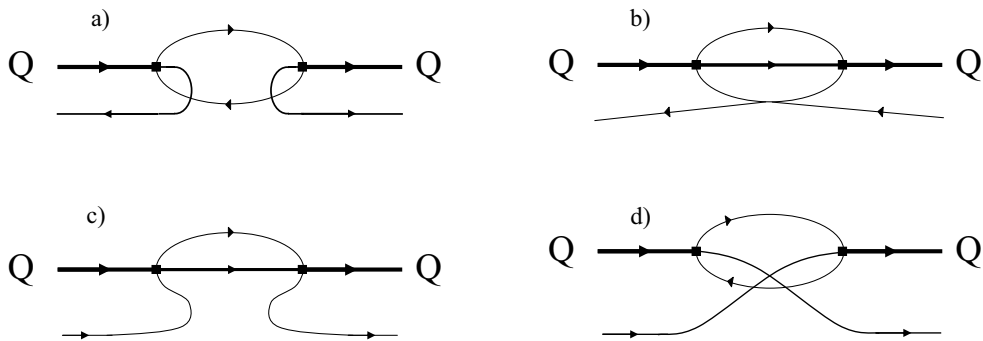}
 \caption{Cutting an internal quark line generates order $1/m_Q^3$ four fermion
operators. a) Weak annihilation in mesonic decays.  d) Pauli Interference in the
decays of a meson.  c)  Weak scattering in baryonic decays.  d)  Pauli
Interference in the decays of a baryon.
    \label{FIG:FOURFER} }
\end{figure}

It has been  
estimated \cite{DSPAPER} that even {\em without} 
WA $D_s$ can actually be
longer   lived than $D^0$, yet by a small amount only due to a
combination   of various ${\cal O}($~\%~$)$
 effects like PI in Cabibbo  
suppressed transitions:  
\beq  
\left. \frac{\tau(D^+_s)}{\tau (D^0)}\right|_{no\; WA} \simeq 1.0 \div 1.07
\eeq  
Furthermore it was stated that WA can modify this ratio  
by about 20\% in {\em either} direction  \cite{DSPAPER}.
 
Data have conclusively  
confirmed that $\tau (D^+_s)$ is moderately longer than $\tau (D^0)$.    
\beq  
\langle \tau(D^+_s)/\tau (D^0)\rangle = 1.22 \pm 0.02
\eeq
 
In summary: the case for PI being the dominant mechanism  
driving lifetime differences among $D$ mesons rests on the following    
facts and observations:  
\begin{enumerate}
\item  
A careful analysis of the HQE and of the  
expectation values of four-quark operators shows that WA contributions are
either helicity suppressed or non-factorizable  and thus suppressed.
Accordingly they are too small for being the  leading effect. At present
this is a purely theoretical argument  although it can be checked in the
future through measurements  of the lepton endpoint spectrum in
semileptonic $B$ or $D$ decays \cite{WA}.  
\item  
The measured $D^+-D^0$ lifetime ratio can be reproduced.  
\item  
The observation of  
\beq  
0.07 \ll |1-\tau(D^+_s)/\tau (D^0)| \ll |1-\tau(D^+)/\tau (D^0)|
\eeq
confirms that WA is nonleading, yet still  
significant. For if WA were the dominant source for the  
$D^+-D^0$ lifetime ratio, one would expect  
$\tau (D^+_s)/\tau (D^0)$ to deviate by a similar amount from  
unity, which is however clearly not the case.  
\end{enumerate}
The observed value for $\tau (D_s)/\tau (D^0)$ is still within range  
of the general estimate  
of what can be accommodated with a modest, yet significant contribution  
from WA \cite{DSPAPER,WA}.
 
It has been suggested recently that there is no need for invoking  
PI and WA to reproduce $\tau(D_s)/\tau(D^0)$ \cite{NUSS2}.  
The authors suggest the following simple minded prescription:  
they pair up all exclusive $D^0$ channels with their $D_s^+$  
counterparts and compute the strength of the latter by taking the  
observed strength of the former and applying simple phase space  
corrections; then they add up all these individual rates and arrive at  
$\tau (D_s)/\tau(D^0) \simeq 1.17$. This is not an unambiguous  
prescription, of course, since the appropriate phase space for  
multibody final states depends on the detailed dynamics of those  
final states. Yet the main problem with this `explanation' is that the  
point is {\em not} whether one can relate classes of hadronic  
decay channels to each other approximately with the help of simple  
prescription. The central theoretical question is whether  
a quark-based, i.e. short distance  treatment  
genuinely inferred from QCD provides an adequate
description of inclusive transitions involving hadrons, which also
addresses the issue of  
quark-hadron  duality and its limitations.  This question  
is not even addressed by an ad-hoc ansatz involving individual hadronic  
modes.
 
Isospin symmetry tells us that the  
semileptonic widths of $D^+$ and $D^+_s$ mesons have to  
be equal up to terms $\sim {\cal O}(\theta_C^2)$;  
therefore the ratio of their semileptonic branching ratios has to  
agree with their lifetime ratio. This is well borne out by the  
data:  
\beq  
\frac{{\rm BR}_{SL}(D^+)}{{\rm BR}_{SL}(D^0)}= 2.50 \pm 0.27  \; vs. \;  
\frac{\tau (D^+)}{\tau(D^0)} = 2.54 \pm 0.01
\eeq  
When one considers the absolute values of these branching ratios  
\beq  
{\rm BR}_{SL}(D^+) = 17.2 \pm 1.9 \% \; , \; 
{\rm BR}_{SL}(D^0) = 6.87 \pm 0.28 \% 
\eeq
there seems to be a fly in the ointment for PI being the main  
reason for the lifetime difference. As already mentioned by reducing  
the $D^+$ nonleptonic width PI {\em enhances}  
BR$_{SL}(D^+)$, while leaving $D^0$ widths largely unaffected.  
WA on the other hand is invoked to enhance the $D^0$  
nonleptonic width and thus {\em reduces} BR$_{SL}(D^0)$.  
Then it comes down to
the  question what one considers to be the `normal' semileptonic $D$
width,  i.e. before hadronization effects differentiating between  
$D^+$ and $D^0$ are included.
 
In the phenomenological models one infers this value from the  
decays of quasifree charm quarks:  
\beq  
{\rm BR}_{SL}(D) \simeq {\rm BR}(c \to \ell \nu s) \sim 16\%
\label{SLBRPARTON}
\eeq
Comparing this expectation with the data would strongly point to  
WA as the dominant mechanism for the lifetime difference.
 
Yet the HQE introduces a new and intriguing twist here:  
as mentioned before in order $1/m_c^2$ it reduces the  
semileptonic width common to $D^+$ and $D^0$ through the  
chromomagnetic moment $\mu_G^2$:  
\beq  
{\rm BR}_{SL}(D^+) \simeq {\rm BR}_{SL}(D^0) +  
{\cal O}(1/m_c^3) \sim 8\%  \; .  
\eeq
The actual numbers have to be taken with quite a grain of salt,  
of course. Yet this effect -- which had been overlooked in the  
original phenomenological analyses -- makes  
the findings of PI being the main engine consistent with the absolute  
values measured for the semileptonic branching ratios.
 
$SU(3)$ symmetry by itself would allow for some still sizeable  
difference in the semileptonic widths for $D_s$ and $D^0$ mesons.  
Yet the HQE yields that $\Gamma_{SL}(D_s)$ and $\Gamma_{SL}(D^0)$  
agree to within a few percent. Therefore one predicts  
\beq  
{\rm BR}_{SL}(D_s) = {\rm BR}_{SL}(D^0) \cdot  
\frac{\tau (D_s)}{\tau (D^0)} \sim 1.2 \cdot {\rm BR}_{SL}(D^0)
\eeq

A detailed analysis of charm baryon lifetimes is not a  
`deja vu all over again'. While it constitutes a complex laboratory  
to study hadronization, it will yield novel lessons:   
(i) There are four weakly decaying baryons: $\Lambda_c$,  
$\Xi_c^{(0,+)}$ and $\Omega_c$. Since we can trust HQE in charm decays on  
a semi quantitative level only, it makes a considerable  
difference whether one can reproduce the pattern of seven rather  
than three hadronic lifetimes.  
(ii) The theoretical challenge is considerably larger here since there  
are more effects driving lifetime differences in order  
$1/m_Q^3$: WS contributions
are certainly not helicity suppressed -- they are actually enhanced by  
QCD radiative corrections and by two-body over three-body phase space;  
also PI can be  constructive as well as destructive. Details can be  
found in Ref.\cite{PRO}.  
(iii) While the light diquark system forms a scalar in
$\Lambda_c$ and  
$\Xi_c$, it carries spin one in $\Omega_c$.  
(iv) The semileptonic widths are
expected to be strongly modified   by constructive PI rather than being
universal.  
(v) There is no unequivocal concept of factorization for the  
{\em baryonic} expectation values of four-fermion-operators.  
Accordingly we have to depend on quark model calculations of those  
matrix elements more than it is the case for mesons. One can entertain
the hope that lattice QCD will provide a reliable handle on those
quantities -- yet that would require unquenched studies.  
The one saving grace is that the wave functions employed can be  
tested by examining whether they can reproduce the observed mass  
splittings among charm baryons and their resonances. 
 
The $\Lambda_c$, $\Xi_c$ and $\Omega_c$ nonleptonic and  
semileptonic widths receive significantly
different  contributions in order $1/m_c^3$ from WS and constructive as
well as destructive PI; for details see the review \cite{PRO}. The 
hierarchy stated in Eq.(\ref{BARHIER}) arises very naturally. 
 
When studying {\em quantitative} predictions we have to be aware of  
the following complexity:  
Since the three {\em non-universal} contributions WS, constructive  
and destructive PI are of comparable size, the considerable  
uncertainties in the individual contributions get magnified when  
cancellations occur between $\Gamma_{WS}$ \&  
$\Delta \Gamma_{PI,+}$ on one side and $\Delta \Gamma _{PI,-}$  
on the other.

\begin{table}
\begin{tabular} {|l|l|l|l|}
\hline
 & $1/m_c$ expect. \cite{PRO}& theory comments & data \\  
\hline  
\hline  
$\frac{\tau (D^+)}{\tau (D^0)}$ &  
$\sim 1+\left( \frac{f_D}{200\; \MeV} \right)^2 \sim 2.4$  
& PI dominant               & $2.54 \pm 0.01$  \\
\hline  
$\frac{\tau (D_s^+)}{\tau (D^0)}$ & 1.0 - 1.07 & {\em without} WA  
\cite{DSPAPER}& \\
 & 0.9 - 1.3 & {\em with} WA \cite{DSPAPER} & $1.22 \pm 0.02$ \\
\hline  
$\frac{\tau (\Lambda _c^+)}{\tau (D^0)}$ & $\sim 0.5$  
& quark model matrix elements          & $0.49 \pm 0.01$ \\  
\hline  
$\frac{\tau (\Xi _c^+)}{\tau (\Lambda _c^+)}$ & $\sim 1.3 \div 1.7$ &  
ditto                                  &  $2.2 \pm 0.1$\\
\hline  
$\frac{\tau (\Lambda _c^+)}{\tau (\Xi _c^0)}$ & $\sim 1.6 \div 2.2$ &  
ditto                                  & $2.0 \pm 0.4$ \\
\hline
$\frac{\tau (\Xi _c^+)}{\tau (\Xi _c^0)}$ & $\sim 2.8$ &  
ditto                                  & $4.5 \pm 0.9$ \\
\hline  
$\frac{\tau (\Xi _c^+)}{\tau (\Omega _c)}$ & $\sim 4$ &  
ditto                                  & $5.8 \pm 0.9$ \\
\hline  
$\frac{\tau (\Xi _c^0)}{\tau (\Omega _c)}$ & $\sim 1.4$ &  
ditto                                  & $1.42 \pm 0.14$ \\
\hline  
\end{tabular}
\centering
\caption{Lifetime ratios in the charm sector}  
\label{TABLECHARM}  
\end{table}

Table \ref{TABLECHARM} summarizes  
the predictions and data. From this comparison one can conclude:  
\begin{itemize}
\item  
The observed lifetime {\em hierarchy} emerges correctly and naturally.  
\item
There is {\em no a priori} justification for a $1/m_Q$ expansion to work already 
for the moderate charm quark mass, only an {\em a posteriori} one. For    
even the quantitative predictions are generally on the mark keeping  
in mind that one has to allow for at least 30\% uncertainties due  
to contributions of higher order in $1/m_c$. For proper appreciation  
one should note that the lifetimes span more than an order of magnitude:
\beq  
\frac{\tau (D^+)}{\tau (\Omega_c)} \sim 20
\eeq
One should also keep in mind that total widths and width {\em differences} (among 
mesons) scale like $m_Q^5$ and $m_Q^2$, respectively. The latter  is thus more 
stable under variations in the value of $m_Q$ than the former. 
It actually has been known for some time that with $m_c \simeq 1.3$ GeV one 
reproduces merely about two thirds of the observed value of $\Gamma (D \to l \nu X)$ 
through order $1/m_Q^3$ retaining factorizable contributions only. 
\item  
One discrepancy stands out, though: the $\Xi_c^+$  
appears to live considerably longer than predicted, namely by  
about 50\%. If one multiplied $\tau (\Xi_c^+)$ by an {\em ad-hoc}   
factor of 1.5, then all the predictions for the baryonic lifetime ratios  
would be close to the central values of the measurements!  
One possible explanation is to attribute it to an anomalously large  
violation of quark-hadron duality induced by the accidental  proximity of
a baryonic resonance with appropriate quantum numbers  near the
$\Xi_c^+$, which interferes destructively with the  usual contributions.
Since the charm region is populated by many  resonances, such `accidents'
are quite likely to happen.
\item  
In computing these ratios one has made an assumption beyond the OPE:  
one has adopted a valence quark ansatz in evaluating four-quark  
operators; i.e.,  
\bea  
\matel{D^+}{(\bar c \Gamma u)(\bar u \Gamma c)}{D^+}  
&=&0= \matel{D^0}{(\bar c \Gamma d)(\bar d \Gamma c)}{D^0}  
\\
\matel{D^{0,+}}{(\bar c \Gamma s)(\bar s \Gamma c)}{D^{0,+}}&=&0  
\eea
etc.  
Such an ansatz cannot be an identity, merely an approximation due to  
the presence of `sea' quarks or quark condensates  
$\matel{0}{\bar qq}{0}$.
The  fact that one obtains the correct lifetime patterns shows a
posteriori  that it is -- maybe not surprisingly -- a good approximation.
Since   an expansion in $1/m_c$ is of limited numerical reliability only,  
there is neither a need nor a value in going beyond this approximation.
\end{itemize}

Isopin symmetry tells us that   
$\Gamma (\Xi_c^+ \to \ell \nu X_s) = \Gamma (\Xi_c^0 \to \ell \nu X_s)$ holds.
Yet it would be highly misleading to invoke $SU(3)_{Fl}$ to   argue for
in general universal semileptonic widths of charm baryons.   On the
contrary one actually expects large differences in the semileptonic
widths mainly due to constructive PI in $\Xi_c$ and  
$\Omega_c$ transitions; the lifetime ratios among the baryons will  
thus not get reflected in their semileptonic branching ratios. One  
estimates \cite{VOLBAR,MELIC}:  
\bea  
BR_{SL}(\Xi_c^0) \sim BR_{SL}(\Lambda_c)  
&\leftrightarrow& \tau(\Xi_c^0) \sim 0.5 \cdot \tau(\Lambda_c) \\
BR_{SL}(\Xi_c^+) \sim 2.5 \cdot BR_{SL}(\Lambda_c)  
&\leftrightarrow& \tau(\Xi_c^+) \sim 1.7 \cdot \tau(\Lambda_c) \\
BR_{SL}(\Omega_c) < 25 \%  
\eea
The conventional way to measure the absolute size of these semileptonic  
branching ratios is to study  
\beq  
e^+e^- \to \Lambda_c \bar \Lambda_c, \; \Xi_c \bar \Xi_c, \;  
\Omega_c \bar \Omega_c  
\eeq
Unfortunately it seems unlikely that the tau-charm factory proposed  
at Cornell University will reach the $\Xi_c$ threshold. Yet the  
spectacular success of the $B$ factories BELLE and BABAR has  
pointed to a novel method for measuring these branching ratios by  
carefully analyzing $B \to \Lambda_c/\Xi_c +X$ transitions. One would  
proceed in two steps: One reconstructs one $B$ mesons  
more or less fully in $\Upsilon (4S) \to B \bar B$ and then exploits  
various correlations between baryon and lepton numbers and strangeness.
 
\noindent  
We have pointed out before that the lifetime ratios between charm  
mesons are much smaller than among kaons. It is curious to note  
that the lifetime ratios between charm baryons differ somewhat  
{\em more} from unity than it is the case for strange baryons:  
\beq  
\frac{\tau (\Xi^0)}{\tau (\Sigma ^+)} \simeq 3.6  
\eeq
The fact that the lifetimes for strange baryons -- unlike those for  
strange mesons -- are comparable is attributed to the fact that  
all baryons can suffer $\Delta I = 1/2$ transitions. One can add that  
even the observed hierarchy  
\beq  
\tau (\Sigma^+) \simeq \tau (\Omega) < \tau (\Sigma^-) \simeq  
\tau (\Xi^-) < \tau (\Lambda) < \tau (\Xi^0)  
\eeq
runs counter to expectations based on considering PI and WS  
contributions as the  engine behind lifetime differences in the  
strange sector.
%
\subsubsection{Future prospects}  
\label{FUTPROSP}
 
When relying on HQE to describe charm hadron lifetimes we have to  
allow for several sources of theoretical uncertainties:  
\begin{enumerate}  
\item  
With the expansion parameter $\Lambda_{NPD}/m_c$ for nonperturbative
dynamics only moderately  smaller than unity, unknown higher order terms
could be sizeable; likewise for higher order perturbative  
corrections.  
\item  
The expectation values of four-fermion operators -- let alone  
of higher-dimensional operators -- are not well known.  
\item  
Terms of the form $e^{-m_c/\Lambda_{NPD}}$ cannot be captured by the  
OPE. They might not be insignificant (in contrast to the case with  
beauty quarks).  
\item  
`Oscillating' terms $\sim \left(\Lambda_{NPD}/m_c\right)^k \cdot  
{\rm sin}(m_c/\Lambda_{NPD})$, $k>0$, likewise are not under good  
theoretical control. They reflect phenomena like hadronic resonances,  
threshold effects etc.
 
\end{enumerate}
Assuming each of these effects to generate a 10\% uncertainty is  
not conservative, and one cannot count on the overall theoretical  
uncertainty to fall below 20 \%. With the exception of the second item in
this list (see its discussion below), the situation is highly  
unlikely to change decisively anytime soon.
 
Any of the sources listed above a priori could have produced uncertainties
of, say, 30\% combining into an overall error of $\sim 100$\%. Our  
description of charm lifetimes would have clearly failed then --  
yet {\em without} causing alarm for the general validity of QCD as  
the theory of strong interactions. However this failure did not happen --  
the expected (and partially even predicted) pattern is in at least  
semi quantitative agreement (except for $\tau (\Xi_c^+)$) with data that
are quite mature now. The general lesson is that the transition from  
nonperturbative to perturbative dynamics in QCD is smoother and more  
regular than one might have anticipated. This insight is bound to enhance  
our confidence that we can indeed treat various aspects in the decays of  
beauty hadrons in a {\em quantitative} way.
 
On the other hand the analysis of charm lifetimes is not meant to yield  
precise quantitative lessons on QCD. With the lifetimes of $D^0$, $D^+$
and $D_s$ now known within  1\% and of $\Lambda_c$ within 3\% measuring
them even more  precisely will not help our understanding of charm
lifetimes, since  that is already limited by theory. One exception to
this general statement is the  search for $D^0 - \bar D^0$ oscillations
to be discussed later.
 
There are two areas where further progress seems feasible:  
\begin{itemize}
\item  
There is an explicit calculation of $\tau (D_s^+)/\tau (D^0)$  
based  on QCD sum rules to estimate the matrix element controlling WA  
\cite{HYC}. It  
finds a ratio that -- while larger than unity -- appears to fall  
well below the data:  
\beq  
\frac{\tau(D_s)}{\tau (D^0)} \simeq 1.08 \pm 0.04
\eeq
It would be premature to view this estimate as conclusive. First  
we have to deepen  
our understanding of WA. Its leading impact is due to the  
expectation value of a four-fermion operators  
$(\bar c_L\gamma_{\mu}q_L)(\bar q_L\gamma_{\mu}c_L)$ and  
$(\bar c_L\gamma_{\mu}\lambda_iq_L)(\bar q_L\gamma_{\mu}\lambda_ic_L)$.
It would provide an interesting benchmark test to extract their sizes  
from data and compare them with the predictions from lattice QCD.  
These quantities will also affect the lepton energy endpoint spectra  
in inclusive semileptonic $D^+$ and $D_s$ decays. Lastly -- and  
maybe most significantly -- their $b$ quark counterparts are expected  
to have a sizeable impact on the lepton energy spectra in $B$ decays  
with the effects being different for $B_d$ and $B^-$  
\cite{WA}. This introduces  
an additional large uncertainty into extracting $V(ub)$ there.  
\item  
With WA driving about 20\% of all $D$ decays, the further  
challenge naturally arises whether footprints of WA can be found in  
{\em exclusive} modes. I.e., can one show that WA --  
rather than modifying all nonleptonic decays in a basically uniform way
-- affects certain channels much more significantly than others. We will  
return to this issue when discussing exclusive decays.

\end{itemize}
 
The data on the lifetimes of the other charm baryons have recently
reached a new maturity level. Yet even so, further improvements  
would be desirable, namely to measure also  
$\tau (\Xi_c^0)$ and $\tau (\Omega_c)$ to within 10\% or even better.
 
We have identified one glaring problem already, namely that the  
$\Xi_c^+$ lifetime exceeds expectation by about 50 \%, while the other  
lifetime ratios are close to expectations. At first sight this might
suggest the accidental presence of a baryonic resonance near the  
$\Xi_c^+$ mass inducing a destructive interference. Since the  
charm region is still populated by light-flavour baryon resonances,  
such an effect might not be that unusual to occur. However  
$\Xi_c^0$ (or $\Lambda_c$) would be more natural `victims' for such an  
`accident' since there the final state carries the quantum numbers that  
an $S=-2$ (or $S=-1$) baryon resonance can possess. It would be
interesting to see whether future measurements leave the ratios  
$\tau (\Lambda_c)/\tau (\Xi_c^0)$ and  
$\tau (\Xi^0_c)/\tau (\Omega_c^0)$ in agreement with expectations.
 
There is another motivation for analysing charm baryon lifetimes.  
For several years now there has been a persistent  
and well publicized problem in the beauty  
sector, namely that the observed beauty baryon lifetime  
falls below HQE predictions:  
\beq  
\left. \frac{\tau (\Lambda_b)}{\tau (B_d)} \right|_{data}  
= 0.797 \pm 0.052 \; \; vs. \; \;  
0.88 \leq \left. \frac{\tau (\Lambda_b)}{\tau (B_d)} \right|_{HQE}  
\leq 1
\eeq  
The data are not conclusive yet, and further insights will be gained by  
measuring $\tau (\Xi_b^0)$ and $\tau (\Xi_b^-)$, which is expected to be
done during the present Tevatron run. Theory might still snatch victory  
of the jaws of defeat.
 
Yet in any case the lifetimes of charm baryons  
can act as important diagnostics also for interpreting either success  
or failures in the HQE predictions for the lifetimes of beauty  
baryons -- as long it makes some sense to treat charm quarks as heavy.
The  weak link in the HQE analysis of baryon widths are the expectation
values  of the four-quark operators, since there size is inferred from  
quark models or QCD sum rules of less than sterling reliability.  
It has been suggested by Voloshin \cite{VOLBAR} and Guberina {\em et al.}
\cite{GUBPATT} to combine HQS, isospin and $SU(3)_{Fl}$ symmetry to  
relate lifetime differences among charm baryons to those among beauty  
baryons. From
\beq  
\frac{\Gamma(\Xi_b^-) - \Gamma(\Xi_b^0)}
{\Gamma(\Xi_c^+) -\Gamma(\Xi_c^0)} = \frac{m_b^2}{m_c^2}  
\frac{|V(cb)|^2}{|V(cs)|^2} \left( 1+ {\cal O}(1/m_c, 1/m_b)\right)  
\eeq  
and using the observed lifetime difference between $\Xi_c^+$ and  
$\Xi_c^0$ one infers  
\beq  
\Gamma(\Xi_b^-) - \Gamma(\Xi_b^0) = ( - 0.14 \pm 0.06) \; ps^{-1} \; ,  
\label{DELTAXIB}
\eeq
which is quite consistent with other predictions. Furthermore to the
degree one can ignore quark masses in the final state one can
approximately equate $\Gamma (\Lambda_b)$ and $\Gamma (\Xi_b^0)$ since
both are subject to WS and to destructive PI, $\Lambda_b$ in the  
$b\to c \bar u d$ and $\Xi_b^0$ in the $b\to c \bar cs$ channels.  
Using the observed $\Lambda_b$ lifetime one infers from  
Eq.(\ref{DELTAXIB}):  
\beq  
\frac{\tau(\Xi_b^-) -\tau(\Lambda_b)}{\tau(\Xi_b^-)} = 0.17 \pm 0.07
\label{DELTALAMBDAXI}
\eeq


\subsection{Masses, weak lifetimes and semileptonic branching
ratios of $C\geq 2$ baryons}  
\label{CGEQ2BARY}

The nonleptonic and semileptonic widths of these baryons are even more  
sensitive probes of the dynamics underlying their structure.
 
The leading contribution is  contained in the quark decay term  
\bea  
\matel{H_{cc}}{\bar cc}{H_{cc}} &=&  
2 - \frac{1}{2} \frac{\mu_{\pi}^2(H_{cc})}{m_c^2} +  
\frac{1}{2} \frac{\mu_{G}^2(H_{cc})}{m_c^2} + {\cal O}(1/m_c^3)  
\nonumber  
\\
\matel{\Omega^{++}_{ccc}}{\bar cc}{\Omega^{++}_{ccc}} &=&  
3 - \frac{1}{2} \frac{\mu_{\pi}^2(\Omega^{++}_{ccc})}{m_c^2} +  
\frac{1}{2} \frac{\mu_{G}^2(\Omega^{++}_{ccc})}{m_c^2} +  
{\cal O}(1/m_c^3)
\; ,  
\label{MECC}  
\eea
where the first term of two [three] reflects the fact that there are  
two [three]  
valence  charm quarks inside $H_{cc}$ [$\Omega_{ccc}$]    
and the leading nonperturbative corrections are expressed through the  
kinetic energy moment $\mu_{\pi}^2(H_{cc})$ and chromomagnetic  
moment $\mu_{G}^2(H_{cc})$. The main differences among the widths of  
the $C=2$ baryons arise in order $1/m_c^3$ due to WS and destructive as  
well as constructive PI, similar to the case of $C=1$ baryons:  
\bea  
\Gamma_{NL} (\Xi_{cc}^{+}) &\simeq& \left[  
\Gamma_{decay,NL}(\Xi_{cc}) + \Gamma_{WS}(\Xi_{cc}^{+})  
\right]  
\nonumber  
\\
\Gamma_{NL} (\Xi_{cc}^{++}) &\simeq& \left[  
\Gamma_{decay,NL}(\Xi_{cc}) - \Delta \Gamma_{PI,-}(\Xi_{cc}^{++})  
\right]  
\nonumber  
\\
\Gamma_{NL} (\Omega_{cc}^{+}) &\simeq& \left[  
\Gamma_{decay,NL}(\Omega_{cc}) + \Delta \Gamma_{PI,+}(\Omega_{cc})  
\right]  
\label{GAMMACC}  
\\
\Gamma_{SL} (\Xi_{cc}^{+}) &\simeq&  \Gamma_{decay,SL}(\Xi_{cc}) , \;
\Gamma_{SL} (\Xi_{cc}^{++}) \simeq  
\Gamma_{decay,SL}(\Xi_{cc})  
\nonumber  
\\  
\Gamma_{SL} (\Omega_{cc}^{+}) &\simeq& \left[  
\Gamma_{decay,SL}(\Omega_{cc}) + \Delta \Gamma_{PI,+}(\Omega_{cc})  
\right] \; .  
\label{SLBRCC}
\eea
The fact that there are two rather than one charm quark that can decay  
with or without PI and undergo WS is contained in the  
size of the expectation values $\matel{H_{cc}}{\bar cc}{H_{cc}}$,  
see Eq.(\ref{MECC}), and  
$\matel{H_{cc}}{(\bar c\Gamma q)(\bar q\Gamma c)}{H_{cc}}$;  
i.e., $\Gamma_{decay}(\Xi_{cc}) = 2\Gamma_{decay}(\Xi_{c})$ to leading  
order in $1/m_c$. Based on the expressions in Eq.(\ref{GAMMACC}) one  
expects substantial lifetime differences with    
\beq  
\tau (\Xi_{cc}^+), \; \tau (\Omega_{cc}) < \tau (\Xi_{cc}^{++})
\eeq
One might ask if the PI contribution to the $\Xi_{cc}^{++}$ width could  
be constructive rather than destructive; in that case the  
$\Xi_{cc}^+$ and $\Xi_{cc}^{++}$ lifetimes would be very similar and both  
short. While the negative sign of PI in the $D^+$ width is not a trivial
matter,  since it depends on properly including QCD radiative
corrections, it is straightforward (though still not trivial) for
baryons. The PI contribution depends on the combination  
$2c_+(2c_- - c_+)$ of the QCD renormalization coefficients $c_{\pm}$:  
since $c_+ < 1 < c_-$ the sign of the effect is stable under radiative  
QCD corrections.
 
With the $D^0$ width given mainly by the decay contribution, it
provides an approximate yardstick for the latter's size. To  
{\em leading} order in $1/m_c$ one has  
\beq  
\frac{1}{2}\Gamma_{decay}(\Xi_{cc})\simeq  
\Gamma_{decay}(\Lambda_c) \simeq
\Gamma_{decay}(D) \simeq \Gamma (D^0)
\eeq
Since the $c$ quarks move more quickly in a double charm than a single  
charm hadron, one expects $\Gamma_{decay}(\Xi_{cc})$ to be actually  
somewhat smaller than $2\Gamma (D^0)$ due to time  
dilatation that enters in order $1/m_c^2$. {\em If} the PI term  
were absent, one would thus have  
$\tau (\Xi_{cc}^{++}) \geq \frac{1}{2}\tau (D^0) \simeq 2\cdot 10^{-13}  
sec$. Including PI, which one confidently predicts to be  
destructive, one expects $\Xi_{cc}^{++}$ to be considerably longer  
lived than this lower bound and possibly even longer lived than  
$D^0$.  The $\Xi_{cc}^+$ lifetime on the other hand will be considerably  
shorter than $\tau (D^0)$. The impact of WS pushing $\tau (\Xi_c^{+})$  
below $2\cdot 10^{-13} sec$ is partially offset by the time dilatation  
effect mentioned above. Nevertheless  
$\tau (\Xi_c^{+}) \sim 10^{-13} sec$ would seem to be a reasonable  
first guess.
 
For more definite predictions one needs to estimate the relevant
$\Xi_{cc}$ expectation values. At present we have to rely on quark
models and QCD sum rules. In the future those could be tested and fine
tuned through their predictions on the mass splittings of the $C=2$
baryons and their  resonances; estimates based on  
lattice QCD might become available as well. The authors of
Refs.(\cite{KISELEV},\cite{GUBDOUBLE}) had the  foresight to take on this
task before there was any experimental  hint for such exotic baryons.  
The two groups follow a somewhat different philosophy in  
choosing the range for $m_c$ and $m_s$.
 
The authors of Ref.\cite{KISELEV} focus on hadrons with two heavy
constituents like the meson $B_c$ and the baryons  
$\Xi_{cc}$, $\Xi_{bc}$ etc.  
They have adopted a rather phenomenological attitude in selecting  
values for $m_c$ (and $m_s$); from $\tau (B_c)$ they infer  
$m_c \sim 1.6$ GeV and find  
$$  
\tau(\Xi_{cc}^{++})  \sim 0.46 \pm 0.05 \; ps \; ; \;  
\tau(\Xi_{cc}^{+})  \sim 0.16 \pm 0.05 \; ps  
$$
\beq  
\tau(\Omega_{cc}^{+}) \sim 0.27 \pm 0.06 \; ps  
\eeq  
\beq  
\Rightarrow  
\frac{\tau(\Xi_{cc}^{++})}{\tau(\Xi_{cc}^{+})} \sim 2.9 \; , \;  
\frac{\tau(\Xi_{cc}^{++})}{\tau(\Omega_{cc}^{+})} \sim 1.7 \; , \;  
\frac{\tau(\Omega_{cc}^{+})}{\tau(\Xi_{cc}^{+})} \sim 1.7
\eeq
 
The authors of Ref.(\cite{GUBDOUBLE}) on the other hand follow a purer  
invocation of the OPE and set $m_c = 1.35$ GeV and $m_s = 0.15$ GeV.  
After some detailed
consideration including even Cabibbo suppressed modes they obtain:  
\bea  
\tau (\Xi_{cc}^{++}) &=& 1.05 \; ps \; , \;  
\tau (\Xi_{cc}^{+}) = 0.20 \; ps \\  
\tau (\Omega_{cc}^{+}) &=& 0.30 \; ps \; , \;  
\tau (\Omega_{ccc}^{++}) = 0.43 \; ps  
\eea
\beq
\Rightarrow  
\frac{\tau(\Xi_{cc}^{++})}{\tau(\Xi_{cc}^{+})} \sim 5.2 \; , \;  
\frac{\tau(\Xi_{cc}^{++})}{\tau(\Omega_{cc}^{+})} \sim 3.5 \; , \;  
\frac{\tau(\Omega_{cc}^{+})}{\tau(\Xi_{cc}^{+})} \sim 1.5
\eeq
and  
\bea
BR_{SL}(\Xi_{cc}^{++}) &=& 15.8 \; \% \; , \;  
BR_{SL}(\Xi_{cc}^{+}) = 3.3 \; \% \\
BR_{SL}(\Omega_{cc}^{+}) &=& 13.7 \; \%  
\eea
Since the semileptonic widths are basically equal for $\Xi_{cc}^{++}$ and  
$\Xi_{cc}^{+}$, the ratio of their semileptonic branching ratios reflects  
the ratio of their lifetimes. On the other hand constructive PI enhances  
the semileptonic $\Omega _{cc}^+$ width.
 
Needless to say, there are substantial differences in these numbers,
less so (as expected) for the ratios. For proper evaluation one has to
take notice of the  following. The authors of Ref.\cite{GUBDOUBLE} use one
value for  baryons and a higher one for mesons for phenomenological
reasons  
\footnote{It is related to the fact mentioned above that with  
$m_{c,kin}(1\; {\rm GeV})\simeq 1.2 - 1.3$ GeV one fails to reproduce  
the observed $\Gamma_{SL}(D)$.}. Yet on theoretical grounds this is  
inadmissible: in the OPE one has to use the same value of $m_c$  
for mesons and baryons alike. Using different values can serve only as  
a {\em temporary crutch} to parameterize an observed difference  
between baryons and mesons one does not understand at all!
 
With neither PI nor WS contributing to $\Omega_{ccc}^{++}$ decays, its  
width is given by the decay of its three charm quarks. Insisting on using  
the same value for $m_c$ as for $D$ mesons, one would predict roughly
the  following numbers:  
\bea  
\tau (\Xi_{cc}^{++}) &=& 0.35 \; ps \; , \;  
\tau (\Xi_{cc}^{+}) = 0.07 \; ps \\
\tau (\Omega_{cc}^{+}) &=& 0.10 \; ps \; , \;  
\tau (\Omega_{ccc}^{++}) = 0.14 \; ps  
\eea
 
Obviously one has to allow for considerable uncertainties in  
all these  
predictions due to the unknown higher order $1/m_c$ contributions  
and our ignorance about the potential controlling the inner dynamics  
of $C=2$ baryons.
 
Yet a certain pattern does emerge, and one can conclude the following:  
\begin{itemize}
\item  
It
is a very  considerable stretch to come up with a $\Xi_{cc}^+$ lifetime as
short as  0.03 ps.
  
\item  
The $\Xi_{cc}^{++}$ lifetime is similar to or even larger than that for  
$D^0$. There appears no way to push the $\Xi_{cc}^{++}$ lifetime
into the "ultrashort" domain $\sim 0.1$ ps.
 
\item  
{\em If} the data forced upon us a scenario with $\tau (\Xi_{cc}^{+})$
well below  
$0.1$ ps and $\tau (\Xi_{cc}^{++})$ near it -- let alone below it --,  we
had to conclude that the successes of the HQE description of  
$C=1$ charm hadron  
lifetimes listed above are quite accidental at least for the $C=1$  
baryons, but probably for the mesons as well. While this is a conceivable  
outcome, it would come with a hefty price for theory, or at least for  
some theorists.  
The HQE offers no argument why $C=1$ hadrons are treatable, while $C=2$ 
are not. Furthermore the double-heavy meson $B_c$ appears to be well 
described by the HQE \cite{PRO}.  
\end{itemize}

%
\section{Leptonic and Rare Decays}  
\label{LEPT}  
%
The simplest final state possible in $D$ decays consists of a lepton  
or a photon pair, namely 
\par
(i) $D_q^+ \to \tau^+\nu$, $\mu^+ \nu$, $e^+\nu$; 
\par
(ii) $D^0 \to \mu^+\mu^-$, $e^+e^-$, $D^0 \to \gamma \gamma$; 
\par
(iii) $D^0 \to e^{\pm}\mu^{\mp}$. 
\par
\noindent  
While the first two transitions can proceed in the SM, albeit at reduced  
or even highly suppressed rates, 
the last one is absolutely forbidden. Transitions (ii) and (iii) thus  
represent clean, yet quite speculative searches for New Physics. The main  
motivation for measuring accurately transitions (i) on the other hand is to
extract the  decay constants $f_D$ and $f_{D_S}$ as one test of our   
theoretical control over nonperturbative QCD. 
\par
There is another simple, yet highly  
exotic final state, namely 
\par
(iv)  $D^+ \to \pi^+ f^0$, $K^+ f^0$
\par
\noindent where $f^0$ denotes the so-called  
familon,
 \index{familon}
  a neutral scalar that could arise as a Nambu-Goldstone  
boson in the  
spontaneous  breaking of a continuous global family symmetry. Although it  
is not a leptonic final state, we will briefly discuss it at the end  
of this Section. 
We will discuss also rare threebody modes 
\par
(v) $D^+, D_s^+ \rightarrow h^+ l^- l^+, \; h^- l^+ l^+, \; h^+e^{\pm}\mu^{\mp}$ 
(with $h = \pi, K $ and $l=e,\mu$), 
\par
\noindent which have some favourable experimental signatures; the theoretical 
interpretation is quite different for the different modes, as explained later. 

\begin{figure}[t]
  \centering 
   \includegraphics[width=10.0cm]{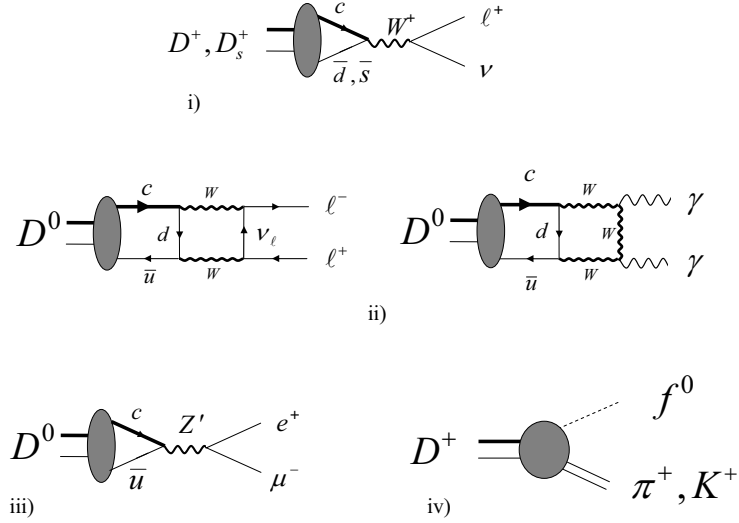} 
 \caption{ Rare and leptonic decay diagrams.
    \label{FIG:LGRAPH} }
\end{figure}

Diagrams for process (i) to (iv) are shown in
Fig.~\ref{FIG:LGRAPH}.

\subsection{Expectations on $D^+_q \to \ell^+ \nu$} 
\label{DLNUTH} 
%
These transitions are driven by the axial 
vector component of the hadronic charged current; the transition 
amplitude  reads:  
\beq 
 {\cal T}(D^+_q \to \ell^+ \nu) =  
\frac{G_F}{\sqrt{2}} \, \langle 0 |A^\mu | D_q \rangle  [\bar \ell 
\gamma_\mu (1-\gamma_5)\nu_\ell ] , \; q=d,s  
\eeq  
The hadronic matrix element is conventionally parametrized by the meson
decay constant: \index{decay constant}  
\beq 
\langle 0 |A^\mu| D_q(p)\rangle = i f_{D_q} p^\mu 
\label{DECCON}  
\eeq 
The total width is then given by  
\beq 
\Gamma(D_q^+\to \ell^+ \nu_l) =  
\frac{G_F^2}{8\pi} f^2_{D_q} |V(cq)|^2 
m_\ell^2 (1-\frac{m_\ell^2}{M_{D_q}^2})^2 M_{D_q} 
\label{DQLNU} 
\eeq  
These transitions are {\em helicity suppressed}
\index{helicity suppression}; i.e., the amplitude  
is proportional to $m_\ell$, the mass of the lepton $\ell$, in complete 
analogy  to $\pi^+ \to \ell^+\nu$. This property holds in 
general  for any (axial)vector charged current -- a point we will return 
to  below. They are also suppressed by $f_{D_q} \ll m_c$. This feature  
can be understood intuitively:   
due to the practically zero range of the weak forces  
the $c$ and $\bar s$ or $\bar d$ quarks have to come together to 
annihilate. The amplitude therefore is proportional to  
$\psi_{c\bar q}(0)$, the $c\bar q$ wave function at zero separation.  
This quantity is related to the decay constant as follows:  
$f^2_{D_q} = 12 |\psi_{c\bar q}(0)|^2/M_{D_q}$ 
\footnote{One should note that  
the description through $f_{D_q}$ as defined by Eq.(\ref{DECCON})  
holds irrespective of the existence of a wave function.}.  
\par
One expects 
the following branching ratios for the $D_s^+$ and $D^+$  
modes, which are Cabibbo allowed and forbidden, respectively:  
\bea  
{\rm BR}(D^+ \to \tau^+ \nu) &=&   
1.0\cdot 10^{-3}\left(\frac{f_D}{220\; {\rm MeV}}\right) ^2  \\   
{\rm BR}(D^+ \to \mu^+ \nu) &=&  
4.6\cdot 10^{-4}\left(\frac{f_D}{220\; {\rm MeV}}\right) ^2  \\  
{\rm BR}(D^+ \to e^+ \nu) &=&  
1.07\cdot 10^{-8}\left(\frac{f_D}{220\; {\rm MeV}}\right) ^2  \\ 
{\rm BR}(D_s^+ \to \tau^+ \nu) &=&   
4.5\cdot 10^{-2}\left(\frac{f_{D_s}}{250\; {\rm MeV}}\right) ^2  \\   
{\rm BR}(D_s^+ \to \mu^+ \nu) &=&  
5.0\cdot 10^{-3}\left(\frac{f_{D_s}}{250\; {\rm MeV}}\right) ^2  \\  
{\rm BR}(D_s^+ \to e^+ \nu) &=&  
1.2\cdot 10^{-7}\left(\frac{f_{D_s}}{250\; {\rm MeV}}\right) ^2   
\eea 
The numbers have been scaled to the lattice estimates of the decay 
constants. 

The fact that there are two different mesons decaying in this way  
represents an extra bonus. Only lattice QCD   
could provide us with predictions for the decay 
constants $f_D$ and $f_{D_s}$ with a theoretical uncertainty not 
exceeding a few percent. Yet one has to be concerned that maybe not all  
lattice artefacts can be brought under such excellent control. These   
systematics should however basically cancel in the ratio $f_{D_s}/f_D$,  
which deviates from unity  only due to $SU(3)_{Fl}$ breaking. 

The motivation for measuring these widths thus lies in extracting  
the decay constants as a precision measure for the theoretical  
control lattice QCD can achieve over certain hadronic quantities.  
Even more importantly lattice QCD successfully passing this test 
constitutes  
an important validation for its prediction on $f_{B_{d,s}}$. It has  
been suggested that even if lattice QCD falls somewhat short of a  
successful prediction for $f_D$, one can use the experimentally  
observed value for $f_D$ together with lattice QCD's prediction for  
$f_B/f_D$ to arrive at a reliable value for $f_B$, which obviously is  
of the greatest benefit for interpreting $B^0-\bar B^0$ oscillations; 
at the very least success in reproducing $f_{D_s}/f_D$ would tightly 
constrain $f_{B_s}/f_B$, which controls $B_s - \bar B_s$ vs. $B_d - \bar B_d$ 
oscillations.

These transitions are sometimes advanced as sensitive probes for  
New Physics in the form of hard pseudoscalar couplings for the charged  
current, which would modify the ratio   
$\Gamma (D_q \to \tau \nu)/\Gamma(D_q \to \mu \nu)$ with respect to the helicity  
factor $(m_{\tau}/m_{\mu})^2$ (plus phase space corrections), see  
Eq.(\ref{DQLNU}). This is an echo of what happened in the early days  
of weak interaction studies when the observation of  
$\Gamma (\pi^+ \to \mu^+\nu) \gg \Gamma (\pi^+ \to e^+\nu)$ pointed to  
the dominance of spin one over spin zero charged currents.

While such couplings are conceivable on purely phenomenological  
grounds, one usually views them as somewhat unlikely  
due to general theoretical arguments. For a spin-zero coupling  
to the $\ell \nu$ pair independent of the lepton mass  
$m_\ell$, would represent a `hard' breaking of chiral invariance.
\index{chiral invariance}
 It would  
be surprising if such a feature had not been noted before. There  
could be Higgs couplings, yet those would again be proportional to  
$m_\ell$ (in amplitude) and thus represent an acceptable soft breaking of 
chiral symmetry. Of course,  the dimensionless numerical coefficient
 in front of 
such a coupling could differ for the different modes. 

\subsection{$D^+_q \to \ell^+ \nu$} 
\label{DLNUEXP} 

Looking at the estimates for the branching ratios one realizes that  
going after $D_q \to e \nu$ is a hopeless enterprise; yet at the same time 
observing it would provide spectacular evidence for New Physics. 
The final state  
$\tau \nu$ has the largest branching ratios, yet poses the highly  
nontrivial problem that the $\tau$ lepton decays typically into  
one-prong final states with additional neutrinos. It then makes sense 
both from an experimental as well as theoretical perspective to make a  
dedicated effort to measure all four channels  
$D_q \to (\tau/\mu) \nu$ as precisely as possible. 
\par
Small decay rate and difficult event topology make these decays a real 
challenge
to the experimenter (Tab.\ref{TAB:LDK}). 
Good lepton identification is required, as well as acceptable tracking
capability, and detector hermeticity to evaluate the neutrino missing energy
which is very large in the case of the $D_s^+ \to \tau^+ \nu_\tau, 
\tau^+\to \mu^+ \nu_\mu {\bar \nu}_\tau$ decay chain.
At fixed-target experiments, the challenge has been so far overwhelming.
Strategies to attempt detecting the $D^+\to \mu^+\nu_\mu$ decay 
by tracking the $D^+$
before its decay have been proposed \cite{Bianco:2003wy}, but the practical
realization of the method has encountered a formidable show-stopper in
unreducible backgrounds processes, such as $D^+\to \pi^0\mu^+\nu_\mu$ with the
undetected neutral pion.
\par
 Model-dependent
determinations based on isospin splittings and hadronic decays 
have also been proposed \cite{Bortoletto:1992rj}.
\par 
For $f_{D^+}$, a measurement is
available from BES, based 
on one event\cite{Bai:cg} $f_D^+ = 300\pm 175 $~MeV from a
branching ratio value of 
$BR(D^+\to \mu^+\nu_\mu)  = 0.08^{+0.17}_{-0.05}$\%. The most recent and most
precise measurement of $f_{D^+_s} = 285 \pm 19 \pm 40$~MeV
 is from ALEPH \cite{Heister:2002fp}, which combines $\mu^+\nu_\mu$ and
 $\tau^+\nu_\tau$ events using relative production rates taken from lattice
 computations.
For a summary of the experimental scenario see 
\cite{Richman:wm,Marinelli02,Artuso03}, and for comparison with pion and 
kaon constants see \cite{Suzuki:dv}.
\par
Great expectations in this field come from CLEO-c and, later, from BES III,
where a projected yield of about 1000 reconstructed leptonic decays should allow
a 2\% error on $f_{D,D_s}$ \cite{Artuso03}. 
%
\subsection{Adagio, ma non troppo}
\label{RARENONTROPPO}
Inclusive or exclusive flavour-changing radiative decays -- $D\to  \gamma X$ or 
$D \to \gamma K^*/\rho/\omega/\phi$, respectively -- reflect very different
dynamics than  
$B\to  \gamma X$ or $B \to \gamma K^*/\rho/\omega/\phi$. While one can {\em
draw}  
Penguin {\em diagrams} \index{Penguin diagram} in both cases, their meaning is
quite different. For the leading  
transition operator in radiative $B$ decays arises from `integrating out' the
top quark in the  
Penguin loop. On the other hand in the Penguin diagram for $c\to u$ the
contribution from an  
internal $b$ quark is negligible due to the almost decoupling of the third quark
family from the  
first two; the contribution from an internal $s$ quark cannot be treated as a
{\em local}  
operator, 
since the $s$ quark is lighter than the external $c$ quark and thus cannot be
integrated out.  
Thus even the inclusive rate $D\to \gamma X$ is not controlled by short-distance
dynamics  
within 
the SM, let alone exclusive transitions $D\to \gamma V$ with $V$ denoting a
vector meson.  
This can be illustrated through  diagrams where weak annihilation
 \index{weak annihilation}
is preceded by photon 
emission 
from the light (anti)quark line.  

The discerning observer can however see a virtue in this complexity. 
Measuring $D \to \gamma K^*/\rho/\omega/\phi$ can
 teach us new lessons on QCD's  nonperturbative dynamics 
in general, provide another test 
challenging lattice QCD's numerical reliability and thus be of 
considerable value for `peeling off' layers of long-distance dynamics  from the
measured rates  
for 
$B \to \gamma K^*/\rho/\omega$ when extracting $V(td)/V(ts)$ from the latter. 
Complementary 
information can be inferred from $B\to \gamma D^*$, which receives no genuine
Penguin  
contribution either. 

Rather detailed analyses exist \cite{BURD95,Fajfer} concerning SM expectations,
which -- not  
surprisingly based on what was said above -- represent order-of-magnitude
predictions only.  
Typical 
numbers are
\bea 
{\rm BR}(D^0 \to \gamma \bar K^{*0}) = (6 \div 36)\cdot 10^{-5} \; , \; 
{\rm BR}(D^0 \to \gamma  \rho^0) = (0.1\div 1)\cdot 10^{-5} 
\nonumber 
\\  
{\rm BR}(D^0 \to \gamma \omega) = (0.1 \div 0.9)\cdot 10^{-5} \; , \; 
{\rm BR}(D^0 \to \gamma \phi ) = (0.1 \div 3.4)\cdot 10^{-5}
\eea 
A more speculative motivation to search for these modes is that in some
nonminimal SUSY  
\index{SUSY}
scenarios induce significant contributions due to local Penguin operators
 for $c\to u$ with 
SUSY fields appearing in the internal loop \cite{GABBIANI}. Since they affect
neither $D\to \gamma K^*$ nor $D \to \gamma \phi$ those two rates provide a good
calibrator for $D \to \gamma \rho$ and 
$D\to \gamma \omega$, 
since several systematic uncertainties will largely drop out from the 
ratios $\Gamma(D^0\to \gamma \rho/\omega)/\Gamma(D^0\to \gamma K^*/\phi)$. 

BELLE has reported the first observation of any of these modes \cite{Abe:2003yv}: 
\beq 
{\rm BR}(D^0 \to \gamma \phi ) = (2.6 ^{+0.70}_{-0.61}(stat.) 
^{+0.15}_{-0.17}(syst.))\cdot 10^{-5}
\eeq
which is consistent with expectations; however 
the latter cover a particularly wide range. 
%
\subsection{Much rarer still: $D^0 \to \mu^+ \mu^-$ and  
$D^0 \to \gamma \gamma$} 
\label{EVENRARER} 
%
The two effects suppressing $D^+ \to \mu^+ \nu$ do likewise in  
$D^0 \to \mu^+\mu^-$:  
(i) Any spin-one coupling of the type  
$[\bar u\gamma_{\mu}\gamma_5c] 
[\bar \mu \gamma_{\mu}(1-\gamma_5)\mu]$ leads to helicity suppression, 
i.e. a transition amplitude proportional to $m_{\mu}$. 
(ii)  Due to the effective zero range also of this coupling the amplitude is  
suppressed by $\sim f_D/m_c \ll 1$. 
On top of that one needs a genuine flavour-changing neutral current:  
(iii) Within the SM such a current does not exist on the tree level -- it  
has to be generated on the one-loop level thus representing a pure quantum 
effect.  

All three effects combine to produce a huge reduction in branching ratio.  
An extremely crude guestimate invokes BR$(D^+\to \mu^+\nu)$ to mirror  
helicity and wavefunction suppression and uses  
$\frac{\alpha_S}{\pi} \cdot (m_s/M_W)^2$ to reflect the second order  
GIM effect; i.e.  
\beq 
{\rm BR}(D^0 \to \mu^+ \mu^-) \sim {\cal O} 
\left( {\rm BR}(D^+ \to \mu^+ \nu) \cdot \frac{\alpha_S}{\pi} \cdot 
\frac{m_s^2}{M_W^2} \right) \sim {\cal O}(10^{-12})  
 \nonumber
\eeq 
It is straightforward to draw one-loop diagrams describing $\Delta C=1$  
neutral currents and computing them as if they  
represent a short-distance effect. However there is no  
a priori reason why $D^0 \to \mu^+\mu^-$ should be controled by simple  
short-distance dynamics.  
A detailed treatment yields \cite{Burdman:2001tf}  
\beq  
{\rm BR}(D^0 \to \mu^+ \mu^-) \simeq 3.0 \cdot 10^{-13}  
  \nonumber
\eeq 
orders of magnitude below the present experimental upper bound:  
\beq  
{\rm BR}(D^0 \to \mu^+ \mu^-)|_{exp} \leq 4.1 \cdot 10^{-6} \; .  
  \nonumber
\eeq 
The authors have employed various prescriptions for estimating possible  
long distance effects analoguous to what one encounters in  
$K_L \to \mu^+\mu^-$. Similar to what happens there they actually find 
that a two-step  transition involving long-distance dynamics 
\index{long distance dynamics}  
provides a large or even  
dominant contribution, namely  
\beq  
{\rm BR}(D^0 \to \gamma \gamma \to \mu^+\mu^-)  
\sim 2.7 \cdot 10^{-5}\cdot  {\rm BR}(D^0 \to \gamma \gamma )  
\sim (0.3 - 1) \cdot 10^{-12}  
   \nonumber
\eeq  
\beq  
{\rm BR}(D^0 \to \gamma \gamma ) \sim (1 - 3.5) \cdot 10^{-8}
   \nonumber
\eeq 
consistent with  the estimates of Ref.\cite{SINGER}. 
The theoretical tools exist for a refined analysis based on the OPE  
that includes long distance dynamics naturally and self-consistently  
through quark condensates. Yet that would appear to be a purely  
academic exercise in view of the immensely tiny branching ratio. For  
it is unlikely that such a refinement could enhance the  
branching ratio by three -- let alone more -- orders of magnitude.  
Later in discussing $D^0 - \bar D^0$ oscillations we will address  
analoguous issues, where they are much more relevant numerically. 

With the SM width so tiny, it could be a promising laboratory to  
search for manifestations of New Physics. The authors of 
Ref.(\cite{Burdman:2001tf}) find that New Physics scenarios  
can produce a wide range in predictions, namely  
\beq  
{\rm BR}(D^0 \to \mu^+ \mu^-)|_{NP} \sim  
10^{-11} / 8 \cdot 10^{-8} / 3.5 \cdot 10^{-6}  
\eeq  
for models with a superheavy $b^{\prime}$ quark/ multi-Higgs sector/   
SUSY with $R$ parity breaking, respectively. 
\par
Since the mode $D^0 \to \mu^+\mu^-$ possesses a clear signature, one can 
entertain the hope to search for it in hadronic collisions. The best limit in
PDG02 actually dates back to 1997 hadroproduction experiment BEATRICE (WA92 at
CERN) \cite{Adamovich:1997wf}.  

 CDF has plans to pursue the study of this decay,  as has BTeV. CDF
showed at Moriond 2003 preliminary results \cite{Korn:2003pt,Acosta:2003ag}
which lower the WA92 best limit by a factor of two (Tab.~\ref{TAB:LDK}. 
About the same sensitivity is expected from B-factories at the present level
 of data collected (about 90 $fb^{-1}$)  \cite{Harrison:1998yr}.
\par
The much more challenging decay  $D^0 \to \gamma \gamma $ requires superb em
calorimetry and high statistics. CLEO possesses both, and recently presented a
first limit \cite{Coan:2002te}. Lower limits are expected from
BABAR and BELLE \cite{Johns03}.

\subsection{The "forbidden" mode: $D^0 \to e^{\pm}\mu^{\mp}$} 
\label{DMUE} 

A transition $D^0 \to e^{\pm}\mu^{\mp}$ violates lepton number  
conservation \index{lepton number violation}; 
observing it manifests unequivocally the intervention  
of New Physics. No signal has been observed so far:  
\beq  
{\rm BR}(D^0 \to \mu^+ e^-)|_{exp} \leq 8.1\cdot 10^{-6} 
\eeq 

Yet as before for $D^0 \to \mu^+\mu^-$ this rate is suppressed by the 
simultaneous factors 
$(f_D/m_c)^2 \sim 0.04$ and $(m_{\mu}/m_c)^2 \sim 0.007$ 
with the latter arising at least  
for a spin-1 component in the underlying $c\bar u$-coupling. 

There are classes of New Physics scenarios that can induce a signal here,  
namely models with Technicolor, nonminimal Higgs dynamics, heavy  
neutrinos, horizontal gauge interactions and SUSY with  R parity breaking. 
The last one again  provides the most promising -- or least discouraging 
-- case allowing for \cite{Burdman:2001tf} 
\beq  
{\rm BR}(D^0 \to \mu^+ e^-)|_{SUSY,\not R} \leq 1.0\cdot 10^{-6} 
\eeq 
One might be inclined to view such searches as bad cases of  
\index{ambulance chasing}
`ambulance chasing'. Yet one should keep in mind that the structure and
strength of flavour-changing neutral currents could be quite different  
in the up- and down-type quark sectors. As far as the latter is 
concerned, very sensitive searches have been and will continue to be 
performed in $K$ and $B$ decays\cite{Isidori02}. 
Yet the   
$u,c$ and $t$ quarks provide very different search scenarios: such  
lepton-number violating flavour-changing neutral currents for $u$ 
quarks can be probed via $\mu - e$ conversion in deep-inelastic  
scattering -- $\mu N \to e X$ -- where the system $X$ might or might  
not contain a charm hadron; top states on the other hand decay as quarks  
before they can hadronize \cite{RAPP} and thus present different 
challenges and promises in such searches. In any case a careful  
study of $D$ decays is thus complementary  
to analyses of $\mu-e$ conversion and top decays rather than a repeat 
effort. 

\subsection{Exotic new physics: $D^+ \to \pi^+/K^+ f^0$} 
\label{FAMILON} 
%
One of the central mysteries of the SM is the replication of  
quark-lepton families, which furthermore exhibit a quite peculiar  
pattern in the masses and CKM parameters. This structure might be  
connected with the existence of some continuous `horizontal' or  
family symmetry that has to be broken. If it is a {\em global} symmetry  
broken {\em spontaneously}, very light neutral Nambu-Goldstone bosons  
have to exist, the `familons' \index{familon}
\cite{FAMTH}. They might actually provide some   
valuable service by en passant solving the `strong CP' problem 
\index{strong CP problem} of the SM  
\footnote{The `strong CP problem' refers to the realization that a priori  
the T-odd operator $G\cdot \tilde G$ can appear in the QCD Lagrangian  
with a coefficient roughly of order unity. Yet bounds on the electric  
dipole moment of neutrons shows this coefficient to be at most of order  
$10^{-9}$. Peccei-Quinn type symmetries have been invoked to provide a  
natural solution to this apparent fine-tuning problem. They imply the  
existence of axions that have turned out to be elusive so far 
\cite{CPBOOK}; familons  
can be their flavour-nondiagonal partners.  
}. 
\par
The effective low-energy interaction of the familon fields $f^a$ with  
fermion fields $\psi_i$ is given by a non-renormalizable derivative 
coupling of the form  
\beq  
{\cal L}_F = \frac{1}{\Lambda_F} \bar \psi _i\gamma _{\mu} 
(V^a_{ij} + A^a_{ij}\gamma_5)\psi_j\partial _{\mu}f^a \; ;  
\eeq 
the mass scale $\Lambda_F$ calibrates the strength of these effective  
forces, while  
$V^a$ and $A^a$ denote matrices containing the vector and axial vector  
couplings of the familons to the different fermions. 

\par
The analogous decays $K^{\pm} \to \pi^{\pm}f^0$ \cite{E787},  
$B^{\pm} \to \pi^{\pm}/K^{\pm}f^0$ and $B_d \to K_Sf^0$ \cite{CLEOFAM} 
have been 
searched for with no signal found yielding 
$\Lambda_F^{(2)}\geq 10^{11}$ GeV and 
$\Lambda_F^{(3)}\geq 10^8$ GeV for the second and third family, 
respectively. 

\par
Surprisingly enough, no 
bounds have been given so far for familons in D decays. This unsatisfactory 
situation (familon couplings could a priori be quite different for up- and 
down-type quarks; searches for $D\to h f$ and $B\to h f$ are thus 
complementary to each other) should be remedied by BELLE, BABAR 
and CLEO-c \footnote{Top quark decays could produce familons as well.}. 
In the $K$ sector a new line of attack can be opened by KLOE at DA$\Phi$NE
$\phi$-factory searching for $K_S^0 \to \pi^0 f^0$. 

\subsection{$D^+, D_s^+ \to h \ell \ell^{\prime}$ with $h = \pi, K$}
\label{Threebody}
Final states with a charged hadron and two charged leptons have in addition to their 
favourable  
experimental signatures some phenomenological advantages as well over 
$D \to l\nu$, $ll$: their amplitudes suffer from neither helicity  
\index{helicity suppression}  nor wavefunction suppression; i.e., the small factors $m_l/m_c$  
and $ f_D/m_c$ are in general absent.  Yet beyond that they represent very different 
dynamical scenarios. 

$D^+, D_s^+ \to h^+ \mu^+e^-$ and $D^+, D_s^+ \to h^+ e^+\mu^-$ require genuine 
flavour changing neutral currents and thus would represent unequivocal signals for 
New Physics; both would actually be natural modes for different 
variants of `horizontal' \index{horizontal gauge interactions} 
or leptoquark \index{leptoquarks} interactions  coupling first and second family 
quarks and leptons to each other. 

The modes $D^+, D_s^+ \to h^- l^+l^+, \; h^-e^+\mu^+$ also require the intervention of 
New Physics; 
yet from a theoretical rather than purely phenomenological perspective they 
appear to represent a rather contrived scenario. 

Finally the rare transitions $D^+, D_s^+ \to h^- l^+l^-$ could certainly be affected 
considerably by New Physics; unfortunately they can proceed also within the SM 
with a strength that cannot be predicted reliably since there they are driven mainly 
by long distance dynamics  \cite{Burdman:2001tf,Fajfer:2001sa,Singer:1996it}, 
which are notoriously difficult to calculate. 
Recent FOCUS results\cite{Link:2003qp} substancially lower previous limits
\cite{Frabetti:1997wp,Aitala:1999db} (Tab.\ref{TAB:LDK}) and already allow one
to exclude SUSY models with R-Parity breaking 
\cite{Burdman:2001tf}. As a technical detail, the FOCUS  limit makes use of a
new bootstrap-based technique to ascertain the limit\cite{Rolke:2000ij}
confidence level,
 which has proved to provide a better {\em coverage} than the standard limit
technique  of Ref.~\cite{Feldman:1997qc}.

\par
\begin{table} 
\begin{center} 
\begin{tabular}{|lrr||lrr|}
\hline
Decay mode     &          & BR \%               & Decay mode   &     & BR \%\\ 
 $D^+\to e^+\nu_\mu$    &             & ---  &  
    $D_s^+\to e^+\nu_\mu$     &       &   --- \\
 $D^+\to \mu^+\nu_\mu$ &  PDG02              &  $0.08^{+0.17}_{-0.05}$ &  
    $D_s^+\to \mu^+\nu_\mu$  &  PDG02        &  $0.50\pm 0.19 $  \\
 \multicolumn{3}{|l|}{$\Rightarrow f_{D^+}= 177 \div 513$ MeV}         &  
 \multicolumn{3}{l|}{$\Rightarrow f_{D_s^+}= 250 \pm 50$ MeV}    \\
 $D^+\to \tau^+\nu_\tau$ &   & ---  &  
    $D_s^+\to \tau^+\nu_\tau$ &   PDG02      &  $6.4\pm 1.5 $  \\
    & & &  
  \multicolumn{3}{l|}{  $\Rightarrow f_{D_s^+}= 298 \pm 37$ MeV }\\
\hline
Decay mode       &         & BR $10^{-6}$              & Decay mode   & 
      & BR $10^{-6}$ \\
  $D^0\to \mu^+\mu^-$  & PDG02      &  $<4.1$ &  
   $D^0\to \mu^+e^-$    &   PDG02     &  $<8.1$  \\
    & \cite{Korn:2003pt,Acosta:2003ag}           &  $<2.4$  &  
                        &           &    \\ 
      $D^0\to \epem$    &PDG02 &$<6.2$ & & & \\                    
 $D^0\to \gamma\gamma$ & \cite{Coan:2002te}           & $<29$  &  
       &                       &   \\  
 $D^+   \to K^+\mu^-\mu^+  $ & PDG02                  & $<44$  &  
 $D^+   \to K^-\mu^+\mu^+  $ & PDG02                  & $<120$    \\ 
  & \cite{Link:2003qp}     & $<9.2$  &  
  & \cite{Link:2003qp}     & $<13$   \\   
 $D^+   \to \pi^+\mu^-\mu^+  $ & PDG02                & $<15$  &  
 $D^+   \to \pi^-\mu^+\mu^+  $ & PDG02                &  $<17$   \\ 
  & \cite{Link:2003qp}   & $<8.8$  &  
  & \cite{Link:2003qp}   & $<4.8$  \\   
 $D_s^+   \to K^+\mu^-\mu^+  $ & PDG02                & $<140$  &  
 $D_s^+   \to K^-\mu^+\mu^+  $ & PDG02                & $<180$   \\ 
  & \cite{Link:2003qp}   & $<36$  &  
  & \cite{Link:2003qp}   & $<13$   \\   
 $D_s^+   \to \pi^+\mu^-\mu^+  $ & PDG02              &  $<140$ &  
 $D_s^+   \to \pi^-\mu^+\mu^+  $ &  PDG02              &  $<82$ \\ 
 & \cite{Link:2003qp} &  $<26$ &  
  & \cite{Link:2003qp} &  $<29$  \\    
\hline
\end{tabular} 
\end{center} 
\caption{Experimental branching ratios for 
charm mesons leptonic and rare decays, and pseudoscalar charm meson decay
constants. World
averages from  PDG02 \cite{Hagiwara:fs}. All limits are
90\% cl. Statistical and systematical errors added in quadrature. } 
\label{TAB:LDK} 
\end{table} 
%
\section{Semileptonic Decays}  
\label{SEMILEPT}  
%
There is no honourable excuse for theory to duck its responsibility  
for treating semileptonic charm decays.
With seven $C=1$ hadrons  
decaying weakly -- and doing that at the Cabibbo allowed  
as well as once forbidden level -- one is certainly facing a complex  
challenge here. Yet there exists strong motivation for a  
dedicated analysis of   
semileptonic charm decays: while there is little hope for New  
Physics to surface there, they constitute a clean laboratory  
for probing  
our quantitative understanding of QCD; this has both intrinsic 
merit and serves as preparation for coming to  grips with hadronization in 
the beauty sector. En passant  one might learn something novel about 
light flavour spectroscopy  as well through a careful analysis  
of final states like $D^+_{(s)} \to (\eta/\eta^{\prime}) \ell \nu $ as  
explained later.  There are several features facilitating a  
theoretical description and making it more interesting at the same  
time:  
\begin{itemize} 
\item  
On each Cabibbo level there is only a single quark level transition  
operator.  
\item  
Factorization holds trivially \index{factorization}; i.e., the amplitudes for semileptonic  
transitions are expressed through the product of a leptonic current and 
the matrix element of a hadronic current. The latter is given by an  
expression bilinear in quark fields. While we have not (yet) established  
full control over quark bilinears, we know a lot about them and in  
any case considerably more than about the matrix elements of four-quark  
operators.  
\item  
On general, though somewhat handwaving grounds one expects semileptonic 
charm widths to be dominated by a handful of exclusive modes. Studying how 
the inclusive width is saturated by exclusive width will 
provide us with some instructive lessons about how duality can emerge  
at relatively low scales.  
\item  
They allow direct access to $|V(cs)|$ and $|V(cd)|$.  
\item  
They constitute a laboratory for studying dynamics similar to those  
shaping semileptonic beauty decays.  
\item  
They provide an important bridge between light flavour and heavy  
flavour dynamics that can be studied by approaching them from  higher mass 
scales using heavy
quark expansions  and from lower ones directly  
through lattice QCD. 
\end{itemize} 

\subsection{Inclusive Transitions}  
\label{SLINCL}
In Sect.\ref{WEAKLIFE} we have already discussed the {\em inclusive} semileptonic  
branching ratios for the various charm hadrons, since they fit  
naturally into the treatment of weak lifetimes. In our discussion  
there we have focussed on {\em ratios}, since there $|V(cs)|$ as well  
as the leading term $\propto m_c^5$ drop out.  

There is still a fly in the ointment. For one might  
become emboldened to predict also the absolute size of $\Gamma _{SL}(D)$. 
The $1/m_Q$ expansion as described in Sect.\ref{WEAKLIFE} yields 
\bea
\Gamma_{\rm SL}(D\to \ell \nu X)&=&\frac{G_{F\,}^2
m_c^5}{192\,\pi^3}\; \raisebox{-.5mm}{\mbox{{\large$|V_{cq}|^2$}}}
\!\left[z_0(r)\!\left(
1\!-\!\frac{\mu_\pi^2\!-\!\mu_G^2}
{2m_c^2}\right)- 2(1\!-\!r)^4
\frac{\mu_G^2}{m_c^2}+ \!{\cal O}(\alpha_{em},\alpha_s,\frac{1}{m_c^3})
\right]\!,
\label{SLwid}
\eea
where $z_0(r)$ is the tree-level phase space factor and
$r\!=\!m_q^2/m_c^2$:
\beq
z_0(r)=1-8r+8r^3-r^4-12r^2\ln{r}\;.
\label{SLwid1}
\eeq
Comparing the measured width with the theoretical expression 
using $|V(cs)|$ and $m_c \sim 1.3$ GeV, as inferred from charmonium  
spectroscopy, one finds that one can reproduce no more than  
two thirds of the observed $\Gamma_{SL}(D)$ with those input values.  
It has been conjectured that {\em non}factorizable contributions might  
make up the deficit. 

Also the  
lepton energy {\em spectrum} of inclusive semileptonic charm hadron  
transitions can teach us valuable lessons. Even among mesons -- with  
practically identical semileptonic widths -- one expects significant  
differences in the spectra of $D^0$ and $D_s^+$ [$D^+$] decays on the  
$c\to s$ [$c\to d$] level due to WA contributions\index{weak annihilation}. 
While one can count  
on no more than a semiquantitative description for the integrated widths,  
even that would be too much to expect for the spectra with their need  
for `smearing' 
\cite{MANIFESTO,BSUVPRL}. Nevertheless in the spirit of Yogi Berra's dicta 
\footnote{Yogi Berra was an outstanding baseball player, manager and coach, who 
was voted onto the All-Century team. He is, however, better known as the founder 
of one of the most popular schools in US philosophy, namely Yogi-ism 
\index{Yogi-ism} characterized by immortal quotes like 
\cite{YOGI}: "You can observe a 
lot by watching"; 
"It's deja vu all over again"; "When you come to a fork in the road ... take it"; "If the 
world were perfect, it wouldn't be"; "The future ain't what it used to be". }
one  
can always start an observation of something by looking at it and  
thus search for evidence of WA in the endpoint spectra; if the observed  
pattern fits the expectations, then one can try to infer the  
$D$ meson expectation value of the four-fermion operator; this quantity  
has both intrinsic and practical interest -- the latter since the  
$B$ meson analogue of this four-fermion operator affects the lepton  
energy endpoint spectra in $B_d$ and $B_u$ transitions  
{\em differently} \cite{WA}. 
OPAL has produced the first measurement \cite{Gagnon:1999nu,Abbiendi:1998ub} of 
the semileptonic branching ratio averaged over all weakly decaying 
charm hadrons produced in $Z^0 \rarr \ccb$ decays. Using  the theoretical 
prediction for $\Gamma(Z^0 \rarr \ccb)$ it finds 
$B(c\rarr \ell) = 0.095\pm0.006^{+0.007}_{-0.006}$, in good 
 agreement with both expectation based on the dominance of $D^0$ over $D^+$ 
 abundance due to $D^*$ 
production and with ARGUS' data at lower energies. 

\subsection{Exclusive Modes} 
\label{SLXCL}
%
There are three main reasons to study exclusive channels: 
\begin{enumerate}
\item 
Exclusive semileptonic decays are easier to measure than inclusive ones, though
more   
difficult to treat theoretically, since they are highly sensitive to  
long distance dynamics. One can turn this vice into a virtue, though: measuring 
these transition rates and comparing them with predictions will extend our 
knowledge and maybe even understanding of nonperturbative dynamics. This is a 
worthy goal in itself -- and of obvious benefit for understanding the
corresponding  
decays of beauty hadrons. It is essential in such an endeavour to have a large
body  
of different well measured modes rather than a few ones. 
\item 
Studying how exclusive modes combine for different charm hadrons to saturate the 
inclusive widths will teach us valuable lessons on how and to which 
degree quark-hadron duality discussed 
in Sect.\ref{QHDUALITY}
 is implemented in charm decays \index{quark-hadron duality}. 
\item 
The final states of semileptonic charm decays can provide novel insights 
into light flavour spectroscopy. 
\end{enumerate}
%
\subsubsection{$H_c \to \ell \nu h$}
\label{HC2ELLNUH}
\begin{figure}[t]
   \centering 
   \includegraphics[width=5.0cm,height=3.0cm]{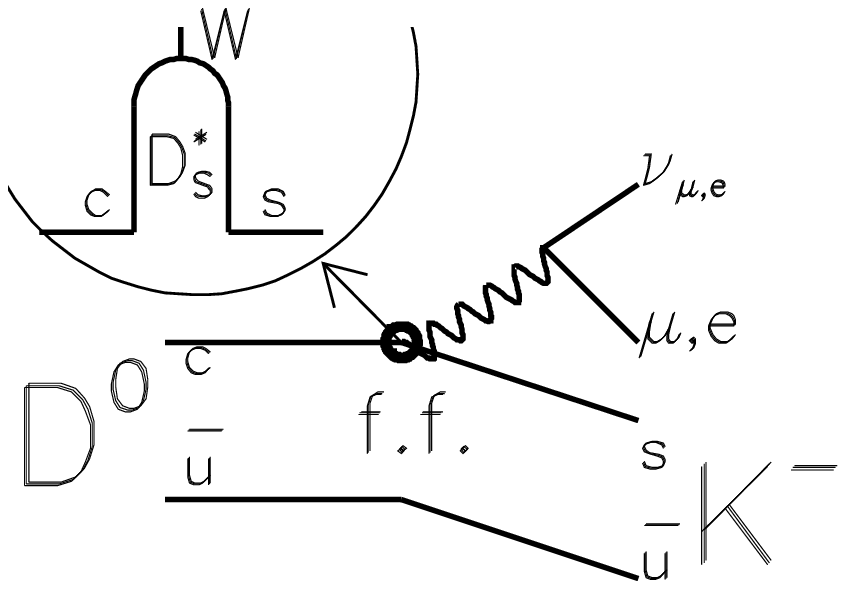} 
   \includegraphics[width=5.0cm,height=3.0cm]{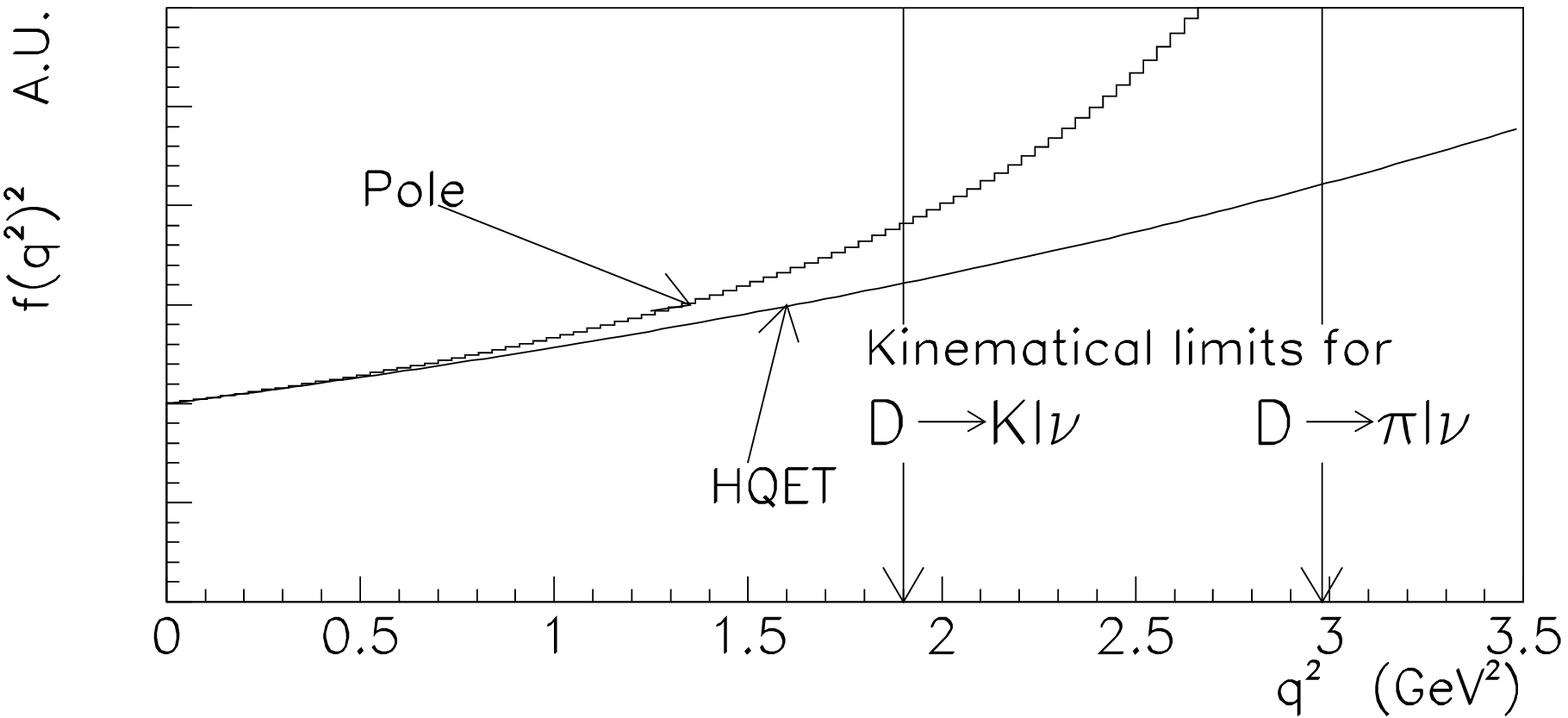} 
 \caption{Semileptonic $D^0$ decay due to the coupling of a virtual $D_s^*(c\bar s)$ 
 vector state in the nearest-pole dominance model (left);  $|f_+(q^2)|^2$ as a
 function of $q^2$ for pole and HQET parameterizations. The kinematic
 limits for  $K\ell\nu$ and $\pi\ell\nu$ are drawn with vertical lines
 (adapted from  \cite{Wiss:1997kb}). 
    \label{fig:slgraph} }
\end{figure}

%
The most tractable case arises when there  
is a single hadron or resonance $h$ in the final state. Long distance  
dynamics can then enter only through hadronic  
form factors,
 \index{form factors}
  which describe how the charm hadron is transformed 
into  
the daughter hadron at the hadron $W$ vertex (Fig.\ref{fig:slgraph}a)  
as a function of the momentum transfer. To 
be more specific: for 
$D$ decays into a final state  with a single pseudoscalar meson $P$, the 
matrix element of the  hadronic weak current can be expressed in terms of 
two  formfactors  
\beq  
\matel{P(p^{\prime})}{J_{\mu}}{D(p)} =  
f_+(q^2)(p + p^{\prime})_{\mu} + f_-(q^2)(p - p^{\prime})_{\mu}\; , \; \;  
q=p - p^{\prime} 
\eeq 
\par
The differential decay 
 rate is given by  
\beq 
 \label{decrate} 
 \frac{d\Gamma(D \to \ell \nu P)}{dq^2} = \frac{G^2_F |V_{cq}|^2 
p_P^3}{24\pi^3} 
 \left\{ |f_+(q^2)|^2+ |f_-(q^2)|^2{\cal O}(m_\ell^2)+... \right\}  
\eeq  
\beq
m^2_\ell \leq q^2 = M_D^2+m_P^2-2M_DE_P \leq 
2M_DE_\ell +2m_P^2E_\ell (2E_\ell-M_D) 
\label{EQ:Q2}
\eeq
where $p_P$ and $E_P$ are the momentum and energy, respectively, of  $P$ in the rest 
frame of  the charmed meson. The dependance on the second  
form factor $f_-$ is proportional to the mass squared of the lepton  
and therefore insignificant in $D$ decays. 
In decays leading to a vector resonance $V$ --  
$D \to \ell \nu V$ -- there are four formfactors  
often denoted by $V$, $A_{1-3}$ and defined by the decomposition of the  
hadronic weak current: 
$$  
\matel{V(p^{\prime}),\lambda}{J_{\mu}}{D(p)} =  
-i(M_D + M_V)A_1(q^2)\epsilon_{\mu}^{*(\lambda)} +  
\frac{iA_2(q^2)}{M_D+M_V}(\epsilon_{\mu}^{*(\lambda)}\cdot p_D) 
(p + p^{\prime})_{\mu}   
$$ 
\beq 
+\frac{iA_3(q^2)}{M_D+M_V}(\epsilon_{\mu}^{*(\lambda)}\cdot p_D) 
(p - p^{\prime})_{\mu} +  
\frac{2V(q^2)}{M_D+M_V}\epsilon_{\mu\nu\rho\sigma} 
\epsilon_{\nu}^{*(\lambda)}p_{\rho}p^{\prime})_{\sigma} \;  ; 
\eeq 
$\lambda$ denotes the polarization of $V$. The $A_3$ contribution  
is suppressed by the lepton mass and thus can be ignored here, similar to  
the $f_-$ term in $D\to \ell \nu P$. The two salient features of the form factors 
are their $q^2$ dependence  
and their normalization at a given value of $q^2$; a typical, though  
not unique, choice is $q^2 =0$. 
Three different strategies have been 
suggested for treating these formfactors theoretically: 
\begin{enumerate} 
\item  
 One uses  
 wavefunctions from a specific quark model to calculate the  
 residue $f_{\pm}(0)$. For the $q^2$ dependance of $f_\pm(q^2)$  
 one assumes a parametrization consistent  
 with the quark model used, like nearest pole 
 dominance \index{nearest pole dominance} \cite{BSW}: 
 \begin{equation} 
  f_\pm(q^2) = f_\pm(0)(1-q^2/M^2_{pole})^{-1} 
  \label{vecmes}   
 \end{equation} 
 In a literal application of this ansatz one chooses the 
 value for $M_{pole}$ as the mass of the nearest charm  
 resonance with the same $J^P$ as the hadronic weak  
 current (Fig.\ref{fig:slgraph}a). This can be an approximation only; 
 one can allow for some flexibility here and fit 
 $M_{pole}$ to the data. A deviation of it from the mass of the 
nearest resonance indicates contributions from higher states. 

  Another model in this class, which is 
 sometimes (although not quite correctly) called the HQET form uses  
 instead  \cite{Scora:1995ty}
 \begin{equation} 
  f_\pm(q^2) =  f_\pm(0)e^{\alpha q^2} 
  \label{wisgur}  
 \end{equation}  
 While the two functional dependences on $q^2$ in Eqs.(\ref{vecmes})  
 and (\ref{wisgur}) look quite different, there is hardly a   
 difference in $D\to K\ell \nu$ due to the small range in $q^2$  
 kinematically allowed there (Fig.\ref{fig:slgraph}b from 
 ref.\cite{Wiss:1997kb}). 
 On the other hand, for the Cabibbo suppressed  channel 
 $D \to \pi\ell \nu$ there is considerable sensitivity to  the differences 
 between the two functions. 
 Such approaches represent theoretical engineering. This is not meant 
 as a put-down. For they are certainly  
 useful: they train our intuition, provide tools for estimating detection  
 efficiencies, yield insights into the strong dynamics and allow 
 extrapolations to the corresponding $B$ decays. Yet such lessons 
 cannot be taken too literally  
 on a quantitative level. For one cannot count on quark models to yield  
 truly reliable error estimates or providing quantitative control over  
 the extrapolations to beauty transitions.  
\item  
 Right after the emergence of HQET there had been considerable optimism  
 about heavy quark symmetry allowing us to reliably evaluate the 
 formfactors for $B\to \ell \nu \pi/\rho$, which are needed to extract 
 $V(ub)$,   by 
 relating them to the ones for $D\to \ell \nu K$, which 
 can be measured (using the known 
 value of $V(cs)$, see our discussion below). 
 Those hopes have largely faded away since unknown 
 corrections of order $1/m_c$ will  
 be quite sizeable. 
 Heavy quark concepts are still useful, yet have to be complemented  
 with other theoretical technologies. One such technology that --  
 unlike quark models -- is directly based on QCD and its field 
 theoretical features employs sum rules \index{sum rules} 
 similar to those introduced by  
 Shifman, Vainshtein and Zakharov and discussed in \ref{SUMRULES}. 
 The particular variant used is referred to as light cone sum rules 
 \cite{LCSR}\index{light cone sum rules}. 
 Ref.\cite{PBALL} finds  
 \bea  
  {\rm BR}(D^0 \to e^+\nu \pi^-) &=&  (1.6 \pm 0.34) \cdot 10^{-3} 
  \label{PREDPI} 
  \\  
  {\rm BR}(D^0 \to e^+\nu \rho^-) &=& (4.8\pm1.4) \cdot 10^{-3} 
 \eea 
 The prediction of Eq.(\ref{PREDPI}) underestimates the data very 
 significantly:  
 \beq  
 {\rm BR}(D^0 \to e^+\nu \pi^-) =  (3.6 \pm0.6 ) \cdot 10^{-3} 
 \eeq 
 The discrepancy appears much larger than the customary 
 20 - 30 \% theoretical uncertainty one has to allow for QCD sum rule as 
 explained in Sect.\ref{SUMRULES}.
  A possible explanation for this failure is that the 
 charm mass is too low a scale for the leading terms to provide an accurate 
 description; such reasoning could not be invoked, though, for a similar 
 failure in $B$ decays. 
\item  
Both the shapes and normalizations of the various form factors listed above 
can be computed based on lattice simulations. Several lattice groups have presented 
what can be viewed as pilot studies of the differential decay width for processes like 
$D \to \pi \ell \nu$ in the {\em quenched} approximation.

In \cite{El-Khadra:2001rv}
 a lattice version of HQET is implemented to determine the relevent form 
factors at different pion energies.  Many sources of error arise in this 
analysis leading to statistical and systematic uncertainties of 
$10-20$\%.  The integrated quantity
\beq
T_D\left(p_{min},p_{max}\right)=\int_{p_{min}}^{p_{max}} dp\,\, p^4\, 
|f_+(E)|^2/E
\eeq
\noindent
can be combined with experimental measurements to obtain $V_{cd}$  using
\beq
|V_{cd}|=\frac{12 \pi 
^3}{G_F^2m_D}\frac{1}{T_D\left(p_{min},p_{max}\right)}\int_{p_{min}}^{p_{max}} 
dp \frac{d\Gamma_{D \to \pi}}{dp}.
\eeq 
Here $p$ is the magnitude of the pion three-momentum in the $D$ meson 
rest frame.  The lower limit $p_{min}$ is chosen to minimize 
extrapolation uncertainties in $p$ and the light quark mass.  The upper 
limit is chosen to reduce statistical and discretization uncertainties.  
With uncertainties of 10-20\% in $T_B$, $|V_{cd}|$ can only be 
determined at the 15\% level using this technique in the quenched 
approximation. 

Other groups have made similar quenched determinations using other 
techniques.  Reference \cite{Aoki:2001rd} uses a lattice version of 
NRQCD, whereas references \cite{Bowler:1999xn,Abada:2000ty}
rely on fermions with light quark normalizations.  Some of these results 
are shown in Fig.~\ref{FIG:LATTFF} demonstrating the agreement between 
the differing techniques and also displaying the overall size of the 
uncertainties.
\begin{figure}[t]
  \centering
   \includegraphics[width=10.0cm]{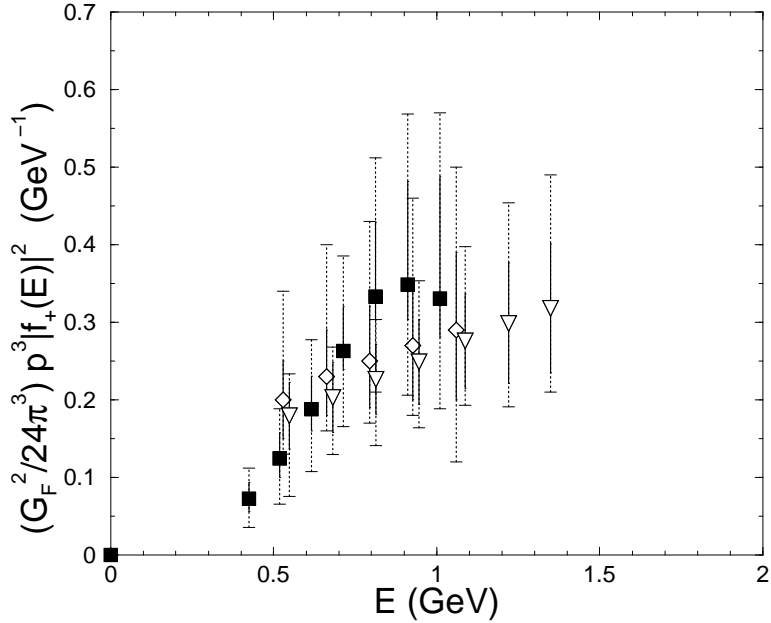}
 \caption{A comparison of various lattice predictions for the 
semi-leptonic differential decay width as a function of the pion energy 
E:  diamonds \cite{Bowler:1999xn},triangles \cite{Abada:2000ty}, and 
squares \cite{El-Khadra:2001rv}.  This plot is taken from reference 
\cite{El-Khadra:2001rv}.
    \label{FIG:LATTFF} }
\end{figure}

The high precision group\cite{LATTICEPREC} has included semi-leptonic 
$D$ meson decays in their list of the "golden-plated" modes for which 
the lattice can provide highly accurate values.  The claim is that with 
{\em un}quenched lattice simulations the overall uncertainty on the 
semi-leptonic form factors can be reduced to level of 1-3\%, which represents 
a noble goal.  There is the promise that this theoretical accuracy level will be 
reached on the same time scale as CLEO-c will provide its precise 
measurements or even before. 
 \end{enumerate}   
\par 
All charm experiments have dedicated programs for studying 
semileptonic decays
 (for extensive reviews see Refs.\cite{Richman:wm,Wiss:1997kb}). 
To measure \q2 in exclusive decays  
one needs to identify all decay products, determine the $H_c$ rest frame, 
and compute \q2 via Eq.(\ref{EQ:Q2}). The determination of this rest frame is not 
unambiguous due to the neutrino escaping detection. 
In experiments with good tracking resolution (typically with a fixed-target geometry) 
the $H_c$ direction can be measured, and the rest frame determined up
  to a quadratic ambiguity. At colliders with an hermetic detector, one can
  study \q2 inclusively by estimating the missing neutrino energy and momentum.
  A must is the capability of selecting semileptonic events by identifying the
  muon, the electron, or both. 
\par
 All charm experiments have contributed to the measurement of semileptonic D
 decays branching ratios, and most of them of formfactors, 
 thanks to generally good
 tracking, vertexing, and lepton identification capabilities. Results new to
 PDG02 (Tab.\ref{TABSLBRSUM} shows 2003 update to \cite{Hagiwara:fs}) 
 have come from FOCUS \cite{Link:2002wg,Link:2002nj}
 and CLEO \cite{Brandenburg:2002eu} (see also the recent
 reviews \cite{Marinelli02,Wiss:2002ju,Johns03}). 
 \par
  A long-standing issue is in $BR(D^+\to K^* \ell \nu)$ relative to $K\pi\pi$,
   where in the electron
  channel the 2002 CLEO measurement is several $\sigma$
 higher than the 1989 E691 value. 
 The PDG03 average $0.61\pm 0.07$ contains a 1.7 error scale factor 
 to reconcile
 CLEO and E691 incompatible measurements.
 Quark models predict values close to CLEO's (see  \cite{Wiss:1997kb}
 for a review).
 The 2002 FOCUS  semimuonic branching ratio is $0.602\pm .022$ ---
 higher that E691, but about 1$\sigma$ lower than CLEO.  
 \par
\begin{figure}
     \centering 
   \includegraphics[width=10.0cm,height=10.0cm]{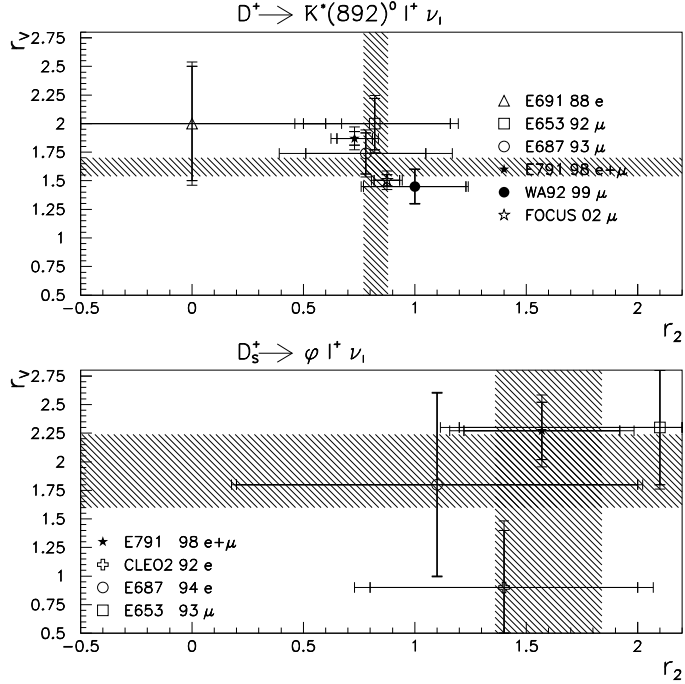} 
 \caption{Compilation of experimental results on semileptonic vector
 formfactor ratios. Error bars show the statistical error, and the
 statistical and systematic added in quadrature.  PDG03 updated averages to
 2002 \cite{Hagiwara:fs} are shown in the boxed area. PDG03 applies a 1.5
 factor on the error of $r_V$ world average for $D^+\rarr {\bar
 K}^{*0}\ell^+\nu_\ell$. 
    \label{FIG:FFRATIO} }
\end{figure}

 Most of the efforts have been concentrated on the precise measurement
   of formfactor ratios (Fig.~\ref{FIG:FFRATIO}). 
 \par
 The measurements of formfactor ratios have generated in the past 
 considerable controversy, with three measurements (E687, E653, E691) in 
 mild disagreement, albeit with large statistical errors 
 (Fig.\ref{FIG:FFRATIO}). A new measurement by FOCUS\cite{Link:2002wg},
  which collected large
 samples  of 
 clean  $D^+\rarr {\bar K}^{*0}\mu^+\nu_\mu$  (fifty-thousand events), 
 determines $r_V, r_2$ with a few-percent error, and confirms the WA92 
 measurement.   SU(3) flavor symmetry 
 between $D_s$ and $D^+$ semileptonic decays would suggest equal $r_V$ and 
 $r_2$ for  $D^+\rarr {\bar
 K}^{*0}\ell^+\nu_\ell$ and $D^+_s\rarr 
 \phi\ell^+\nu_\ell$, where a spectator $\bar d$ quark is replaced by a
 spectator $\bar s$ quark. This indeed is the case for $r_V$, while $r_2$
 appears inconsistent at the $3\sigma$ level. 
  \par
The  FOCUS measurement\cite{Link:2002ev} featured the observation of a
broad  S-wave amplitude component in the $K\pi$ resonant substructure, which
evidently   accompanies the $K^*(890)$ amplitude in the decay $D^+\rarr K^-
\pi^+\mu^+\nu_\mu$.
 This claim is based on  the observation of a relevant forward-backward
 asymmetry in  $\cos \theta_V$ (the angle between the pion and the $D$ direction
 in the $K^-\pi^+$ rest frame of decay), for events with $K^-\pi^+$ mass below
 the ${\bar K}^{*0}$ pole. FOCUS interprets this as the interference of a broad
 s-wave amplitude and the ${\bar K}^{*0}$ amplitude. The asymmetry is well
  represented by a $0.36\exp(i\pi/4)(GeV^{-1})$ s-wave, with a strength of about
 $7$\% relative to ${\bar K}^{*0}$. 
 \par
The S wave component has important consequences in the computation of 
formfactors, and in the measurement of relative semileptonic branching ratios. 
The new measurement by FOCUS of branching ratios does include the 
effect of such a s-wave component 
and it 
significantly changes the PDG02 average. 
\par
It is desirable to have these findings checked by other experiments. If
confirmed,  
one should study whether this broad S wave structure has some connection to 
 \index{$\kappa$ resonance in charm decays}
the broad scalar $\kappa$ state that has emerged recently in a Dalitz plot 
analysis by E791,  see Sect.\ref{DALITZRESULTS}. 
\par
There are also consequences for $B$ decays, namely a proper interpretation 
of the rare $B \to (K\pi)\gamma$, $(K\pi)\ell ^+\ell ^-$ and
 $B_s \to (K\pi) \ell \nu$ transitions. 
FOCUS will soon provide a measurement of formfactor
  ratios for $D^+_s\rarr  \phi\ell^+\nu_\ell$ from  a ten-thousand events
  sample.
  From this measurement it will be interesting to see whether a $\phi-f_0$
  interference  phenomenon in the $K^-K^+\mu\nu$ decay can be found, analogous
  to the broad S-wave reported in $K^-\pi^+\mu\nu$. FOCUS
   have also announced
  \cite{Wiss:2002ju}
  a large sample of clean 
  $D\to \pi \ell \nu$ signals,
  whose relevance in extracting the \q2 dependance has been discussed
  previously. BELLE and BABAR have already accumulated much larger statistics 
  than CLEO and shall present new results soon. 
\par
 Results from B-factories are also highly expected on 
$D^+_{(s)} \to (\eta/\eta^{\prime}) \ell \nu$.
Only two measurements do exist in PDG02, from
 CLEO and the neutrino experiment E653. Their rates can shed light on the 
 relative weight of decay mechanisms 
 -- spectator vs. WA -- and on the strange vs. nonstrange vs. 
 {\em non-}$q\bar q$ 
 components of the $\eta$ and $\eta ^{\prime}$ wavefunctions. 
 Very preliminary new results from CLEO have been very recently
announced\cite{Johns03}. 
\begin{table}
\begin{center}
\begin{tabular}{|llrrr|}
\hline
           & Norm. Mode & Rel BR & ~$^{(1)}$Abs. BR  & Scale \\
           &            &        &          \% & factor \\	   
	  \hline\hline
 $D^+\to$ &              &        &         &              \\
   $\bar K^0 e^+\nu_e$ &
               & 
              & 
    $6.5\pm 0.9$       &
                              \\
                      &
    $\bar K^0 \pi^+$           & 
    $ 2.39\pm 0.31$           & 
    $$       &
                              \\
   $$ &
    $K^-\pi^+\pi^+$           & 
    $ 0.73\pm 0.09$           & 
    $$       &
                              \\
   $\bar K^0 \mu^+\nu_\mu$ &
               & 
                & 
    $7.0^{+3.0}_{-2.0}$       &
                              \\
   $\bar K^-\pi^+e^+\nu_e$ &
    $$           & 
    $$           & 
    $4.4^{+0.9}_{-0.7}$       &
                              \\
   $\bar K^*(892)^0e^+\nu_e$ &
    $$           & 
    $ $           & 
    $3.2\pm 0.33$       &
                              \\
   $K^-\pi^+\mu^+\nu_\mu$ &
    $$           & 
    $ $           & 
    $3.79\pm 0.33$       &
     1.1                         \\
   $\bar K^* (892)^0\mu^+\nu_\mu$ &
    $$           & 
    $ $           & 
    $3.0\pm 0.4$       &
                              \\
   ${K^-\pi^+\mu^+\nu_\mu}_{(NR)}$ &
    $$           & 
    $ $           & 
    $0.31\pm 0.12$       &
                              \\
   $$ &
    $K^-\pi^+\mu^+\nu_\mu$           & 
    $0.083\pm 0.029$           & 
    $$       &
                              \\
   $\pi^0\ell^+\nu_\ell$ &
    $$           & 
    $ $           & 
    $0.31\pm 0.15$       &
                              \\
   $$ &
    $\bar K^0\ell^+\nu_\ell$           & 
    $ 0.046\pm 0.014\pm 0.017$           & 
    $$       &
                              \\
   $\bar K^*(892)^0e^+\nu_e$ &
    $$           & 
    $ $           & 
    $5.3\pm 0.7$       &
     1.5                         \\
   $$ &
    $K^-\pi^+e^+\nu_e$           & 
    $ 1.21^{+0.21}_{0.24}$           & 
    $$       &
                              \\
   $$ &
    $K^-\pi^+\pi^+$           & 
    $0.60\pm 0.07 $           & 
    $$       &
     1.7                         \\
   $\bar K^*(892)^0\mu^+\nu_\mu$ &
    $$           & 
    $ $           & 
    $5.2\pm 0.$       &
     1.1                         \\
   $$ &
    $K^-\pi^+\pi^+$           & 
    $ 0.590\pm 0.023$           & 
    $$       &
    ~$^{(2)}$1.1                          \\
   $\rho^0 e^+\nu_e$ &
    $$           & 
    $ $           & 
    $0.24\pm0.09$       &
                              \\
   $$ &
    $\bar K^*(892)^0 e^+\nu_e$           & 
    $ 0.045\pm 0.014\pm 0.009$           & 
    $$       &
                              \\
   $\rho^0 \mu^+\nu_\mu$ &
    $$           & 
    $ $           & 
    $0.32\pm 0.08$       &
                              \\
   $$ &
    $\bar K^*(892)^0\mu^+\nu_\mu$           & 
    $ 0.061\pm0.014$           & 
    $$       &
                              \\
			    \hline
 $D^0\to$ &              &        &         &              \\			    
   $K^-e^+\nu_e$ &
    $$           & 
    $$           & 
    $3.58\pm 0.18$       &
      1.1                        \\
   $$ &
    $K^-\pi^+$           & 
    $  0.94\pm0.04$           & 
    $$       &
                              \\
   $K^-\mu^+\nu_\mu$ &
    $$           & 
    $ $           & 
    $3.19\pm 0.17$       &
                              \\
   $$ &
    $K^-\pi^+$           & 
    $0.84\pm 0.04 $           & 
    $$       &
                              \\
   $$ &
    $\mu^+X$           & 
    $0.49\pm 0.06 $           & 
    $$       &
                              \\
   $K^-\pi^0e^+\nu_e$ &
    $$           & 
    $ $           & 
    $1.1^{+0.8}_{-0.6}$       &
     1.6                         \\
   $\bar K^0\pi^-e^+\nu_e$ &
    $$           & 
    $ $           & 
    $1.8\pm 0.8$       &
     1.6                         \\
   $\bar K^*(892)^-e^+\nu_e$ &
    $$           & 
    $ $           & 
    $1.43\pm 0.23$       &
                              \\
   $$ &
    $K^-e^+\nu_e$           & 
    $0.60\pm 0.10 $           & 
    $$       &
                              \\
   $$ &
    $\bar K^0\pi^+ \pi^-$           & 
    $ 0.36\pm 0.06$           & 
    $$       &
                              \\
   $\pi^+e^+\nu_e$ &
    $$           & 
    $ $           & 
    $0.36\pm 0.06$       &
                              \\
   $$ &
    $K^-e^+\nu_e$           & 
    $ 0.101\pm 0.017$           & 
    $$       &
                              \\
			    \hline
 $D^+_s\to$ &              &        &         &              \\	
   $\phi\ell^+\nu_\ell$ &
    $$           & 
    $ $           & 
    $2.0\pm 0.5$       &
                              \\
   $$ &
    $\phi\pi^+$           & 
    $0.55\pm 0.04 $           & 
    $$       &
                              \\
   \multicolumn{2}{|l}{$\eta\ell^+\nu_\ell + \eta^\prime(958)\ell^+\nu_\ell$}  & 
    $ $           & 
    $3.4\pm 1.0$       &
                              \\
   $$ &
    $\phi\ell^+\nu_\ell$           & 
    $1.72\pm 0.23 $           & 
    $$       &
                              \\
   $\eta\ell^+\nu_\ell$ &
    $$           & 
    $ $           & 
    $2.5\pm 0.7$       &
                              \\
   $$ &
    $\phi\ell^+\nu_\ell$           & 
    $ 1.27\pm 0.19$           & 
    $0.89\pm 0.33$       &
                              \\
   $\eta^\prime(958)\ell^+\nu_\ell$ &
    $$           & 
    $ $           & 
    $$       &
                              \\
   $$ &
    $\phi\ell^+\nu_\ell$           & 
    $ 0.44\pm 0.13$           & 
    $$       &
                              \\			    			    			    			    			    			    			    			    			    			    			    			    			    			    			    			    			    			    			    
\hline
\end{tabular}
\end{center}
\caption{Summary of semileptonic branching ratios
 for $D^0$, $D^+$ and $D_s^+$ \cite{Hagiwara:fs}. Limits are not shown. 
 Notes:
 $^{(1)}$ Absolute
 BR are either results of PDG fit, or direct measurements at $\epem$ (see
 \cite{Hagiwara:fs} for details). 
 $^{(2)}$ This PDG03 new fit includes the FOCUS measurement that finds a small
 S-wave $K^-\pi^+$ amplitude along with the dominant $K^*$. Fitting the FOCUS
 result together with the previous measurements should be taken {\it cum grano
 salis}.}
\label{TABSLBRSUM}
\end{table}
In Table \ref{TABSLBR} we compare predictions from quark models with the data 
from a few modes. 
\begin{table}
\begin{center}
\begin{tabular}{|l|c|c|c|c|c|}
\hline
\vspace*{-.1cm}
SL Channel & ISGW2 \cite{Scora:1995ty} & Jaus\cite{Jaus:np} & MS 
\cite{Melikhov:2000wg} & WWZ \cite{Wang:2002zb} &  Exp \\
\hline
\hline
$\Gamma(D \to K)$ & 10.0 & 9.6 & 9.7 & 9.6 &  8.7 \\
$\Gamma(D \to K^*)$ &  5.4 & 5.5 & 6.0 & 4.8 & 5.2 \\
$\Gamma(D \to \pi)$ & 0.47 & 0.80 & 0.95 & 0.73 &  0.88 \\
$\Gamma(D \to \rho)$ & 0.24 & 0.33 & 0.42 & 0.34 &  0.26 \\
$\Gamma(D_s \to K)$ & 0.43 &  & 0.63 & &  \\
$\Gamma(D_s \to K^*)$ & 0.21 &  & 0.38 & &  \\
\hline
\end{tabular}
\end{center}
\caption{Predictions for various semileptonic decay rates in $10^{10} s^{-1}$. }
\label{TABSLBR}
\end{table}
The predictions show some significant variations, which are -- not surprisingly -- 
very sizeable for $D \to \pi$. 
\par 
 We conclude this subsection  with the discussion of a few selected topics in
 charm {\em baryon} semileptonic decays.
 If both initial and final state hadrons possess heavy flavour -- like in $b\to
 c \ell \nu$ --  
 then the semileptonic decays of baryons would exhibit some simpler features 
 than those of mesons. Since in $Qqq^{\prime}$ the light diquark $qq^{\prime}$ 
 carries 
 spin zero, in the heavy quark limit there is a {\em single} spin-1/2
 groundstate in the  
 baryon sector rather than 
 the spin-0 and -1 meson states. Likewise the three form factors 
 that in general describe 
 $\Lambda _{Q_1} \to \Lambda _{Q_2} \ell \nu$ can be expressed 
 through a single independent 
 Isgur-Wise function. In $\Lambda_c \to \Lambda \ell \nu$ the strange 
 baryon cannot 
 be treated as a heavy flavour object. Nevertheless it would be interesting 
 to study to which 
 degree these heavy quark symmetry expectations hold in $\Lambda_c$ 
 decays and the total $\Lambda_c$ semileptonic width is saturated by 
 $\Lambda_c \to \Lambda \ell \nu$. 
 \par
 The experimental scenario for charmed baryons semileptonic decays is at its
 infancy. Only an handful of branching ratio measurements are reported in PDG02,
 for decays $\Lambda_c\to \Lambda \ell \nu, \Xi_c\to \Xi \ell \nu$. 
 New results have recently come from CLEO\cite{Ammar:2002pf} and
 BELLE\cite{Chistov03} on $\Omega_c\to \Omega \ell \nu$ 
 and are reviewed in \cite{Johns03}, where new findings from CLEO
  on $\Lambda_c$
 form factors are also presented.  
 These studies mark just the beginning and
 we look forward to exciting news.

\subsubsection{Saturating the inclusive width}
\label{SATSL}

As explained in Sect.\ref{QHDUALITY} quark-hadron duality
\index{quark-hadron duality} 
represents a powerful concept at the heart of countless theoretical 
arguments. With 
charm being just `beyond the border' of heavy flavours it can provide 
us with important 
insights about the transition to the duality regime. This is true in 
particular for semileptonic 
decays. More specifically one can study not only to which degree a 
quark-gluon based 
treatment can describe the total semileptonic width of the various charm 
hadrons, see 
Sect.\ref{OPE}, but also how the latter is built up by {\em exclusive} channels. This is not 
`merely'  
of intrinsic intellectual interest, but might lead to practical lessons helping our understanding 
of $D^0 - \bar D^0$ oscillations and even $B\to \ell \nu X_u$. 

Of course, the quantitative features are quite different in 
$D\to \ell \nu X_{S=-1,0}$ and 
in $B\to \ell \nu X_u$, let alone $B\to \ell \nu X_c$, and lessons can 
therefore not be drawn 
mechanically. Nevertheless  it will be instructive to analyze the 
precise substructure -- 
resonant 
as well as nonresonant -- in final states like
 $\ell \nu K\pi$, $\pi \pi$, $\eta \pi$, 
$K\pi \pi$, $3\pi$ etc. and the relative weight of semileptonic final 
states with more than 
two hadrons. A detailed measurement of $D \to \ell \nu \pi \pi$ 
provides a lab for studying 
$\pi-\pi$ interactions in a different kinematical regime than in $K_{e4}$
 decays. 

\subsubsection{Light flavour spectroscopy in semileptonic decays.}
\label{SPECTSL} 

Theoretical predictions for $\Gamma (H_c \to \ell \nu h)$ depend 
also on the wavefunction of 
the hadron $h$ in terms of quarks (and even gluons). Considering the 
spectator diagram one finds that in $D^+_s \to \ell \nu \eta/\eta^{\prime}$ 
the 
$\eta/\eta^{\prime}$ are excited via their $\bar ss$ components, whereas in 
$D^+ \to \ell \nu \eta/\eta^{\prime}$ it is done through $\bar dd$. 
Measuring these four 
widths accurately will provide novel information on the relative
 weight of their 
strange and nonstrange $\bar qq$ components. Comparing the predictions 
from two models for $D^+_s \to \ell \nu \eta/\eta^{\prime}$ with the 
data one finds 
\begin{table}
\begin{center}
\begin{tabular}{|l|c|c|c|c|c|}
\hline
\vspace*{-.1cm}
SL Channel & ISGW2 \cite{Scora:1995ty} & Jaus\cite{Jaus:np} & MS 
\cite{Melikhov:2000wg} & WWZ \cite{Wang:2002zb} &  Exp \\
\hline
\hline
$\Gamma(D_s \to \eta)$ & 3.5 &  & 5.0 & & 5.3 \\
$\Gamma(D_s \to \eta^{'})$ & 3.0 &  & 1.85 & & 1.9 \\
\vspace{-.1cm}
$\Gamma(D_s \to \phi)$ & 4.6 &  & 5.1 & & 4.1 \\
\hline
\end{tabular}
\end{center}
\caption{Predictions for $D_s$ semileptonic decay rates in $10^{10} s$. }
\end{table}

\begin{figure}
     \centering 
   \includegraphics[width=5cm]{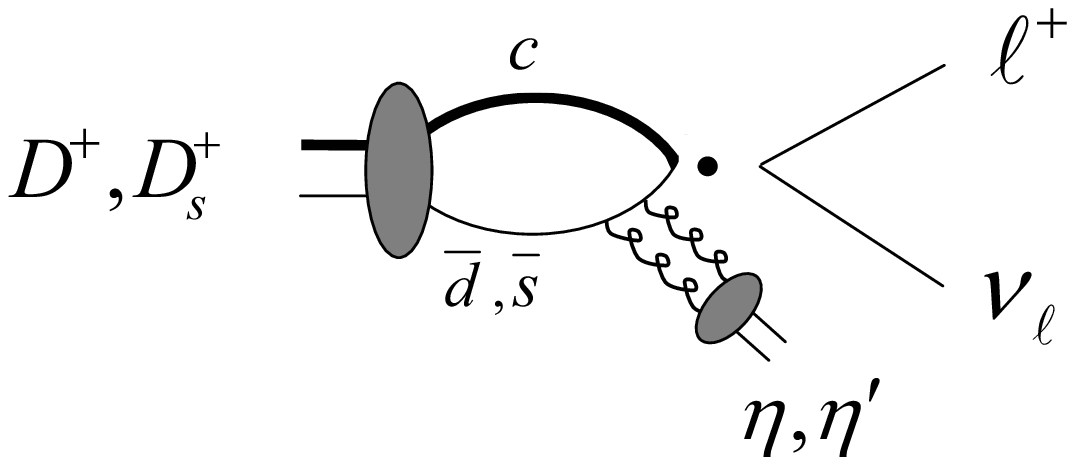} 
 \caption{Diagrams for $(D^+,D_s^+) \to (\eta,\eta^\prime) \ell \nu$
    \label{FIG:DETASL} }
\end{figure}

Yet there is another layer of complexity to it: while WA provides no more than a 
nonleading contribution to inclusive rates, it could affect exclusive modes very 
considerably. This can be visualized through the diagram of
Fig.~\ref{FIG:DETASL},  
where the $c$ and $\bar s$ or $\bar d$, before they annihilate into a virtual 
$W$ emit two gluons generating the $\eta$ or $\eta^{\prime}$ through a $gg$ 
component in the latter's wavefunctions \cite{OLDPAPER}. It could conceivably 
affect also $D_s^+ \to \ell \nu \phi$. 
\par
This mechanism 
can give rise also to the unusual mode $D_s^+ \to l^+\nu \pi^+\pi^-$. 
There had even been speculation \cite{OLDPAPER} that a glueball can be produced 
in such a way, if the latter is sufficiently light.  
\par
To conclude this subsection, we remind another example of unexpected surge of
connection between charm semileptonics decays, and light flavour hadronic
physics (see Sect.\ref{HC2ELLNUH}). The recent evidence 
for a broad S-wave resonant
component in $D^+\to (K\pi) \ell \nu$ can possibly be related to the debated
observation of the $\kappa$ resonance in Dalitz plot analysis of charm
decays (Sect.\ref{DALITZ}).
%
\subsection{$V(cs)$ \& $V(cd)$} 
\label{VCS} 
%
The CKM parameters constitute fundamental quantities of the SM. It is 
believed that they are shaped by dynamics at high scales -- a conjecture  
strengthened by the seemingly non-trivial pattern in their values, 
a point we 
will return to in our discussion of CP violation.  
Therefore one would like to determine 
them as precisely as possible.  
At first sight it might seem that of the three CKM parameters  
involving charm quarks -- $V(cd)$, $V(cs)$ and $V(cb)$ -- the first  
two are already known with excellent precision, as stated in PDG '00:  
$$  
|V(cd)| = 0.222 \pm 0.002\; , \; |V(cs)| = 0.9742 \pm 0.0006  
$$ 
\beq  
|V(cb)| = 0.040 \pm 0.002 
\label{00UNIT} 
\eeq 
Yet in arriving at these numbers one has {\em imposed} three-family  
unitarity\index{three-family unitarity}. While there is no evidence 
for a fourth family and we  
actually know the $Z^0$ decays into three neutrino types only, we  
should not limit ourselves to a three-family scenario.   
Listing the values obtained from the data 
{\em without} three-family unitarity one has according to PDG '98:  
$$  
|V(cd)| = 0.224 \pm 0.016\; , \; |V(cs)| = 1.04 \pm 0.16  
$$  
\beq  
|V(cb)| = 0.0395 \pm 0.0017 
\eeq 
The uncertainties are considerably larger now, in particular for  
$|V(cs)|$. The latter is obtained primarily from $D \to \ell \nu K$ decays,  
the description of which suffers from sizeable theoretical  
uncertainties, as discussed before. On the other hand  
$|V(cd)|$ is  derived mainly from the observed charm production rate in 
deep inelastic neutrino scattering is much better known -- as is 
$|V(cb)|$  extracted from semileptonic $B$ decays.

With three-family unitarity PDG '02 lists:  
$$ 
|V(cd)|_{(3)} = 0.223 \pm 0.002\; , \; |V(cs)|_{(3)} = 0.9740 \pm 0.0006  
$$ 
\beq  
|V(cb)|_{(3)} = 0.041 \pm 0.002 \; ;  
\label{02UNIT} 
\eeq 
i.e., no significant change relative to Eq.(\ref{00UNIT}).  
Allowing  
for the existence of more families these numbers are relaxed to:  
$$  
|V(cd)|_{(>3)} = 0.218 \pm 0.006\; , \;  
|V(cs)|_{(>3)} = 0.971 \pm 0.003  
$$ 
\beq  
|V(cb)|_{(>3)} = 0.041 \pm 0.002 \; ;  
\label{02MORE} 
\eeq 
i.e., consistent values, yet again a larger uncertainty in particular for 
$|V(cs)|$. Relying on direct determinations PDG '02  lists  
\footnote{Comparing the '98 and '00 errors for  
$|V(cb)|$ shows that PDG values should not  
be treated as gospel. Yet the uncertainties stated now are more 
credible than previously quoted ones.}   
$$  
|V(cd)| = 0.224 \pm 0.016\; , \; |V(cs)| = 0.996 \pm 0.013  
$$  
\beq  
|V(cb)| = 0.0412 \pm 0.0020 
\eeq  
There is a remarkable reduction in the uncertainty for  
the directly determined value of $|V(cs)|$ relative to the status in  
1998. 
This is  due to the usage of the method of extracting it from W leptonic
branching fraction $B(W\to \ell \bar{\nu}_\ell)$, a novel method which comes 
from a
combined analysis of about 60~000 WW events collected by four LEP experiments
\cite{BARBERIOCKM}.  The leptonic branching
fraction of the W is related to the CKM matrix elements without top quark 
\beq
 \frac{1}{B(W\to \ell \bar{\nu_\ell})}=
    3 \Bigl[ 1 + 
        \Bigl[
	    1 +  \frac{\alpha_s(M^2_W)}{\pi} 
	\Bigr]
	 \sum_{i=u,c;j=d,s,b}|V_{ij}|^2 
      \Bigr]
\eeq 
The above given value of $|V_{cs}|$ is obtained assuming 
knowledge of $V_{ud,us,ub,cd,cb}$. 
\par
It has also been pointed out recently\cite{BARBERIOCKM}  that $|V{cs}|$ can
be extracted independently of other elements by measuring the $W\to cs$
branching fraction, which requires the reconstruction of   
$W \to charm \; {\rm jet}\; +  \; strange \; {\rm jet}$. 
Charm (or beauty) jets
can be tagged easily based on peculiar characteristics, such as long lifetimes. 
Tagging of $s$ jets is performed by tagging high-momentum kaons.The
identification of kaons requires relevant particle identification technology,
which at LEP was possessed only by DELPHI. They attempted to measure $V_{cs}$
directly
\cite{Abreu:1998ap}
 using 120 hadronic W decays, and
found $|V(cs)| = 0.97 \pm 0.37$.
\par
While the PDG02 value  $|V(cs)| = 0.996 \pm 0.013$ seems to be the limit
accuracy to-date, it is pointed out\cite{BARBERIOLC} 
 how at a Linear Collider with
$5\cdot 10^6$ W decays one could reach a precision of about 0.1\%.
\par 
Optimism has been expressed that lattice QCD will be able to compute   
the form factors for $D \to \ell \nu K/\pi$ very considerably in the next 
very    few years. It will be a tall order, though, to reduce them down to  
even the 5 \% level, since it requires a fully unquenched calculation.  
%
\section{Exclusive nonleptonic decays}  
\label{NONLEPT}  
In Sect.\ref{LIFE}
 we have already discussed a number of exclusive decays of hidden and 
open charm hadrons. This was done mainly in the context of spectroscopy 
and of lifetime measurements. We will treat exclusive modes now in their own
right.  
\subsection{The $\rho-\pi$ puzzle}
\label{RHOPIPUZZLE}
%
\index{$\rho-\pi$ puzzle}
The annihilation amplitudes for $\jp$ and 
$\pspr$ are proportional to the wavefunction at the origin 
$\psi(\bf{r}=0)$ as discussed in Sect.\ref{LIFE}. The decay widths  
for these S-wave states to a particular final hadronic state $f$ consisting 
of light mesons then depends on $|\psi(0)|^2$. The decay width to $e^+e^-$ 
also depends on this quantity, so one expects the following universal ratio:
\beq
Q_h=\frac{BR(\pspr \to f)}{BR(\jp \to f)} \approx \frac{BR(\pspr \to 
e^+e^-)}{BR(\jp \to e^+e^-)} = (12.3 \pm 0.7) \%
\eeq
This relationship is indeed found to be satisfied for many hadronic 
final states like $p\bar p \pi^0$, $2(\pi^+ \pi^-)\pi^0$, and 
 $\pi^+ \pi^- \omega$. However, in 1983 the Mark II Collaboration found a 
startling violation of this `12\% rule' in decays to certain 
vector-pseudoscalar final states \cite{Franklin:1983ve}. They observed ratios 
$Q_{\rho \pi}<0.6$\% and $Q_{K^* K}<2$\%. These numbers have recently 
been confirmed by the BES Collaboration: $Q_{\rho \pi}<0.23$\%, 
$Q_{K^{*+} K^-}<0.64$\%, and $Q_{K^{*0} \bar K^0}=(1.7 \pm 0.6)$\% 
\cite{Harris:1999wn}. The suppression of the ratio $Q_h$ in these 
states is often referred to as the $\rho-\pi$ puzzle.
\par
Twenty years after the first signal of this puzzle, the issue is still 
unresolved theoretically. There are several recent discussions of the 
theoretical situation in the literature \cite{Tuan:1999ig,Gu:1999ks}. 
Below we will give a brief description of a few of the proposed 
solutions and their drawbacks.
\par
One of the earliest proposed solution \cite{Hou:1982kh,Brodsky:1987bb} 
to the $\rho-\pi$ utilized the vector gluonium state, ${\it O}$, proposed 
earlier by Freund and Nambu \cite{Freund:1975pn}. This vector glueball 
would decay predominantly to vector pseudoscalar states like $\rho\pi$ 
and $K^*K$. If it is sufficiently close in mass to the $\jp$ and narrow, it can 
mix significantly with the $\jp$, but not the $\pspr$ and thus enhance the 
$\jp$ decay rate into these final states relative to the three gluon rate. 
The current 
data require $|m_{{\it O}}-m_{\jp}|<80$ MeV and $4 \; {\rm 
MeV}<\Gamma_{{\it O}}<50 \; {\rm MeV}$\cite{Harris:1999wn}. BES has 
searched for the state and found no evidence\cite{Bai:1996rd}. Lattice 
QCD simulations suggest that these states should be several hundred MeV 
heavier\cite{Bali:1993fb}.

It has also been suggested that the $\rho$ meson may contain a large 
intrinsic charm component\index{intrinsic charm} \cite{Brodsky:1997fj}. 
The radial wavefunction 
of the $c\bar c$ component of the $\rho$ is argued to be nodeless and 
thus to have a much larger overlap with the $\jp$ wavefunction than  
with the wavefunction of the $\pspr$ again enhancing the $\jp$ rate. So 
a sizeable intrinsic charm component in the $\rho$ meson could produce 
an enhancement in the $\jp$ rate to $\rho \pi$ that is not seen in the 
corresponding $\pspr$ rate. Similar excitations of intrinsic charm in 
light mesons can be tested using $\eta_c(1S)$ and $\eta_c(2S)$ decays.

Another proposal \cite{Chen:1998ma} uses the framework of NRQCD to show 
that the decay of the $\jp$ can proceed predominantly through a higher 
Fock state in which the $c \bar c$ pair is in a color octet which decays 
via $c \bar c \to q \bar q$. For $\pspr$ decays this mechanism is 
suppressed by a dynamical effect arising from the proximity of the 
$\pspr$ mass to the $D \bar D$ threshold. So for $\pspr$ the decay 
through three gluons is expected to dominate. If the color octet 
contribution is sizable, we again see a mechanism that could generate an 
enhancement of the $\jp$ rate to states like $\rho \pi$.

The above ideas all implicitly assume that hadron helicity is conserved 
in these processes. Perturbative QCD requires that decays to states in 
which the mesons have nonzero helicity are suppressed by at least 
$1/M_{\jp}$ \cite{Brodsky:1981kj}. For the decay to $\rho \pi$, there 
exists only one possible Lorentz-invariant quantity that can describe 
the $\jp-\rho-\pi$ coupling and the structure of this form factor 
requires the $\rho$ to have helicity $\pm 1$ violating hadron helicity 
conservation. For the above analyses, subleading processes are 
introduced that should otherwise be negligible were it not for the 
suppression of the leading helicity violating amplitude.

There does exist evidence that the assumption of hadron helicity 
conservation is questionable for $\jp$ and $\pspr$. The BES 
Collaboration finds that the rate for $\pspr \to \pi^0 \omega$ is not 
suppressed relative to the $\jp$ rate, i.e. 
$Q_{\pi^0 \omega}=(9.3 \pm 5.0)$\% \cite{Harris:1999wn}. This 
process violates hadron helicity 
conservation but does not exhibit the behaviour that would be expected 
if any of the above models were correct. Measuring the decay rates to 
the hadron helicity conserving final state $\pi^+ \pi^-$ should provide 
insight into the validity of this assumption.

There are many more models that attempt to resolve this intriguing 
puzzle, many of which do not invoke hadron helicity conservation. 
Unfortunately, there is no current model which can satisfactorily 
describe all of the relevant data. Clearly this is a problem that 
requires more attention both theoretically and experimentally.

%
\subsection{Other charmonium decays}
\label{ONIUMDK}
%
There are four classes of decays: 
\begin{enumerate}
 \item 
   Electromagnetic decays of higher mass charmonia to lower mass ones -- 
   $[\bar cc]_N \to [\bar cc]_{n<N} + \gamma$ -- or of charmonia into on- or 
   off-shell photons: $[\bar cc]_N \to \gamma^* \to l^+l^-$, $\gamma \gamma$. 
 \item 
   Radiative  decays into one or several light-flavour hadrons: 
   \beq 
     [\bar cc] \to \gamma + \{ h_{light} \} 
   \eeq 
 \item 
   Hadronic transitions of higher mass to lower mass charmonia:  
   \beq 
     [\bar cc]_N \to [\bar cc]_{n<N} + \{ h_{light} \} 
   \eeq 
 \item 
   Hadronic decays into light-flavour hadrons 
   \beq 
     [\bar cc] \to  \{ h_{light} \} 
   \eeq
\end{enumerate}
A wealth of new data is flowing in thanks to $\epem$ experiments 
(B-factories BABAR and BELLE, upgraded versions of experiments CLEO and BES),
and to final results from established $\bar p p $ formation experiment E835 at
Fermilab. 
\par
An important restructuring of the information contained in the PDG occurred in
2002, and it affected
relevantly (up to  about 
 30\%) the values of some branching ratios of $\pspr$
and $\chi_{c0,1,2}$. Details are discussed  in \cite{Hernandez-Rey:dj}:
with the global fit provided in the 2002 edition, the  correlations and vicious
circles introduced through the use of self-referencing relative branching ratios
should have been corrected.
\par
Great interest is focussed on the $\rho\pi$ puzzle discussed in the previous
section. Other recent 
topical issues are a new determination of $\Gamma(\chi_{c0,2}\to
\gamma\gamma)$ by E835, and the complementary measurement of cross-section 
 for charmonium production in two-photon processes
 $\epem \to \epem \gamma \gamma \to \epem R$
  from DELPHI
   $(R=\eta_c)$  and BELLE $(R=\chi_{c0})$ \cite{Abe:2002va};
 experiment E835 also presented the first evidence for the decay 
 $\chi_{c0}\to \pi^0\pi^0, \eta\eta$ \cite{E835:2003sk}.
  First evidence for $\chi_{c0,1,2}\to
 \Lambda \bar \Lambda$ was reported by BES \cite{Bai:2003ds}.
\par
New results are too numerous  to be covered here, 
the reader is addressed to recent
reviews \cite{Cester02,Pastrone:2003wt,Mahlke-Krueger03} for results on the four
classes above, and to the comments already discussed on 
 selected topics interlaced with charmonium spectroscopy 
  in Sect.\ref{SPECONIA};
 here we will add comments 
 on class (2).
%
%
\subsubsection{$[\bar cc] \to \gamma + \{ h_{light} \}$}
\label{RADTOGLUBALL}
The search for glueball candidates\index{glueballs} is a central motivation in 
studying these transitions -- in particular from the $J/\psi$ --, which to 
lowest order 
are driven by 
$[\bar cc] \to \gamma +gg$: any kinematically accessible glueball or other 
hadron with a sizeable gluon component in its wave function should figure 
prominently in the hadronic final state. The most direct way is to measure the 
{\em inclusive} hadronic recoil spectrum  
$d\sigma (J/\psi\to \gamma X)/dM_X$ and search 
for peaks there.  
Alternatively one can analyze {\em exclusive} hadronic final states 
for the presence 
of new states outside the usual $\bar qq$ multiplets. 

 Experiments at $\epem$  colliders accumulating large data samples at the $\jp$
 have made of glueball search one of the top priorities. No conclusive signal
 has been found there to-date. The scalar glueball is predicted around 1700~MeV,
 and the lowest tensor glueball at about 2220~MeV.
 In 1996 Mark~III \cite{Dunwoodie:1997an} claimed evidence for the lightest
 scalar glueball,  
 indicating the  $f_0(1700)$ after a Partial Wave Analysis. It was advocated
 later\cite{Close:2001zp} how this candidacy should be compared to the E791
 Dalitz plot studies of  
 $D_s\to KK\pi$ (Sect.\ref{DALITZ}). 
 BES claimed candidates for the tensor glueball (the $f_J(2220)$) in 1996, and 
 they are now performing detailed searches of several radiative final states 
 \cite{Shen:2002nv} with their total sample of 58 million $\jp$ decays.
 No true glueball candidate is  supposed to be observed in $\gamma\gamma$
 production, and therefore CLEO and LEP experiments have performed a search,
 setting limits. Finally, searches for the $f_J(2220)$ have been carried over at
 CLEO\cite{Masek:2002yb} in $\Upsilon\to \gamma f_J$ decays.
 \par
 The experimental table is evidently very rich, however dinner is not ready yet.
 We may very well need to await for the  honour guest --- CLEO-c.
\subsection{On absolute charm branching ratios} 
\label{ABSBR} 
Attempts to measure absolute charm  
branching ratios date back to the very early stages of the charm 
adventure at accelerators. In principle one has "merely" to observe all 
decays of a certain hadron, and then count
how often a specific final state appears. In 
reality things are of course less straightforward. 
The first attempts 
\cite{PERUZZI77} relied on estimating the charm cross section at 
the $\psi(3770)$ to convert the observed signal 
into D branching fractions. It was assumed that the 
$\d0d0$ and $D^{\pm}D^{\mp}$ partial width were the same and that they 
saturate $\Gamma (\psi(3770))$. 

Such analyses suffer from  the systematic uncertainty on the exact size 
of the relevant production rate. To overcome this handicap, several 
approaches have been pursued over the past two decades. 
\begin{enumerate} 
 \item 
  The Mark III collaboration at SPEAR\index{SPEAR at Stanford} first exploited the method 
  of {\em tagged decays}\index{tagged decays}. Since the $\psi(3770)$ 
  is just barely above charm threshold, its decays into charm cannot produce more than 
  $D^0\bar D^0$ and $D^+D^-$ pairs. Absolute branching ratios 
  for $D$ mesons can then be obtained  \cite{BALTRU} 
  by comparing the number of single  
  and double tags, i.e., fully reconstructed $\ddb$ events . 
  Single tag events were more abundant, but double tags benefitted from
   additional 
  kinematic constraints, such as beam energy and momentum conservation, etc. 
  This method does {\em not} have to assume that the total $\psi(3770)$ width is saturated 
  by $\Gamma (\psi(3770)) \to D \bar D)$. 
  Typical errors were of order 10\%. 
 \item 
   The method pionereed by ALEPH \cite{DECAMP}  and HRS \cite{ABACHI} 
  exploits the fact that the pion from the strong decay 
  $D^{*+}\to D^0\pi^+$ is very soft due to the small Q-value. In particular, the 
  $D^0$ is not reconstructed, but its direction is well approximated by 
   the thrust axis of the event, and the soft pion characteristic kinematical 
   signature tags the event\index{$D^*$ tag}. 
\item 
  ACCMOR\cite{Barlag:ww} and LEBC-EHS\cite{Aguilar-Benitez:1987cv} 
   measured the ratio of $D^0\to K^-\pi^+$ 
  relative to the total number of  even-prong decays. 
  The absolute branching
   ratio 
  is then calculated using topological branching fractions 
  to correct for missing zero-prong decays.
\item 
  ARGUS\cite{Albrecht:1993gr} and then CLEO\cite{Artuso:1997mc}
  determined  
  the inclusive number of $D^0$'s  by partial reconstruction of 
   $\bar B^0 \to D^{*+} \ell^- \bar \nu$ with $D^{*+} \to D^0 \pi^+$ 
   where only the soft pion and the lepton are detected. The $D^*$ direction 
  is, as always, approximated by the direction of the soft pion. 
  The missing mass squared 
  $(E_B-E_\ell-E_{D^*})^2-|\vec{P}_B-\vec{P}_\ell-\vec{P}_{D^*}|^2$  
  shows a prominent peak at zero: the number of 
   events in the peak provides the 
  normalizing  
  factor of the number of $D^*$ decays.  
 \item  
  For the absolute branching ratios of the $D_s^+$ meson,  
  E691\cite{Anjos:1990cm} and later E687\cite{Frabetti:1993ih} determined it by
  measuring $\Gamma(D_s^+\to \phi \mu^+\nu)/\Gamma(D_s^+\to \phi\pi^+)$, by
  using the $D_s^+$ lifetime, and assuming the theoretical expectation that
  $\Gamma(D_s^+\to \phi\mu^+\nu)\sim\Gamma(D^+\to \bar K^{*0}\mu^+\nu)$.
 \item 
   Also for the $D_s^+$ meson, low-statistics 
  samples  
  were collected by BES\cite{Bai:1994cv}  by using double-tag  
  $D^+_sD^-_s$ pairs exclusively produced in $\epem$ at energy just below the  
  $D_s^*$ production threshold. 
 \item 
  In the charmed baryon sector, the only measurements available for absolute 
   branching ratios refer to the $\Lambda_c^+$. A concise but complete review is    
  found in \cite{Burchat:de}.  No model-independent measurements exist.
  ARGUS\cite{Albrecht:1988an} 
  and
  CLEO\cite{Crawford:1991at} measure $B(\bar B \to \Lambda_c^+X)\cdot
  B(\Lambda_c^+\to 
  pK^-\pi^+)$  and, assuming that $\Lambda_c X$ channel saturate the B meson
  decays into baryons and that $\Lambda_c^+X$ final states other than
  $\Lambda^+_c\bar N X$ can be neglected, they also measure $B(\bar B \to
  \Lambda_c^+ X)$. Hence, $B(\Lambda_c^+\to pK^-\pi^+)$ is extracted.
  ARGUS\cite{Albrecht:1991bu} 
  and CLEO\cite{Bergfeld:1994gt} also measure 
  $\sigma(\epem \to \Lambda_c^+
  X)\cdot B(\Lambda_c^+ \to \Lambda\ell^+ \nu_\ell)$. The PDG group combines 
  these
  measurements with $\sigma(\epem \to \Lambda_c^+X)\cdot B(\Lambda_c^+ \to
  p K^- \pi^+)$, also estimating $B(\Lambda_c^+\to pK^-\pi^+)$. The
  model-dependent systematic error estimated is of order 30\%.
  A different approach is attempted by \c2
  \cite{Jaffe:2000nw},  that
  tags charm events with the semielectronic decay of a 
  $D^*$-tagged $\bar{D}$, and the $\Lambda^+_c$ production with a 
  $\bar{p}$. 
  The assumption here is that having a $\bar D$ meson in one hemisphere, and a
  $\bar p$ in the opposite hemisphere, is a tagging for a $\Lambda_c^+$ in the
  hemisphere of the antiproton.
  Their final value is $B(\Lambda_c^+\rarr p 
  K^-\pi^+)=(5.0 \pm 0.5 \pm 1.2)$~\%, which coincides with the PDG02 average
  from the two older measurements described above. 
\item 
  A novel method analyzes  
  $B$ meson decays into charm hadrons and then extracts the latter's  
  branching ratio by relying on various correlations in the overall  
  $B$ decay. This method has been pioneered by CLEO. 
   The measurement\cite{Artuso:1996xr} is  performed 
     using partially reconstructed  
  decays $\bar B^0 \to D^{*+}D_s^{*-}$. This decay is peculiar since it 
  contains
  a    soft pion from $D^*$ decay, and a soft photon from the $D^*_s$ decay.
  Another example has been discussed in Sect.\ref{ABSBR}:
   utilizing the huge 
  data samples to be accumulated by the $B$ factories 
  one reconstructs one $B$ meson in $\Upsilon(4S) \to B \bar B$ and 
  then exploits correlations between baryon and lepton numbers and 
  strangeness among the decay products of the other $B$ meson to infer 
the absolute semileptonic branching ratios of charm baryons \cite{BIGISLBR}. 
\end{enumerate} 
 Tab.\ref{TAB:ABSBR} 
 shows the present world averages for the measured absolute branching 
 ratios \cite{Hagiwara:fs} 
 along with the average $\chi^2$ as computed by the PDG group. 
 The 20\%   
 error on the $\Lambda_c$ is clearly unsatisfactory, and it should be noted that there 
 are no absolute branching ratios for the other charmed baryons. 
\begin{table} 
\begin{center} 
\begin{tabular}{|l|r|r|}
\hline\hline
 Abs. BR                  & PDG02                 
      & Error scale factor \\ 
\hline\hline
$D^+\to$  & & \\ 
 $\bar K^0 \pi^+$            & $0.0277 \pm 0.0018$  &  
  --- \\
 $K^-\pi^+\pi^+ $            & $0.091\pm 0.006$     &  
  --- \\
 $\bar K^0\pi^+\pi^0 $       & $0.097\pm 0.030$     &  
  1.1 \\
 $ K^-\pi^+\pi^+\pi^0 $       & $0.064\pm 0.011$     &  
  --- \\
 $\bar K^0\pi^+\pi^+\pi^- $  & $0.070\pm 0.009$     &  
  --- \\
 $K^*(892)^-\pi^+\pi^+ $     & $0.021\pm 0.009$     &  
  --- \\
 $\phi\pi^+\pi^0 $           & $0.023\pm 0.010$     &  
  --- \\
 $K^+K^-\pi^+\pi^0 $         & $0.015\pm 0.007$     &  
  --- \\
 $K^*(892)^+\bar K^*(892)^0 $& $0.026\pm 0.011$    &  
  --- \\
	\hline
$D^0\to$                 &                 &    \\	
 $ K^-\pi^+ $                 & $0.0380\pm 0.0009$   &  
 --- \\
 $ \bar K^0 \pi^+\pi^-$       & $0.0592 \pm 0.0035$  &  
  1.1 \\
 $ K^-\pi^+\pi^0 $            & $0.131\pm 0.009$     &  
  1.3 \\
 $ \bar K^0 \pi^+\pi^-\pi^0$  & $0.108\pm 0.013$     &  
  --- \\
\hline 
 $D_s^+\to \phi \pi^+ $      & $0.036\pm 0.009$     &   --- \\
\hline
 $\Lambda_c^+\to pK^-\pi^+$ & $0.050 \pm 0.013$    &    --- \\
\hline 
\end{tabular} 
\end{center} 
\caption{World averages for charm mesons and baryons absolute branching ratios
from \cite{Hagiwara:fs}.} 
\label{TAB:ABSBR} 
\end{table}  

It is legitimate to ask why one wants to measure charm branching ratios  
so accurately, why, say, a 10\% accuracy is not satisfactory, when  
one has to allow for about 30\% uncertainties in theoretical predictions.  
There are several motivations of somewhat different nature.  
\begin{itemize} 
\item  
The absolute values of the $D^{0,+}$, $D_s$, $\Lambda_c$ etc. branching 
ratios are an  important `engineering' input for a sizeable number of  
$B$ decay analyses like the following:    
\begin{itemize} 
\item  
When extracting  
the CKM parameter $V(cb)$ from the semileptonic $B$ 
width  and from the formfactor at zero recoil for  
the exclusive transitions  
$B \to \ell \nu D^*$ and $B\to \ell \nu D$ they are needed to translate the 
observed  rate for, say, $B\to \ell \nu (K\pi)_D\pi$ into a width for  
$B \to \ell \nu D^*$ etc.  
\item  
They are also of very direct importance for  
evaluating the charm content in nonleptonic $B$ decays
\index{charm content in $B$ decays} 
in particular and to  
compare the $b\to c \bar ud$ and $b\to c \bar cs$ widths of $B$ mesons.  
It had been pointed out first almost ten years ago that the observed 
semileptonic $B$ branching ratio is somewhat on the low side of 
theoretical expectation \cite{BAFFLING}. This could be understood if the  
$b \to c \bar cs$ width were larger than expected. The charm content of 
$B$ decays, i.e. the average number of charm hadrons in the final 
states of $B$ mesons is thus an observable complementary to  
BR$(B \to \ell \nu X_c)$. Its determination obviously depends on the 
absolute size of $D^0$, $D^+$, $D_s$ and $\Lambda_c$ branching ratios. 
\end{itemize} 
It turns out that the accuracy, with which absolute charm branching 
ratios are known, is about to become -- or will be in the foreseeable 
future -- a major bottle neck in the analysis of beauty decays. 
\item  
As discussed in Sect.\ref{HQET2} the HQE predicts quite spectacular variations  
in the semileptonic widths and branching ratios of charm baryons.  
\item  
The one-prong decays $D^+,\; D_s^+ \to \mu ^+ \nu$, $\tau^+\nu$ are  
measured to extract the decay constants $f_D$ and $f_{D^*}$ from the 
{\em widths} and 
compare the results with predictions in particular of lattice QCD, see 
Sect.\ref{LEPT}.  
\item  
Analogous statements apply to exclusive semileptonic charm decays.  
It is an often repeated promise of lattice QCD that it will calculate  
the semileptonic form factors accurately both in their normalization 
and $q^2$ dependence. 
\end{itemize} 
Fortunately help is in sight: (i) The new tau-charm factory CLEO-c will measure 
the absolute scale of $D$ and $D_s$ absolute branching ratios 
with a 1-2\% error, 
and possibly the  $\Lambda_c$ absolute branching 
ratio significantly more accurately as well. 
(ii) Novel methods have been proposed to get at the branching ratios for 
charm baryons \cite{BIGISLBR}.  

\subsection{Two-body modes in weak nonleptonic decays} 
\label{TWOBODY} 
Nonleptonic {\em weak} decays pose a much stiffer challenge to a  
theoretical description than semileptonic ones: there are more colour 
sources and sinks in the form of quarks and antiquarks, more different  
combinations for colour flux tubes to form, and the energy reservoir is  
not depleted by an escaping lepton pair. There are only two classes  
of nonleptonic decays where one can harbour reasonable hope of some 
success at least, namely fully inclusive transitions like lifetimes etc.  
already discussed and channels with a two-body final state. Once one goes  
beyond two hadrons in the final state, the degrees of freedom and the  
complexities of phase space increase in a way that pushes them beyond  
our theoretical control. 

Therefore we will discuss only two-body channels in detail including 
those with resonances.

For two-body final states the phase space is trivial and the number of  
formfactors quite limited. Yet even so such transitions present a 
formidable  theoretical challenge, since they depend on long-distance  
dynamics in an essential way. As 
an optimist one might point to some mitigating  factors:  
\begin{itemize} 
\item  
Two-body final states allow for sizeable momentum transfers thus  
hopefully reducing the predominance of long-distance dynamics.  
\item  
It is not utopian to expect lattice QCD to treat these transitions  
some day in full generality. Such results will however be reliable  
only, if obtained with incorporating fully dynamical fermions  
-- i.e. without "quenching" -- and without relying on a  
$1/m_c$ expansion. 

\item  
\index{Watson's theorem}
Watson's theorem can still provide some useful guidance.  
\end{itemize} 
In addition there are motivations for taking up this challenge:  
\begin{itemize} 
\item  
Carefully analysing branching ratios and Dalitz plots can teach us 
novel lessons on  light-flavour hadron spectroscopy, like on 
characteristics of some resonances like the $\sigma$ or on the  
$\eta$ and $\eta^{\prime}$ wavefunctions and a possible  
non-$\bar qq$ component in them.  
\item  
Analogous $B$ decay modes are being studied also as a mean to extract 
the complex phase of $V(ub)$. One could hope that $D$ decays might serve as a  
validation analysis. For honesty's sake one has to add that in the  
two frameworks presently available for treating such $B$ decays -- usually 
refered  to as "QCD factorization" and "PQCD" -- such a connection cannot  
be exploited largely due to technical reasons. For contrary to the HQE 
treatment of {\em inclusive}   rates, where the leading nonperturbative 
corrections arise only in  order $1/m_Q^2$, these frameworks for  
{\em exclusive} widths allow for corrections already 
$\sim {\cal O}(1/m_Q)$. Those, which for   
$D$ decays potentially are very large, cannot be controled.  
\item  
As discussed later a most sensitive probe for New Physics is provided  
by searches for CP violation in the decays of charm hadrons.  
Two-body nonleptonic modes provide good opportunities for such searches.  
Signals of {\em direct} CP violation typically require the intervention  
of FSI in the form of phaseshifts. For designing search strategies and 
for properly interpreting a signal (or the lack of it) one needs 
independent information on such FSI phases. A comprehensive analysis of  
two-body charm decays can provide such information. 

\item  
It is not unreasonable to ask whether New Physics -- rather than  
novel features of SM dynamics -- could manifest itself also by enhancing  
or suppressing some exclusive widths, as discussed recently  
\cite{LIPKINCLOSE}. However one is facing a `Scylla and Charybdis' 
dilemma here: due to our limited theoretical control only very  
sizeable deviations from SM expectations can be viewed as  
significant. But then one has to wonder why such a discrepancy has not  
been noticed before in other transitions. Doubly Cabibbo suppressed modes  
presumably offer the best `signal-to-noise' ratio. 

\end{itemize}  
We can easily see that FSI produce generally large phase shifts by  
using isospin decompositions of decay amplitudes. Consider the modes  
$D\to K\pi$: there are three channels -- $D^0 \to K^-\pi^+$,  
$D^0 \to \bar K^0\pi^0$ and $D^+ \to \bar K^0\pi^+$ -- yet only two  
independent amplitudes, namely with isospin 1/2 and 3/2 in the final 
state, $T_{\frac{1}{2}}$ and $T_{\frac{3}{2}}$, respectively. Thus there 
has to be a  relation between the three (complex) transition amplitudes; 
i.e., they  have to form a triangle relation, which is easily worked out:  
\bea  
\nonumber  
T(D^0 \to K^-\pi^+) &\equiv& T_{-+} = \frac{1}{\sqrt{3}}\left(  
\sqrt{2}T_{\frac{1}{2}} + T_{\frac{3}{2}}\right) \\ 
\nonumber  
T(D^0 \to \bar K^0\pi^0) &\equiv& T_{00} = \frac{1}{\sqrt{3}}\left(  
-T_{\frac{1}{2}} + \sqrt{2} T_{\frac{3}{2}}\right) \\ 
T(D^+ \to \bar K^0\pi^+) &\equiv& T_{0+} = \sqrt{3} T_{\frac{3}{2}} 
\label{IDECOMP} 
\eea 
and thus  
\beq  
T_{-+} + \sqrt{2}T_{00} - \frac{2}{3}T_{0+} = 0 \; .  
\eeq 
Using the most recent PDG values for the branching ratios one infers  
\bea 
\nonumber 
|T_{\frac{1}{2}}|&=& (3.05 \pm 0.06) \times 10^{-3} MeV\\ 
\nonumber 
|T_{\frac{3}{2}}|&=& (7.67\pm 0.25) \times 10^{-4} MeV\\ 
\delta_{\frac{3}{2}} - \delta_{\frac{1}{2}}&=& (95.6 \pm 6.3)^{0}. 
\eea 
where $T_{\frac{j}{2}} \equiv  
|T_{\frac{j}{2}}|e^{i\delta_{\frac{j}{2}}}$, $j=1,3$. I.e., the  
phase shift $|\delta_{\frac{3}{2}} - \delta_{\frac{1}{2}}|$ is indeed  
very large in this case.

While this turns out as expected and can encourage us to search for  
direct CP asymmetries in $D$ decays, it also implies that describing  
nonleptonic two-body modes will be challenging, to put it  
euphemistically. 

\subsubsection{Early phenomenology} 
\label{BSW} 

The first two pieces of the "charm puzzle" \index{charm puzzle}, i.e. of  
evidence that charm decays do not proceed  quite to the {\em original} 
expectations actually emerged in $D^0$  two-body modes  
\footnote{The third piece was the observation that  
$\tau (D^+)$ exceeds $\tau (D^0)$ considerably.}:  
\bea  
\left. \frac{\Gamma (D^0 \to K^+K^-)}{\Gamma (D^0 \to \pi^+ \pi^-)} 
\right|_{\rm data} &\sim 3&  \; \; \; vs. \; \; \sim 1.4  \; \;  
\; \; {\rm "originally \; expected"} 
\label{KKPIPI1} \\ 
\left. \frac{\Gamma (D^0 \to \bar K^0\pi^0)} 
{\Gamma (D^0 \to K^-\pi^+)} 
\right|_{\rm data} &\sim \frac{1}{2}&  \; \; \; vs. \; \; \ll 1  \; \;  
\; \; {\rm "originally \; expected"}  
\label{K0PI0} 
\eea  
Subsequently it was suggested \cite{FUKUGITA} that another channel  
would be very telling:  
\beq 
{\rm BR}(D^0 \to \bar K^0\phi) \simeq 0 \; \; \; \; \;  
{\rm "naively \; expected \; without \; WA"} 
\label{KPHI} 
\eeq 
The success or failure of a theoretical description will depend on its  
ability to reproduce or even predict the {\em whole pattern} of decay  
modes. Nevertheless the three observable ratios of  
Eqs.(\ref{KKPIPI1},\ref{K0PI0},\ref{KPHI}) provide a first orientation  
for evaluating how well a certain model does in describing the data.

To understand what underlies the "expectations" and what  
one learns from these discrepancies we have to describe  
how one arrives at such predictions. In doing so we will  
{\em not} follow the historical sequence.

The starting point is always provided by the effective weak  
$\Delta C=1$ Lagrangian, as discussed in Sect.\ref{EFFWEAK}.  
For our subsequent discussion it is instructive to write it in terms  
of the multiplicatively renormalized 
operators\index{multiplicatively renormalized operators}.   
For Cabibbo 
allowed modes we have:  
\beq   
{\cal L}_{eff}^{\Delta C=1}(\mu = m_c) = (4G_F\sqrt{2}) 
V(cs)V^*(ud) \cdot  
[c_- O_- + c_+ O_+]  
\label{EFFWEAKLAG} 
\eeq 
\beq  
O_{\pm}=\frac{1}{2}\left[ 
(\bar s_L\gamma_{\nu}c_L)(\bar u_L\gamma_{\nu}d_L)] \pm  
(\bar u_L\gamma_{\nu}c_L)(\bar s_L\gamma_{\nu}d_L)\right]  
\eeq 
\beq 
 c_- \simeq 1.90 \; , \; c_+ =  0.74  
\label{EFFWEAKOP} 
\eeq  
Both operators $O_{\pm}$ carry isospin $(I,I_3) = (1,1)$, yet are  
distinguished by their $V$ spin quantum numbers, as described in  
Sect.\ref{EFFWEAK}: the $\Delta V=0$ $O_-$ is enhanced in ${\cal L}_{eff}^{\Delta C=1}$, 
the $\Delta V=1$ $O_+$ reduced. This explains why the {\em 
width} for  
$D^+ \to \bar K^0\pi^+$ is reduced relative to $D^0 \to K^- \pi^+$.  
For $D^+$ carrying $V=0$ can decay via $O_-$ only into $V=0$  
final state. Bose statistics, however, tells us that the two  
$V$ spinors $\bar K^0$ and $\pi^+$ have to be symmetric under exchange  
and thus form a $V=1$ configuration. Yet the latter can be reached  
only through $O_+$. $D^0 \to K^- \pi^+$ on the other hand can be  
driven by $O_-$.

\begin{figure}[t]
	\centering
	\epsfysize=3.5cm
      \epsffile{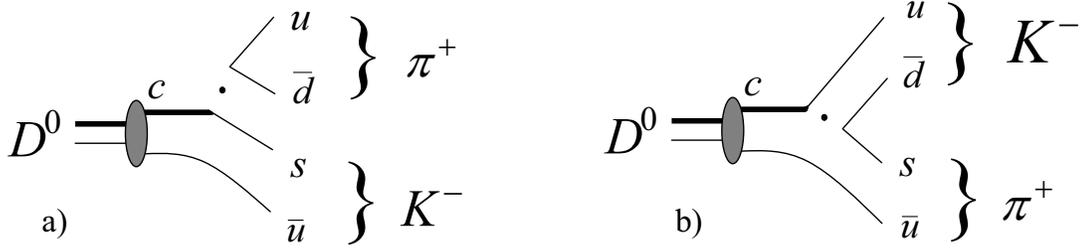} 
 \caption{a) The external emission of the quark-antiquark pair forms a colour
 singlet.  b)  Internal emission of the quark-antiquark pair does not result in
 a colour singlet. 
     \label{FIG:INTEXT} }
\end{figure}

The quark level diagrams for nonleptonic and semileptonic decays do not 
look so different. Yet with having two quark and two antiquark fields  
in the final state rather than one each there are more different routings 
for the colour flux tubes between the fields and ways in which those 
arrange themselves into hadrons.  
To illustrate this point consider again $D^0 \to K^- 
\pi^+$: Fig.\ref{FIG:INTEXT}~a [\ref{FIG:INTEXT}~b]  
shows the diagram for this transition driven by the operator  
$O_1$ [$O_2$] through a topology usually called "external  
[internal] $W$ emission".  
\index{internal vs. external $W$ emission} The mode  
$D^0 \to \bar K^0\pi^0$ is driven  by diagrams with the same topologies, 
yet with 
$O_1$ and $O_2$ having  switched places. For inclusive decays this 
distinction between  "internal" and "external" $W$ emission does not 
exist, yet for  exclusive ones it does.

Drawing quark diagrams is one thing, attaching numerical values to them  
quite another. For the different fields interact strongly with each 
other. On the one hand this is a blessing since these strong interactions 
generate a large number of possible hadronic final states from the 
limited set of quark level configurations. On the other hand it is quite 
difficult to make rigorous predictions regarding the evolution from quarks 
and gluons to asymptotic hadronic states. The problem lurks in calculating 
hadronic matrix elements,  since those are controled by long-distance 
dynamics. Lattice QCD is  widely expected to yield reliable answers -- 
someday;  
$1/N_C$ expansions provide a compact classification scheme, but  
not quantitative answers as explained below; $1/m_Q$ cannot make reliable  
statements on exclusive charm decays. Thus we have a situation, where  
quark models are called to the front as tools of last resort. 

Stech and coworkers tackled the problem not unlike  
Alexander the Great did the Gordian knot. They relied on an unabashedly 
phenomenological approach where they invoked simplifying assumptions  
as much as practically possible while allowing for complexities only  
when unavoidable. Their approach turned out to be quite successful. In 
describing it we will point out its differences to earlier attempts.

Their prescription involves the following rules:  
\begin{itemize} 
\item  
One employs the effective weak Lagrangian of Eq.(\ref{EFFWEAKLAG}),  
yet leaves the coefficients $c_{1,2}$ as free parameters at first.  
\item  
One ignores WA diagrams completely.  
\item  
One allows for contributions both from external and internal $W$ 
emission. In the former, Fig.(\ref{FIG:INTEXT}~a), the quark-antiquark pair
already   
forms a colour singlet, in the latter, Fig.(\ref{FIG:INTEXT}~b), it does not.
Accordingly   
one assigns a colour factor $\xi$ to the matrix element of the latter  
relative to the former. Naively, i.e. by just counting the number  
of the different colour combinations, one would guestimate  
$\xi \simeq 1/N_C = 1/3$. Here instead one leaves the numerical value of  
$\xi$ a priori completely free, yet maintains it to possess a  
{\em universal} value for all channels, whatever it is. This is a 
critical assumption, to which we will return.  
\end{itemize} 

The first consequence of these three rules is that all decay amplitudes  
can be expressed as linear combinations of two terms:  
\beq  
T(D \to f) \propto a_1\matel{f}{J_{\mu}^{(ch)}J_{\mu}^{(ch)\prime}}{D}  
+ a_2\matel{f}{J_{\mu}^{(neut)}J_{\mu}^{(neut)\prime}}{D} 
\eeq 
\beq  
a_1 = c_1 + \xi c_2 \; , \; a_2 = c_2 + \xi c_1  
\label{A1A2} 
\eeq 
where $J_{\mu}^{(ch)}$ and $J_{\mu}^{(ch)\prime}$  
[$J_{\mu}^{(neut)}$ and $J_{\mu}^{(ch)\prime}$] denote the charged  
[neutral] currents appearing in Fig.\ref{FIG:INTEXT}~a [\ref{FIG:INTEXT}~b]. 
It should be kept in  
mind that the quantities $c_1$ and $c_2$ on one hand and $\xi$ on the  
other are of {\em completely different} origin despite their common  
appearance in $a_1$ and $a_2$: while $c_{1,2}$ are determined by 
short-distance dynamics, $\xi$ parametrizes the impact of long distance  
dynamics on the size of matrix elements.  
\begin{itemize} 
\item  
One adopts a {\em factorization} ansatz \index{factorization ansatz}  
for evaluating the matrix 
elements, i.e. one approximates the matrix element for $D \to M_1M_2$  
by the products of two simpler matrix elements:  
\beq  
\matel{M_1M_2}{J_{\mu}J_{\mu}}{D} \simeq  
\matel{M_1}{J_{\mu}}{0}\matel{M_2}{J_{\mu}}{D}\; , \;  
\matel{M_2}{J_{\mu}}{0}\matel{M_1}{J_{\mu}}{D} 
\eeq 
\item  
For $M_{1,2}$ being pseudoscalar mesons $P$  one has  
\beq 
\langle P(p)|A_{\mu}|0\rangle =-i f_{P} p^{\mu}  
\eeq 
$$ 
\langle P(p_{P})|V^{\mu}|D(p_{D})\rangle =(p_{P}^{\mu} + 
p_{D}^{\mu} - \frac {M_{D}^{2} - M_{P}^{2}}{q^{2}} 
q^{\mu})f_{+}(q^{2})  
$$  
\beq  
+ \frac {M_{D}^{2} - M_{P}^{2}}{q^{2}} 
q^{\mu})f_{0}(q^{2}) \; ; \; q=p_D - p_P \; , \; f_{+}(0)=f_{0}(0) 
\label{DPFF} 
\eeq 
The decay constants are known for the low mass mesons. For the  
formfactors $f_{0,+}$ BSW assumed dominance by the nearest  
t-channel pole:  
\beq 
f_{+,0}=\frac{h_{+,0}}{1-\frac{q^{2}}{M^{*}}}. 
\eeq 
For the values of the residues $h_{i}'s$ one can rely on quark model  
calculations or take them from the available data on semi-leptonic decays.

Similar, though lengthier, expressions apply, when vectormesons are  
involved.  
\item  
FSI in the form of rescattering, channel mixing and conceivably even  
resonance enhancements are bound to affect charm decays. An optimist  
can entertain the hope that a large part or even most of such effects  
can be lumped into the quantity $\xi$, which reflects long-distance  
dynamics; however there is no good reason why such a 
{\em global} treatment of FSI  
should work even approximately. Stech et al. included residual FSI 
effects on a case-by-case basis; in the spirit of Ockham's razor FSI 
effects (in the form of phase shifts and absorption) were included only 
as much as really needed.  
\item  
It has to be kept in mind that certain final states cannot be produced  
by such factorizable contributions. One example is $D^0 \to \bar 
K^0\phi$.  
\end{itemize} 
One can distinguish three classes of two-body modes \cite{BSW}:  
\begin{itemize} 
\item  
class   I:   $D^0 \to M_1^+M_2^-$  
\item  
class  II:   $D^0 \to M_1^0M_2^0$ 
\item  
class III:   $D^+ \to M_1^+M_2^0$ 
\end{itemize} 
Class I[II] transitions receive contributions from $a_1$ [$a_2$]  
amplitudes only, whereas class III modes involve the interference between  
$a_1$ and $a_2$ amplitudes. It is therefore the latter that allow us  
to determine the relative sign between $a_1$ and $a_2$. 

In summary: the BSW model put forward almost twenty years ago contains two 
free parameters -- $a_1$ and $a_2$ -- plus some considerable degree of 
poetic license in the  size assumed for the residue factors $h_i$ and the 
amount of explicit  FSI that has been included.

With such limited freedom BSW were able to obtain a 
decent  fit to twenty-odd two-body modes of $D^0$, $D^+$ and $D_s^+$ 
mesons. This is quite remarkable even keeping in mind that their  
description was helped by the sizeable errors in most measurements then.  
Their fit to the data yielded  
\beq  
a_1(exp) \simeq 1.2 \pm 0.1 \; , \; a_2(exp) \simeq - 0.5 \pm 0.1  
\label{A1A2FIT} 
\eeq 
Obviously one wants to compare this with what one might infer 
theoretically from  Eq.(\ref{A1A2}):   
\beq  
a_1(QCD) \simeq 1.32 - 0.58 \xi \; , \;  
a_2(QCD) \simeq - 0.58 + 1.3 \xi  
\label{A1A2QCD} 
\eeq 
It is again remarkable that these numbers are in the right `ballpark' --  
and even more so that $\xi \simeq 0$ brings the numbers in  
Eq.(\ref{A1A2QCD}) in full agreement with those in Eq.(\ref{A1A2FIT}). 

As already stated, {\em naively} one guestimates  
$\xi \simeq 1/N_C = 1/3$ leading to $a_1\simeq 1.1$ and  
$a_2 \simeq -0.15$, which would greatly reduce expectations for  
class II and III transition rates. This and other related issues can be  
illustrated by considering $\Gamma (D^0 \to \bar K^0\pi^0)$ vs.  
$\Gamma (D^0 \to K^-\pi^+)$. Ignoring QCD effects -- in particular  
setting $c_1=1$, $c_2=0$ -- one expects  
$\Gamma (D^0 \to \bar K^0\pi^0)/\Gamma (D^0 \to K^-\pi^+) \simeq  
\left( \frac{1}{\sqrt{2}}\frac{1}{3}\right)^2 = \frac{1}{18}$. Using  
$c_1 \simeq 1.3$ and $c_2 \simeq - 0.6$ reduces this ratio  
even further. However one should not have trusted this result in 
the first place! For the suppression of $T(D^0 \to \bar K^0\pi^0)$  
is due to an accidental cancellation of the $T_{I_f=1/2}$ and  
$T_{I_f=3/2}$ amplitudes in the simple prescription used, see  
Eq(\ref{IDECOMP}). Any FSI is 
quite  likely to vitiate this cancellation. To say it differently:  
the mode $D^0 \to \bar K^0\pi^0$ is particularly sensitive to the  
intervention of FSI. The BSW fit can reproduce the observed ratio only  
by invoking sizeable FSI as inferred from the observed $K\pi$ phase 
shifts. It is also amusing to note that this ratio was at first seen as 
clear evidence for WA playing a dominant role in all $D$ decays.  
For WA produces a pure $I=1/2$ final state where  
$\frac{\Gamma (D^0 \to \bar K^0\pi^0)}{\Gamma (D^0 \to K^-\pi^+)}=  
\frac{1}{2}$ has to hold. 

The mode $D_s^+\to \pi^+\pi^0$, which might be seen as a signal for WA,  
is actually forbidden by isospin invariance: due to Bose statistics the  
pion pair with charge $1$ has to carry $I=2$, which cannot be reached  
from the isoscalar $D_s^+$ via a $\Delta I=1$ Lagrangian. The analogous  
Cabibbo forbidden $D^+ \to \pi^+\pi^0$ can be reached via the  
$\Delta I=3/2$ Lagrangian, yet {\em not} by WA. 

Adding up the widths for all these (quasi-)two-body channels results in a 
number close to the total nonleptonic widths for the charmed mesons. 
I.e., $D$ decays largely proceed through the hadronization of two  
quark-antiquark pairs in the final state.This means also that the  
destructive PI mechanism prolonging the $D^+$ lifetime has to emerge  
in the two-body $D^+$ modes as well. It should be noted that this  
correlation does {\em not} exist for $B$ decays, where two-body channels  
represent merely a small fraction of all $B$ decays. It is then quite  
conceivable -- and it has indeed been observed -- that two-body  
$B^-$ modes exhibit {\em constructive} PI {\em contrary} to its  
full width! 

\subsubsection{Cabibbo forbidden channels} 
\label{CABFORB} 

Several further complications and complexities arise on the once Cabibbo  
forbidden level. First, there are two transition operators before  
QCD corrections are included, namely for $c \to s \bar s u$ and  
$c\to d \bar d u$. The multiplicatively renormalized operators are  
\bea  
O_{\pm}^{(ss)} &=& \frac{1}{2}  
\left[ (\bar s_L\gamma_{\nu}c_L)(\bar u_L\gamma_{\nu}s_L)  
\pm (\bar s_L\gamma_{\nu}s_L)(\bar u_L\gamma_{\nu}c_L) \right]  
\label{OODDEVENSS} 
\\ 
O_{\pm}^{(dd)} &=& \frac{1}{2}  
\left[ (\bar d_L\gamma_{\nu}c_L)(\bar u_L\gamma_{\nu}d_L)  
\pm (\bar u_L\gamma_{\nu}c_L)(\bar d_L\gamma_{\nu}d_L) \right]  
\label{OODDEVENDD} 
\eea  
The operator $O_{-}^{(ss)}$ carries $\Delta V=1/2$ only,  
$O_{+}^{(ss)}$ also $\Delta V=3/2$; both are isospinors.  
The quantum numbers of $O_{\pm}^{(dd)}$ are quite analogous to those  
driving strange decays: $O_{-}^{(dd)}$ is purely $\Delta I=1/2$,  
whereas $O_{+}^{(dd)}$ contains also $\Delta I=3/2$. Their 
renormalization coefficients $c_{\pm}$ are as described in  
Sect.\ref{EFFWEAK}. One should note that to a very good approximation one 
has $V(cd) \simeq - V(us)$. In addition one has  the Penguin-like 
operator  
\beq  
O_P = (\bar u_L\gamma_{\nu}c_L) 
\left[ \sum_{q=u,d,s}(\bar q_L\gamma_{\nu}q_L) +  
\sum_{q=u,d,s}(\bar q_R\gamma_{\nu}q_R)\right]  
\eeq  
One should note that the usual Penguin diagram does not yield  
a local operator since the internal quarks $d$ and $s$ are lighter  
than the external $c$ quark and not even a short-distance operator  
since $m_s < \Lambda_{NPD}$. This is in marked contrast to the  
situation with Penguin diagrams for $s$ and $b$ decays. 

In the limit of $SU(3)_{Fl}$ symmetry the rates for the channels  
$D^0 \to K^+K^-$, $\pi^+\pi^-$ have to coincide. Pure phase  
space favours $D^0 \to \pi^+\pi^-$ over $D^0 \to K^+ K^-$ somewhat.  
Yet this is expected to be more than compensated by the larger  
form factors and decay constants for the $K\bar K$ final state. Within  
the factorization ansatz one finds  
\beq  
\frac{\Gamma (D^0 \to K^+K^-)}{\Gamma (D^0 \to \pi^+ \pi^-)} \simeq  
\left| \frac{f_K}{f_{\pi}}\right|^2 \simeq 1.4 \; ,  
\eeq 
which goes in the right direction, yet not nearly far enough. It has been 
suggested that Penguin operators can close the gap since they contribute  
constructively to $D^0 \to K^+K^-$, yet destructively to  
$D^0 \to \pi^+\pi^-$ because of $V(us) \simeq - V(cd)$. Yet, as indicated  
above, the Penguin operator is shaped essentially by long distance  
dynamics, this suggestion has remained a conjecture. 

The stand-by culprit for discrepancies between predictions and data are  
FSI. The preponderance of $D^0 \to K^+K^-$ over $D^0 \to \pi^+\pi^-$  
could a priori be due to a large fraction of $\pi^+\pi^-$ being  
rescattered into $\pi^0\pi^0$ final states. Yet this is clearly not the  
case, see Table \ref{tab3}. There is a simple intuitive argument why  
$\Gamma (D^0 \to K^+ K^-)$ should exceed $\Gamma (D^0 \to \pi^+ \pi^-)$:  
Two kaons eat up already a large fraction of the available phase 
space -- unlike two pions; the probability for another quark-antiquark 
pair to be created should be lower in the former than in the latter 
case. To say it differently: $D^0 \to K\bar K$ represents a larger  
fraction of  $D^0 \to K\bar K + n\; \pi 's$ than  
$D^0 \to \pi\pi$ of $D^0 \to n \; \pi 's$. This feature has indeed  
emerged in the data:  
\bea  
BR(D^0 \to K^+K^-\pi^+\pi^-) &=& (2.52\pm 0.24)\cdot 10^{-3} \\  
BR(D^0 \to \pi^+\pi^-\pi^+\pi^-)&=&(7.4\pm 0.6)\cdot 10^{-3} 
\eea 
again exhibit very large $SU(3)_{Fl}$ symmetry breaking  
\beq  
\frac{BR(D^0 \to K^+K^-\pi^+\pi^-)}{BR(D^0 \to \pi^+\pi^-\pi^+\pi^-)}  
\simeq 0.34 \pm 0.04 
\eeq  
yet in  a direction opposite to what happens in the two-body  
modes. Adding these two- and four-body rates one obtains  
\beq  
\frac{BR(D^0 \to K^+K^-,K^+K^-\pi^+\pi^-)} 
{BR(D^0 \to \pi^+\pi^-, \pi^+\pi^-\pi^+\pi^-)}  
\simeq 0.8 \pm 0.1 
\label{KKNPIONS} 
\eeq  
These numbers illustrate the general feature that individual exclusive  
modes can exhibit very large symmetry violations that average out when  
summing over exclusive rates. The ratio in Eq.(\ref{KKNPIONS}) is  
actually fully consistent with approximate $SU(3)_{Fl}$ invariance, as  
expected for the inclusive widths $\Gamma(D_q \to s \bar s u\bar q)$ vs.  
$\Gamma(D_q \to d \bar d u\bar q)$. These issues will be addressed  
again in our discussion of $D^0 - \bar D^0$ oscillations below.

There is a subtlety in relating the observable rate for 
$D^0 \to K_SK_S$ to that for $D^0 \to K^0\bar K^0$. Since 
$|K^0\bar K^0\rangle$ is equivalent to 
$|K_SK_S\rangle - |K_LK_L\rangle$ we have 
\beq 
\Gamma (D^0 \to K^0\bar K^0) = 2 \Gamma (D^0 \to K_SK_S) 
\eeq
rather than $\Gamma (D^0 \to K^0\bar K^0) = 4 \Gamma (D^0 \to K_SK_S)$.

The dynamics of {\em doubly Cabibbo suppressed decays} simplifies again  
since there is only one operator before QCD corrections are included:  
$c_L \to d_L \bar s_L u_L$. The multiplicatively renormalized operators  
yet again are  
\beq  
O_{\pm}^{\Delta C=\Delta S =-1} = \frac{1}{2}  
\left[ (\bar d_L\gamma_{\nu}c_L)(\bar u_L\gamma_{\nu}s_L)  
\pm (\bar u_L\gamma_{\nu}c_L)(\bar d_L\gamma_{\nu}s_L) \right] 
\eeq 
with $O_{-}^{\Delta C=\Delta S =-1}$ being purely $\Delta I=0$ and  
$O_{+}^{\Delta C=\Delta S =-1}$ $\Delta I=1$. 

The relative weight to the corresponding Cabibbo allowed modes  
is controlled by $(tg\theta_C)^4 \simeq 2.3\cdot 10^{-3}$  
{\em on average}. Channel-by-channel there can be considerable  
deviations from this number due to differences in formfactors and FSI. 
For Cabibbo allowed modes are purely $\Delta I =1$, while  
doubly Cabibbo suppressed channels are mainly $\Delta I=0$.  
Furthermore doubly Cabibbo suppressed $D^+$ channels are {\em not} 
suppressed by PI.  Therefore they are {\em enhanced on average} by 
the lifetime ratio  
$\tau (D^+)/\tau (D^0)$. Doubly Cabibbo suppressed $D^+_s$ rates  
on the other hand are reduced by PI; therefore they are {\em further  
decreased} by about the same factor on average.

The first evidence of the DCS  decay $D^+\rarr K^+K^-K^+$ had   
been reported  by 
 FOCUS
 \cite{Link:2002iy}, which measures  
\beq 
  \Gamma(D^+\rarr K^+K^-K^+)/\Gamma(D^+\rarr  K^-\pi^+\pi^+)=(9.5\pm 2.2) 
  \times  10^{-4} 
\label{BRKKK} 
\eeq

Such a decay can proceed via the quark decay reactions only if coupled  
with a FSI like the rescattering $\bar K^0 K^0\Rightarrow K^+K^-$,  
which is quite conceivable. Comparing the branching ratio in  
Eq.(\ref{BRKKK}) with the naive guestimate of  
$\Gamma_{DCSD}/\Gamma_{CF} \propto \tan^4\theta_C \simeq 2  \times 
10^{-3}$ shows there is ample room for such a rescattering to take place  
and still reproduce the observed width.  

\subsubsection{The $1/N_C$ ansatz} 
\label{BURAS} 
%
The fit result $\xi \simeq 0$ lead to the intriguing speculation that  
these weak two-body decays can be described more rigorously through  
$1/N_C$ expansions sketched in Sect.\ref{1OVERN}  
\index{$1/N_C$ expansions}\cite{GERARD}. They are invoked to calculate hadronic  
matrix elements. The  procedure is the following: One employs the 
effective weak transition  operator ${\cal L}_{eff}(\Delta C=1)$  
given explicitly in Eq.(\ref{EFFWEAKLAG}); since it describes short  
distance dynamics, one has kept $N_C=3$ there. Then one expands the  
matrix element for a certain transition driven by this operators in  
$1/N_C$:  
\beq  
T(D\to f) = \matel{f}{{\cal L}_{eff}(\Delta C=1)}{D} =  
\sqrt{N_C}\left( b_0 + \frac{b_1}{N_C} +  
{\cal O}(1/N_C^2)   
\right) 
\eeq  
There are straightforward rules for determining the colour weight of the 
various quark-level diagrams:  
(i) assign a colour weight $N_C$ to each {\em closed} quark loop and  
$1/\sqrt{N_C}$ to each quark-gluon coupling;  
(ii) treat every gluon line as a quark and antiquark line in colour  
space;  
(iii) normalize the meson wave functions in colour space,  
which amounts to another factor $1/\sqrt{N_C}$ per meson in the  
final or initial state. 

Quark-gluon dynamics is treated here in a nonperturbative way, as can be  
seen by considering a quark loop: any "ladder" diagram -- i.e. any planar  
diagram where gluon lines form the rungs between the quark and antiquark  
lines -- has the same colour factor since every such gluon line creates  
a new loop in colour space, rule (ii), and thus a factor $N_C$, rule (i),  
which is compensated by the two quark-gluon couplings, rule (i). 

Using these rules it is easy to show that the following simplifying  
properties hold for the contributions leading in $1/N_C$:  
\begin{itemize} 
\item  
one has to consider {\em valence} quark wave functions only;  
\item  
{\em factorization} holds;  
\item  
{\em WA} has to be ignored as have {\em FSI}. 
\end{itemize} 
To leading order in $1/N_C$ only the term $b_0$ is retained; then one 
has effectively $\xi =0$ since $\xi \simeq 1/N_C$ represents a higher 
order contribution.  However the next-to-leading term $b_1$ is in 
general beyond theoretical  control. $1/N_C$ expansions therefore do not 
enable us to decrease  the uncertainties {\em systematically}.

The $N_C \to \infty$ prescription is certainly a very compact one with  
transparent rules, and it provides not a bad first approximation --  
but not more. One can ignore neither FSI nor WA completely. 

\subsubsection{Treatment with QCD sum rules} 
\label{BLOK} 
%
In a series of papers \cite{BLOKSHIFMAN} Blok and Shifman developed  
a treatment of $D_q \to PP$, $PV$ decays based on a judicious application  
of QCD sum rules\index{QCD sum rules}. They analyzed four-point 
correlation functions between the weak Lagrangian  
${\cal L}_{weak}(\Delta C=1)$ and three currents -- one a pseudoscalar  
one generating $D_q$ mesons and two axialvector or vector currents  
for the mesons in the final state. As usual an OPE \index{OPE} is applied to the 
correlation function in the  Euclidean region;   nonperturbative dynamics is 
incorporated  through condensates  
\index{condensates} $\matel{0}{m\bar qq}{0}$, $\matel{0}{G\cdot G}{0}$  
etc., the  numerical values of which are extracted from other  
{\em light}-quark systems. Blok and Shifman extrapolate their results to the  
Minkowskian domain through a (double) dispersion relation  
\index{dispersion relations}. They succeed in finding a stability  
range for matching it with phenomenological hadronic expressions;  
hence they extract the decay amplitudes. For technical reasons their 
analysis does not extend to $D_q \to VV$ or axialvector resonances. 

 Their analysis has some nice features: 
 
\noindent $\oplus$ It has a clear basis in QCD, and includes, 
 in principle at least, nonperturbative dynamics in a well-defined way.

\noindent $\oplus$ It incorporates different quark-level processes  
-- external and internal $W$ emission, WA and PI --  
in a natural manner. 

\noindent $\oplus$ It allows to include nonfactorizable contributions  
systematically. 

In practice, however, it suffers from some shortcommings: 

\noindent $\ominus$ The charm scale is not sufficiently high that one 
could have full confidence in the various extrapolations undertaken. 

\noindent $\ominus$ To make these very lengthy calculations at all  
managable, some simplifying assumptions had to be made, like  
$m_u=m_d=m_s=0$ and $SU(3)_{Fl}$ breaking beyond $M_K > m_{\pi}$  
had to be ignored; in particular $\matel{0}{\bar ss}{0}$ =  
$\matel{0}{\bar dd}{0}$ = $\matel{0}{\bar uu}{0}$ was used. Thus  
one cannot expect $SU(3)_{Fl}$ breaking to be reproduced correctly. 

 \noindent $\ominus$ {\em Prominent} FSI that vary rapidly with the  
energy scale -- like 
effects due to narrow  resonances - cannot be described in  
this treatment; for an extrapolation from the Euclidean to the  
Minkowskian domain amounts to some averaging or `smearing' over  
energies. 

A statement that the predictions do not provide an excellent fit to  
the data on about twenty-odd $D^0$, $D^+$ and $D_s^+$ modes -- while  
correct on the surface, especially when $SU(3)_{Fl}$ breaking  
is involved -- misses 
the main point:  
\begin{itemize} 
\item  
No a priori model assumption like factorization had to be made.  
\item  
The theoretical description does not contain any free parameters in  
principle, though in practice there is leeway in the size of some  
decay constants. 
\end{itemize} 

 {\em In summary}:  
Overall a decent phenomenological description was achieved, yet  
realistically this framework provides few openings for {\em systematic} 
improvements.

\subsubsection{Status of the data} 
\label{DATATWOBODY} 
%
All charm
experiments have important analyses projects ongoing, which
address both twobody and multibody nonleptonic decays.

 \par
 An often overlooked, yet relevant source of systematic errors arises 
 from the --- at times poor --- knowledge of {\em absolute} branching 
 ratios of the {\em normalizing} modes.
 The status of measurements of {\em absolute} branching ratios was 
discussed in Sect.\ref{ABSBR}. 
In Table \ref{TAB:RECENTHAD2BGT2B}  
we list some twobody and multibody modes together with their 
normalizing mode, recent data on the {\em relative} rates and the PDG02 values; 
in Table \ref{TAB:RECENTABSGT2B}  
we give values of their 
{\em absolute} branching ratios including 
our estimate of the uncertainty arising 
from the error in the branching ratio for the normalizing mode. 
 \par
\begin{table}
\begin{center} 
\begin{tabular}{|lllrr|}
\hline
 Exp.            &
   Decay Mode  & 
   Norm. Mode    & 
   $\frac{BR(f)}{BR(f_{norm})} $  &
   $\frac{BR(f)}{BR(f_{norm})} $  \\
                 & 
  $f$             & 
  $f_{norm} $     & 
   exp    \%           & 
    PDG02 \%		           \\
  \hline
  \multicolumn{5}{|c|}{{\bf 2 BODY}} \\
                 &
  $D^0 \to$      &
  $D^0 \to$      &
                 &
                    \\
 FOCUS  \cite{Link:2002hi}& 
  $K^-K^+$ &
  $K^-\pi^+$  &
  $9.93\pm 0.20$     &
  $10.83\pm 0.27$                           \\ 
 FOCUS \cite{Link:2002hi} & 
  $\pi^-\pi^+$ &
  $K^-\pi^+$  &
  $3.53\pm0.13$     &
  $3.76\pm 0.17$                      \\
 CDF  \cite{Furic03}& 
  $K^-K^+$ &
  $K^-\pi^+$  &
  $9.38\pm 0.20$     &
                          \\ 
 CDF prel.\cite{Furic03} & 
  $\pi^-\pi^+$ &
  $K^-\pi^+$  &
  $3.686\pm0.084$     &
                       \\
 BELLE  \cite{Abe:2002ke} & 
  $K^+\pi^-$ &
  $K^-\pi^+$  &
  $0.372\pm0.027$     &
  $0.39\pm 0.06$                 \\  
 BELLE \cite{Abe:2001si}& 
  $K^0_L\pi^0$ &
  $K^0_S \pi^0$  &
  $0.88\pm 0.13$     &
  --                      \\ 
                 &
  $\Lambda_c^+ \to$      &
  $\Lambda_c^+ \to$      &
                 &
                    \\
 FOCUS  \cite{Link:2002zx}& 
  $\Sigma^+K^{*0}(892)$ &
  $\Sigma^+\pi^+\pi^-$  &
  $7.8\pm2.2$     &
  --                     \\   
 FOCUS  \cite{Link:2002zx}& 
  $\Sigma^+ \phi$ &
  $\Sigma^+\pi^+\pi^-$  &
  $8.7\pm1.7$     &
  $8.7\pm1.6$                        \\ 
 FOCUS  \cite{Link:2002zx}& 
  $\Xi(1690)^0K^+$ &
  $\Sigma^+\pi^+\pi^-$  &
  $2.2\pm0.8$     &
  $2.3\pm 0.7$                          \\  
  \hline     
  \multicolumn{5}{|c|}{{\bf $>$ 2 BODY}} \\
                 &
  $D^+ \to$      &
  $D^+ \to$      &
                 &
                    \\		   
 FOCUS  \cite{Link:2002mm}& 
  $K^-3\pi^+\pi^-$ &
  $K^-2\pi^+$  &
  $5.8\pm 0.6$     &
  $8.0\pm 0.9$                         \\
 FOCUS  \cite{Link:2002mm}& 
  $3\pi^+2\pi^-$ &
  $K^-3\pi^+\pi^-$  &
  $29.0\pm 2.0$     &
  --                   \\
 FOCUS  \cite{Link:2002mm}& 
  $K^+K^- 2\pi^+\pi^-$ &
  $K^-3\pi^+\pi^-$  &
  $4.0\pm 2.1$     &
  --        \\
FOCUS \cite{Link:2002iy} & 
  $K^-K^+K^+$ &
  $K^-\pi^+\pi^+$  &
  $(9.49\pm2.18)\, 10^{-2}$     &
  --                         \\  
                 &
  $D_s^+ \to$      &
  $D_s^+ \to$      &
                  &
                     \\
 FOCUS  \cite{Link:2002mm}& 
  $3\pi^+2\pi^-$ &
  $K^+K^-\pi^+$  &
  $14.5\pm1.4$     &
  $15.8\pm 5.2$                 \\  
 FOCUS  \cite{Link:2002mm}& 
  $K^+K^-2\pi^+\pi^-$ &
  $K^+K^-\pi^+$  &
  $15.0\pm3.1$     &
  $18.8 \pm 5.4$                         \\  
 FOCUS  \cite{Link:2002mm}& 
  $\phi 2\pi^+\pi^-$ &
  $\phi \pi^+$  &
  $24.9\pm 3.2$     &
  $33\pm 6$        \\      
FOCUS \cite{Link:2002iy}  & 
  $K^+K^-K^+$ &
  $K^+K^-\pi^+$  &
  $0.895\pm 0.310$     &
  $<1.6$                            \\   
                 &
$\Lambda_c^+ \to$      &
  $\Lambda_c^+ \to$      &
                 &
                    \\ 
 FOCUS  \cite{Link:2002zx}& 
  $\Sigma^+K^+K^-$ &
  $\Sigma^+\pi^+\pi^-$  &
  $7.1\pm 1.5$     &
  $8.1\pm 1.0$                         \\  
 FOCUS  \cite{Link:2002zx}& 
  $\Sigma^-K^+\pi^+$ &
  $\Sigma^+K^{*0}(892)$  &
  $<35$(90\% cl)     &
  --                        \\ 
 FOCUS  \cite{Link:2002zx}& 
  $(\Sigma^+K^+K^-)_{nr}$ &
  $\Sigma^+\pi^+\pi^-$  &
  $<2.8$ (90\% cl)    &
  $<1.8$                          \\  
 CLEO  \cite{Cronin-Hennessy:2002we} & 
  $\Lambda \pi^+ \omega$ &
  $pK^-\pi^+$  &
  $24\pm 8$     &
  --                       \\ 
 CLEO  \cite{Cronin-Hennessy:2002we} & 
  $\Lambda \pi^+ \eta$ &
  $pK^-\pi^+$  &
  $41\pm 20$     &
  $35\pm 8$                          \\  
 CLEO  \cite{Cronin-Hennessy:2002we} & 
  $\Lambda \pi^+ (\pi^+\pi^-\pi^0)_{nr}$ &
  $pK^-\pi^+$  &
  $<13$ (90\% cl)     &
  --                        \\  
 CLEO  \cite{Cronin-Hennessy:2002we} & 
  $\Lambda \pi^+(\pi^+\pi^-\pi^0)_{tot}$ &
  $pK^-\pi^+$  &
  $36\pm 13$     &
  --                          \\       
\hline
\end{tabular} 
\end{center} 
\caption{Recent results not included in PDG02 on 2-body and $>$2-body 
charm mesons and
baryons branching ratios. Errors are added in quadrature. }
\label{TAB:RECENTHAD2BGT2B} 
\end{table}  

\begin{table}
\begin{center} 
\begin{tabular}{|llrr|}
\hline
 Exp.            &
   Decay Mode    & 
   $BR_{exp}$ \% &
    $\frac{\sigma(f_{norm})}{\sigma(f)+\sigma(f_{norm})}$    \\
\hline
  \multicolumn{4}{|c|}{{\bf 2 BODY}} \\
                 &
  $D^0 \to$      &
                 &
                    \\
 FOCUS  \cite{Link:2002hi}& 
  $K^-K^+$ &
 $0.394\pm 0.013$     &
  1.2                        \\ 
 FOCUS \cite{Link:2002hi} & 
  $\pi^-\pi^+$ &
  $0.138\pm 0.005$     &
  0.7                         \\
 BELLE  \cite{Abe:2002ke} & 
  $K^+\pi^-$ &
  $0.014\pm 0.001$     &
  0.3                         \\
 BELLE  \cite{Abe:2003yv} & 
  $\phi\pi^0$ &
  $0.0801\pm 0.0052$     &
         0.3                 \\     
 BELLE  \cite{Abe:2003yv} & 
  $\phi\eta$ &
  $0.0148\pm 0.0048$     &
         0.1                 \\			  
                 &
  $\Lambda_c^+ \to$      &
                 &
                    \\
 FOCUS  \cite{Link:2002zx}& 
  $\Sigma^+K^{*0}(892)$ &
  $0.28\pm 0.10$     &
  1.0                         \\  
 FOCUS  \cite{Link:2002zx}& 
  $\Sigma^+ \phi$ &
  $0.31\pm 0.08$     &
  1.5                         \\ 
 FOCUS  \cite{Link:2002zx}& 
  $\Xi(1690)^0K^+$ &
  $0.08\pm 0.03$     &
  0.8                         \\   
\hline
  \multicolumn{4}{|c|}{{\bf $>$ 2 BODY}} \\
                 &
  $D^+ \to$      &
                 &
                    \\		   
 FOCUS  \cite{Link:2002mm}& 
  $K^-3\pi^+\pi^-$ &
  $0.53\pm 0.06$     &
  0.7                         \\
 FOCUS  \cite{Link:2002mm}& 
  $3\pi^+2\pi^-$ &
  $0.212\pm 0.036$     &
  1.4                         \\
 FOCUS  \cite{Link:2002mm}& 
  $K^+K^- 2\pi^+\pi^-$ &
  $0.029\pm 0.017$     &
  0.2    \\
FOCUS \cite{Link:2002iy} & 
  $K^-K^+K^+$ &
  $(0.86\pm 0.19)\, 10^{-2}$     &
  0.3                         \\  
                 &
  $D_s^+ \to$      &
                 &
                    \\
 FOCUS  \cite{Link:2002mm}& 
  $3\pi^+2\pi^-$ &
  $0.64\pm 0.18$     &
  2.7                        \\  
 FOCUS  \cite{Link:2002mm}& 
  $K^+K^-2\pi^+\pi^-$ &
  $0.66\pm 0.22$     &
  1.3                         \\  
 FOCUS  \cite{Link:2002mm}& 
  $\phi 2\pi^+\pi^-$ &
  $0.90\pm 0.26$     &
  2.0                         \\      
FOCUS \cite{Link:2002iy}  & 
  $K^-K^+K^+$ &
  $0.039\pm 0.016$     &
  0.8                         \\   
                 &
  $\Lambda_c^+ \to$      &
                 &
                    \\  
 FOCUS  \cite{Link:2002zx}& 
  $\Sigma^+K^+K^-$ &
  $0.281\pm 0.069$     &
  1.3                         \\ 
 FOCUS  \cite{Link:2002zx}& 
  $(\Sigma^+K^+K^-)_{nr}$ &
  $<0.06$ (90\% cl)    &
                           \\  
 CLEO  \cite{Cronin-Hennessy:2002we} & 
  $\Lambda \pi^+ (\pi^+\pi^-\pi^0)_{tot}$ &
  $1.80\pm 0.64$     &
  0.7                       \\ 
 CLEO  \cite{Cronin-Hennessy:2002we} & 
  $\Lambda \omega \pi^+$ &
  $1.20\pm 0.43$     &
  0.8                       \\ 
 CLEO  \cite{Cronin-Hennessy:2002we} & 
  $\Lambda \eta \pi^+$ &
  $1.75\pm 0.49$     &
  0.5                         \\  
 CLEO  \cite{Cronin-Hennessy:2002we} & 
  $\Lambda \pi^+ (\pi^+\pi^-\pi^0)_{nr}$ &
  $<0.65$ (90\% cl)     &
                          \\       
\hline
\end{tabular}  
\end{center} 
\caption{Recent results not included in PDG02 on 2-body and
 $>$2-body charm mesons and
baryons absolute branching ratios. Last column shows 
ratio of relative errors on branching fractions of normalizing mode 
and decay mode.}
\label{TAB:RECENTABSGT2B} 
\end{table} 
%
\subsubsection{Modern models} 
\label{MODERN} 
%
As the data improved, the BSW prescription became inadequate, however  
almost every subsequent attempt to describe two-body nonleptonic decays  
in the $D$ system uses the assumption of naive factorization as a  
starting point. 

At this point the following question arises: Why not just wait for 
lattice QCD (or some other calculational breakthrough) to gain us  
theoretical control over exclusive decays? For refining quark model  
predictions could be viewed somewhat unkindly like  
adding epicycles to a Ptolemaic system: while producing more accurate  
numbers, it would not deepen our understanding.

There are several reasons for not waiting idly:  
\begin{itemize} 
\item  
The wait might be quite long considering a quenched (or even partially  
unquenched) approximation is inadequate to include FSI.  
\item  
Even a merely phenomenological description of exclusive modes will be of 
great help for estimating the strength of $D^0 - \bar D^0$ oscillations 
in the SM, as explained later.  
\item  
As already stated, information on the strong phases due to FSI is  
instrumental for understanding {\em direct} CP violation.  
\end{itemize} 

Improvements (hopefully) and generalizations of the BSW description are  
made in three areas:  
\begin{enumerate} 
\item  
Different parametrizations for the $q^2$ dependence of the form factors  
are used and different evaluations of their normalization. This is 
similar to what was addressed in our discussion of exclusive semileptonic  
decays. One appealing suggestion has been to use only those expressions  
for form factors that asymptotically -- i.e. for $m_c$, $m_s \to \infty$ 
-- exhibit heavy quark symmetry.  
\item  
Contributions due to WA and Penguin operators have been included.  
\item  
Attempts have been made to incorporate FSI more reliably. 

\end{enumerate} 

One can easily see that the form factors of BSW do not agree 
with QCD's HQS in the heavy quark limit. In this limit, the various form 
factors can be expressed in terms of one universal form factor. 
For example, Eq.~(\ref{DPFF}) then reads as follows 
$$ 
\langle P(p_{P})|V^{\mu}|D(p_{D})\rangle = 
\xi(v \cdot v')(v+v')^{\mu}  
$$ 
with $v[v']$ refering to the velocity of the $D[P]$ meson. This 
expression allows us to relate the form factors $F_1$ and $F_0$: 
\beq 
\xi(v \cdot v')=\frac{2\sqrt{M_D M_P}}{M_D+M_P}F_1(q^2)= 
\frac{2\sqrt{M_D M_P}}{M_D+M_p}\frac{F_0(q^2)}{1-q^2/(M_D+M_P)^2} 
\label{FFRELAT} 
\eeq 
\noindent 
Assuming a simple pole dependence for both $F_1$ and $F_0$ is 
clearly inconsistent with the preceding expression.  
Yet it turns out that the $q^2$ dependence of the form factors -- whether 
it is described by a pole, double pole or exponential -- has a rather  
limited impact on the results -- not surprisingly, since in $D$ (unlike 
$B$) decays the $q^2$ range is quite limited. This is displayed in Table 
\ref{FORMFACS}.

\begin{table}[ht]
\begin{center}
\begin{tabular}{l c c c c c c}
& $F_0^{DK}(m_{\pi}^2)$ & $F_0^{D\pi}(m_K^2)$ & $F_1^{DK}(m_\rho^2)$ & $F_1^{D\pi}(m_{K^*}^2)$ & $A_0^{DK^*}(m_\pi^2)$ & $A_0^{D\rho}(m_K^2)$ \\ 
\hline
pole & 0.76 & 0.72 & 0.84 & 0.79 & 0.74 & 0.69 \\
multipole & - & - & 0.91 & 0.92 & 0.74 & 0.72\\
exponential & 0.75 & 0.78 & 0.85 & 0.94 & 0.83 & 0.96 \\
\end{tabular}
\end{center}
\caption{Numerical values of pole, double pole and exponential form factors}
\label{FORMFACS}
\end{table}

The much greater challenge is provided by items 2 and 3. As discussed in 
Sect.\ref{WEAKLIFE} WA is not the dominant engine driving the $D^0-D^+$ 
lifetime  difference. It had been predicted \cite{DSPAPER} to contribute  
about 10 - 20 \% of the overall 
$D^0$ and $D^+_s$ widths, and their observed lifetime ratio shows 
that it indeed does. However it does not generate a $\sim$ 10 - 20 \%  
contribution {\em uniformly} to all channels. Actually it could substantially  
enhance or suppress the widths of individual channels or even dominate  
them \cite{MIRAGE}. It would be interesting to see whether such 
`exclusive footprints'  of WA could be identified in the data.

There is a further complication beyond WA's strength varying greatly  
from channel to channel. WA always produces non-exotic final states,  
i.e. those with quantum numbers possible for a  
$\bar q_1 q_2$ [$q_1q_2q_3$] combination in $D$ [$\Lambda_c$]  
decays. Exactly those channels are sensitive to the most prominent FSI, 
namely resonance effects. As far as {\em fully inclusive} widths,  
which are not subject to FSI, are concerned, WA has a well-defined  
meaning as an independent process (and enters as an independent  
$1/m_Q^3$ term in the HQE). Yet for exclusive nonleptonic modes the  
distinction between WA and FSI becomes blurred. Till one has established  
full theoretical control over FSI, its contributions and those of WA are  
indistinguishable in an {\em individual} channel. Only by carefully analyzing 
the whole pattern can one hope to arrive at some meaningful conclusions.

There exists a wide variety of ans\"atze in the literature differing in 
their parametrizations, treatment of non-factorizable contributions etc.  
The first detailed model of the post-BSW generation was presented in  
Ref.\cite{BUCCELLA}. It is based on a modified (??) factorization 
prescription, fits the colour suppression factor $\xi$ to the data and 
allows for prominent WA terms. FSI are implemented by allowing for  
rescattering among final states belonging to the same $SU(3)$ representation.

Using these assumption the authors fitted the then available data.  
The results are shown in Tables~(\ref{tab1}-\ref{tab4}). We give there the data 
as listed in PDG96 (relevant for the time the analysis of Ref.\cite{BUCCELLA} was 
performed) and PDG02. On the experimental side it is intriguing to see 
that the quoted errors do not always decrease in time and that changes in the 
central values by two sigma do occur. 

On the theoretical side we would like to note that the overall 
agreement is quite good. However a comparison of the second and third columns 
shows that the inclusion of well chosen FSI is essential for this success; 
for proper perspective one should keep in mind that including FSI at 
present involves a considerable amount of `poetic license'. There is a good side 
to the prominence of FSI as well: they are a conditio sine qua non for direct 
CP violation revealing itself in partial width asymmetries. Anticipating the later discussion 
of CP violation we have listed also the asymmetries predicted by Ref.\cite{BUCCELLA} 
for Cabibbo forbidden modes in Tables~(\ref{tab2},\ref{tab3}). However there are 
some glaring discrepencies. In particular,  
the predicted branching ratio for $D^+_s \to \rho^+ \eta$ is well above the {\em presently} 
measured value, when FSI are included. The predicted value for 
BR$(D^+_s \to \rho^+ \eta^{\prime})$ on the other hand appears well below the data. 

\begin{table}
\begin{center}
\begin{tabular}{|l|c|c|c|c|c|}
\hline
\vspace*{-.1cm}
Decay Channel &\hspace{-.4cm} & $BR_{th}$(no FSI)              & 
$BR_{th}$(FSI)  & $BR_{exp} $  & Exp Ref\\
              &\hspace{-.4cm} &                 $\times 10^2$ & 
$\times 10^2$ & $ \times 10^2$  & \\
\hline
\hline
$D^0 \to K^- \pi^+ $ & \hspace{-.4cm} & 5.35 & 3.85 & $3.83 \pm 0.12$ & 
PDG96\\
 & & & & $3.80 \pm 0.09$ & PDG03\\
$D^0 \to K_s \pi^0 $ &\hspace{-.4cm} & 0.44 & 0.76 & $1.05 \pm 0.10$ & 
PDG96\\
 & & & & $1.14 \pm 0.11$ & PDG03\\
$D^0 \to K_s \eta $ & \hspace{-.4cm} & 0.13 & 0.45 & $0.35 \pm 0.05$ & 
PDG96\\
 & & & & $0.38 \pm 0.05$ & PDG03\\
$D^0 \to K_s \eta' $ &\hspace{-.4cm} & 0.54 & 0.80  & $0.85 \pm 0.13$ & 
PDG96\\
 & & & & $0.94 \pm 0.14$ & PDG03\\
$D^0 \to \bar K^{*0} \pi^0 $ &\hspace{-.4cm} & 1.66 & 3.21  & $3.10 \pm 
0.40$ & PDG96\\
 & & & & $2.8 \pm 0.4$ & PDG03\\
$D^0 \to \rho^0 K_s $ & \hspace{-.4cm} & 1.01 & 0.45  & $0.60 \pm 0.085$ 
& PDG96\\
 & & & & $0.74 \pm 0.15$ & PDG03\\
$D^0 \to K^{*-} \pi^+ $ &\hspace{-.4cm} & 1.12 & 4.66  & $5.0 \pm 0.4$ & 
PDG96\\
 & & & & $6.0 \pm 0.5$ & PDG03\\
$D^0 \to \rho^+ K^- $ & \hspace{-.4cm} & 9.62 & 11.2  & $10.8 \pm 1.0$ & 
PDG96\\
 & & & & $10.2 \pm 0.8$ & PDG03\\
$D^0 \to \bar K^{*0} \eta $ &\hspace{-.4cm} & 1.21 & 0.47  & $1.90 \pm 
0.50$ & PDG96\\
 & & & & $1.8 \pm 0.9$ & PDG03\\
$D^0 \to \bar K^{*0} \eta' $ & \hspace{-.4cm} & 0.001 & 0.004  & $<0.11$ 
& PDG96\\
 & & & & $<0.11$ & PDG03\\
$D^0 \to \omega K_s $ &\hspace{-.4cm} & 0.26 & 0.97  & $1.05 \pm 0.20$ & 
PDG96\\
 & & & & $1.1 \pm 0.2$ & PDG03\\
$D^0 \to \phi K_s $ & \hspace{-.4cm} & 0.059 & 0.414  & $0.43 \pm 0.05$ 
& PDG96\\
 & & & & $0.47 \pm 0.06$ & PDG03\\
$D^+ \to K_s \pi^+$ & \hspace{-.4cm} & 1.14 & 1.35  & $1.37 \pm 0.15$ & 
PDG96\\
 & & & & $1.39 \pm 0.09$ & PDG03\\
$D^+ \to \bar K^{*0} \pi^+ $ & \hspace{-.4cm} & 1.46 & 2.00  & $1.92 \pm 
0.19$ & PDG96\\
$D^+ \to \rho^+ K_s $ & \hspace{-.4cm} & 1.71 & 5.82  & $3.30\pm 1.25$ & 
PDG96\\
$D^+_s \to K^+  K_s$  & \hspace{-.4cm} & 1.35 & 2.47  & $1.80 \pm 0.55$ 
& PDG96\\
$D^+_s \to \pi^+ \eta$ & \hspace{-.4cm} & 4.53 & 1.13  & $2.0 \pm 0.6$ & 
PDG96\\
 & & & & $1.7 \pm 0.5$ & PDG03\\
$D^+_s \to \pi^+ \eta'$ & \hspace{-.4cm} & 2.60 & 5.44  & $4.9 \pm 1.8$ 
& PDG96\\
 & & & & $3.9 \pm 1.0$ & PDG03\\
$D^+_s \to \rho^+ \eta$ & \hspace{-.4cm} & 4.42 & 8.12  & $10.3\pm 3.2$ 
& PDG96\\
 & & & & $3.80 \pm 0.09$ & PDG03\\
$D^+_s \to \rho^+ \eta'$ & \hspace{-.4cm} & 1.08 & 2.46  & $12.0 \pm 
4.0$ & PDG96\\
 & & & & $10.8 \pm 3.1$ & PDG03\\
$D^+_s \to \phi \pi^+$ & \hspace{-.4cm} & 2.51 & 4.55  & $3.6 \pm 0.9$ & 
PDG96\\
 & & & & $3.80 \pm 0.09$ & PDG03\\
$D^+_s \to \bar K^{*0} K^+$ & \hspace{-.4cm} & 5.27 & 4.81  & $3.4 \pm 
0.9$ & PDG96\\
 & & & & $3.6 \pm 0.9$ & PDG03\\
$D^+_s \to K^{*+} K_s$ & \hspace{-.4cm} & 0.87 & 1.10  & $2.15 \pm 0.7$ 
& PDG96\\
$D^+_s \to \rho^0 \pi^+$ & \hspace{-.4cm} & 0.012 & 0.01  & $<0.29$ & 
PDG96\\
 & & & & $<0.07$ & PDG03\\
$D^+_s \to \rho^+ \pi^0$ & \hspace{-.4cm} & 0.023 & 0.01  & $-\!\!-$ & \\
$D^+_s \to \omega \pi^+$ & \hspace{-.4cm} & 0.023 & 0.20  & $0.27 \pm 
0.12$ & PDG96\\
\vspace{-.1cm}
 & & & & $0.28 \pm 0.11$ & PDG03\\
\hline
\end{tabular}
\end{center}
\caption{Theoretical predictions and experimental values for the 
branching ratios of various Cabibbo-allowed decays.}
\label{tab1}
\end{table}

\begin{table}
\begin{center}
\begin{tabular}{|l|c|c|c|c|c|c|}
\hline
\vspace*{-.1cm}
$D^+ \to$ & \hspace{-.3cm} & $BR_{th}$(no FSI) & 
$BR_{th}$(FSI)  & $BR_{exp} $ & Exp Ref & $a_{CP} $ \\
   & \hspace{-.3cm} & $\times 10^2$ & 
   $\times 10^2$ & $\times 10^2$ &  & $(10^{-3})$ \\
\hline
\hline
$ \pi^+ \pi^0 $ & \hspace{-.3cm} & $0.186$ & $0.185$ & $0.25 \pm 0.07$ & 
PDG96 &  $-\!\!-$ \\
$ \pi^+ \eta$ & \hspace{-.3cm} & $0.38$ & $0.38$ & $0.75 \pm 0.25$ & 
PDG96 & -0.77 \\
 & & & & $0.3 \pm 0.06$ & PDG03 & \\
$ \pi^+ \eta' $ & \hspace{-.3cm} & $0.058$ &  $0.768$ & $<0.9$ & PDG96 & 
+0.90 \\ 
 & & & & $0.5 \pm 0.1$ & PDG03 & \\
$ K^+ \bar K^0 $ & \hspace{-.3cm} & $1.49$ &  $0.763$ & $0.72 \pm 0.12$ 
& PDG96 & -0.52 \\
 & & & & $0.58 \pm 0.06$ & PDG03 & \\
$ \pi^+ \rho^0 $ & \hspace{-.3cm} & $0.012$ & $0.104$ & $<0.14$ & PDG96 
& -1.96 \\
 & & & & $0.104 \pm 0.018$ & PDG03 & \\
$ \rho^+ \pi^0$ & \hspace{-.3cm} & $0.208$ &  $0.451$ & $-\!\!-$ &  & 
+0.89 \\
$ \rho^+ \eta $ & \hspace{-.3cm} & $0.695$ & $0.064$ & $<1.2$ & PDG96 & 
-1.60 \\
 & & & & $<0.7$ & PDG03 & \\ 
$ \rho^+ \eta' $ & \hspace{-.3cm} & $0.004$ & $0.122$ & $ <1.5$ & PDG96 
& $\sim$ 0 \\
 & & & & $<0.5$ & PDG03 & \\
$ \pi^+ \omega $ & \hspace{-.3cm} & $0.124$ &  $0.038$ & $<0.7$ & PDG96 
& -0.60 \\
$ \pi^+ \phi$ & \hspace{-.3cm} & $0.146$ &  $0.619$ & $0.61 \pm 0.06$ & 
PDG96 & -0.09 \\
$ K^+ \bar K^{*0} $ & \hspace{-.3cm} & $1.99$ &  $0.436$ & $0.42 \pm 
0.05$ & PDG96 & +0.68\\
$ K^{*+} \bar K^{0} $ & \hspace{-.3cm} & $1.52$ & $0.86$ & $3.0 \pm 1.4$ 
& PDG96 & -0.19\\
\vspace{-.1cm}
 & & & & $3.1 \pm 1.4$ & PDG03 & \\
\hline
\hline
\vspace{-.1cm}
$D^+_s \to$ & \hspace{-.3cm} & $BR_{th}$(no FSI) & 
$BR_{th}$(FSI)  & $BR_{exp} $ & Exp Ref & $a_{CP} $ \\
   & \hspace{-.3cm} & $\times 10^2$ & 
   $\times 10^2$ & $\times 10^2$ &  & $(10^{-3})$ \\
\hline
\hline
$ K^+ \pi^0 $ & \hspace{-.3cm} & $0.222$ & $0.146$ & $-\!\!-$ &  & +1.07\\
$ K^+ \eta$ & \hspace{-.3cm} & $0.046$ & $0.299$ & $-\!\!-$ &  & -0.05\\
$ K^+ \eta'$ & \hspace{-.3cm} & $0.318$ &  $0.495$ & $-\!\!-$ &  & -0.64\\
$ \pi^+ K^0$ & \hspace{-.3cm} & $0.586$ & $0.373$ & $<0.8$ & PDG96 & 
+0.48\\
$ K^+ \rho^0 $ & \hspace{-.3cm} & $0.952$ & $0.198$ & $<0.29$ & PDG96 & 
+0.25\\
$ K^0 \rho^+$ & \hspace{-.3cm} & $0.384$ & $1.29$ & $-\!\!-$ &  & +0.36\\ 
$ K^{*+} \pi^0$ & \hspace{-.3cm} & $0.0004$ & $0.076$ & $-\!\!-$ &  & 
-0.92\\
$ K^{*0} \pi^+$ & \hspace{-.3cm} & $0.191$ & $0.444$ & $0.65 \pm 0.28$ & 
PDG96 & -0.75\\
$ K^{*+} \eta$ & \hspace{-.3cm} & $0.200$ & $0.146$ & $-\!\!-$ &  & -0.41\\
$ K^{*+} \eta'$ & \hspace{-.3cm} & $0.044$ & $0.029$ & $-\!\!-$ &  & 
-0.09\\
$ K^+ \omega $& \hspace{-.3cm} & $0.252$ &  $0.178$ & $-\!\!-$ &  & -0.34\\
\vspace{-.1cm}
$ K^+ \phi $ & \hspace{-.3cm} & $0.103$ & $0.008$ &  $<0.05$ & PDG96 & 
+1.79\\
\hline
\end{tabular}
\end{center}
 \caption{Branching ratios for Cabibbo-forbidden decays of $D^+$ and 
$D^+_s$ mesons along with predictions for CP
asymmetries}
\label{tab2}
\end{table}

\begin{table}
\begin{center}
\begin{tabular}{|l|c|c|c|c|c|c|}
\hline
\vspace*{-.1cm}
$D^0 \to$ & \hspace{-.3cm} & $BR_{th}$(no FSI) & 
$BR_{th}$(FSI)  & $BR_{exp} $ & Exp Ref & $a_{CP} $ \\
   & \hspace{-.3cm} & $\times 10^2$ & 
   $\times 10^2$ & $\times 10^2$ &  & $(10^{-3})$ \\
\hline
\hline
$ \pi^+ \pi^- $ & \hspace{-.2cm} & $0.505$ & $0.152$ & $0.152 \pm 0.011$ 
& PDG96 & $-0.10$\\
 & & & & $0.143 \pm 0.007$ & PDG03 & \\
$ \pi^0 \pi^0 $ & \hspace{-.2cm} & $0.106$ &  $0.115$ & $0.084 \pm 
0.022$ & PDG96 &
+0.51\\
$ K^+ K^- $ & \hspace{-.2cm} & $0.589$ & $0.427$ & $0.433 \pm 0.027$ & 
PDG96 & -0.10\\
 & & & & $0.412 \pm 0.014$ & PDG03 & \\
$ K^0 \bar K^0 $ & \hspace{-.2cm} & $0$ & $0.108$ & $0.13 \pm 0.04$ & 
PDG96 &
+0.26\\
 & & & & $0.071 \pm 0.019$ & PDG03 & \\ 
$\pi^0 \omega$ & \hspace{-.2cm} & $0.013$ & $0.003$ & $-\!\!-$ &  & -0.01\\
$ \pi^0 \phi $ & \hspace{-.2cm} & $0.127$ & $0.105$ & $<0.14$ & PDG96 & 
-0.04 \\
$\eta \phi $ & \hspace{-.2cm} & $0.080$ & $0.080$ &$ <0.28$ & PDG96 & 
-0.15\\
$ \pi^0 \phi $ & \hspace{-.2cm} &  &  & $0.0801\pm0.0052$ & 
\cite{Abe:2003yv} &  \\
$\eta \phi $ & \hspace{-.2cm} &  &  &$0.0148\pm 0.0048$ & 
\cite{Abe:2003yv} & \\
$ K^0 \bar K^{*0}$ & \hspace{-.2cm} & $0.031$ & $0.052$ & $<0.16$ & 
PDG96 & -0.56\\
 & & & & $<0.17$ & PDG03 & \\ 
$ \bar K^0 K^{*0}$ & \hspace{-.2cm} & $0.031$ & $0.062$ & $<0.08 $ & 
PDG96 & -0.65\\
 & & & & $<0.09$ & PDG03 & \\
$ K^- K^{*+} $ & \hspace{-.2cm} & $0.542$ & $0.431$ & $0.35 \pm 0.08$ & 
PDG96 &-0.04\\
 & & & & $0.38 \pm 0.08$ & PDG03 & \\
$ K^+ K^{*-} $ & \hspace{-.2cm} & $0.178$ & $0.290$ & $0.18 \pm 0.08$ & 
PDG96 &0.27\\
 & & & & $0.20 \pm 0.11$ & PDG03 & \\ 
$ \pi^0 \eta $ & \hspace{-.2cm} & $0.055$ & $0.054$ & $-\!\!-$ &  & -1.44\\
$ \pi^0 \eta'$ & \hspace{-.2cm} & $0.174$ & $0.175$ & $-\!\!-$ &  & +0.89\\
$ \eta \eta$ & \hspace{-.2cm} & $0.171$ & $0.093$ & $-\!\!-$ &  & -0.51\\
$ \eta \eta' $ & \hspace{-.2cm} & $0.011$ & $0.186$ & $-\!\!-$ &  & -0.31\\
$ \eta \rho^0$ & \hspace{-.2cm} & $0.010$ & $0.020$ & $-\!\!-$ &  & -0.53\\
$ \eta' \rho^0$ & \hspace{-.2cm} & $0.007$ & $0.008$ & $-\!\!-$ &  &+0.01\\
$ \eta \omega$ & \hspace{-.2cm} & $0.002$ & $0.209$ & $-\!\!-$ &  & -0.02\\
$ \eta' \omega $& \hspace{-.2cm} & $0.017$ & $0.0002$ & $-\!\!-$ &  & 
-3.66\\
$ \pi^0 \rho^0 $ & \hspace{-.2cm} & $0.199$ &  $0.216$ & $-\!\!-$ &  
&-0.01\\
$ \pi^+ \rho^-$ & \hspace{-.2cm} & $0.442$ & $0.485$ & $-\!\!-$ &  & 
-0.43\\
\vspace{-.1cm}
$ \pi^- \rho^+$ & \hspace{-.2cm} & $1.45$ &  $0.706$ & $-\!\!-$ &  & 
+0.34\\
\hline
\end{tabular}
\end{center}
\caption{Branching ratios for Cabibbo suppressed decays of $D^0$ 
mesons along with predictions for CP asymmetries}
\label{tab3}
\end{table}

\begin{table}
\begin{center}
\begin{tabular}{|l|c|c|c|c|c|}
\hline
\vspace*{-.1cm}
Decay Channel &\hspace{-.2cm} & $BR_{th}$(no FSI)$\times 10^2$ & 
$BR_{th}$(FSI) $\times 10^2$ & $BR_{exp} \times 10^2$ & Exp Ref \\
\hline
\hline
$D^0 \to K^+ \pi^- $ & \hspace{-.2cm} & 0.028 & 0.033 & $0.029 \pm 
0.014$ & PDG96 \\
 & & & & $0.0148 \pm 0.0021$ & PDG03 \\
$D^0 \to K^{*0} \pi^0 $ &\hspace{-.2cm} & 0.004 & 0.004  & $-\!\!-$ & \\
$D^0 \to K^{*+} \pi^- $ &\hspace{-.2cm} & 0.033 & 0.038 & $-\!\!-$ & \\
$D^+ \to K^{+} \pi^0 $ &\hspace{-.2cm} & 0.044 & 0.055 & $-\!\!-$ & \\
$D^+ \to K^{*+} \pi^0 $ &\hspace{-.2cm} & 0.054 & 0.057 & $-\!\!-$ &\\
$D^+ \to K^{*0} \pi^+ $ &\hspace{-.2cm} & 0.040 & 0.027 & $<0.019$ & 
PDG96 \\
 & & & & $0.036 \pm 0.016$ & PDG03 \\
$D^+ \to \phi K^{+} $ &\hspace{-.2cm} & 0.003 & 0.003 & $<0.013$ & PDG96\\
$D^+ \to \rho^0 K^{+} $ &\hspace{-.2cm} & 0.027 & 0.029 & $<0.06$ & PDG96\\
\vspace{-.1cm}
 & & & & $0.025 \pm 0.012$ & PDG03 \\
\hline
\end{tabular}
\end{center}
\caption{Theoretical predictions and experimental values for the 
branching ratios of various Double-Cabibbo forbidden decays.}
\label{tab4}
\end{table}

A less ambitious 
approach has also been undertaken \cite{ROSNER,CHENG}. One can 
decompose the 
various amplitudes for a given class of decays into four 
{\em topological} amplitudes. 
These diagrams account for final state interactions and are not 
actual Feynman graphs. 
The four diagrams are: (1) a colour-favoured external W-emission 
tree diagram T, (2) a colour-suppressed internal W-emission tree diagram 
C,(3) an exchange amplitude E, and (4) an annihilation amplitude A. 

Using the available data, 
these amplitudes can be fitted for the Cabibbo allowed modes. At 
first glance, this may appear to be an empty analysis. However, by making 
various phenomenological assumptions one can begin to discuss 
nonfactorizable effects and final state interactions. For example, in 
Ref.~(\cite{CHENG}) the various topological amplitudes are defined in 
a similar vein to the BSW approach, e.g. 
\beq 
T=\frac{G_F}{\sqrt{2}}V_{ud}V_{cs}^* a_1 f_\pi 
\left(M_D^2-M_K^2)\right) F_0^{DK}\left(M_\pi ^2\right) 
\eeq 
Here $a_1$ is related to the coefficient of the same name 
discussed above. It consists of the naive factorization piece along with a 
piece describing nonfactorizable contributions. 
\beq 
a_1=c_1+c_2\left(\frac{1}{N_c}+\chi_1\right) 
\eeq 
This analysis is again consistent with $\chi_1=-1/N_C$ or $\xi 
\approx 0$. Final state rescattering is also discussed in this 
approach. Using the same resonance based rescattering model of 
Ref.\cite{BUCCELLA}  
\footnote{There is a disputed factor of 2 between 
these analyses in the definition of the rescattering phase.}, it is found 
that the fit value for the exchange topology E is consistent with a 
vanishing quark level exchange contribution as expected from helicity 
suppression arguments. 

This analysis can be extended to include singly and doubly 
Cabibbo-suppressed modes. In Ref. \cite{Chiang} it was shown that 
by simply rescaling the Cabibbo allowed topological amplitudes by the 
relevent CKM factor one finds relatively good agreement with the current 
experimental data. A notable exception is the mode $D^+ \to \bar{K}^0 
K^{*+}$ discussed below. A large annihilation contribution is required to 
reach the current experiment value. Such a sizable contribution appears 
to be ruled out by bounds from other processes.  

\subsubsection{On manifestations of New Physics} 
\label{NEWPHYS} 

Up to now we have focussed on two motivations for detailed studies of 
charm decays:  
\begin{itemize} 
\item  
They help prepare us to search for New Physics in  
$D^0-\bar D^0$ oscillations, the CP phenomenology in charm decays and in 
$B$ decays.  
\item  
They can shed a novel light on the formation of low-mass hadronic 
states, as described in Sect.\ref{DALITZ}.  
\end{itemize} 
However one can raise the question whether exclusive $D$ decays can by 
themselves reveal the intervention of New Physics. The rarer the  
mode, the better the a priori chance for such an effect. Doubly Cabibbo  
suppressed channels would offer the best chance, once Cabibbo  
suppressed ones the next best one. Of course we have to be cognizant of 
our limited  theoretical control over hadronization and not jump to  
conclusion.

This debate has recently been joined by two groups  
\cite{LIPKINCLOSE,LUSIGNOLIRECENT}. Close and Lipkin start by pointing  
at the "anomalously high branching ratios" for two Cabibbo suppressed  
transitions  
\bea  
{\rm BR}(D^+ \to K^*(892)^+\bar K^0) &=& 3.2 \pm 1.5 \%  
\nonumber  
\\  
{\rm BR}(D^+ \to K^*(892)^+\bar K^*(892)^0) &=& 2.6 \pm 1.1 \% \; ,  
\label{LICL1} 
\eea 
which are amazingly similar to the corresponding Cabibbo allowed  
branching ratios:  
\bea  
{\rm BR}(D^+ \to \rho^+\bar K^0) &=& 6.6 \pm 2.5 \%  
\nonumber  
\\  
{\rm BR}(D^+ \to \rho^+\bar K^*(892)^0) &=& 2.1 \pm 1.3 \% \; .  
\label{LICL2} 
\eea 
In $D^0$ decays on the other hand the expected Cabibbo hierarchy is  
apparent; e.g.,   
\bea  
{\rm BR}(D^0 \to K^*(892)^+ K^-) &=& 0.35 \pm 0.08 \%  
\nonumber  
\\  
{\rm BR}(D^0 \to \rho^+ K^-) &=& 10.8 \pm 0.9 \% \; .  
\label{LICL3} 
\eea 

A note of skepticism is appropriate here: while the central values in  
Eq.(\ref{LICL1}) are certainly large, the size of the error bars  
does not allow a firm conclusion. But it is intriguing to follow the  
authors of Ref.\cite{LIPKINCLOSE} and speculate `what if' the error bars  
were to shrink substantially, yet the central value to stay basically the  
same. It can serve as a case study for how such arguments go.

Branching ratios also are not the best yardstick here. For it  
was pointed out more than twenty years ago  
\cite{COLETTE} that $D^+$ modes generated by $c \to s \bar su$  
-- like $D^+ \to \bar K^0 K^+$ -- are {\em less} than Cabibbo suppressed 
relative to the corresponding $c \to s \bar du$ modes  
-- like $D^+ \to \bar K^0 \pi^+$ --  
since the former in contrast to the  latter (and to $c \to d 
\bar du$ modes like $D^+ \to \bar \pi^0 \pi^+$) do {\em not} 
experience  destructive PI. With such PI the dominant mechanism driving  
$\tau (D^+)/\tau (D^0) \simeq 2.5$ the branching branching {\em ratios} 
for the  channels in Eq.(\ref{LICL1}) are enhanced by a factor $\sim 2.5$  
due to `known' physics. To have a fairer comparison of $D^+$ and $D^0$  
branching ratios, one should `recalibrate'  
$D^+ \to \bar K^{(*)}K^{(*)}$ branching ratios {\em downward} by a  
factor 2.5, which puts it around the 1\% level for the two modes in  
Eq.(\ref{LICL1}). Even then they are still larger than the branching 
ratios for the corresponding $D^0$ modes and also for some similar  
$D^+$ modes, namely  
\beq  
{\rm BR}(D^+ \to \bar K^0 K^+) =  0.58 \pm 0.06 \%  
\eeq   
The latter is indeed considerably larger than  
BR$(D^+ \to \bar \pi^0 \pi^+) = 0.25 \pm 0.07$, yet not by as much as one  
would expect based on comparing quark-level diagrams. The fact that some  
{\em individual} branching ratios fall outside the general pattern 
should be seen as prima facie evidence for the intervention of FSI in 
the form of resonance effects. It is hard to see how truly New Physics 
could effect $D^0$ and $D^+$ decays very differently, let alone 
individual $D^+$ modes on the same Cabibbo level.

If the central values of the branching ratios in Eq.(\ref{LICL1})  
were substantially confirmed, the by far most likely explanation would  
be to `blame' it on somewhat accidentally strong FSI. The authors of  
Ref.\cite{LUSIGNOLIRECENT} arrive at the same conclusion based on their  
detailed and quite successful model for nonleptonic $D$ decays. The  
reader can be forgiven if she feels that theorists tend to call on  
FSI as a very convenient `deus ex machina' to bail them out of trouble.  
Yet the burden of the proof has to rest on the shoulders of those 
arguing in favour of New Physics. It is a worthwhile effort, though, to  
probe for violations of G parity in $D$ decays as suggested in  
Ref.\cite{LIPKINCLOSE}, more to further our education on QCD's dynamics  
than to establish New Physics. 

\subsection{Light-flavour spectroscopy from charm hadronic decays} 
\label{LIGHTFLAVOUR} 

Charm decays can be utilized also as a novel probe of {\em light}-flavour  
spectroscopy\index{spectroscopy}. We have already touched upon this point 
in our discussion of semileptonic decays. The situation is both richer and 
more complex in nonleptonic decays requiring special tools.

\subsubsection{Dalitz plot techniques} 
\label{DALITZ} 

Dalitz plots were invented in 1953 \cite{DALITZPAPERS} for treating the 
full complexity of decays into three body final states, specifically 
$K^+ \to \pi^+\pi^- \pi^+$. The main goal originally was to infer 
the spin and parity of the decaying particle from its decay products. 
Dalitz' work constituted an essential step in resolving the 
`$\tau -\theta$' puzzle, i.e. that $\tau^+ \to \pi^+\pi^-\pi^+$ and 
$\theta^+ \to \pi^+\pi^0$ represented decays of the same particle, 
namely the $K^+$ -- at the `price' that parity was violated in 
weak decays \cite{DALITZMEM}. Dalitz showed that three body 
final states could conveniently 
be described through a two-dimensional scatter plot 
\beq
 d\Gamma (P \to d_1d_2d_3)\propto 
 \frac{1}{M_P^3} |{\cal M}|^2  ds_{12} ds_{23} \; , \; 
 s_{ij} \equiv (p_{d_i} + p_{d_j})^2 \; .  
\eeq 
The density
of the Dalitz plot will be constant if and only if the matrix element 
${\cal M}$ is constant; any variation in ${\cal M}$ reveals a {\em dynamical} 
rather than kinematical effect. 

Charm decays proceed in an environment shaped by many resonances and 
virulent FSI. 
Our previous discussion has already benefitted handsomely from Dalitz 
plot studies when  
listing branching ratios for $D\to PV$ channels like $D \to \bar K^*\pi$,  
$\bar K\rho$ etc. and comparing them with theoretical predictions. Yet 
the benefits -- real and potential -- extend further still. For  
better appreciation we sketch the relevant analysis tools. 
Two main approaches are used in charm physics to formalize the Dalitz plot fit function
 parametrization, namely the {\em isobar model}\index{isobar model} and the 
 {\em K-matrix model}\index{K-matrix model}. Both have 
 been strongly criticized by theorists, as we shall discuss later. 
\par
 In the {\em isobar model} the overall amplitude is parametrized via 
 a coherent sum of Breit-Wigner amplitudes for modelling the interfering
 resonances, each with amplitude ${\cal A}^{BW}$, and of a constant amplitude
 $\cal  A_{NR}$ for the nonresonant amplitude
\beq
 {\cal A} = a_{NR}e^{i\delta_{NR}} + \sum^n_{j=1}a_je^{i\delta_j}{\cal A}_j^{BW}
  \label{EQ:DPISO}
\eeq
 where the $a$ parameters give the various relative contributions, and the
 phases $\delta$ account for FSI.
 Measurements of the
 phase shifts between different resonant components allow one to gauge the role
 of FSI and thus to shed some light onto the underlying weak decay dynamics.
At tree level, the weak amplitudes are treated as real; 
phase
 shifts in the decay are due to FSI.
 In this picture, the decay to quasi-two body final states is viewed as an 
 s-channel process (Fig.~\ref{FIG:DPCART}), whose propagator is represented by
 the complex amplitude Eq.~(\ref{EQ:DPISO}). 
 The explicit appearance of resonant amplitudes, as well as details of fit
 strategies,  vary with the experimental
 group. The E791 collaboration (see, as instance, \cite{Aitala:2002kr})
 uses a combination of  Breit-Wigner functions, 
 complemented by momentum-dependent form factors which reflect the non-pointlike
 nature of the D meson and of the resonance, and an angular momentum term.
 An additional complication arises when the j-th resonance
 is kinematically allowed to
 decay to different channels. This is the typical case of $f_0$ which can decay
 to 
 both $\pi\pi$ and $KK$. In this case one uses a coupled-channel Breit-Wigner
 (Flatte' formula)
 \cite{WA76}. 
\par
  The isobar model is the tool normally used for extracting information from
  the Daliz plots, and a wealth of updated reviews can be found 
  in the literature
\cite{Wiss:1997kb,reisfs02,mirandavietri02}.

  The use of isobar models for Dalitz plots in charm decays has, however, been subjected
to multiple criticisms: doubts have been cast whether overlapping broad resonances can 
be represented correctly (which is closely connected  to the issue of formulating the unitarity 
constraint in three and four-body final states), on how to relate the observations
from charm to those from scattering, to formulate a coupled-channel treatments, etc.
\par
 Recently the use of {\em K-matrix}-inspired approaches has been advocated, in close
 connection 
 to the experimental puzzle of the observation of the scalar $\sigma$ resonance 
 \index{$\sigma$ resonance in charm decays}
 in $D_s\to \pi \pi  \pi$ (for an experimentalist's point of view see
 \cite{reisfs02,mirandavietri02,Malvezzi:2003jp}).
\par
\index{K matrix vs. isobar formalism in Dalitz plots}
\index{isobar vs. K matrix  formalism in Dalitz plots}
 The K-matrix is a representation of the scattering matrix $S$, where
 the resonances are defined as poles of $S$. 
 The K-matrix, originally developed in the context of scattering problems
 can be extended to cover the case of more complex resonance formation through
 the P-vector approach \cite{PVECTOR}. The propagator of Fig.\ref{FIG:DPCART} is written 
 $  (I-iK\cdot\rho)^{-1}$
 where $K$ is the matrix for scattering of particles 1 and 2, $I$ the identity
 matrix, and $\rho$ the phase space matrix.
 Consequently, the 
 amplitude (Eq.~\ref{EQ:DPISO}) is written as a coherent sum of a nonresonant
 term, Breit-Wigner terms for narrow, well isolated resonances, and K-matrix
 terms for broad overlapping resonances 
\beq
 {\cal A} = a_{NR}e^{i\delta_{NR}} + \sum^n_{j=1}a_je^{i\delta_j}{\cal A}_j^{BW}
  + \sum^m_{k=n+1}a_ke^{i\delta_k}{\cal A}_k^K
  \label{EQ:DPKMA}
\eeq 
 Experimental information on poles is taken from $\pi\pi$ scattering data, and
 this allows one to treat coupled-channel decays such as $f_0\to \pi\pi, KK$
straightforwardly. 

 \subsubsection{Results from Dalitz analyses}
 \label{DALITZRESULTS}
  {\em The recent experimental studies of charm decays have opened up
a new experimental window for 
understanding light meson spectroscopy and especially the controversial
scalar mesons, which are copiously produced in these decays} 
Ref.~\cite{Close:2002zu}. 
This statement may be regarded as overly emphatic and excessively
optimistic, but the plain truth is that the study of charm mesons hadronic
decays via Dalitz plot formalisms is a field hosting  a rich 
`ecosystem' of diverse concepts involving heavy and light quarks, gluonia, 
hybrids or hermaphrodites  etc. enlivened by seemingly endless debates 
on the proper formalism to treat the data.
\par
For the light quark {\it aficionado},  charm decays have unique features that in
principle make them ideally 
 suited for light quark spectroscopy, i.e., a J=0, well defined D meson
 initial state, small nonresonant component, small background, large coupling to
 scalars, and independence from isospin and parity conservation
 \cite{reisfs02,mirandavietri02}.
For the charm quark zealot, the study of nonleptonic quasi-two body decays via
Dalitz plots is an essential tool to study the extent of final-state interaction
effects. This is accomplished by means of the study of the interference among
amplitudes which describe conflicting resonant channels. 
Furthermore finding a state in 
vastly different environments like charm decays and low energy hadronic  
collisions or photoproduction with consistent values for its mass and  
width strengthens considerably its claim for being a genuine  
resonance rather than a `mere' threshold enhancement. 
\par 
 \begin{figure}[t] 
  \centering 
   \includegraphics[width=3.0cm,height=3.0cm]{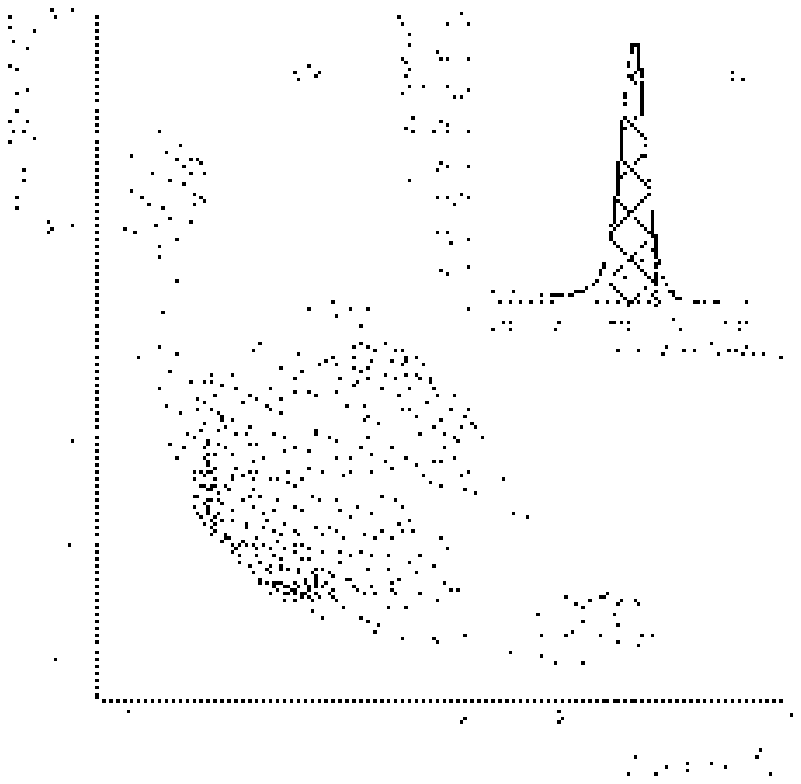} 
   \includegraphics[width=3.0cm,height=3.0cm]{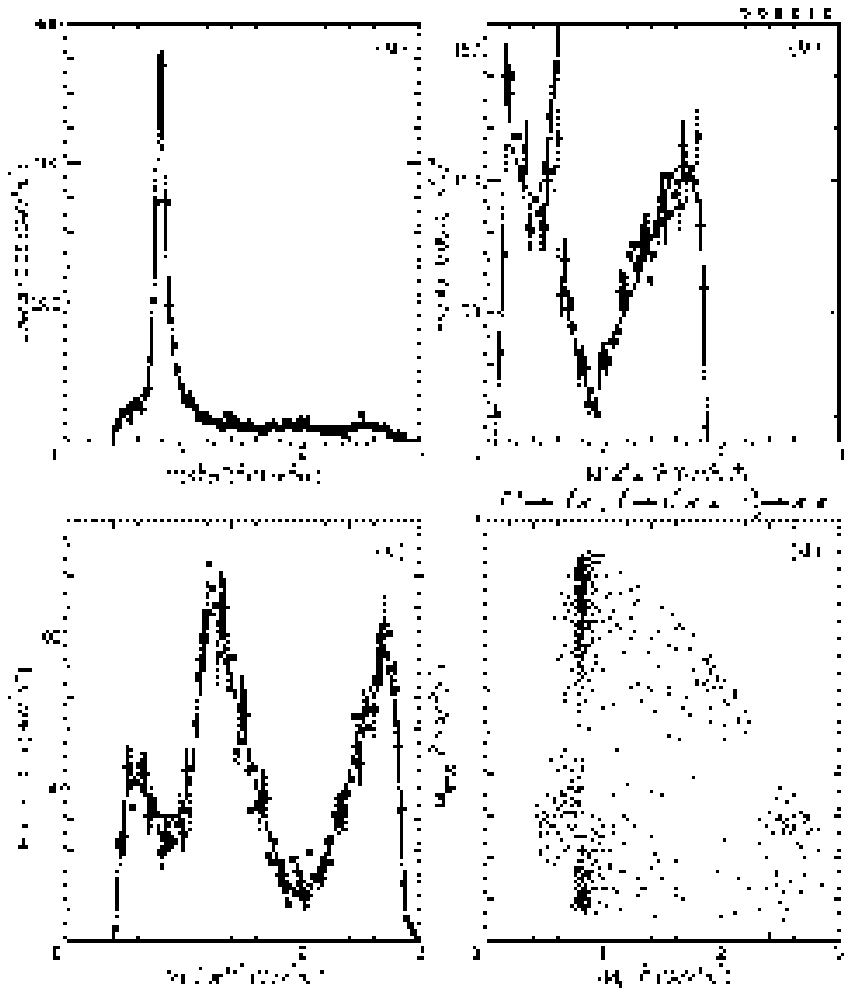}
   \includegraphics[width=3.0cm,height=3.0cm]{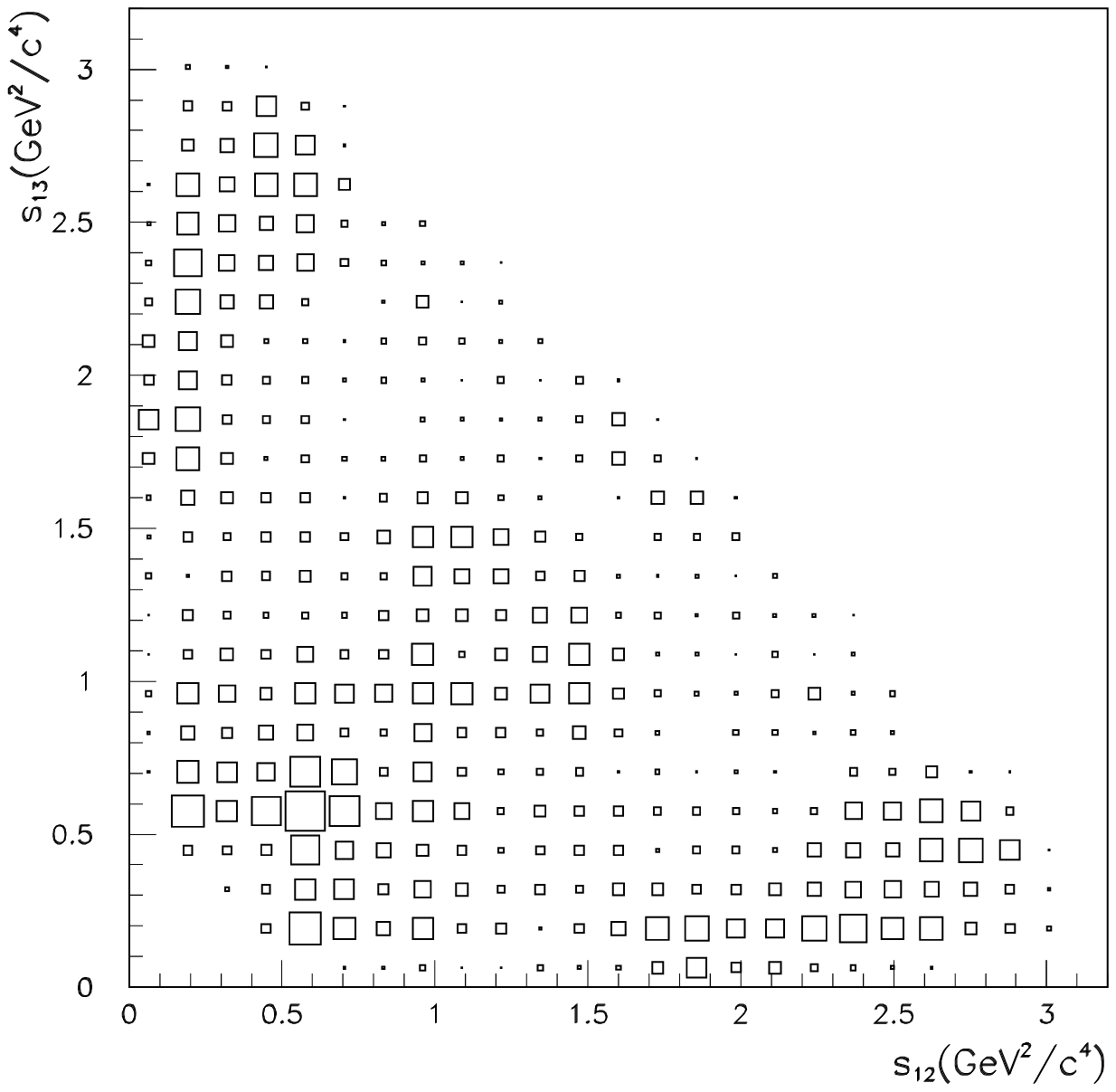}
   \includegraphics[width=3.0cm,height=3.0cm]{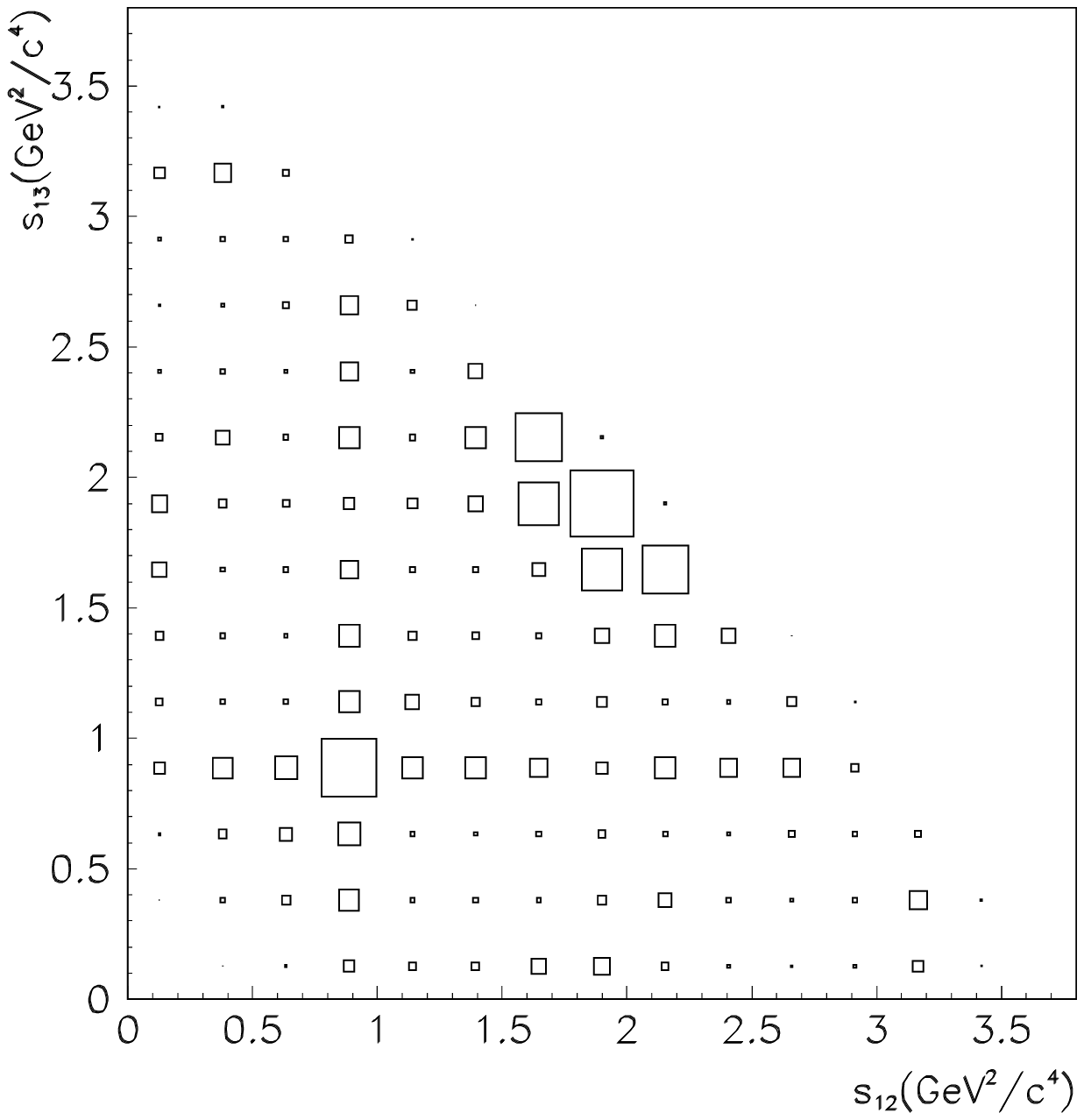}
   \includegraphics[width=3.0cm,height=3.0cm]{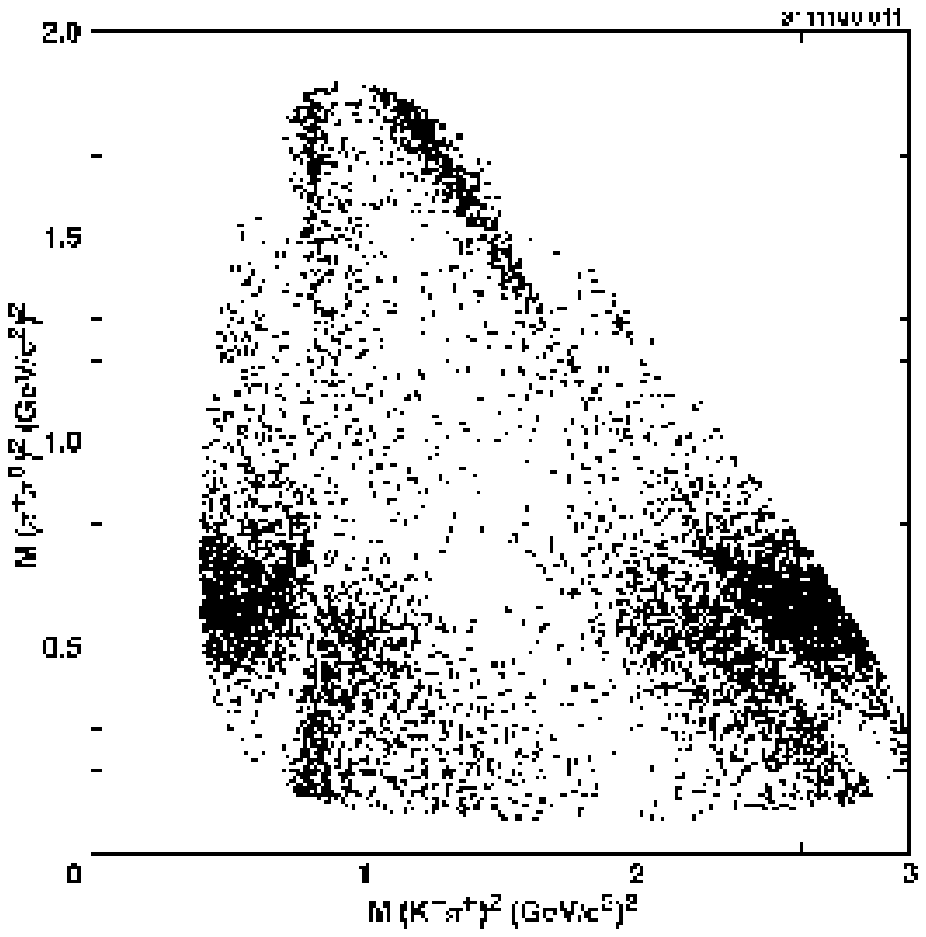}
   \includegraphics[width=3.0cm,height=3.0cm]{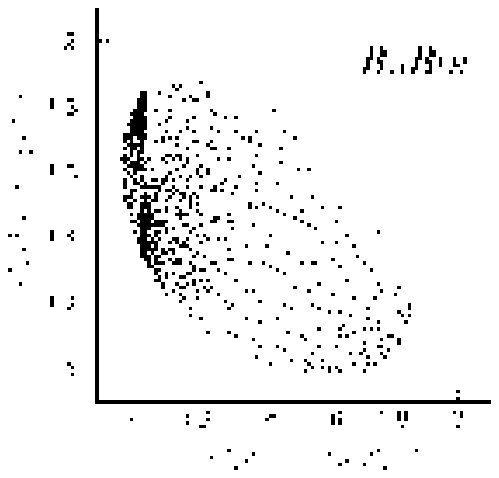} 
   \includegraphics[width=3.0cm,height=3.0cm]{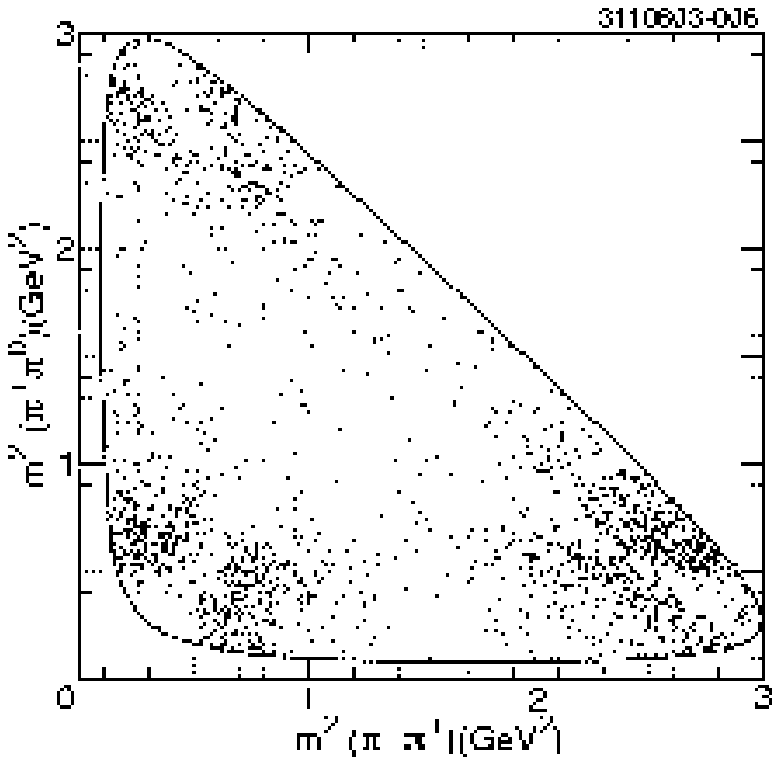}      
     
 \caption{\it 
 A selection of Dalitz plots of  charm meson decays:
   (a) E791 $D^+   \to K^- \pi^+\pi^+$      \cite{Aitala:2002kr}  ;  
   (b) CLEO $D^0   \to K^0_s \pi^+ \pi^-$   \cite{Muramatsu:2002jp}   ;
   (c) E791 $D^+   \to \pi^- \pi^+\pi^+$    \cite{Aitala:2000xu} ;
   (d) E791 $D^+_s \to \pi^- \pi^+\pi^+$    \cite{Aitala:2000xt} ;
   (e) CLEO $D^0   \to K^-  \pi^+\pi^0$     \cite{Kopp:2000gv};
   (f) BABAR  $D^0 \to K^0_s  K^+K^-$       \cite{Aubert:2002yc}; 
   (g) CLEO $D^0\to \pi^-\pi^+\pi^0  $      \cite{Frolov:2003jf}.
    \label{FIG:DPCOMP} } 
\end{figure} 
\par 
 \begin{figure}[t] 
  \centering 
   \includegraphics[width=5.0cm,height=3.0cm]{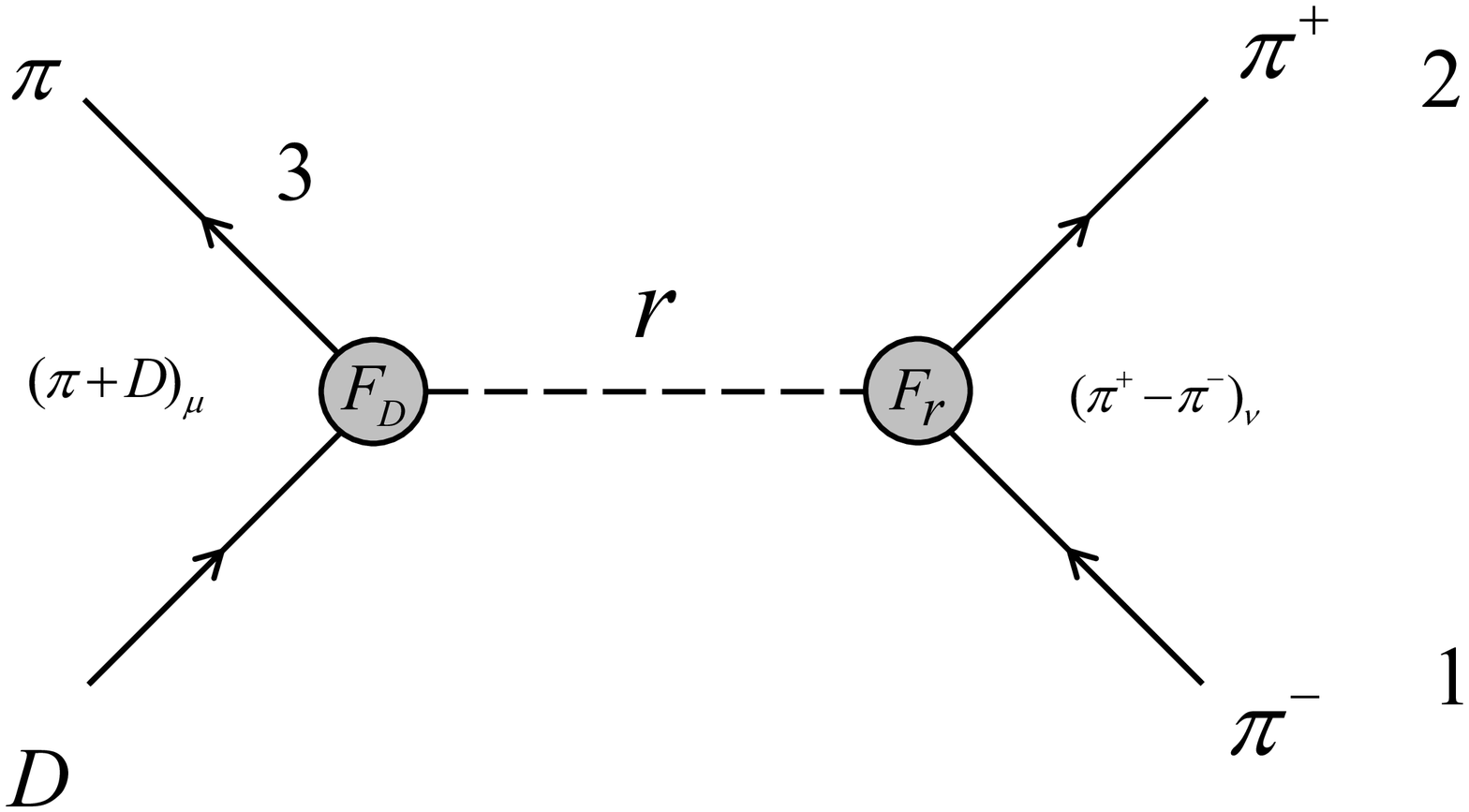} 
 \caption{\it 
 Charm meson decay represented via an s-channel process via resonance propagator
 $r$.
    \label{FIG:DPCART} } 
\end{figure} 
\par
 The data scenario on Dalitz 
 plots analyses is quite rich, with contributions
 from all high-resolution, high-statistics 
 charm experiment both at fixed target
 (E687, E791, FOCUS) and at low-energy 
 $\epem$ colliders (CLEO, BABAR, BELLE).
 We shall outline the main features of the 
 experimental scenario for $K\pi\pi$,
 $KK\pi$ and $\pi\pi\pi$, reporting a 
 summary of  measurements in Tables
 \ref{TAB:DPAMPHD1},\ref{TAB:DPAMPHD2},\ref{TAB:DPAMPHDS}.
\par 
\begin{table}
\begin{center} 
\begin{tabular}{|l|r|r|r|}
\hline
 Decay Mode   &  Fit Fraction \% & Amplitude & Phase ~$^o$  \\ 
\hline
 \multicolumn{4}{|c|}{$D^+\to K^-\pi^+\pi^+$ E791 2002 \cite{Aitala:2002kr}} \\
 $\bar{K}^*(892)^0\pi^+$                &
     $12.3\pm 1.3$             &
     $1.0$ fixed                     &
     $0$ fixed                               \\
 $NR$                   &
     $13.0\pm 7.3 $             &
     $1.03\pm 0.34 $                     &
     $-11 \pm 16  $                               \\
 $\kappa\pi^+  $                   &
     $47.8\pm 13.2$             &
     $1.97\pm 0.37 $                     &
     $187 \pm  20 $                               \\
 $\bar{K}_0^*(1430)^0\pi^+  $                   &
     $12.5\pm 1.5 $             &
     $1.01\pm 0.13 $                     &
     $48  \pm 12  $                               \\   
 $ \bar{K}_2^*(1430)^0\pi^+ $                   &
     $0.5 \pm 0.2 $             &
     $0.20\pm 0.06$                     &
     $-54 \pm  11 $                               \\
 $ \bar{K}^*(1680)^0\pi^+ $                   &
     $2.5 \pm 0.8$             &
     $0.45\pm 0.16 $                     &
     $ 28 \pm 0.20$                               \\   
 \multicolumn{4}{|c|}{$D^0\to K_s^0\pi^+\pi^-$ E687 1992 \cite{Frabetti:1992we}}  \\
 $K^{*-}\pi^+  $                   &
     $64\pm 9 $             &
     $1.0$ fixed                      &
     $0$ fixed                                 \\
 $NR  $                   &
     $26\pm  9$             &
     $0.41\pm 0.09 $                     &
     $-252\pm 34  $                               \\
 $\bar{K}^0\rho^0  $                   &
     $20 \pm 7$             &
     $0.39\pm 0.07 $                     &
     $-275 \pm 57 $                               \\ 
 \multicolumn{4}{|c|}{$ D^0\to K_s^0\pi^+\pi^-$ E687 1994 \cite{Frabetti:1994di}}  \\                
 $ K^{*-}\pi^+ $                   &
     $62.5\pm 4.1 $             &
     $ $                     &
     $ 0 $ fixed                               \\ 
 $K_0^*(1430)^-\pi^+  $                   &
     $10.9\pm 3.9 $             &
     $ $                     &
     $ -166\pm 11 $                               \\
 $\bar K^0 \rho^0 $                   &
     $35.0\pm 7.3 $             &
     $ $                     &
     $ -136\pm 6 $                               \\
 $\bar K^0 f_0(975)  $                   &
     $6.8\pm 2.3 $             &
     $ $                     &
     $38\pm 11  $                               \\
 $\bar K^0 f_2(1270)  $                   &
     $3.7\pm 2.2 $             &
     $ $                     &
     $ -174\pm 23 $                               \\ 
 $ \bar K^0 f_0(1400) $                   &
     $7.7\pm 3.6 $             &
     $ $                     &
     $ -45\pm 24 $                               \\
 \multicolumn{4}{|c|}{$ D^+\to K^-\pi^+\pi^+$ E687 1994 \cite{Frabetti:1994di}}  \\                
 $\bar K^{*0}\pi^+  $                   &
     $13.7\pm 1.0 $             &
     $ $                     &
     $ 48\pm 2 $                               \\ 
 $ \bar K^*_0(1430)^0\pi^+ $                   &
     $28.4\pm 3.9 $             &
     $ $                     &
     $ 63\pm 4 $                               \\
 $ \bar K^* (1680)^0 \pi^+ $                   &
     $4.7\pm 0.6 $             &
     $ $                     &
     $ 73\pm 17 $                               \\
 $ NR $                   &
     $99.8 \pm 5.9 $             &
     $ $                     &
     $ 0 $ fixed                               \\
 \multicolumn{4}{|c|}{$ D^0\to K^-\pi^+\pi^0$ E687 1994 \cite{Frabetti:1994di}}  \\                
 $ K^-\rho^+ $                   &
     $76.5\pm 4.6 $             &
     $ $                     &
     $0  $ fixed                               \\ 
 $ K^{*-}\pi^+ $                   &
     $14.8\pm 5.6 $             &
     $ $                     &
     $ 162\pm 12 $                               \\
 $ \bar K^{*0} \pi^0 $                   &
     $ 16.5\pm 3.3$             &
     $ $                     &
     $ -2 \pm 26 $                               \\
 $NR  $                   &
     $ 10.1\pm 4.4$             &
     $ $                     &
     $ -122\pm 23 $                               \\
 \multicolumn{4}{|c|}{$D^+\to K^-\pi^+\pi^+$ E691 1993 \cite{Anjos:1992kb}}  \\     
 $NR  $                   &
     $83.8 $             &
     $1.0$ fixed                      &
     $ 0$  fixed                                \\
 $ \bar{K}^*(892)^0\pi^+ $                   &
     $17.0\pm 0.9 $             &
     $0.78\pm 0.02 $                     &
     $-60\pm 3  $                               \\
 $\bar{K}_0^*(1430)^0\pi^+   $                   &
     $24.8\pm 1.9 $             &
     $0.53\pm 0.2 $                     &
     $132\pm 2  $                               \\ 
 $  \bar{K}^*(1680)^0\pi^+  $                   &
     $3.0\pm 0.4 $             &
     $0.47\pm 0.03 $                     &
     $-51\pm 4  $                               \\
 \multicolumn{4}{|c|}{$D^0\to K^-\pi^+\pi^0$ E691 1993 \cite{Anjos:1992kb} }  \\          
 $NR  $                   &
     $3.6 $             &
     $1.0$ fixed                     &
     $0$ fixed                            \\
 $  \bar{K}^*(892)^0\pi^0  $                   &
     $14.2\pm 1.8 $             &
     $3.19\pm 0.20 $                     &
     $167\pm 9  $                               \\
 $   \bar{K}^*(892)^-\pi^+$                   &
     $ 8.4\pm 1.1$             &
     $ 2.96\pm 0.19 $                     &
     $ -112\pm 9 $                               \\
 $ K^-\rho^+ $                   &
     $64.7\pm 3.9 $             &
     $8.56\pm 0.26 $                     &
     $40\pm 7  $                               \\
 \multicolumn{4}{|c|}{$D^0\to K^0_s\pi^+\pi^-$ E691 1993 \cite{Anjos:1992kb} }  \\          
 $NR  $                   &
     $26.3 $             &
     $1.0$ fixed                    &
     $0$ fixed                                \\ 
 $K^*(892)^-\pi^+  $                   &
     $48.0\pm 9.7 $             &
     $2.3\pm 0.23 $                     &
     $109\pm 9  $                               \\ 
 $\bar{K}^0\rho^0  $                   &
     $21.5\pm 5.1 $             &
     $1.59\pm 0.19 $                     &
     $-123\pm 12  $                               \\ 
 \multicolumn{4}{|c|}{$D^+\to \pi^-\pi^+\pi^+$ E791 2001 \cite{Aitala:2000xu} }  \\          
 $ \rho^0(770)\pi^+ $                   &
     $33.6\pm 3.9 $             &
     $1.0$ fixed                    &
     $0 $ fixed                               \\ 
 $\sigma\pi^+  $                   &
     $46.3\pm 9.2 $             &
     $1.17\pm 0.14 $                     &
     $205.7\pm 9.0  $                               \\ 
 $NR  $                   &
     $7.8\pm 6.6 $             &
     $0.48\pm 0.20 $                     &
     $57.3\pm 20.3  $                               \\ 
 $f_0(980)\pi^+  $                   &
     $6.2\pm 1.4 $             &
     $0.43\pm 0.05 $                     &
     $165\pm 11  $                               \\
 $f_2(1270)\pi^+  $                   &
     $19.4\pm 2.5 $             &
     $0.76\pm 0.07$                     &
     $57.3\pm 8.0  $                               \\
 $f_0(1370)\pi^+  $                   &
     $2.3\pm 1.7 $             &
     $0.26\pm 0.09 $                     &
     $105.4\pm 17.8  $                               \\
 $\rho^0(1450) \pi^+  $                   &
     $0.7\pm 0.8 $             &
     $0.14\pm 0.07 $                     &
     $319.1\pm 40.5 $                               \\
  \multicolumn{4}{|c|}{$D^+\to K^-K^+\pi^+$ E687 1995 \cite{Frabetti:1995sg}}\\     
 $ \bar K^*(892)^0 K^+ $                   &
     $30.1\pm 3.2 $             &
     $ $                     &
     $ 0  $   fixed                            \\
 $ \phi \pi^+ $                   &
     $ 29.2\pm 4.3$             &
     $ $                     &
     $ -159\pm 14 $                               \\
 $ \bar K_0^*(1430)^0 K^+ $                   &
     $37.0\pm 3.9 $             &
     $ $                     &
     $70\pm 8  $                               \\ 
     \hline 
\end{tabular} 
\end{center} 
\caption{Dalitz plot coherent analyses results on $K\pi\pi$ and 
$\pi\pi\pi$ decays of $D$. Errors
are summed in quadrature.} 
\label{TAB:DPAMPHD1} 
\end{table}  
\begin{table}
\begin{center} 
\begin{tabular}{|l|r|r|r|}
\hline
 Decay Mode   &  Fit Fraction \% & Amplitude & Phase ~$^o$  \\ 
\hline       
 \multicolumn{4}{|c|}{$D^+\to \pi^-\pi^+\pi^+$ E687 1997 \cite{Frabetti:1997sx}}
  \\       
 $NR  $                   &
     $58.9\pm 13.3 $             &
     $1$   fixed                  &
     $0 $ fixed                               \\
 $\rho^0(770)\pi^+  $                   &
     $28.9\pm 8.0$             &
     $0.70\pm 0.11 $                     &
     $27\pm 18  $                               \\
 $f_0(980)\pi^+  $                   &
     $2.7\pm 4.9$             &
     $0.22\pm 0.13 $                     &
     $197\pm 37  $                               \\
 $f_2(1270)\pi^+  $                   &
     $5.2\pm 4.9 $             &
     $0.30\pm 0.11 $                     &
     $207\pm 17  $                               \\
  \multicolumn{4}{|c|}{$D^0\to \bar K^0\pi^+\pi^-$ CLEO 2002 
  \cite{Muramatsu:2002jp} } \\
 $K^*(892)^+\pi^-  $                   &
     $0.34\pm 0.22 $             &
     $(11\pm 4)\, 10^{-2} $                     &
     $321\pm 14  $                               \\
 ${\bar K}^0 \rho^0  $                   &
     $26.4\pm 1.3 $             &
     $1.0 $      fixed               &
     $0  $ fixed                                \\
 $ {\bar K}^0 \omega $                   &
     $0.72\pm 0.20 $             &
     $(37\pm 7) \,10^{-3} $                     &
     $114\pm 9  $                               \\            
 $K^*(892)^-\pi^+  $                   &
     $65.7\pm 3.1 $             &
     $1.56\pm 0.11 $                     &
     $150\pm 4  $                               \\  
 $ {\bar K}^0 f_0(980) $                   &
     $4.3\pm 1.0 $             &
     $0.34 \pm 0.05 $                     &
     $188\pm 9  $                               \\  
 $ {\bar K}^0 f_2(1270)  $                   &
     $0.27\pm 0.29 $             &
     $0.7\pm 0.5 $                     &
     $308\pm 42  $                               \\
 $ {\bar K}^0 f_0(1370)  $                   &
     $ 9.9\pm 3.1$             &
     $ 1.8\pm 0.2$                     &
     $ 85\pm 20 $                               \\
 $ K_0^*(1430)^-\pi^+ $                   &
     $7.3\pm 2.2 $             &
     $2.0\pm 0.3 $                     &
     $3\pm 11  $                               \\     
 $ K_2^* (1430)^-\pi^+$                   &
     $1.1\pm 0.5 $             &
     $1.0\pm 0.24 $                     &
     $155\pm 17  $                               \\
 $ K^*(1680)^-\pi^+ $                   &
     $ 2.2\pm 1.6$             &
     $ 5.6\pm 4$                     &
     $ 174\pm 17 $                               \\
 $ NR $                   &
     $0.9\pm 1.2 $             &
     $1.1\pm 0.9 $                     &
     $160\pm 57  $                               \\               
  \multicolumn{4}{|c|}{$D^0\to K^-\pi^+\pi^0$ MARK III
  1987 \cite{Adler:1987sd}}  \\ 
 $K^-\rho^+  $                   &
     $ 81\pm 7$             &
     $ $                     &
     $ 0 $ fixed                               \\
 $K^{*-}\pi^+  $                   &
     $12\pm 4 $             &
     $ $                     &
     $ 154\pm 11 $                               \\
 $\bar K^{*0}pi^0  $                   &
     $13\pm 4 $             &
     $ $                     &
     $7\pm 7  $                               \\           
 $NR  $                   &
     $9\pm 4 $             &
     $ $                     &
     $52\pm 9  $                               \\
   \multicolumn{4}{|c|}{$D^0\to \bar K^0\pi^+\pi^-$ MARK III
  1987 \cite{Adler:1987sd} }\\  
 $\bar K^0 \rho^0  $                   &
     $12\pm 7 $             &
     $ $                     &
     $93\pm 30  $                               \\
 $K^{*-}\pi^+  $                   &
     $56\pm 6 $             &
     $ $                     &
     $0  $ fixed                               \\           
 $NR  $                   &
     $33\pm 11 $             &
     $ $                     &
     $--$                               \\
   \multicolumn{4}{|c|}{$D^+\to \bar K^0\pi^+\pi^0$ MARK III
  1987 \cite{Adler:1987sd} }  \\       
 $\bar K^0 \rho^+  $                   &
     $68\pm 14 $             &
     $ $                     &
     $ 0 $ fixed                               \\
 $ \bar K^{*0}\pi^+ $                   &
     $19\pm 8 $             &
     $ $                     &
     $43\pm 23  $                               \\  
 $NR  $                   &
     $13\pm 11 $             &
     $ $                     &
     $250\pm 19  $                               \\
   \multicolumn{4}{|c|}{$D^+\to \bar K^-\pi^+\pi^+$ MARK III
  1987 \cite{Adler:1987sd}   }  \\      
 $ \bar K^{*0}\pi^+ $                   &
     $13\pm 7 $             &
     $ $                     &
     $ 105\pm 8 $                               \\           
 $ NR $                   &
     $79\pm 16 $             &
     $ $                     &
     $ 0.0 $                               \\             
    \multicolumn{4}{|c|}{$D^0\to  K^0K^-\pi^+$ BABAR 2002 \cite{Aubert:2002yc}}  \\
 $ {\bar K}_0^*(1430)^0 K^0 $                   &
     $4.8\pm 2.1 $             &
     $ $                     &
     $52\pm 27  $                               \\
 $ {\bar K}_1^*(892)^0 K^0 $                   &
     $0.8\pm 0.5 $             &
     $ $                     &
     $175\pm 22  $                               \\
 $ {\bar K}_1^*(1680)^0 K^0 $                   &
     $6.9\pm 1.6 $             &
     $ $                     &
     $-169\pm 16  $                               \\
 $ {\bar K}_2^*(1430)^0 K^0 $                   &
     $2.0\pm 0.6 $             &
     $ $                     &
     $ 51\pm 18 $                               \\
 $  K_0^*(1430)^+ K^- $                   &
     $13.3\pm 5.2 $             &
     $ $                     &
     $ -41\pm 25 $                               \\
 $  K_1^*(892)^+ K^-  $                   &
     $63.6\pm 5.7 $             &
     $ $                     &
     $0   $ fixed                               \\
 $ K_1^*(1680)^+ K^- $                   &
     $15.6\pm 3.3 $             &
     $ $                     &
     $-178\pm 10  $                               \\
 $ K_2^*(1430)^+ K^-$                   &
     $13.8\pm 8.3 $             &
     $ $                     &
     $-52\pm 7  $                               \\
 $ a_0(980)^- \pi^+ $                   &
     $2.9\pm 2.4 $             &
     $ $                     &
     $ -100\pm 13 $                               \\
 $  a_0(1450)^- \pi^+ $                   &
     $ 3.1\pm 2.1$             &
     $ $                     &
     $ 31\pm 16 $                               \\
 $ a_2(1310)^- \pi^+  $                   &
     $ 0.7\pm 0.4$             &
     $ $                     &
     $ -149\pm 27 $                               \\
 $ NR $                   &
     $ 2.3\pm 5.6$             &
     $ $                     &
     $ -136\pm 23 $                               \\ 
\multicolumn{4}{|c|}{$D^0\to \pi^+\pi^-\pi^0$ CLEO 2003  \cite{Frolov:2003jf}
 }  \\     
 $ \rho^+ \pi^- $                   &
     $ 76.5\pm 5.1$             &
     $1 $  fixed                   &
     $ 0 $   fixed                            \\   
 $ \rho^0 \pi^0 $                   &
     $ 23.9\pm 5.0$             &
     $0.56\pm 0.07$                     &
     $ 10\pm 4 $                               \\ 
 $ \rho^- \pi^+ $                   &
     $ 32.3\pm 3.0$             &
     $0.65 \pm 0.05$                     &
     $ -4\pm 5 $                               \\ 
 NR                  &
     $ 2.7\pm 1.9$             &
     $1.03\pm 0.35$                     &
     $ 77\pm 14 $                               \\                                                                                                                                                                                        
\hline 
\end{tabular} 
\end{center} 
\caption{Dalitz plot coherent analyses results on $K\pi\pi$, $K K\pi$ and 
$\pi\pi\pi$ decays of $D$. Errors
are summed in quadrature.} 
\label{TAB:DPAMPHD2} 
\end{table}  
\begin{table}
\begin{center} 
\begin{tabular}{|l|r|r|r|}
\hline
 Decay Mode   &  Fit Fraction \% & Amplitude & Phase ~$^o$  \\ 
\hline
    \multicolumn{4}{|c|}{$D^0\to  {\bar K}^0K^+\pi^-$ BABAR 2002 
    \cite{Aubert:2002yc}}  \\
 $ K_0^*(1430)^0\bar K^0 $                   &
     $26\pm 16.3 $             &
     $ $                     &
     $-38\pm 22  $                               \\
 $ K_1^*(892)^0\bar K^0 $                   &
     $2.8\pm 1.5 $             &
     $ $                     &
     $ -126 \pm 19 $                               \\
 $ K_1^*(1680)^0\bar K^0 $                   &
     $15.2 \pm 12.1 $             &
     $ $                     &
     $161\pm 9  $                               \\
 $ K_2^*(1430)^0\bar K^0 $                   &
     $ 1.7\pm 2.5$             &
     $ $                     &
     $ 53\pm 38 $                               \\
 $ K_0^*(1430)^- K^+ $                   &
     $2.4 \pm 8 $             &
     $ $                     &
     $ -142\pm 115 $                               \\
 $ K_1^*(892)^- K^+ $                   &
     $ 35.6\pm 8$             &
     $ $                     &
     $ 0 $ fixed                               \\
 $ K_1^*(1680)^- K^+ $                   &
     $ 5.1\pm 6$             &
     $ $                     &
     $ 124\pm 27 $                               \\
 $ K_2^*(1430)^- K^+ $                   &
     $ 1\pm 1$             &
     $ $                     &
     $ -26 \pm 38 $                               \\
 $ a_0^+(980) \pi^- $                   &
     $ 15.1\pm 12.1$             &
     $ $                     &
     $ -160\pm 42 $                               \\
 $ a_0^+(1450) \pi^- $                   &
     $ 2.2\pm 2.9$             &
     $ $                     &
     $148\pm 25  $                               \\
   NR                   &
     $ 37\pm 26$             &
     $ $                     &
     $ -172\pm 13 $                               \\                 
     \multicolumn{4}{|c|}{$D^0\to  {\bar K}^0K^+K^-$ 
     BABAR 2002 \cite{Aubert:2002yc}}  \\ 
 $ \bar K^0 \phi $                   &
     $ 45.4\pm 1.9$             &
     $ $                     &
     $ 0 $ fixed                               \\
 $ \bar K^0 a_0(980)^0 $                   &
     $61\pm 15 $             &
     $ $                     &
     $ 109\pm 15 $                               \\
 $  \bar K^0 f_0(980) $                   &
     $ 12.2\pm 9.1$             &
     $ $                     &
     $ -161\pm 14 $                               \\
 $  a_0(980)^+K^-$                   &
     $34.3\pm 7.5 $             &
     $ $                     &
     $ -53\pm 4 $                               \\
 $ a_0(980)^-K^+ $                   &
     $ 3.2\pm 3.9$             &
     $ $                     &
     $ -13\pm 15 $                               \\
   NR                 &
     $ 0.4\pm 0.8$             &
     $ $                     &
     $ 40\pm 44 $                               \\      
 \multicolumn{4}{|c|}{$D^+_s\to \pi^-\pi^+\pi^+$ E791 2000 \cite{Aitala:2000xt}}  \\          
 $ f_0(980)\pi^+ $                   &
     $54.1\pm 4.0 $             &
     $1.0  $ fixed                     &
     $0$ fixed                               \\ 
 $\rho^0(770)\pi^+  $                   &
     $11.1\pm 2.5$             &
     $0.45\pm 0.06 $                     &
     $81\pm 15 $                               \\ 
 $NR  $                   &
     $5.0\pm 3.8 $             &
     $0.30\pm 0.12 $                     &
     $149\pm 25  $                               \\ 
 $f_2(1270)\pi^+  $                   &
     $20.8\pm 3.0 $             &
     $0.62\pm 0.05 $                     &
     $124\pm 11  $                               \\
 $f_0(1370)\pi^+  $                   &
     $34.7\pm 7.2$             &
     $0.80\pm 0.11 $                     &
     $159\pm 14  $                               \\
 $\rho^0(1450) \pi^+  $                   &
     $0 $             &
     $ 0$                     &
     $ 0 $                               \\
 \multicolumn{4}{|c|}{$D^+_s\to \pi^-\pi^+\pi^+$ E687 1997 
 \cite{Frabetti:1997sx}}  \\      
 $f_0(980)\pi^+  $                   &
     $107.4\pm 14.6 $             &
     $1$ fixed                     &
     $0 $ fixed                               \\
 $NR  $                   &
     $12.1\pm 12.3$             &
      $0.34\pm 0.14 $                     &
     $235\pm 22 $                               \\
 $\rho^0(770)\pi^+  $                   &
     $2.3\pm 2.9 $             &
     $0.15\pm 0.09$                     &
     $53\pm 45 $                               \\
 $f_2(1270)\pi^+  $                   &
     $12.3\pm 5.9$             &
     $0.34\pm 0.09$                     &
     $100\pm 19 $                               \\
 $  S(1475) \pi^+$                   &
     $ 27.4\pm11.5$             &
     $0.50\pm 0.13 $                     &
     $234\pm 15  $                               \\          
  \multicolumn{4}{|c|}{$D^+_s\to K^-K^+\pi^+$ E687 1995 \cite{Frabetti:1995sg}
 }  \\ 
   $ \bar K^*(892)^0 K^+ $                   &
     $47.8\pm 6.1  $             &
     $ $                     &
     $ 0 $ fixed                               \\
 $ \phi \pi^+ $                   &
     $39.6 \pm 5.7 $             &
     $ $                     &
     $178\pm 31  $                               \\
 $ f_0(980) \pi^+ $                   &
     $ 11.0 \pm 4.4$             &
     $ $                     &
     $ 159\pm 27 $                               \\
 $ f_J(1710)\pi^+ $                   &
     $3.4\pm 4.2 $             &
     $ $                     &
     $110\pm 26  $                               \\ 
 $ \bar K_0^*(1430)^0 K^+ $                   &
     $ 9.3\pm 4.5$             &
     $ $                     &
     $ 152\pm 56 $                               \\         
\hline 
\end{tabular} 
\end{center} 
\caption{Dalitz plot coherent analyses 
results on  $KK\pi$ and $\pi\pi\pi$ decays of
$D_s$. Errors are summed in quadrature.} 
\label{TAB:DPAMPHDS} 
\end{table}  

 {\bf (i)} Among the major channels, the mode $D^+\to K^-\pi^+\pi^+$  
 is the only  with an important nonresonant component.
  The fit fraction was also shown to largely exceed 100\%, thus signalling very
  large constructive interference. A recent result from E791 claims that very
  much of the nonresonant fraction can be better represented by a broad scalar
  resonance, which they interpret as the $\kappa$.
 A possible link can be thought with the recent observation by FOCUS 
 \cite{Link:2002ev} of an
 anomalous interference effect in $D^+$ semileptonic decay which could be
 interpreted with the interference of a broad scalar (the $\kappa$ ?) with the
 $K^*(890)$. Recent measurement by CLEO \cite{Muramatsu:2002jp},
 however, fails in finding support for
 the $\kappa$ hypothesis in Dalitz plot. 
 \par 
 {\bf (ii)} The mode $D^0 \to K_S\phi$ has played an interesting role 
 in our efforts  
to identify the mechanisms driving $D$ decays. Its existence was 
predicted with a branching ratio of $\sim 0.1 - 0.5$~\% based on the  
assumption that WA creates most of the excess in the $D^0$ over the  
$D^+$ width \cite{FUKUGITA}. It was claimed that finding it would amount 
to a `smoking gun' evidence supporting this assumption. Afterwards it 
was indeed found -- with close to the predicted branching ratio:  
BR$(D^0 \to K_S\phi) = (0.47 \pm 0.06)$~\%. In retrospect, however, it 
should be viewed as evidence for the impact FSI can have. 
\par 
There is, though, some lesson of future relevance we can learn. The state 
$K_S\phi$ is an odd CP eigenstate; as such it provides intriguing ways  
to search for CP violation in $D^0 \to K_S\phi$ as explained later.  
The corresponding beauty mode $B_d \to K_S\phi$ is under active study 
for similar reasons at the $B$ factories. Yet when extracting   
$K_S\phi$ from $K_SK^+K^-$ final states one has to contend with the  
scalar $\bar KK$ resonance $f_0(980)$ close to the $\phi$ mass. This  
problem is aggravated by the fact that $K_Sf_0(980)$ has the {\em opposite}   
CP parity of $K_S\phi$. A CP asymmetry in $D^0 \to K_SK^+K^-$ coming  
from $D^0 \to K_S\phi$ would then be (partially) cancelled by one due  
to $D^0 \to K_Sf_0(980)$. Likewise for $B_d \to K_SK^+K^-$. 
\par
 {\bf (iii)} The $D^+\to KK\pi$  Dalitz plot shows the presence of various $K^*$
 states, with 
 the asymmetry between $\bar K^*$ `lobes' being interpreted \cite{Wiss:1997kb}
  as an interference of a broad S=0 resonance with the $\bar K^{*0}K^+$. The
  $D_s^+\to KK\pi$  Dalitz plot is strongly dominated by the $\phi(1020)$. 
 However we know that $D_s^+\to \pi\pi\pi$ is dominated by $f_0(980)$ 
 (see below), which
 also decays to $KK$. It is clear that the $\phi$ contribution overlaps
 with the $f_0(980)$  contributions, and the two should be disentangled. This
 has 
 far reaching consequences and implications, also in B physics, as we shall
 discuss later on in this section.
\par
{\bf (iv)} The issues involved here -- resonance vs. threshold enhancement,
Breit-Wigner vs.  
{\em non}-Breit-Wigner form of the excitation curve, isobar vs. K matrix model
\index{K matrix vs. isobar formalism in Dalitz plots}
\index{isobar vs. K matrix  formalism in Dalitz plots}
--  
 \index{$\sigma$ resonance in charm decays}
are being  debated with particular passion in connection with the role of 
the $\sigma$ resonance in $D^+, D_s^+\to 3\pi$ modes. 
\par
The $\sigma$ has 
a checkered past: after being   
introduced as an s-wave scalar resonance in $\pi\pi$ scattering its mass 
and width -- even its name -- have 
changed over time. It has been never {\em conclusively} observed in $\pi\pi$
scattering, 
 and its relevance is connected to the role of Higgs-like particle in
 Nambu--Jona-Lasinio models. Evidence for the $\sigma$'s existence was recently
obtained by BES in $\jp\to \omega\pi$ data. 
Finding it in $D_s \to 3\pi$ would provide a 
nontrivial boost to its resonance status. 
\par
\begin{figure}[t] 
  \centering 
   \includegraphics[width=5.0cm,height=3.0cm]{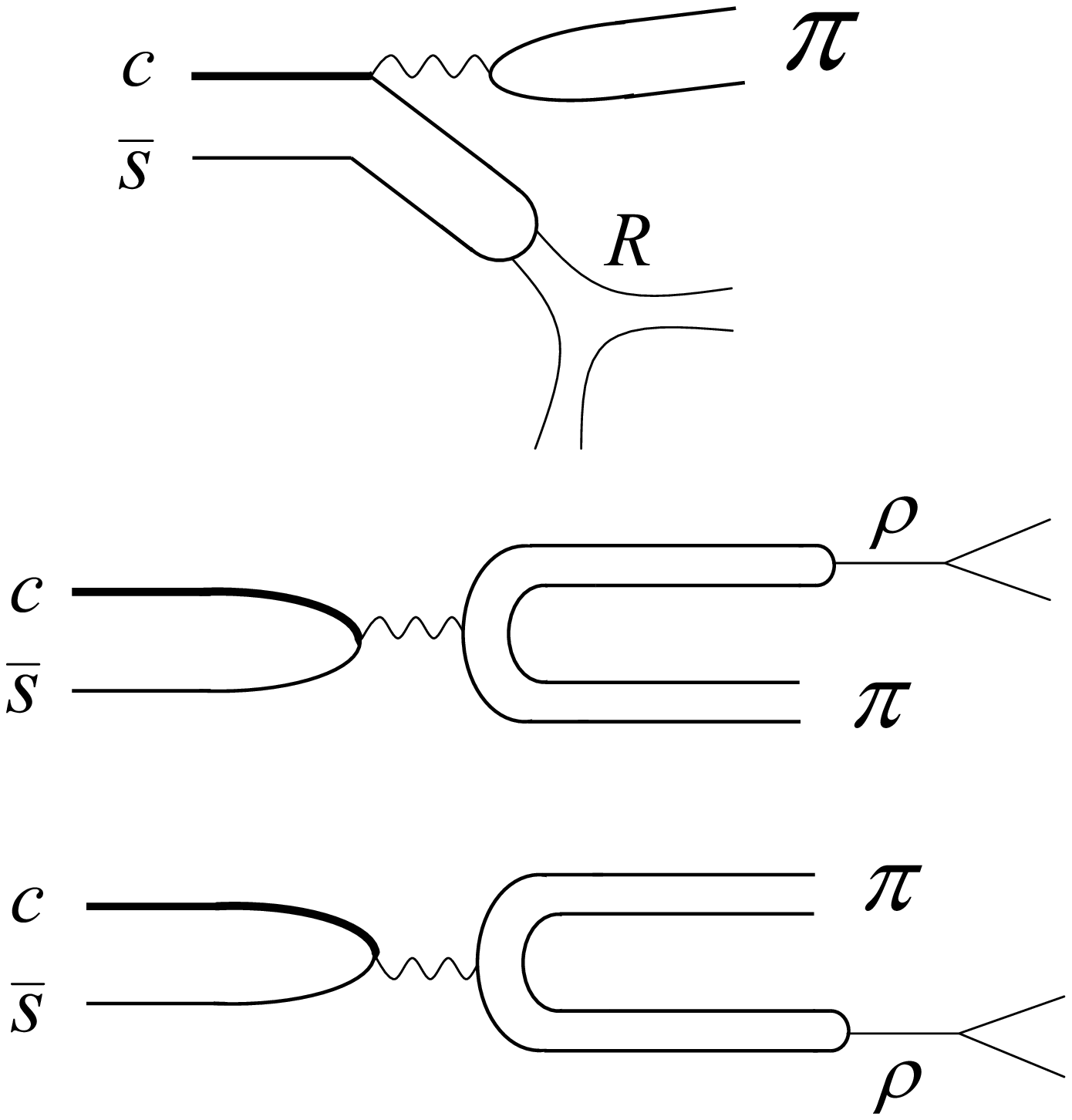} 
 \caption{Diagrams for $D_s^+\to \pi\pi\pi$
    \label{FIG:DS3PI} } 
\end{figure}

 
One has to keep in mind that the dynamical stage for $D_s^+ \to 3\pi$ 
and $D^+ \to 3\pi$ is actually  
quite different even beyond the fact that the former is Cabibbo allowed 
and the latter Cabibbo suppressed:  
\begin{itemize} 
\item  
$D_s^+ \to 3\pi$ can be generated by WA as well by the leading decay process  
coupled with $K\bar K \pi \Rightarrow 3\pi$ rescattering. Its final state  
carries $I=1$.  
\item  
$D^+\to 3 \pi$ on the other hand can proceed via  
(i) WA,  (ii) the $\Delta I=1/2$ and  (iii) $\Delta I=3/2$ components of the quark 
decay process.   
In the first two cases the final state isospin is $1$, in the last one  
$I=2$.  
\end{itemize} 
These general facts can be translated into more channel-specific statements. 
 Given the relevant strange
 quark content of $D_s$, resonances involved in its decays should couple to both $KK$ and
 $\pi\pi$, and obvious candidates are the $f_0$'s.
 Figure \ref{FIG:DS3PI} shows the diagrams such a decay can proceed through.
 The
 spectator-like diagram involves the resonance able to couple simultaneously to
 a couple of strange quarks, and to a couple of pions. The other two processes
 are WA-like; while those are reduced in inclusive rates, 
 they can be quite strong in some exclusive modes. 
 Evidence of such a WA component has
 been recently put forward by E791 which measures a relative fit fraction of
 about 10\% for $\rho\pi$. Interference between the WA processes has also been
 discussed. In general, the $D_s^+\to \pi\pi\pi$ is dominated by the $f_0(980)$
 narrow resonances, with very little nonresonant component left over.
 On the other hand,  the 
 $D^+\to \pi\pi\pi$ is characterized by a very large either nonresonant, or
 broad resonant component. 
 Thus one expects to find a quite different Dalitz plot population  
in $D_s^+ \to 3\pi$ and $D^+ \to 3\pi$ -- and this is indeed borne out  
by the data. 
They reveal   for D and $D_s$ charm mesons a rich
wealth of resonance substructures (Fig.~\ref{FIG:DPCOMP}). The
evidence can be visualised by the cartoon of a quasi-two body charm meson decay
proceeding effectively via a $D\to \pi^+ $ current (Fig.\ref{FIG:DPCART}),
interacting
to a $P_1P_2$ current (such as $K^+K^-$, $\pi^+\pi^-$, $K^-\pi^+$)  through a
strongly-decaying resonance propagator. 
The interpretation is, however, still in dispute. E687 was able to satisfactorily fit the Dalitz plot
 area with a 80\% fit fraction of nonresonant. E791 claims the observation 
 of a large $\sigma\pi^+$ component in $D^+\to \pi\pi\pi$. Analyses based on the 
 K-matrix ansatz do
 recognize the broad, low-mass excess, but claim to be able to explain it purely
 via $\pi\pi$ scattering amplitudes \cite{Anisovich:2002ij}.
\par
There are several possible reasons 
for this discrepancy: 
\begin{itemize}
\item 
It might be due to the excess in the E791 and BES data simply 
reflecting an attractive enhancement (rather than a full fledged resonance) in 
 $\pi\pi$ scattering. 
 \item 
The K-matrix approach might be of limited validity at
 small invariant energies. 
 \item 
 General caveats have been put forward towards 
 extracting resonance parameters from charm
 decay data when referring to scalars. Only the low energy
  tails of the resonance phase shifts  are visible in charm data and noone  
  has ever observed a complete $\sigma$
   \index{$\sigma$ resonance in charm decays}
 \index{$\kappa$ resonance in charm decays}
 nor $\kappa$  Breit-Wigner phase motion through $180^o$. 
 Novel approaches
 for testing the phase shift over the Dalitz plot area have been 
 suggested\cite{Bediaga:2002au},
 but so far  
 they are not applicable due to severe statistics
 limitations of datasets. We also remind the reader that Breit-Wigner parameters and pole
 positions  can fairly easily shift by hundred of MeVs, in particular for broad
 resonance candidates such as $\sigma$ or $\kappa$. 
 \item 
 A Breit-Wigner resonance form is an approximation of varying accuracy; in particular  
for scalar di-meson resonances a lot of information exists from data that  
a Breit-Wigner ansatz provides in general a poor description  
\cite{ULF}. More work is needed here, and theorists are asked to spend  
some quality time on this problem. 
 \end{itemize}
The last item has an important consequence also for $B$ decays \cite{ULF}:  
 Whatever the structure of $\sigma$ is, either a genuine resonance or a strong attractive S 
wave enhancement at low di-pion  masses, it will affect the important mode 
$B\to \rho \pi$ to be used for extracting the angle $\phi_2/\alpha$ in the unitarity 
triangle for the scalar form factor that one should adopt to represent $\pi\pi$
 resonances such 
as the $\sigma$ is very different from the normally used Breit-Wigner 
 parametrization with a running width. 
\par
 Finally, a far-reaching effect of the $\sigma$ puzzle 
 is under the eyes of low-energy hadronic community, and it regards the recent
 measurement of radiative decay $\Gamma(\phi\to\gamma f_0)$ at the DA$\Phi$NE
 $\phi$-factory \cite{Aloisio:2002bt} in the $\pi^0\pi^0 \gamma$ final state. 
 In this analysis the presence of a $\sigma$
 is essential ingredient to the fit of data. A recent analysis
 \index{K matrix vs. isobar formalism in Dalitz plots}
\index{isobar vs. K matrix  formalism in Dalitz plots}
 \cite{Boglione:2003xh} refits the KLOE data by using the K-matrix formalism and
 thus avoiding to employ explicity the $\sigma$, finding an order of magnitude
 smaller branching ratio.
 \par
\par
There are also ulterior reasons for a detailed understanding of Dalitz plots: 
as discussed in Sect. \ref{CPV} larger CP asymmetries might surface in them than 
in fully integrated widths.

\subsection{Baryon decays} 
\label{BARYONNL} 
%
All the issues addressed above for meson decays have their analogue  
in the decays of charm baryons.  
\begin{itemize} 
\item 
One wants to know absolute branching ratios for $\Lambda_c$ and  
$\Xi_c$ for the purpose of {\em charm counting}  
\index{charm counting} in $B$ decays.  
\item  
Knowing the relative importance of $\Lambda_c \to \Lambda +X$ vs.  
$\Lambda_c \to N \bar K +X$ and $\Xi_c \to \Xi +X$ vs.  
$\Xi_c \to \Lambda \bar K +X$ is of great help in identifying  
$\Lambda_b$ and $\Xi_b$ decays.  
\item  
The final states in $\Lambda_c$ and $\Xi_c$ decays can shed new light on 
the spectroscopy of light flavour baryons. 
\end{itemize} 
Quasi-two-body modes of $\Lambda_c$ and $\Xi_c$ can be described with the 
same tools as $D$ decays, namely quark models, QCD sum rules and lattice  
QCD. Yet the baryon decays pose even more formidable theoretical 
challenges. There are more types of contributions all on the same  
level, namely destructive as well as constructive PI and even WS with the latter 
{\em not} suffering from helicity suppression\index{helicity suppression}. 
Evaluating matrix elements of the relevant operators for baryons is even harder 
than for mesons. 

Let us merely comment here on attempts to search for footprints of WS in  
certain exclusive channels. An early example is provided by  
$\Lambda_c^+ \to \Delta^{++}K^-$. Having an only  
slightly reduced branching ratio, it might be cited \cite{WSDELTA} 
as specific evidence 
for WS since that mechanism produces it naturally, whereas quark decay  
does not lead to it {\em directly}. Yet one encounters the usual problem 
with interpreting a transition as due to WS:   
it can be produced by the quark decay reaction followed by FSI. Until we  
gain full control over the nonperturbative dynamics in exclusive 
transitions we cannot distinguish the two explanations on a case-by-case 
is. 

A similar comment applies to the Cabibbo suppressed (CS) mode $\Xi^+_c \rarr p
K^-   \pi^+$ 
first seen by SELEX \cite{Jun:1999gn} (confirmed shortly 
thereafter \cite{Link:2001rn}):
\beq 
B(\Xi_c^+\rarr pK^-\pi^+)/B(\Xi_c^+   \rarr \Sigma^+(pn)\, K^- \pi^+)= 
  0.22\pm0.06\pm 0.03  \; ; 
\eeq 
once  corrected for phase space this number is compatible with the branching ratio for the 
only other CS decay well   measured, $\Lambda^+_c  \rarr p K^-  K^+$, relative to three--
body CF decay $\Lambda^+_c  \rarr p K^-  \pi^+$. 

%
\subsection{Resume} 
\label{NLRES} 
%
A decent theoretical description of nonleptonic (quasi-)two-body modes  
in $D$ decays has emerged. Yet we owe this success not completely --  
maybe not even mainly -- to our ingenuity: nature was kind enough to  
present us with a relatively smooth dynamical environment  
considering how disruptive FSI could have been. On the other hand we  
seem to have reached a point of quickly diminishing returns. Treatments  
based on the quark  model and on QCD sum rules probably have been pushed  
as far as they go without any qualitatively new theoretical insights on  
the limits of factorization, on how to go beyond it and on FSI. One is  
entitled to appeal to lattice QCD for more definitive answers, yet it 
might be quite a while before lattice QCD can deal with exclusive 
nonleptonic decays involving pions and kaons {\em without} the quenched 
approximation. 
\par
There are two main motivations beyond professional pride 
why we would like to do better in our understanding of nonleptonic 
charm decays: (i) It is desirable to obtain a reliable 
estimate for the strength of $D^0 - \bar D^0$ oscillation generated by SM dynamics 
to interprete experimental findings; 
(ii) for similar reasons we want to obtain reliable predictions on CP asymmetries in charm 
decays due to SM and New Physics dynamics. These issues will be discussed in more 
detail later on. 
\par
A less profound, yet still interesting question to study is which fraction of the 
weak decays of a given hadron 
are of the two-body or quasi-two-body type. This is however much easier said than done 
due to interference effects between different resonant final states.  If one merely adds 
up the quoted widths for the different contributions to a final state, one typically either 
over- or under-saturates the overall widths, because the different components in general 
induce destructive or constructive interference between them. Examples can be 
found in Tables \ref{TAB:DPAMPHD1} - \ref{TAB:DPAMPHDS}.  
There is some discussion 
on this point in \cite{Browder:1996af,FRABETTI95}.  

It would make sense for PDG to list not merely the branching ratios, but 
also the amplitudes including their phases for Dalitz plot analyses. 
A way of addressing the issue, if not of solving it, is to analyze for  
each final state and resonant structure the magnitude of interference.
\par  
\begin{table} 
 \caption{Fraction (\%) of partial widths for charmed particles identified in 
exclusive modes, and fraction of exclusive modes identified as two or 
quasitwo body (hadronic) and three or quasithree (semileptonic). Data from 
PDG2002. No absolute branching fractions have been measured for 
$\Xi_C^+$. No branching fractions have been measured for $\Xi_C^0$ nor 
$\Omega_C^0$. 
  \label{tab:ratios} 
 } 
 \footnotesize 
 \begin{center} 
 \begin{tabular}{|l|l|l|l|l|l|} \hline 
  & inclusive    & excl. fract. & excl. fract. & excl. fract.    &excl. fract.  
\\ 
  &  $(e+\mu+h)X$          & (semi)leptonic & 3B or          & nonleptonic       &2B or  
        \\ 
  &            &                &       quasi-3B & nonleptonic       &   
quasi-2B\\
 \hline 
   $D^+$&   106  $\pm$ 21  &21$\pm$ 3 & 24$\pm$ 3    & 42$\pm$ 3  &40$\pm$ 4 \\ 
 $D^0$&  113  $\pm$ 23   &10 $\pm$ 1   & 9.3$\pm$ 0.4 & 54 $\pm$ 2     & 57
 $\pm$ 3  \\  
    $D^+_s$& 160 $\pm$ 70    &12$\pm$ 2   & 5$\pm$ 1  & 64 $\pm$ 8   & 57 $\pm$
    6 \\  
   $\Lambda_C^+$ &  150   $\pm$  50  & 4.2$\pm$ 0.8 & 4.2$\pm$ 0.8 & 44 $\pm$ 9&
   9$\pm$2 \\  
\hline 
 \end{tabular} 
  \vfill 
 \end{center}  
\end{table} 
%
\section{$D^0 - \bar D^0$ Oscillations}  
\label{DDOSC}  
%
%
%
Very often the two notions `oscillations' and `mixing' are employed in an 
interchangable way. Both are concepts deeply embedded in quantum mechanics, 
yet we want to distinguish them in our review\index{mixing vs. oscillations}. 
{\em Mixing} means that classically 
distinct states are not necessarily so in quantum mechanics and therefore can interfere. 
For example in atomic physics the weak neutral currents induce a `wrong' parity 
component into a wavefunction, which in turn generate parity odd observables. Or mass 
eigenstates of quarks (or leptons) contain components of different flavours  
giving rise to a non-diagonal CKM (or PMNS) matrix. Such mixing creates a plethora 
of observable effects. The most intriguing ones arise when the violation of a certain flavour 
quantum number leads to the {\em stationary} or {\em mass}  eigenstates not being 
{\em flavour} eigenstates. This induces 
{\em oscillations}, i.e. particular transitions, of which those between  
neutral mesons and anti-mesons of a given flavour (as discussed below), neutrinos of different 
flavours (including neutrino-antineutrino transitions due to Majorana terms) and neutron and 
antineutrons are the most discussed examples: a beam initially containing only one flavour 
`regenerates' other flavours at later times through oscillatory functions of time, namely sine 
and cosine terms. One furthermore 
distinguishes between {\em spontaneous} regeneration or oscillations in vacuum and 
regeneration when the oscillating state has to transverse matter. 
\par
Mixing is thus a necessary, though not sufficient condition for oscillations with their peculiar 
time dependence to occur.  
\par
Flavour-changing 
$\Delta S,\Delta C,\Delta B \neq 0$ weak interactions have a two-fold 
impact on neutral heavy mesons. In addition to driving their decays, they 
induce  $K^0 - \bar K^0$, $D^0-\bar D^0$, 
$B^0-\bar B^0$ oscillations. I.e., the two mass eigenstates of 
neutral heavy flavour mesons are linear combinations of the flavour 
eigenstates; while no longer carrying a definite flavour quantum 
number, they possess {\em non}degenerate masses and lifetimes.  
Its most striking signature is that an initially pure beam of, say,  
$B^0$ mesons will not only lose intensity due to $B^0$ decaying away,  
but also change its flavour identity over to $\bar B^0$, then go back to  
$B^0$ and so on. CP violation in the underlying dynamics can manifest 
itself in a variety of ways as described later. These  
generic statements apply to $K^0-\bar K^0$, $B^0-\bar B^0$ as well as  
$D^0 - \bar D^0$.  

On the practical there are large differences, though. Unlike for 
strange and $B$ mesons, $D^0 - \bar D^0$ oscillations are predicted to 
proceed quite slowly  within the Standard Model. Searching for them has 
been recognized  for a long time as a promising indirect probe for New 
Physics.  Yet with the experimental sensitivity having reached the few 
percent  level, we have to carefully evaluate the reliability of 
our  theoretical treatment of charm transitions. For on one  
hand charm hadrons -- in contrast to kaons -- are  
too heavy to have only a few decay channels;   
on the other hand -- unlike $B$ mesons -- we cannot be  
confident of the applicability of the  
Heavy Quark Expansion (HQE) to charm decays, although it has  
been proposed. This problem is particularly  
serious for $D^0 - \bar D^0$ oscillations; over the last  
twenty years vastly differing predictions have appeared  
in the literature   
\cite{NELSON}.  
A model-independent treatment becomes highly desirable  
even if it provides us with a mainly qualitative understanding  
rather than precise numbers. 

\subsection{Notation} 
\label{DDNOT} 

The time evolution of neutral $D$ mesons is obtained from  
solving the (free) Schr\"odinger equation  
\beq  
i\frac{d}{dt} \left(  
\begin{array}{ll} 
D^0 \\ 
\bar D^0 
\end{array}   
\right)  = \left(  
\begin{array}{ll} 
M_{11} - \frac{i}{2} \Gamma _{11} &  
M_{12} - \frac{i}{2} \Gamma _{12} \\  
M^*_{12} - \frac{i}{2} \Gamma ^*_{12} &  
M_{22} - \frac{i}{2} \Gamma _{22}  
\end{array} 
\right)  
\left(  
\begin{array}{ll} 
D^0 \\ 
\bar D^0 
\end{array}   
\right)  
\label{SCHROED}  
\eeq 
CPT invariance imposes  
\beq  
M_{11}= M_{22} \; \; , \; \; \Gamma _{11} = \Gamma _{22} \; .  
\label{CPTMASS} 
\eeq 
The mass eigenstates obtained through diagonalising this matrix  
are given by (for details see \cite{LEE,CPBOOK})  
\bea 
 |D_L\rangle &=&  
\frac{1}{\sqrt{|p|^2 + |q|^2}} \left( p |D^0 \rangle +  
q |\bar D^0\rangle \right)   
\nonumber 
\\   
|D_H\rangle &=&  
\frac{1}{\sqrt{|p|^2 + |q|^2}} \left( p |D^0 \rangle -   
q |\bar D^0\rangle \right)  
\label{P1P2}  
\\ 
\frac{q}{p} &=&   
\sqrt{\frac{M_{12}^* - \frac{i}{2} \Gamma _{12}^*} 
{M_{12} - \frac{i}{2} \Gamma _{12}}} 
\label{Q/P}  
\eea 
with differences in mass and width:  
\bea  
\Delta M_D &\equiv& M_H - M_L =  
-2 {\rm Re} \left[ \frac{q}{p}(M_{12} -  
\frac{i}{2}\Gamma _{12})\right]  \\ 
\Delta \Gamma _D  &\equiv& \Gamma _L -  
\Gamma _H =  
-2 {\rm Im}\left[ \frac{q}{p}(M_{12} -  
\frac{i}{2}\Gamma _{12})\right]  
\label{DELTAEV}  
\eea 
The labels $H$ and $L$ are chosen such that $\Delta M_D >0$.  
Once this {\em convention}  
has been adopted, it becomes a sensible question  
whether  
\beq  
\Gamma _H > \Gamma _L  \; \; \; {\rm or} \; \; \;  
\Gamma _H < \Gamma _L  
\eeq  
holds, i.e. whether the heavier state is shorter or  
longer lived.  
Note that the subscripts $H$ and $L$ have been swapped in  
going from $\Delta M_D$ to $\Delta \Gamma_D$; this is done to   
have the analogous quantities positive for kaons.

The time evolution of the `stationary' states $|D_{H,L}\rangle$  
is then as follows  
\beq  
|D_{H,L}(t)\rangle = 
e^{-\frac{1}{2}\Gamma_{H,L}t}e^{-iM_{H,L}t}|D_{H,L}\rangle  
\eeq 
The probability to find a $\bar D^0$ at time $t$ in an initially  
pure $D^0$ beam is given by  
\beq  
|\langle \bar D^0|D^0(t)\rangle |^2 =  
\frac{1}{4} \left| \frac{q}{p}\right| ^2 e^{-\Gamma_Ht}  
\left( 1+ e^{-\Delta \Gamma_D t} - 2e^{-\frac{1}{2}\Delta \Gamma_D t} 
{\rm cos}\Delta M_D t\right)  
\eeq 
While the flavour of the initial meson is tagged by its production,  
the flavour of the final meson is inferred from its decay.

There are two dimensionless ratios describing the interplay between 
oscillations and decays:   
\beq  
x_D = \frac{\Delta M_D}{\bar \Gamma_D} \; , \; \;  
y_D = \frac{\Delta \Gamma_D}{2\bar \Gamma_D} \; ,  \quad {\rm with }
\bar \Gamma_D \equiv \frac{1}{2}(\Gamma_1 + \Gamma_2)  
\eeq  
In the limit of CP invariance $\frac{q}{p} =1$ and  
the mass eigenstates are CP eigenstates as well; we can then  
ask whether the heavier state is CP odd (as for kaons)  
or even. With the {\em definitions}   
$CP|D^0\rangle = |\bar D^0\rangle$ and  
$CP|D_{\pm}\rangle = \pm |D_{\pm}\rangle$ we have  
\beq  
|D_{\pm}\rangle = \frac{1}{\sqrt{2}}  
\left(  
|D^0\rangle \pm |\bar D^0\rangle 
\right)  
\eeq 
\beq  
M_{odd} - M_{even} = M_H - M_L =  -2  
{\rm Re} \left[ \frac{q}{p} 
\left( M_{12} - \frac{i}{2} \Gamma _{12} \right) 
\right] =  
- 2 M_{12}  
\label{DM-+} 
\eeq 
It is instructive to follow how expressions change, yet  
observables remain the same, when different conventions for  
$q/p$ and the $CP$ operator are adopted \cite{CPBOOK}. 

\subsection{Phenomenology} 
\label{DDPHENO} 

$D^0 - \bar D^0$ oscillations can reveal themselves through  
four classes of observables:  
\begin{enumerate} 
\item  
`wrong-sign' decays, i.e.  
a `global' or 'time integrated' violation of a selection rule in the  
final state of $D$ decays;  
\item  
$D^0$ and $\bar D^0$ after being produced exclusively in  
$e^+e^- \to D^0 \bar D^0$ decaying into two seemingly 
identical final states;  
\item  
different lifetimes in different channels;  
\item  
non-exponential decay rate evolutions. 

\end{enumerate} 
We now discuss the four cases.
\par
{\bf 1.} The final states of $D$ decays are characterized by certain  
selection rules like $\Delta C=\Delta S= - \Delta Q_l$ for semileptonic  
and $\Delta C=\Delta S$ for Cabibbo allowed nonleptonic transitions.  
Having established `correct sign' modes  
\footnote{We prefer the term `correct sign' over `right sign', since we 
do not share the often unreflected preference of the right over the  
left.}, which satisfy these selection rules, one can then search for  
corresponding `wrong sign' channels violating these rules and compare 
the two rates 
\beq  
r_{WS}^D(f) = \frac{\Gamma (D^0 \to f_{WS})}{\Gamma (D^0 \to f_{CS})}  
\label{RDDEF}
\eeq  
A related quantity of phenomenological convenience is 
\beq 
\chi_{WS}^D(f) \equiv \frac{\Gamma (D^0 \to f_{WS})}
{\Gamma (D^0 \to f_{CS}) + \Gamma (D^0 \to f_{WS})} 
\label{CHIDEF}
\eeq
Within the SM a clean signal for $D^0 - \bar D^0$ oscillations  
is the emergence of `wrong-sign' leptons:  
\beq  
r_D \equiv   
\frac{\Gamma (D^0 \to l^-X)}{\Gamma (D^0 \to l^+X)} =   
\frac{|q/p|^2(x_D^2 + y_D^2)}{2 +x_D^2 - y_D^2} \simeq  
\frac{1}{2}(x_D^2 + y_D^2) \; ,  
\label{RDOSC} 
\eeq  
where we have used the usual short-hand notation  
$r_D = r_{WS}^D(l^{\mp})$; we have also  
anticipated that $x_D^2$, $y_D^2$ $\ll 1$ and  
$|q|^2 \simeq |p|^2$ hold. A signal is produced by either  
$x_D \neq 0$ or $y_D \neq 0$ or both. One should note that while  
$r_D \neq 0$ is an unambiguous sign of New Physics within the SM, it  
does not require very specific {\em time} features of $D^0 - \bar D^0$ 
oscillations {\em per se}, as illustrated by the next example with   
 `wrong-sign' kaons. For there is a  
SM background -- doubly Cabibbo suppressed decays (DCSD). Thus there  
are three classes of contributions, namely purely from DCSD,  
from $\d0d0$ oscillations followed by a Cabibbo allowed  
transition and from the interference between the two:  
$$   
\tilde r_{WS}^D(K^{\pm}\pi^{\mp}) \equiv  
\frac{\Gamma (D^0 \to K^+ \pi ^-)}{\Gamma (D^0 \to K^-\pi^+)} =  
\frac{|T(D^0\to K^+\pi^-)|^2}{|T(D^0\to K^-\pi^+)|^2}  \cdot  
$$ 
\beq   
\left[ 1 + \frac{1}{2} \frac{x_D^2 +y_D^2}{{\rm tg}^4\theta _C} 
|\hat \rho(K\pi)|^2   
+ \frac{y_D}{ {\rm tg}^2\theta _C}  
{\rm Re}\left( \frac{q}{p}\hat \rho _{K\pi} 
\right)   
+    
\frac{x_D}{{\rm tg}\theta _C^2} 
{\rm Im}\left(  
\frac{q}{p}\hat \rho _{K\pi }    
\right)  \right]\; ,  
\label{DKPIOSCGLOBAL} 
\eeq  
where  
\beq   
\frac{1}{-{\rm tg}^2\theta _C} \hat \rho _{K\pi} \equiv  
\frac{T(\bar D^0 \to K^+\pi^-)}{T(D^0 \to K^+\pi^-)}  
\eeq 
denotes the ratio of {\em instantaneous} transition amplitudes with  
$\hat \rho _{K\pi} \sim {\cal O}(1)$.  
We have retained terms of first and second order in the small  
parameters $x_D$, $y_D$ that are enhanced by  
$1/{\rm tg}^2\theta_C$ and $1/{\rm tg}^4\theta_C$,  
respectively.  
$|T(D^0\to K^+\pi^-)|^2/|T(D^0\to K^-\pi^+)|^2$ is controled by  
tg$^4\theta_C \sim 3\cdot 10^{-3}$. The quark decay process actually  
enhances the DCS mode by $(f_K/f_{\pi})^2 \sim 1.5$, and thus one  
expects $|T(D^0\to K^+\pi^-)|^2/|T(D^0\to K^-\pi^+)|^2$ to be in the  
range of ${\rm few}\cdot 10^{-3}$; yet one has to allow for some significant  
uncertainty in this prediction. Therefore one can infer  
$D^0-\bar D^0$ oscillations from the time integrated ratio  
$\tilde r_{WS}^D(K^{\pm}\pi^{\mp})$ only if it is relatively sizable.

{\bf 2.}   
Producing charm in $e^+e^-$ annihilation on the vector meson resonance  
$\psi ^{\prime \prime}(3770)$ leads to an exclusive particle-antiparticle 
pair. Consider the case where both charm mesons decay into seemingly 
identical final states:  
\beq  
e^+e^- \to \psi ^{\prime \prime}(3770) \to D^0 \bar D^0 \to  
f_D f_D 
\eeq 
It might appear that  
even without $D^0-\bar D^0$ oscillations such final states are possible. 
E.g., $f_D = K^-\pi^+$ could  arise due to the Cabibbo allowed  
[doubly Cabibbo suppressed] mode $D^0 [\bar D^0]\to K^-\pi^+$; likewise  
for the CP conjugate channel $f_D=K^+\pi^-$. Or $f_D=K^+K^-$ would be  
driven by the Cabibbo suppressed channels $D^0[\bar D^0]\to K^+K^-$. 
However the two charm mesons have to form a P-wave and C odd 
configuration;  
Bose-Einstein statistics then does not allow them to 
decay into identical  final states. This is one example of  
exploiting EPR correlations \index{EPR} named after Einstein,  
Podolsky and Rosen, who first pointed them out as a consequence  
of quantum mechanics' intrinsically {\em non-}local features  
\cite{EPRPAPER}.

The situation becomes more complex in the presence of $D^0-\bar D^0$ 
oscillations. Bose-Einstein statistics still forbid the original $\d0d0$ 
pair to evolve into a $D^0D^0$ or  
$\bar D^0\bar D^0$ configuration at one time $t$. The EPR 
correlation tells us that if one 
neutral $D$ meson reveals itself as a, say, $D^0$, the other one has to 
be a $\bar D^0$ at {\em that} time. Yet at {\em later} times it can 
evolve into a  
$D^0$, since the coherence between the original $D^0$ and $\bar D^0$  
has been lost. Let us consider specifically $f_D = K^{-}\pi^{+}$.  We then find 
\beq  
\frac{\sigma (e^+e^- \to D^0\bar D^0 \to  
(K^{\mp}\pi^{\pm})_D (K^{\mp}\pi^{\pm})_D)} 
{\sigma (e^+e^- \to D^0\bar D^0 \to  
(K^{\mp}\pi^{\pm})_D (K^{\pm}\pi^{\mp})_D)} = r_D \; ; 
\eeq
i.e., this process can occur only through oscillation -- $r_D \neq 0$ -- 
and the EPR correlation forces the pair of charm mesons to act   
like a single meson.

The requirements of  Bose-Einstein statistics are satisfied in a subtle 
way, as best seen when  describing the reaction in terms of the {\em mass} 
eigenstates  $D_H$ and $D_L$:  
\beq  
e^+e^- \to \psi ^{\prime \prime}(3770) \to D_H D_L  \to  
(K^-\pi^+)_{D_H} (K^-\pi^+)_{D_L} \; ;   
\eeq 
i.e., the two decay final states $K^-\pi^+$ are {\em not} truly identical -- 
their energies differ by $\Delta M_D$. This difference is of course much 
too tiny to be measurable directly.  

In principle the same kind of argument can be applied to more complex  
final states like $f_D = K^{\mp}\rho^{\pm}$, however it is less 
conclusive there, in particular when the $\rho$ is identified merely  
by the dipion mass.   

The final state $f_D = K^+K^-$ can occur in the presence of  
$D^0-\bar D^0$ oscillations, but only if CP invriance is violated. For   
the CP parity of the initial $J^{PC}=1^{--}$ state is even, yet odd for 
the final state $[(K^+K^-)_D(K^+K^-)_D]_{l=1}$. We will return to this  
point later on.  

The fact that the $D^0\bar D^0$ pair has to form a coherent $C=-1$ 
quantum state in $e^+e^- \to \psi ^{\prime \prime}(3770) \to D^0 \bar D^0$ 
has other subtle consequences as well\index{coherent production of $D$ pairs}. 
Since the EPR effect (anti)correlates 
the time evolutions of the two neutral $D$ mesons as sketched above, they 
act like a single charm meson as far as {\em like}-sign dileptons are concerned, i.e., 
\beq
\frac{\sigma (e^+e^- \to D^0 \bar D^0 \to l^{\pm}l^{\pm}\, X)}
{\sigma (e^+e^- \to D^0 \bar D^0 \to l\, l\, X)} = 
\chi_{WS}^D(ll) 
\eeq
rather than the expression found for an {\em in}coherent pair of $D^0\bar D^0$
\beq 
\frac{\sigma (D^0 \bar D^0|_{in{\rm coh}} \to l^{\pm}l^{\pm}\, X)}
{\sigma (D^0 \bar D^0|_{in{\rm coh}} \to l\, l\, X)} = 
2\chi_{WS}^D(ll) [1- \chi_{WS}^D(ll)] 
\eeq

{\bf 3.}  
Lifetime measurements in different $D^0$ channels  
represent a direct probe for $y_D\neq 0$. In the limit of CP  
invariance -- which holds at least approximately in $D$ decays,  
see below -- mass eigenstates are CP eigenstates as well. There  
are three classes of final states: CP even and odd states --  
$D^0 \to \pi \pi, \; K\bar K. ...$ and  
\index{$D^0\to K_s\phi$ and $B^0\to K_s\phi$} 
$D^0 \to K_S\eta, \; K_S \phi  
\footnote{One has to keep in mind that $D^0 \to K_S[K^+K^-]_{\phi}$  
has to be distinguished against $D^0 \to K_S[K^+K^-]_{f_0}$   
since the latter in contrast to the former is CP even.} 
, ...$, respectively -- and mixed  
ones -- $D^0 \to l^+\nu K^-, \, K^-\pi^+, ...$ with   
$\tau _{D^0 \to l^+X} 
\simeq \left( \tau _{D_+} + \tau _{D_-} \right)/2$.

{\bf 4.}  
The most direct signal for oscillations is the observation that decay  
rate evolutions in time do not follow a strictly exponential law. For  
semileptonic transitions we have  
\beq  
{\rm rate}(D^0(t) \to l^{\pm}\nu X) \propto  
e^{- \Gamma_H t}+e^{-\Gamma_L t} \pm 2e^{\bar \Gamma t} 
{\rm cos}\Delta M_Dt  
\eeq  
The most dramatic demonstration is provided by  
$D^0 \to K^+ \pi^-$ which without oscillations is doubly  
Cabibbo suppressed:  
$$  
\frac{{\rm rate}(D^0(t) \to K^+ \pi ^-)} 
{{\rm rate}(D^0(t) \to K^- \pi ^+)} =    
\frac{|T(D^0\to K^+\pi^-)|^2}{|T(D^0\to K^-\pi^+)|^2}  \cdot  
$$  
$$  
\cdot \left[ 1 +  
\frac{(\Delta m_Dt)^2 +\frac{1}{4}(\Delta \Gamma _Dt)^2} 
{4{\rm tg}^4\theta _C} 
\left| \frac{q}{p}\right|^2 |\hat \rho _{K\pi }|^2    
+ \frac{\Delta \Gamma _Dt} 
{2{\rm tg}^2\theta _C} 
{\rm Re}\left(  
\frac{q}{p}\hat \rho _{K\pi }    
\right)   
+ \right.   
$$  
\beq 
\left. + \frac{\Delta m_Dt}{{\rm tg}^2\theta _C}  
{\rm Im}\left(  
\frac{q}{p}\hat \rho _{K\pi }    
\right)   
\right]  
\label{DKPIOSCLOCAL} 
\eeq 
where we have used the notation of Eq.(\ref{DKPIOSCGLOBAL}). We see that  
the dependence on the (proper) time of decay 
$t$ differentiates between  DCSD, $\d0d0$ oscillations and their 
interference.

\subsection{Theory expectations} 

Within the SM two structural reasons combine to make $x_D$ and $y_D$  
small in contrast to the situation for $B^0 - \bar B^0$ and  
$K^0 - \bar K^0$ oscillations:  
\begin{itemize} 
\item  
The amplitude for $D^0 \leftrightarrow \bar D^0$ transitions is twice  
Cabibbo suppressed and therefore  
$x_D$, $y_D$ $\propto {\rm sin}^2 \theta _C$. The amplitudes for  
$K^0 \leftrightarrow \bar K^0$ and $B^0 \leftrightarrow \bar B^0$  
are also twice Cabibbo and KM suppressed -- yet  
so are their decay widths.  
\item  
Due to the GIM mechanism\index{GIM mechanism} 
one has $\Delta M = 0 = \Delta \Gamma$ in  
the limit of flavour symmetry. Yet $K^0 \leftrightarrow \bar K^0$ is  
driven by $SU(4)_{Fl}$ breaking characterised by $m_c^2 \neq m_u^2$,  
which represents no suppression on the usual hadronic scales. In  
contrast $D^0 \leftrightarrow \bar D^0$ is controled by  
$SU(3)_{fl}$ breaking typified by $m_s^2 \neq m_d^2$  
(or in terms of hadrons $M_K^2 \neq M_{\pi}^2$), which on the  
scale $M_D^2$ provides a very effective suppression. 
\end{itemize}  
These general considerations can be illustrated by 
considering transitions to two pseudoscalar mesons, which
are common to 
$D^0$ and $\bar D^0$ decays and can thus communicate between 
them:   
\beq 
D^0 \; \; \stackrel{CS\,} \Rightarrow \; \; 
K^+ K^- ,\; 
\pi ^+ \pi ^- \; \; \stackrel{CS\,} \Rightarrow \; \; 
\bar D^0 \; , 
\label{DKK}
\eeq 
\beq  
D^0 \; \; \stackrel{CA\,} \Rightarrow \; \;  
K^- \pi ^+ \; \; \stackrel{DCS} \Rightarrow \; \; 
\bar D^0 \;  \;  {\rm or} \; \; 
D^0 \; \; \stackrel{DCS} \Rightarrow \; \;  
K^+ \pi ^- \; \; \stackrel{CA\,} \Rightarrow \; \; \bar D^0 
\; . 
\label{DKPI}
\eeq 
where $CA$, $CS$ and $DCS$ denotes the channel as 
Cabibbo allowed, Cabibbo suppressed and doubly Cabibbo 
suppressed, respectively. 
We have used the symbol "$\Rightarrow$" to indicate that these transitions 
can be real on-shell ones -- for $\Delta \Gamma_D$ -- as well as virtual 
off-shell ones -- for $\Delta m_D$. 
Since  
\bea 
T(D^0 \Rightarrow K^-\pi^+/K^+\pi^- \Rightarrow \bar D^0) 
&\propto&  -{\rm sin}^2\theta_C {\rm cos}^2\theta_C 
\nonumber 
\\
T(D^0 \Rightarrow K^-K^+/\pi^+\pi^- \Rightarrow \bar D^0) 
&\propto&  
{\rm sin}^2\theta_C {\rm cos}^2\theta_C
\label{D2PSBARD} 
\eea
one obviously has in the $SU(3)$ limit  
$\Delta \Gamma (D^0 \to K\bar K,\pi \pi , K\pi , 
\pi \bar K) = 0$;  for the amplitudes for 
Eqs.\,(\ref{DKPI}) would then be equal in size and opposite in sign 
to those of Eq.\,(\ref{DKK}).  
Yet the measured branching ratios \cite{Hagiwara:fs}
\bea 
{\rm BR}(D^0 \to K^+K^-) &=& (4.12 \pm 0.14) \cdot 10^{-3}  , \; 
{\rm BR}(D^0 \to \pi ^+ \pi ^-) = (1.43 \pm 0.07) \cdot 10^{-3} 
\nonumber 
\\ 
{\rm BR}(D^0 \to K^- \pi ^+) &=& (3.80 \pm 0.09) \cdot 10^{-2}  , \; 
{\rm BR}(D^0 \to K^+ \pi ^-) = (1.48 \pm 0.21) \cdot 10^{-4}
\nonumber
\eea
show very considerable $SU(3)$ breakings:  
\bea 
\frac{{\rm BR}(D^0 \to K^+K^-)}{{\rm BR}(D^0 \to \pi^+\pi^-)} 
&\simeq& 2.88 \pm 0.18  
\label{KKPIPI2}   
\\ 
\frac{{\rm BR}(D^0 \to K^+ \pi ^-)}{{\rm BR}(D^0 \to K^- \pi ^+)} 
&\simeq& (1.5 \pm 0.2) \cdot \tan^4 {\theta _C} 
\label{RATIO2B} 
\eea  
compared to ratios of unity and 
$\tan^4{\theta_C}$, 
respectively, in the symmetry limit. 

One would then conclude that the 
$K\bar K,\pi \pi , K\pi , 
\pi \bar K$ contributions to $\Delta \Gamma$  
should be merely Cabibbo suppressed with flavor $SU(3)$ 
providing only moderate further reduction -- similar to the 
general expectation of Eq.\,(\ref{GENERAL}): 
\beq 
\left. \frac{\Delta \Gamma }{\Gamma }\right| 
_{D \to K\bar K,\pi \pi , K\pi , 
\pi \bar K} \sim {\cal O}(0.01) \; .
\eeq 
Yet despite these large $SU(3)$ breakings an almost complete 
cancellation
takes place  between their contributions to 
$D^0 \!- \!\bar D^0$ oscillations 
\footnote{Since Eq.\,(\ref{PP})  is meant 
only as a qualitative 
illustration of our general argument 
we have ignored that $SU(3)$ {\em breaking} final state 
interactions can generate a strong phase shift 
$\delta_{K\pi}$  
between $D^0 \to K^- \pi ^+$ and $D^0 \to K^+ \pi ^-$, which 
would induce  a factor $\cos{\delta _{K\pi}}$ in the 
last interference term.}: 
$$  
{\rm BR}(D^0 \to K^+K^-) + {\rm BR}(D^0 \to \pi ^+ \pi ^-) 
- 2 \sqrt{{\rm BR}(D^0 \to K^- \pi ^+) 
{\rm BR}(D^0 \to K^+ \pi ^-)} \simeq 
$$ 
\beq \left(  - 8 ^{+12}_{-10} \right) \cdot 10^{-4} 
\label{PP}
\eeq
to be compared to 
$$  
{\rm BR}(D^0 \to K^- \pi ^+) + 
{\rm BR}(D^0 \to K^+K^-) + {\rm BR}(D^0 \to \pi ^+ \pi ^-) 
+ {\rm BR}(D^0 \to K^+ \pi ^-) \simeq 
$$ 
\beq 
(4.46 \pm 0.01 ) \cdot 10^{-2} 
\eeq 
Having two Cabibbo suppressed classes of decays one concludes 
for the overall oscillation strength:  
\beq  
\frac{\Delta M_D}{\bar \Gamma_D}, \; \Delta \Gamma_D \sim  
\; SU(3)\; {\rm breaking}  
\times 2 {\rm sin}^2\theta_C   
\eeq 
The proper description of $SU(3)$ {\it breaking} thus becomes the  
central issue -- see our discussion in Sect.\ref{TWOBODY}.  
The lesson to be drawn from the example given above 
is that one 
{\em cannot count} on the GIM mechanism to reduce the 
$D^0 \Rightarrow \bar D^0$ transition by more than a factor of 
three (in particular for $\Delta \Gamma_D$), yet cannot rule it out either. 
Thus  
\beq  
\frac{\Delta M_D}{\Gamma _D} \; \lsim \;   
\frac{\Delta \Gamma _D}{\Gamma _D}   
\lsim \;\frac{1}{3} \;  
\times 2 \sin^2{\!\theta_C} \;  \sim \; 
{\rm few} \times 0.01 
\label{GENERAL}  
\eeq 
represents a conservative bound -- for $\Delta M_D$  
maybe overly conservative -- based on general features of the SM.

The vastness of  
the literature on $D^0 - \bar D^0$ oscillations makes it difficult  
to track where a certain idea originated. There can be little doubt  
that many people knew about the statement that the  
$D^0 \leftrightarrow \bar D^0$ amplitude is at least  
${\cal O}(m_s^2)$,  
i.e. of second order in $SU(3)$ breaking, and  
even mentioned it in their papers. Without any claim to originality  it is given in  
a few lines following Fig.~2 in Ref.~\cite{DD} together with a  
concise proof. We want to repeat this elementary reasoning  
since it will elucidate subsequent points. 

The $m_s$ dependence of the $D^0\to\bar{D}^0$ transition amplitude  
can be inferred by considering $U$ spin, which relates $s$ and  
$d$ quarks. For its analysis it is  
advantageous to work in the basis of  
{\em interaction} eigenstates  
\beq 
s'= s\cos{\theta_c} - d \sin{\theta_c}, \qquad  
d'= d\cos{\theta_c}+ s 
\sin{\theta_c} \; ,   
\label{16} 
\eeq 
Their  
corresponding quantum numbers `current strangeness'  
$S'$ and  
`current downness'' $D'$ are conserved by the  
strong and electromagnetic 
forces. $D^0$ and $\bar{D}^0$ carry $C= +1$ and $-1$,  
respectively, and both  
have $S'=D'=0$. Since the relevant  
transition operator  
$(\bar uc)(\bar s'd')$  
has the {\it exact\,} selection rules  
$\Delta C= -\Delta S' = \Delta D'=1$, any amplitude $D^0\to\bar{D}^0$ has 
$\Delta S'=2$. For $m_s=m_d$ QCD dynamics strictly conserves $S'$. The 
only term violating the conservation of $S'$ or $D'$ is the mass  
term $\delta {\cal H}= \sin{\theta_c} cos{\theta_c} (m_s\!-\!m_d)  
\bar d' s' + {\rm h.c.}\,$ and   
{\it any} $D^0\to\bar{D}^0$ amplitude involves a second iteration of 
$\delta {\cal H}$; q.e.d. 
One should note that this reasoning is based on  
$U$ spin considerations alone rather than full flavour $SU(3)$: it holds  
irrespective of the mass of the $u$ quark, whether it is light or heavy,  
degenerate or not with $d$ or $s$.

Nothwithstanding this observation, there can be contributions to  
$T(D^0 \to \bar D^0)$ that are of first order in $m_s$ \cite{DD} -- a point  
claimed by the authors of Ref.~\cite{g4} to be wrong. The main point  
to note is that conventional perturbation theory breaks down when  
transitions can occur between {\em degenerate} states; for the energy  
denominators become singular then. Such degeneracies arise here  
due to the presence of the pseudogoldstone bosons (PGB)  
$K^0$, $\pi^0$ and $\eta$. A contribution $\sim {\cal O}(m_s)$  
emerges due to an IR singularity in the PGB loop. In the SM with  
purely left-handed charged currents such effects cancel out, yet  
they are present in a more general theory.  
\footnote{In a footnote in Ref.~\cite{g4}  
it is claimed that such an effect cannot arise  
``because the $\pi$, $K$, and $\eta$ are coupled derivatively''.}
Numerical estimates have usually been obtained as follows:  
({\bf i}) Quark-level contributions are estimated   
by the usual quark box diagrams; they yield only insignificant 
contributions  to $\Delta M_D$ and $\Delta \Gamma _D$ (see below).  
({\bf ii}) Various schemes employing contributions of selected 
hadronic states are invoked to estimate  
the impact of long distance  
dynamics; the numbers typically resulting are  
$x_D \, , \; y_D \,\sim \, 10^{-4} - 10^{-3}$  
\cite{Burdman:2001tf,BUCCELLA}.   
({\bf i\hspace*{-.2mm}i\hspace*{-.2mm}i}) These findings  
lead to the following widely  
embraced conclusions:   
An observation of $x_D > 10^{-3}$ would reveal the intervention of New  
Physics beyond the SM,    
while $y_D \simeq \left. y_D \right|_{SM} \leq 10^{-3}$ has  
to hold since New Physics has hardly a chance to enhance it. 

Both $\Delta M_D$ and  
$\Delta \Gamma _D$ have to vanish in the $SU(3)$ limit, yet the dynamics  
underlying them have different features:  
$\Delta M_D$ receives contributions from {\em virtual}    
intermediate states whereas $\Delta \Gamma$ is generated by  
{\em on}-shell transitions. Therefore the former  
represents a more robust quantity than the latter; actually it  
has often been argued that quark diagrams {\em cannot} be relied upon  
to even estimate $\Delta \Gamma$. 
Yet despite these  
differences there is no {\em fundamental} distinction   
in the theoretical treatment of $\Delta M_D$ and $\Delta \Gamma _D$:   
{\em both can be described through an 
OPE\index{operator product expansion} in terms of the expectation values  
of local operators and condensates\index{condensates} incorporating short distance 
as well as long distance dynamics}. Only the  
numerical aspects differ between $\Delta M_D$ and $\Delta \Gamma _D$, 
as does their sensitivity to New Physics. Finally  
the evaluation relies on local quark-hadron duality  
for both  
$\Delta M_D$ and $\Delta \Gamma _D$; the latter is, however, more 
vulnerable to limitations to duality since it involves less  
averaging \cite{DD}.

The formally leading term in the OPE for $\Delta C\!=\!-2$ 
transitions comes from 
dimension-6 four-fermion operators of the generic form 
$(\bar{u}c)(\bar{u}c)$  with the   
corresponding Wilson coefficient receiving contributions from  
different sources; it coresponds to the quark box diagram.

({\bf a}) Effects due to intermediate $b$ quarks are   
evaluated in a straightforward way since 
they are far off-shell: 
\beq 
\Delta M_D^{(b \bar b)} \simeq - \frac{G_F^2 m_b^2}{8\pi ^2}  
\left| V^*_{cb}V_{ub}\right| ^2   
\frac{\matel{D^0}{(\bar u \gamma_{\mu} (1 \!-\! \gamma _5)c) 
(\bar u \gamma_{\mu} (1 \!- \!\gamma _5)c)}{\bar D^0}}{2M_D}\; ;   
\label{B2} 
\nonumber  
\eeq 
however they are highly suppressed by 
the tiny CKM parameters. Using factorization to 
estimate the  matrix element one finds:  
\beq 
x_D^{(b\bar b)} \sim \:{\rm few} \times 10^{-7}\;. 
\label{B3} 
\eeq 
Loops with one $b$ and one light quark likewise are suppressed. 

({\bf b}) 
For the light intermediate quarks -- $d,s$ -- the momentum scale is set by 
the {\em external} mass $m_c$. However, it is highly GIM suppressed 
\footnote{  
This contribution is obviously saturated at the momentum scale 
$\sim \!m_c$, and thus refers to the Wilson coefficient of the $D\!=\!6$ 
operator. Therefore they are not  
long-distance contributions despite being proportional to $m_s^2$.}  
$$   
\Delta M_D^{(s,d)} \simeq    
- \frac{G_F^2 m_c^2}{8\pi ^2}  
\left| V^*_{cs}V_{us}\right| ^2  
\frac{\left( m_s^2 \!-\! m_d^2\right) ^2}{m_c^4} \times  
$$  
\beq   
\frac{\matel{D^0}{(\bar u \gamma_{\mu} (1 \!-\! \gamma _5)c) 
(\bar u \gamma_{\mu} (1 \!-\! \gamma _5)c) +  
(\bar u (1 \!+ \!\gamma _5)c) 
(\bar u  (1 \!+ \!\gamma _5)c)}{\bar D^0}}{2M_D} 
\;. 
\label{B4} 
\eeq 
The contribution to $\Delta \Gamma_D$ from the bare quark box is  
greatly suppressed by a factor $m_s^6$. The GIM mass insertions yield  
a factor $m_s^4$. Contrary to the claim in Ref.~\cite{g4} the additional  
factor of $m_s^2$ is {\em not} due to helicity suppression -- the  
GIM factors already take care of that effect; it is of an accidental  
nature: it arises because the weak currents are purely $V-A$ and only  
in four dimensions. Including radiative QCD corrections to the  
box diagram yields contributions $\propto m_s^4\alpha_S/\pi$.   
Numerically one finds:  
\beq  
\Delta \Gamma_D^{\rm box} <  
\Delta M_D^{\rm box} \sim {\rm few} \; \times 10^{-17} \; {\rm GeV}  
\; \; \hat = \; \; x_D^{\rm box} \sim {\rm few} \; \times 10^{-5}  
\label{B6}  
\eeq

With the leading Wilson coefficient so highly suppressed, 
one has to consider also formally non-leading contributions 
from higher dimensional 
operators. It turns out that the $SU(3)$ GIM suppression is in 
general not as severe as $(m_s^2 \!-\! m_d^2)/m_c^2$ per fermion line: it 
can be merely $m_s/\mu_{\rm hadr}$ if the fermion line is soft \cite{DD}. In 
the so-called practical version of the OPE
 this is described by condensates\index{condensates}  
contributing to higher orders in $1/m_c$. To be more specific: having a 
condensate induces a suppression factor $\sim \mu_{\rm hadr}^3/m_c^3$; 
yet the  GIM suppression now becomes only $m_s/\mu_{\rm hadr}$ 
yielding altogether a factor of order $\mu_{\rm hadr}^2/(m_s m_c)$ which can 
actually result in an enhancement, since $\mu_{\rm hadr}/m_c$ is not much 
smaller than unity. 

Analyzing the contributions coming from higher-dimensional operators  
with the help of condensates one estimates \cite{DD}  
\footnote{The huge lifetime ratio $\tau (K_L)/\tau (K_S) \sim 600$  
is due to the accidental fact that the kaon mass is barely above the  
three pion threshold.}  
\beq  
x_D\; , \; y_D \; \sim {\cal O}(10^{-3})  
\eeq 
with the realization that these estimates involve high powers of the  
ratio of comparable scales implying considerable numerical  
uncertainties that are very hard to overcome almost as a matter of  
principle. 

Yet despite the similarities in numbers for $x_D$ and  
$y_D$ the dynamics driving these two  
$\Delta C=2$ observables   
are quite different:  
\begin{itemize}  
\item  
$\Delta m_D$ being generated by contributions from virtual states  
is sensitive to New Physics which could raise it to the  
percent level. At the same time it necessarily involves an  
integral over energies thus making it rather robust against  
violations of local duality.  
\item  
$\Delta \Gamma _D$ being driven by on-shell transitions  
can hardly be sensitive to New Physics. At the 
same time, however,  it is very vulnerable to violations of local 
duality: a nearby  narrow resonance could easily wreck any GIM 
cancellation and raise  the value of $\Delta \Gamma _D$ by an order 
of magnitude!    
\end{itemize} 
The authors of Ref.~\cite{g4} claim  
to have shown in a model-independent way that the SM indeed generates  
$x_D$, $y_D \sim 1$\%. Yet even a simple minded model -- they use  
basically a phase space ansatz -- is still a model. Their analysis  
can be viewed as illustrating that $x_D$ and $y_D$ indeed could reach  
the $1$\% level -- but certainly no proof.

If data revealed $y_D \ll 1$\%$ \leq x_D$, we would have an intriguing case 
for the presence of New Physics. Yet considering the theoretical 
uncertainties basing the case for New Physics solely on the observation of  
$D^0 - \bar D^0$ oscillations cannot be viewed as conservative.   

If data revealed $y_D \ll x_D \sim 1$\% we would have a strong case to  
infer the intervention of New Physics. If on the other hand  
$y_D \sim 1$\% -- as hinted at by the FOCUS data -- then two scenarios  
could arise:  
if $x_D \leq {\rm few}\times 10^{-3}$ were found, one would infer  
that the $1/m_c$ expansion within the SM yields a correct  
semiquantitative result while blaming the "large" value for  
$y_D$ on a sizeable and not totally surprising violation of  
duality. If, however, $x_D \sim 0.01$ would emerge, we would face a  
theoretical conundrum: an interpretation ascribing this to  
New Physics would hardly be convincing since $x_D \sim y_D$.  

\subsection{Experiments and data} 
\label{OSCDATA}  
%
Traditionally, $\d0d0$ oscillations have been searched for by  means of 
event-counting techniques, i.e., measurements of the wrong vs. 
right sign events\index{wrong vs. right sign events} for a given final state in 
$D$ decays, 
which are then related to the parameter 
$r^D_{WS}(f)$ (or $\chi^D_{WS}(f)$) defined in  Eq.(\ref{RDOSC}). 
The first search was undertaken in 
deep inelastic  neutrino nucleon scattering. {\em Opposite-}sign dimuon 
events reflect  charm production:  
$\nu N \to \mu^- D^0 X \to \mu^-\mu^+ X^{\prime}$.  
{\em Like-}sign dimuon events then are 
a signature  of charm production accompanied by $\d0d0$ oscillations:  
$\nu N \to \mu^- D^0 X \Rightarrow \mu^- \bar D^0 X\to \mu^-\mu^- 
X^{\prime}$. Background due to associated charm production is suppressed 
and furthermore leads to a very different spectrum for the  
secondary muon:  
$\nu N \to \mu^- D \bar D X \to \mu^- \mu^- DX^{\prime}$.   
Tantalizing evidence for $\d0d0$ oscillations was seen by MARK III
\cite{Schindler:1985pm}, where one event consistent with 
\beq 
 e^+e^- \to \psi ^{\prime \prime}(3770) \to (K^+\pi^-)_D (K^+ \pi^- \pi^0)_D 
\label{MARKIIIDD1}
\eeq  
and two events consistent with
\beq 
e^+e^- \to \psi ^{\prime \prime}(3770) 
 \to (K^+\pi^-\pi^0)_D (K^+ \pi^- \pi^0)_D 
\label{MARKIIIDD2}
\eeq
emerged from a 162 events sample. Background due to 
{\em doubly}-misidentified decays was estimated in $0.4\pm 0.2$
 events. The event in Eq.(\ref{MARKIIIDD1}) could be attributed to DCS decay 
 $D^0\to K^+\pi^-$.
Due to the fact that an analysis of $K (n\pi)$ resonant substructure showed that
the nonresonant 
component was very small, i.e., all threebody $D$ decays effectively lead to
quasi-twobody pseudoscalar-vector final states, the $K^+ \pi^- \pi^0$
combinations could reasonably be assumed as  coming from $K^{*0}\pi^0$ and, as
such, could not be attributed to DCSD. 
Instead, they were considered as 
candidates for $\d0d0$ mixing, although with a vanishing
statistical significance:
taking into
account the estimated  background, removing the DCS-compatible decay and
accounting for fluctuations of the nonzero background in the invariant mass
spectrum, one was left with a little more than a 
single signal event, corresponding to an alleged
mixing rate of $1.6\,$\%.
 Alas -- oscillations could not be established with one event. 
\par
Immediately after the MARK III claim, 
on the other side of the ocean, ARGUS at DESY \cite{Albrecht:1987ei}
was able to set a
$<1.4$\%(90\% cl) limit on charm mixing or DCS decay rates, 
based on a sample of 162 correct-sign events, and
zero wrong-sign events. 
\par
In these searches the important ingredient missing was
the ability to vertex $D^0$ decays and to determine the time evolution. 
Without such capability, one can distinguish between DCS decays and genuine oscillations 
only by employing quantum correlations\index{quantum correlations} as described above, 
which requires more statistics. 
\par
Recent advances in event statistics as well as vertexing $D^0$ decays have 
allowed to look for the specific time signature of oscillations and to probe 
$x_D$ and $y_D$ separately. 
First measurements exploiting lifetime information came from 
fixed-target experiment E691 \cite{E691osc} and from ALEPH at LEP \cite{Barate:1998uy}. 
With further
improvements in detector performances, and most importantly with the geometrical
increase in reconstructed charm sample size, attempts were performed in
measuring mixing via the $y$ parameters, i.e., by measuring lifetime asymmetries
of CP-conjugate eigenstates. See Tab.\ref{TAB:MIXSYN} 
for a synopsis of mixing results, and \cite{Cheunghf9,Grothe:2003zg} 
for excellent
recent reviews.
Present results from B-factories have pushed the limits on mixing parameter $r$ 
down to the
0.1\% limit, and the sensitivity on y beyond the 1\% level. 
 \par
 We now discuss briefly the experimental techniques for unveiling oscillation 
 parameters and conclude with a discussion of the data presently available, and an 
outlook of the future.
%
\subsubsection{Wrong sign vs right sign counting }  

\begin{figure}[t] 
  \centering 
   \includegraphics[width=10.0cm]{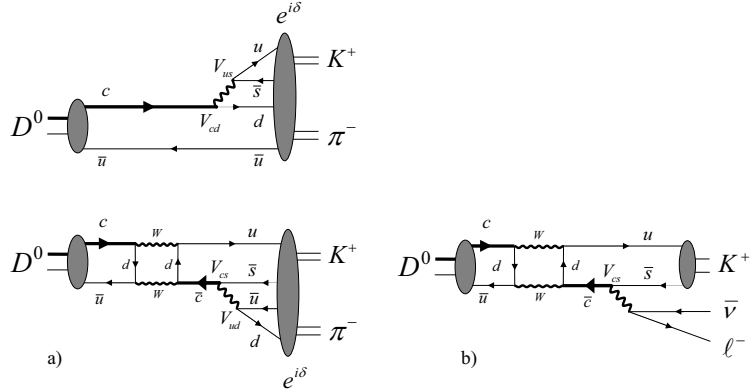} 
  \caption{\it 
  Cartoon which explains 
  at a quark-level (a) the two routes (DCS decays and mixing)
  to get to $K^+\pi^-$ starting from a $D^0$; (b) the one route (mixing) which
  is possible for semileptonic final state. 
    \label{FIG:RMIX} } 
\end{figure}

Searches for $r_{WS}^D(f)$, Eq.(\ref{RDDEF}), benefit greatly from the 
`$D^*$ tag trick' \index{$D^*$ tag},  which allows to tag the flavour of the 
neutral $D$ meson originating from a $D^*$ decay by the accompanying `slow' pion; 
the flavour at decay is tagged by a charged kaon or lepton: 
 \bea 
   D^{*+} & \rarr & D^0 \pi^+     \nonumber        \\ 
          &       & D^0    \Rightarrow\bar{D}^0  \nonumber        \\ 
          &       &   \bar{D}^0 \rarr  K^+\pi^-, 
          K^+\pi^-\pi^+\pi^-, K^+\ell^- \bar{\nu}_\ell  
\eea 
 In the case of nonleptonic decays, the interpretation of the data is  
complicated by the fact that the selection rule $\Delta C = \Delta S$ can 
be violated also by doubly Cabibbo  
suppressed transitions, whose relative branching ratio is given by  
$\tan^4\theta_C \sim 3 \cdot 10^{-3}$:    
\bea 
  D^{*+} & \rarr & D^0 \pi^+     \nonumber        \\ 
         &       &   D^0 \rarr  K^+\pi^-, 
         K^+\pi^-\pi^+\pi^- 
\eea 
 Fig.\ref{FIG:RMIX} shows a pictorial quark level illustration 
of the interplay between oscillations and DCSD processes. The   
wrong-sign events receive contributions  
from DCSD, $\d0d0$ oscillations (followed by a Cabibbo 
allowed decays) and the interference  between the 
two \cite{Blaylock:1995ay,BIGIBERKELEY} and that those can be  
distinguished and thus measured separately  by their dependence on the 
(proper) time of decay, as stated above in Eq.(\ref{DKPIOSCLOCAL}):  
\beq
\frac{{\rm rate}(D^0(t) \to K^+\pi^-)}{{\rm rate}(D^0(t)\to K^-\pi^+)} 
 =  \frac{|T(D^0\to K^+\pi^-)|^2}{|T(D^0\to K^-\pi^+)|^2}  \cdot  
\label{EQ:WS}
\eeq
\beq   
[X_{K\pi}+Y_{K\pi}(t\Gamma_D) +  
Z_{K\pi}(t\Gamma_D)^2)]  
\eeq  
\bea  
X_{K\pi} & \equiv & 1   
\nonumber \\ 
Y_{K\pi} & \equiv &  
\frac{y_D}{{\rm tg}^2\theta_C }  
{\rm Re}\left(\frac{q}{p}\hat \rho_{K\pi}  
\right)  + \frac{x_D}{{\rm tg}^2\theta_C}  
{\rm Im}\left(\frac{q}{p}\hat \rho_{K\pi}  
\right)    
\nonumber\\  
 Z_{K\pi} & \equiv & \left| \frac{q}{p}\right|^2  
\frac{x_D^2 +y_D^2}{4{\rm tg}^4\theta_C} |\hat \rho _{K\pi}|^2   
\eea  
$X$ and $Z$ represent the DCSD and  $\d0d0$ terms, respectively, and 
$Y$ their interference.  The latter receives a nonzero contribution from  
Im$\left(\frac{p}{q}\frac{\hat \rho_{K\pi}}{|\hat \rho_{K\pi}|}\right)$,  
if there is a {\em weak} phase, which leads to CP violation as discussed  
later, and/or if a {\em strong} phase is present due to different FSI  
in $D^0\to K^+\pi^-$ and $\bar D^0\to K^+\pi^-$. One has to allow for  
such a difference since the latter is a pure $\Delta I=1$ transition, 
while the former is given by a combination of an enhanced  
$\Delta I=0$ and a suppressed $\Delta I=1$ amplitude.

Assuming CP conservation, i.e. the absence of weak phases, implies  
$T(D^0 \to K^+\pi^-) = T(\bar D^0 \to K^-\pi^+)$ and  
$|q/p|=1$. As explained later the phase of $q/p$ is actually unphysical  
and can be absorbed into the definition of $\bar D^0$ in  
$\frac{q}{p}\frac{T(\bar D^0 \to K^+\pi^-)}{T(D^0 \to K^+\pi^-)}$;  
the latter quantity can then be written as  
\beq  
\frac{q}{p}\frac{T(\bar D^0 \to K^+\pi^-)}{T(D^0 \to K^+\pi^-)} =  
e^{i\delta}\frac{1}{{\rm tg}^2\theta_C} \hat \rho(K\pi)  
\eeq 
leading to the simplified expression  
\beq  
r_{WS}(t) = \frac{|T(D^0\to K^+\pi^-)|^2}{|T(D^0\to K^-\pi^+)|^2}  
\nonumber
\left[  
1 + \frac{(x_D^2+y_D^2)(\Gamma_Dt)^2} 
{\frac{|T(D^0\to K^+\pi^-)|^2}{|T(D^0\to K^-\pi^+)|^2}} -  
\frac{y_D^\prime(\Gamma_Dt)} 
{\frac{|T(D^0\to K^+\pi^-)|}{|T(D^0\to K^-\pi^+)|}} 
\right]  
\eeq 
where  
\beq 
y_D^\prime \equiv y_D\cos\delta-x_D\sin\delta \, ,\; x_D^\prime\equiv 
x_D\cos\delta+y_D\sin\delta \;  
\label{EQ:DELTA} 
\eeq 
with $x_D^2 + y_D^2 \equiv (x_D^\prime)^2 + (y_D^\prime)^2$. The observable  
ratio of wrong- to correct-sign events as a function of the (proper)  
time of decay $t$ is thus expressed in terms of the branching ratio for 
the doubly Cabibbo suppressed mode, the oscillation parameters  
$x_D$, $y_D$ and the strong phase $\delta$.  

Finding $Y_{K\pi}$ and/or $Z_{K\pi}$ to  
differ from zero unequivocally establishes the presence of oscillations.  
It would also allow us to extract $x_D^2+y_D^2$ and  
$y_D^\prime$ -- yet not $x_D$ and $y_D$ separately due to our ignorance  
concerning the strong phase $\delta$.  One can extract $\delta$ in a clean way, 
namely by exploiting   
the {\em coherence} of a $D^0\bar D^0$ pair produced in $e^+e^-$ 
annihilation  close to threshold. We already stated that  
\beq  
\frac{\sigma (e^+e^- \to D^0 \bar D^0 \to  
(K^{\pm}\pi^{\mp})_D (K^{\pm}\pi^{\mp})_D} 
{\sigma (e^+e^- \to D^0 \bar D^0 \to  
(K^{\pm}\pi^{\mp})_D (K^{\mp}\pi^{\pm})_D} \simeq  
\frac{x_D^2 + y_D^2}{2} \; ;  
\eeq 
i.e., this transition can occur only due to $\d0d0$ oscillations.  
EPR correlations produce a quite different ratio between these final  
states when the underlying reaction is  
$e^+e^- \to D^0\bar D^0 \gamma$ (due to $D^{0*}\bar D^0$,  
$\bar D^{0*} D^0$ $\to D^0\bar D^0 \gamma$), since now the  
$D^0\bar D^0$ pair forms a {\em even} configuration:  
$$  
\frac{\sigma (e^+e^- \to D^0 \bar D^0 \gamma \to  
(K^{\pm}\pi^{\mp})_D (K^{\pm}\pi^{\mp})_D\gamma } 
{\sigma (e^+e^- \to D^0 \bar D^0 \gamma \to  
(K^{\pm}\pi^{\mp})_D (K^{\mp}\pi^{\pm})_D\gamma } \simeq 
$$ 
\beq  
\frac{3}{2}\left( x_D^2 + y_D^2\right) +  
4\left|  
\frac{T(D^0 \to K^+\pi^-)}{T(D^0 \to K^-\pi^+)} 
\right|^2  + 8 y_D {\rm cos} \delta \left|  
\frac{T(D^0 \to K^+\pi^-)}{T(D^0 \to K^-\pi^+)} 
\right| 
\label{DELTAEX1} 
\eeq 
Alternatively one can measure  
$$  
\sigma (e^+e^- \to D^0 \bar D^0 \to  
(K^{\pm}\pi^{\mp})_D (K^{\mp}\pi^{\pm})_D)  
\propto  
$$  
\beq  
\left(  
1-\frac{x_D^2 - y_D^2}{2} 
\right) \left(  
1 - 2{\rm cos}\delta \left|  
\frac{T(D^0 \to K^+\pi^-)}{T(D^0 \to K^-\pi^+)} 
\right|^2  
\right)  
\label{DELTAEX2} 
\eeq 
Eq.(\ref{DELTAEX1}) or Eq.(\ref{DELTAEX2}) coupled with the 
previously described analyses of single $D$ decays allow to extract  
$x_D$, $y_D$ and $\delta$.  
 \par 
\par 
In the search for oscillationns from wrong-sign events,  the 
event selection procedure 
at fixed-target experiments requires a good candidate secondary vertex 
consistent with the $D^0$ mass, a suitable primary vertex consisting of a  
minimum 
number of other tracks, and well isolated from the secondary vertex. The main 
background to the WS signal is due to  
doubly misidentified $K^+\pi^-$ pairs from $D^0$ decays which form a broad  
peak 
directly under the $D^0$ signal in $K^+\pi^-$ and  a narrow peak in the  
$D^*-D$ 
mass difference  signal region. The mass difference background is 
indistinguishable  from the real WS tagged signal. To eliminate this 
background, the $K\pi$ invariant mass is computed with the kaon and pion 
particle hypotheses swapped. Any candidate whose swapped mass is within 
some distance (a few $\sigma$'s) of the $D^0$ mass is subjected to a cut on  
the sum of the 
$K\pi$ separations for both tracks.  
Finally, all tracks in the production vertex are tested as potential 
$\pi$ candidates, and are accepted if within a narrow (typically, $\pm 
50\,{\rm MeV}/c^2$ ) 
window of the nominal $D^*-D^0$ mass difference, and if they satisfy a loose 
particle identification cut.  
\par 
Experiments at $\epem$ colliders follow similar reconstruction strategies,  
with 
the important difference being the superior resolution on the $D^*-D^0$ mass 
difference (typically 200~keV). A cut on the angle of the pion candidate in  
the 
$D^0$ rest frame with respect to the $D^0$ boost rejects asymmetric $D^0$ 
decays, where the pion candidate has low momentum. Another relevant difference 
with fixed-target experiments is the fact that the primary vertex is reconstructed over 
blocks of data as the centroid of the luminous $\epem$ interaction region. The 
vertical extent of such region is only about $10\mu m$, and this permits to 
reconstruct the proper time $t$ using only the vertical component of the  
flight 
distance of the $D^0$ candidate. 
\par 
For both FT and $\epem$ experiments, the WS mass peak is fitted in the Q-M  
plane, i.e., the 
scatter plot $M(D^*-D^0) \quad vs \quad M(D^0)$. Lifetime information is 
extracted on the proper time distribution of WS candidates within a few sigmas 
from the CFD signal value. Due to the superior lifetime resoultion, the 
distribution of proper times is an exponential at FT, and an exponential 
convoluted with a gaussian resolution function for $\epem$ experiment. 
The dominant contribution to the systematic error comes from uncertainties in 
the shapes and acceptances of backgrounds. 
\par 
The alternative option in counting techniques is to use semileptonic, 
 $D^*$-tagged final states $K \ell \nu$; they do not suffer from DCSD pollution, but are 
 harder experimentally. For semileptonic final states, Eq.(\ref{EQ:WS}) reduces  
to 
\bea 
 r_{WS}^D(lX) \propto \frac{r_D}{2}t^2 e^{-t}     
 \label{EQ:WSSL} 
\eea 
Event selection and vertex reconstruction techniques follow guidelines common 
 to all semileptonic studies, where the presence of an undetected neutrino 
 prevents one from reconstructing the $D^0$ momentum directly. By exploiting the 
 information on the primary and secondary vertices, the $K$ and lepton momenta, 
 and assuming a $D^0$ parent mass, one can reconstruct the neutrino 
 momentum modulo a two-fold ambiguity. Finally the invariant mass of $D^*$ and 
 proper decay time of $D^0$ are computed, and events are selected in the Q-M 
 plane in analogy to the hadronic case. 
Feedthrough of hadronic modes is mainly by $K^-\pi^+\pi^0$ with an undetected 
 $\pi^0$ faking the neutrino, and a pion is misidentified as a lepton. 

\subsubsection{Lifetime difference measurements}  
\label{LIFEDIFF}

In the presence of $D^0 - \bar D^0$ oscillations a single lifetime  
does not suffice to describe all transitions. The situation can most 
concisely be discussed when CP invariance is assumed --   
$[H_{\Delta C\neq 0},CP]=0$ -- since the mass eigenstates are then CP  
eigenstates as well (CP violation will be addressed in the next Section). 
Transitions $D\to K^+K^-$, $\pi^+\pi^-$ are  
controlled by the lifetime of the CP even state \cite{LIU} and  
$D\to K_S\pi^0$, $K_S\eta$ by that of the CP odd state. 
\par
The 
channel $D^0 \to K_S\phi$ has a CP odd final state; yet in 
$D^0 \to K_SK^+K^-$ one has to disentangle the $\phi$ and 
$f^0$ contributions in $K^+K^-$, since the latter leads to a 
CP even state.  This complication and its implications for 
$B \to K_S \phi$ will be addressed again in our discussion 
of CP violation.
\index{$D^0\to K_s\phi$ and $B^0\to K_s\phi$} 
As an example of the problem, it has been shown \cite{Pedrini02Frontier},
 out of
photoproduction data, 
 the fraction of $f_0$ component in  $D^0 \to K_SK^+K^-$ being as large as
  $37.8\pm 3.0$\% vith a relatively loose mass cut
   $M(KK)^2 < 1.1 GeV^2$, which is reduced 
 to 8\% with a narrower mass cut $(1.034 < M^2 < 1.042)\,GeV^2$,
 with the obvious penalty of    a very large reduction in statistics.
      BABAR has shown      \cite{Aubert:2002yc}
       preliminary results where the $f_0/\phi$ ratio is about 25\% integrated
       over the entire range of $M^2$.
\par
The final state  
$D\to K^-\pi^+$ on the other hand has no definite CP parity. Up to terms  
of order $(\Delta \Gamma /\bar \Gamma)^2$ its time dependance is 
controlled by the average lifetime $\bar \tau = \bar \Gamma ^{-1}$\cite{LIU} and 
thus  
\beq  
y_D \simeq y^D_{CP} = \frac{\Gamma _{K^+K^-} - \Gamma _{K^{\mp}\pi^{\pm}}} 
{\Gamma _{K^{\mp}\pi^{\pm}}} =  \frac{\tau(D\rarr K\pi)}{\tau(D\rarr KK)} -1
\eeq  
\begin{table} 
\caption{Synopsis of recent $\d0d0$ mixing results. CPV phase is 
$\varphi$,  strong 
 phase is $\delta$, interference angle is 
$\phi=\arg(ix+y)-\varphi-\delta$. As instance, $\varphi \neq 0$ stands for no CP
conservation is assumed.
  \label{TAB:MIXSYN} 
 } 
 \footnotesize 
 \begin{center}  
\begin{tabular}{|lllrr|} \hline 
                      & Assumptions    & Mode      &$N_{RS}$&Result (\%)  
\\  
\hline  
{\bf E691 88}     \cite{Anjos:1987pw}      & 
             $\varphi=0,\cos\phi=0$               &
		  $K\pi$           & 
		  1.5k              & 
  $r< 0.5$                   \\
     {\it (95\%~CL)}   & 
                       &
		  $K\pi\pi\pi$           & 
		             & 
  $r< 0.5$                   \\  
                      &
                                &              
                                   & 
        &
	combined: $r<0.37$  \\  
	\hline
{\bf ALEPH}\cite{Barate:1998uy}   & No mix         & $K\pi$    & 
                      1.0k 
                      &$r_{DCS}=1.84\pm .59 \pm .34$                      
\\  {\it (95\%~CL)}       & $\varphi=0,\cos\phi=0$ &           & 
                      &$r<0.92$                                           
\\ 
                      & $\varphi=0,\cos\phi=+1$ &          & 
                      &$r<0.96$                                          
\\ 
                      & $\varphi=0,\cos\phi=-1$ &          & 
                      &$r<3.6$                                            
\\  
\hline  
{\bf E791}\cite{Aitala:1996vz}    & $\varphi=0$    
  & $K\ell\nu$   & 
2.5k 
                      & 
  $r=0.11^{+0.30}_{-0.27}$              \\ 
{\it (90\%~CL)}       &                
   &                 & 
                   &                                                  
        $ r<0.50$\\  
\hline  
{\bf E791}\cite{Aitala:1998fg}    & No mix         & $K\pi$    & 5.6k 
                      & $r_{DCS}=0.68^{+0.34}_{-0.33}\pm 0.07$           
\\  {\it (90\%~CL)}       & No mix         & $K3\pi$   & 3.5k 
                      & $r_{DCS}=0.25^{+0.36}_{-0.34}\pm 0.03$           \\ 
                      &$\varphi\neq 0$ in $y$&           & 
                      & $r=0.39^{+0.36}_{-0.32}\pm 0.16$    \\ 
                      &                 &           & 
                      & $r<0.85$    \\ 
\hline  
{\bf E791}\cite{Aitala:1999nh}    & $\varphi=0,\delta=0$ & $KK$    & 6.7k 
                      & $\Delta\Gamma=0.04\pm0.14\pm0.05\,{\rm 
ps}^{-1}$   \\  {\it (90\%~CL)}       &                &    $K\pi$  & 60k 
                   & $(-0.20<\Delta\Gamma<0.28)\,{\rm ps}^{-1}$         
\\ 
                     &                &              & 
                      & $y=0.8\pm 2.9 \pm 1.0 \quad (-4<y<6)$ \\  
\hline  
{\bf CLEO 00}\cite{Godang:1999yd}       & 
                 No mix            &
		  $K\pi$           & 
		  14k              & 
  $r_{DCS}=0.332^{+0.063}_{-0.065} \pm 0.040$                  \\
    {\it (95\%~CL)}                &
     $\varphi\neq 0,\delta\neq0$       &              
                                   & 
       \multicolumn{2}{l|}{ $r_{DCS}=0.47^{+0.11}_{-0.12} \pm 0.01$}      \\  
             $9\,{\rm fb}^{-1}$    &
	       &               
	                           & 
 \multicolumn{2}{l|}{ $y^\prime =-2.3^{+1.3}_{-1.4}\pm 0.3 \,(-5.2 < y^\prime
 < 0.2)$ }           \\          
                                    & 
                                    &
                                   & 
 \multicolumn{2}{l|}{ $x^\prime =0\pm 1.5 \pm 0.2 \, (-2.8< x^\prime
 < 2.8) $}       \\
                                    & 
                                    &
                                   & 
 \multicolumn{2}{l|}{ $(x^\prime)^2/2<0.038$}       \\  
 \hline
{\bf FOCUS 00}\cite{Link:2000cu}           &
              $\varphi=0,\delta=0$ & 
	      $K\pi$             &
	       120k               &                   \\  
                             &
                                   &  
		       $KK$        & 
		       10k        &
 $y=3.42\pm 1.39\pm 0.74 $                    \\  
\hline
{\bf FOCUS 01}\cite{Link:2001cx}           &
              $\varphi\neq 0$ & 
	      $K\pi$             &
	       36.7k               & 
	       $-12.4<y^\prime<-0.6$                  \\  
          prelim.                   &
                                   &  
		          & 
		               &
 $|x^\prime|<3.9 $                    \\ 

\hline
{\bf FOCUS 02}\cite{MalvezziICHEP02}           &
              $\varphi\neq 0$ & 
	      $K\ell \nu$             &
	       60k               & 
	stat.err. only       $r<0.12$                  \\  
          prelim.                   &
                                   &  
		             & 
		            &
                    \\ 
\hline
{\bf CLEO 02}\cite{Csorna:2001ww}    & 
 $\varphi=0,\delta=0$               & 
 $\pi\pi$                           & 
  710                              &    
      $A_{CP}(KK)=0.0\pm 2.2\pm 0.8$             \\
			   &
                                    &
	    $KK$                    & 
	    1.9k                    &
	 $A_{CP}(\pi\pi)=1.9\pm 3.2\pm 0.8$                   \\  
          $9\,{\rm fb}^{-1}$       & 
	                             & 
	   $K\pi$                    &
	    20k                      &                                                   
               $y=-1.2\pm 2.5 \pm 1.4$                             \\  
 \hline
{\bf BELLE 02}\cite{Abe:2001ed}           &
              $\varphi=0,\delta=0$ & 
	      $K\pi$             &
	       214k               & 
	                                       \\  
        $23.4\,{\rm fb}^{-1}$                     &
                                   &  
		       $KK$        & 
		       18k        &
 $y=-0.5\pm 1.0^{+0.7}_{-0.8} $                    \\
 \hline
{\bf BABAR 03}\cite{Aubert:2003pz}    & 
 $\varphi=0,\delta=0$               & 
 $\pi\pi$                           & 
  13k                              &   
                       $y=0.8\pm 0.4^{+0.5}_{-0.4}$             \\
			   &
                                    &
	    $KK$                    & 
	    26k                    &
 $\Delta y=-0.8\pm 0.6 \pm 0.2$       
	            \\  
          $91\,{\rm fb}^{-1}$       & 
	                             & 
	   $K\pi$                    &
	    265k                      &                                                   
                                            \\
\hline
{\bf BABAR 03}\cite{Aubert:2003ae}       & 
                 No mix            &
		  $K\pi$           & 
		  120k              & 
  $r_{DCS}=0.357\pm 0.022  \pm 0.027$                  \\
    {\it (95\%~CL)}                &
                                &              
                                   & 
       \multicolumn{2}{l|}{  $A_{CP}(K^+\pi^-)= 9.5\pm 6.1\pm 8.3 $}     \\  
             $57.1\,{\rm fb}^{-1}$    &
	Nomix, $\varphi\neq 0,\delta\neq0$        &               
	                           & 
 \multicolumn{2}{l|}{ $ r_{DCS}=0.359\pm 0.020  \pm 0.027  $ }           \\        
                &
	$\varphi\neq 0,\delta\neq0$        &               
	                           & 
 \multicolumn{2}{l|}{ $-5.6 < y^\prime < 3.9$ }           \\   
                &
	      &               
	                           & 
 \multicolumn{2}{l|}{ $x^{\prime 2}<0.22, \quad r<0.16$ }           \\
                &
	$\varphi =0,\delta\neq0$        &               
	                           & 
 \multicolumn{2}{l|}{ $-2.7 < y^\prime < 2.2$ }           \\   
                &
	      &               
	                           & 
 \multicolumn{2}{l|}{ $x^{\prime 2}<0.2, \quad r<0.13$ }           \\
 \hline
{\bf BELLE 03}\cite{YABSLEY03}    & 
 $\varphi=0,\delta=0$               & 
 $KK$                           & 
  36.5k                              &   
                       $y=1.15\pm 0.69 \pm 0.38$             \\
          $158 \,{\rm fb}^{-1}$       & 
	                             & 
	   $K\pi$                    &
	    448k k                      &                                                   
                                            \\					    
\hline   
\end{tabular} 
  \vfill 
 \end{center}  
\end{table} 
Several sources of systematic errors common to $KK$ and $K\pi$ 
final states cancel 
in the ratio of their lifetimes and thus in $y^D_{CP}$. The strategy common to  
fixed target and $\epem$ experiments is to select high-statistics, clean 
modes, for which $D \to K^+K^-$ and $K^-\pi ^+$ are prime examples. 
For the latter requirement, a $D^*$-tag 
\index{$D^*$ tag}
is of paramount importance.  
Many of the analysis cuts and the fitting strategies have close analogies with  
those employed in lifetime measurements (see Sect.~\ref{LIFE}). In all  
measurements 
 of $y^D_{CP}$, the systematic error only refers to the lifetimes asymmetry and 
 not to the absolute lifetime of the CP eigenstates. As such, lifetimes of, 
 i.e.,   $\tau(D\rarr K\pi)$ have a statistical error only.  

{\em Measuring $y$ at fixed target experiments:}  
  The cuts used to obtain a clean signal are designed to produce a nearly 
  flat efficiency in reduced proper time $t^\prime \equiv (\ell - N 
  \sigma)/(\gamma\beta c )$. The charm secondary is selected by means of a 
  candidate driven algorithm, with stringent requests on particle 
  identification, as well as requiring a minimum $\sigma_\ell$ detachment 
  between primary and secondary vertex. To select a clean sample, 
  either a  $D^*$ tag is required, or a set of 
  more stringent cuts, such as more stringent Cerenkov requirements on kaons 
  and pions, momenta of decay particles balancing each other, 
  primary vertex inside the target material, and resolution of proper time 
  less than 60~fs. The $D^*$ tagged sample has a better signal-to-noise ratio, 
  while the inclusive sample accommodates larger sample size. Both samples are 
  treated as systematics checks. 
\par 
\begin{figure}
\begin{center} 
       \includegraphics[width=6cm]{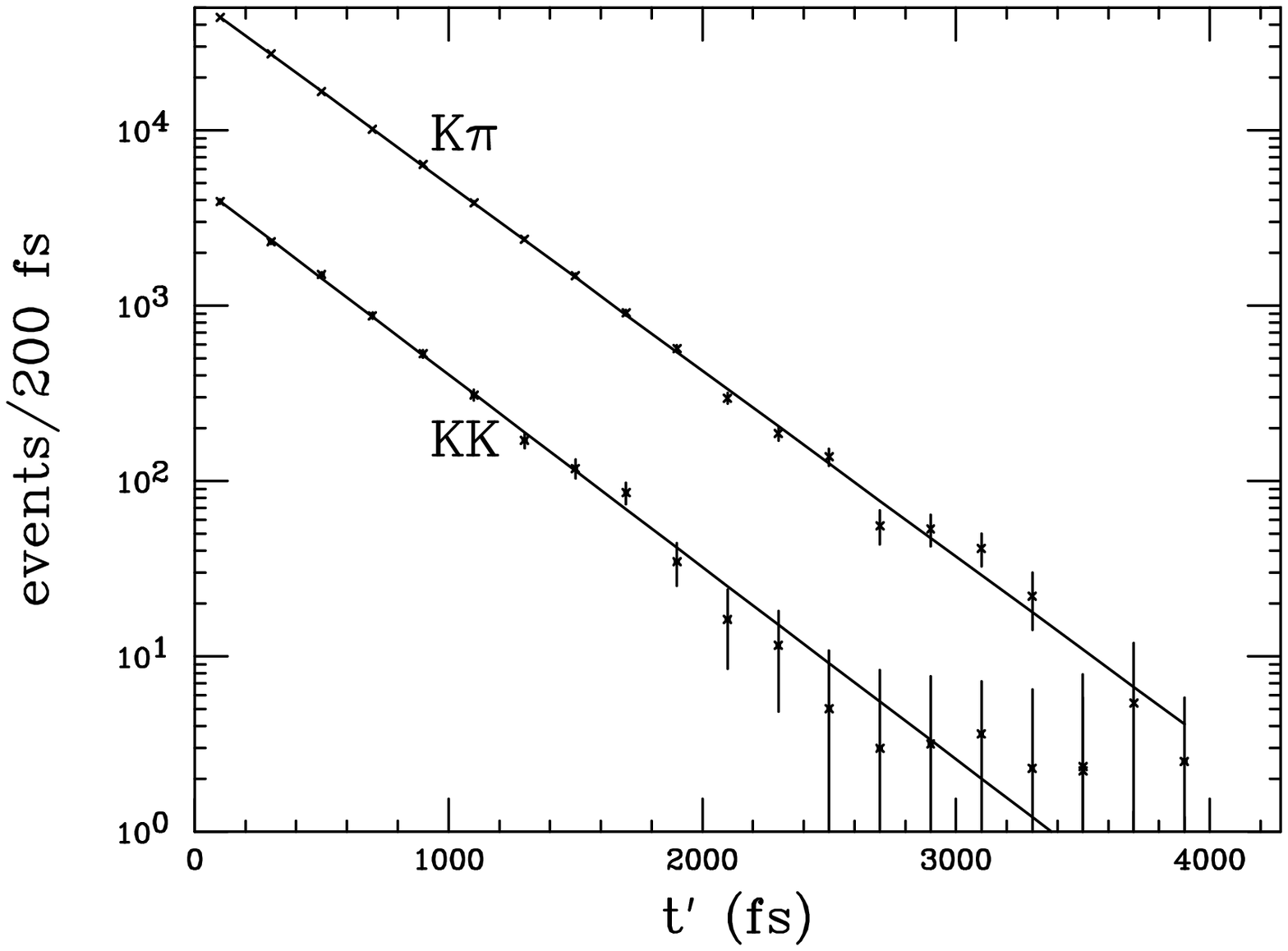} 
     \includegraphics[width=6cm]{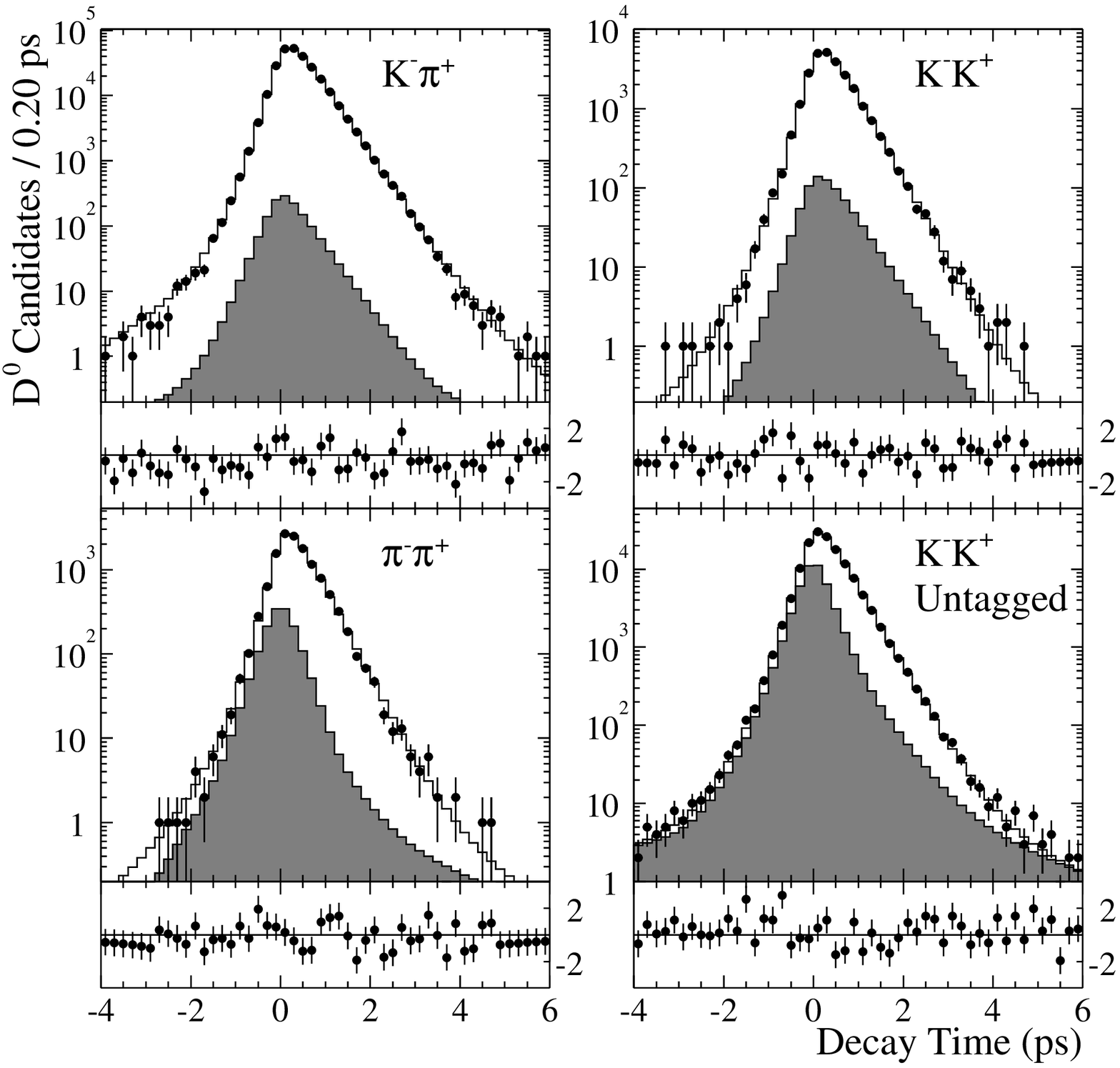} 
\end{center} 
\caption{ 
 Lifetime distributions used to extract lifetime asymmetry $y_{CP}$ (left)
 Fixed-target experiment FOCUS \cite{Link:2000cu} (right) $\epem$
experiment BABAR \cite{Aubert:2003pz}.
} 
\label{FIG:SIGNALS} 
\end{figure} 
\par
The $D^0\rarr K^-K^+$ sample is characterized by a prominent reflection 
background coming from misidentified $D^0\rarr K^-\pi^+$ decays 
(Fig.~\ref{FIG:SIGNALS}). 
 The amount of  $D^0\rarr K^-\pi^+$ reflection is obtained  
by a mass fit to the $K^-K^+$ sample and the shape of the reflection is 
deduced from a high-statistics montecarlo sample. It is assumed that the time 
evolution of the reflection is described by the lifetime of $D^0\rarr 
K^-\pi^+$ and a fit is performed of the reduced proper time distributions of  
the 
$D^0\rarr K^-\pi^+$ and $D^0\rarr K^-K^+$ samples at the same time. The fit 
parameters are the $D\rarr K\pi$ lifetime, the lifetime asymmetry $y_{CP}$, 
and the number of background events under each  
$D^0\rarr K^-\pi^+$ and $D^0\rarr K^-K^+$ signal region. The signal 
contributions for the $D^0\rarr K^-\pi^+$, $D^0\rarr K^-K^+$ and the 
reflection from the misidentified $D^0\rarr K^-\pi^+$ in the reduced proper 
time histograms are described by  a term 
 $f(t^\prime)\, \exp(-t^\prime/\tau) $
in the fit likelihood function. The function $f(t^\prime)$, determined from  
montecarlo,  covers any  
deviation of the reduced proper time distribution from a pure exponential 
due to acceptance.
 The background yield parameters are either left floating, or fixed to the 
number of events in mass sidebands using a Poisson penalty term in the fit 
likelihood function.  
\par
The systematic error on the lifetime asymmetry is determined by calculating the shifts
in $y_{CP}$ for a set of detachment cuts, kaon identification cuts, background
normalizations, and lifetime fit ranges.
\par
\begin{figure}
 \begin{center} 
       \includegraphics[width=10cm]{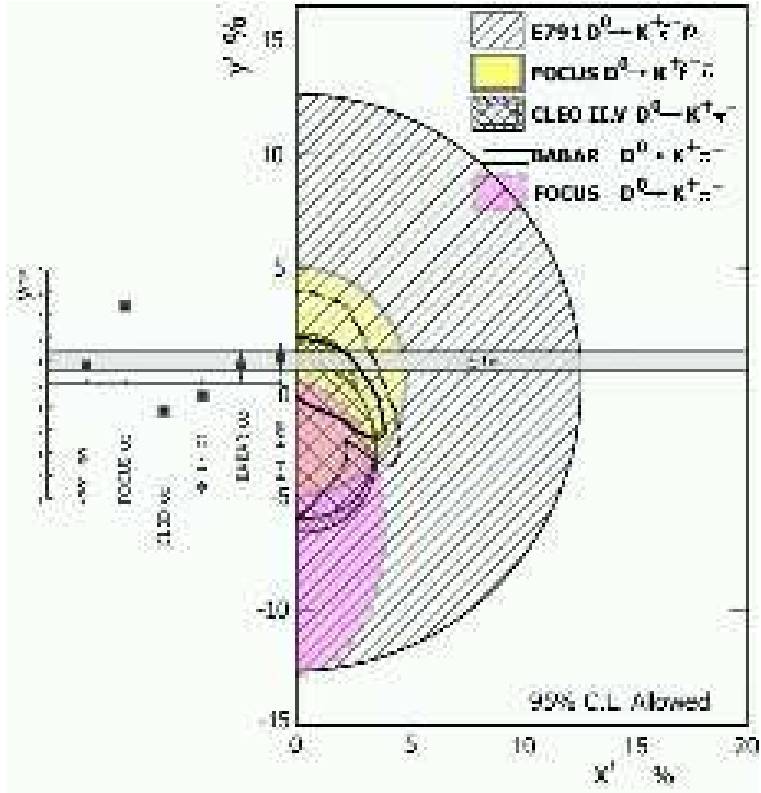} 
 \end{center}    
 \caption{Summary of $x', y', y_{CP}$ measurements. The BABAR
  limits (solid contour: CP conservation assumed; dashed contour: CP
  conservation not assumed)  were scaled to 
 $x^\prime$ from Ref.\cite{Aubert:2003ae},
 and are superimposed for qualitative comparison.} 
 \label{FIG:YVSX} 
\end{figure} 
{\em Measuring $y_{CP}$ at $\epem$ $B$ factories:} 
Event selection and fitting are carried out by CLEO, BABAR and BELLE in a very similar 
fashion to each other and also to fixed-target experiments. A $D^*$-tag is normally used to 
select a clean sample of $D^0$ decays, with the noteworthy exception of BELLE which 
manages to get a sample of equivalent purity thanks to claimed superior particle identification.
\par
Lifetime fits are performed via unbinned maximum likelihood method, where the fit function 
contains proper time signal and background terms. An additional penalty term in the likelihood 
function proportional to the difference to the nominal $D^0$ invariant mass is also used.
\par
Main sources of systematic error on the lifetime asymmetry are the uncertainty in the MC 
correction functions $f(t^\prime)$, and the background contribution to the signals.  

 \subsection{Where do we stand today, and what next?}
The oscillation industry is being revamped by a steady flux of results
(for recent reviews, see \cite{Cheunghf9,Grothe:2003zg}). 
All results so far are consistent with no oscillations. 
FOCUS had stirred up excitement with a $\sim 2\sigma$ evidence
 for positive $y_{CP}$. After several ups and downs (Tab.\ref{TAB:MIXSYN}),
 $y_{CP}$ is dominated by the 2003 preliminary 
 measurements \cite{Aubert:2003pz,YABSLEY03}
\beq
 y_{CP}^{BABAR03}=0.8\pm 0.6  \pm 0.2 \% \qquad 
 y_{CP}^{BELLE03}=1.15\pm 0.69 \pm 0.38  \%
 \nonumber
\eeq
 We 
average the  results in Tab.\ref{TAB:MIXSYN} and we get the world average
\beq
<y_{CP}>^{2003}= (1.0 \pm 0.5)  \% 
\nonumber
\eeq 
In year 2000 CLEO  published
 a limit on WS counting rates of  $r<0.041$\% based on
$9\,fb^{-1}$ data set. BABAR's new limit is about three times higher,
although it is
based on a tenfold dataset (Fig.\ref{FIG:YVSX}). Besides, BABAR contour limits
have same size as CLEO contour limits.
 Possible reasons for such inconsistent
results have been discussed \cite{Grothe:2003zg}; they are  likely due to 
differences in fit and limit computation techniques. 
Particularly intriguing is
the effect of positive $y^\prime$ fit results providing larger limits than those
provided by negative $y^\prime$ fit results.
\par
Although everything is consistent with zero, $y_{CP}$ seems
 to have a tendency to
prefer positive values, while  hadronic mixing measurements sort of favour
laying in the 
 $y^\prime <0$ semiplane. 
Very important measurements will be performed in counting techniques using
semileptonic channels, where a preliminary FOCUS result shows a $r<0.12$\% limit
and results from B-factories should be coming up.
A critical issue is how we can determine experimentally the strong phase shift $\delta$. 
We have already described how it can be extracted from 
$e^+e^- \to \psi(3770)\to D^0 \bar D^0$ as will be studied by CLEO-c. 
Alternatively one can infer it from $D\to K_L\pi$ decays \cite{GOLOPAK}. 
\par
The future will bring a rich harvest of new data.  $B$ factories plan to
multiply by five their 
 dataset by 2006. The statistical error on $y_{CP}$ will then be negligible with
 respect to the systematic error, and a 0.2\% precision will be reached.
 Similarly for hadronic mixing which should get to below the 0.5\% level. At
 CLEO-c oscillation will be investigated under novel coherence conditions, where
 $\delta$ can be measured independently \cite{TAUCHARM,GGR}.

\subsection{Resume} 
\label{RESDDOSC} 

Table \ref{TAB:MIX} summarizes our present experimental 
knowledge on meson-antimeson 
oscillations. In the second column we have indicated the relative weight 
{\em within the SM} between contributions to $\Delta M(P^0)$ from 
short distance (SD) and long distance (LD) dynamics. We have also translated 
$x$ into $\Delta M(P^0)$ for an absolute yardstick.   
$K^0 -\bar K^0$ and $B_d - \bar B_d$ oscillations 
have been established and well measured. The observed values for 
$\Delta M(K)$ and $\Delta M(B_d)$ can be reproduced by 
SM dynamics within reasonable theoretical uncertainties. 
\par
Experimental evidences in nonleptonic $D^0$  decays can bear on the discussion
of the experimental determination of strong phase $\delta$ in hadronic $\d0d0$
mixing. Isospin decomposition of  $D^0\to KK, \pi\pi$ decays is used to
determine the phase shift between isospin amplitudes\cite{Link:2002hi}. Phase
shift is large, this being a signal for large FSI. 
A further attempt is made by authors of \cite{Link:2002hi} to estimate the
relative importance of elastic and inelastic FSI. They find that the elastic FSI
cannot account for all the discrepancy found between experiment and theory
predicted value of the branching ratio, and they argue that the explanation may
be the presence of an inelastic FSI component acting in 
the transition $KK\to \pi\pi$. The inelastic FSI component is a smoking gun for
a sizable strong angle $\delta$ which affects also the $K\pi$
 final states 
 \par
The experimental sensitivity for $D^0 - \bar D^0$ oscillations 
has been increased considerably  in the last few years 
with no signal established. Yet for proper perspective one should note 
that $D^0$ is the only meson with up-type quarks in this list and that 
furthermore the 
upper bound on $\Delta M_D$ is not much smaller than 
$\Delta M_K$ and $\Delta M(B_d)$.  The best theoretical estimate yields 
$y_D$, $x_D$ $\sim {\cal O}(10^{-3})$; yet a value $\sim 0.01$ cannot be 
ruled out nor is there a clear prospect for reducing the theoretical uncertainties 
significantly. Yet rather then resigning ourselves to `experimental nihilism' 
we advocate exercising sound judgment. It is 
important to measure $x_D$ and $y_D$ separately as accurately as possible. 
Finding, say, $y_D \sim 10^{-3}$ together with $x_D \sim 0.01$ would represent a 
tantalizing  `prima facie' case for the presence of New Physics; on the 
other hand $x_D \sim y_D \sim 0.01$ would suggest that the OPE treatment 
involving a $1/m_Q$ expansion can provide little quantitative guidance here;  
$x_D \sim 10^{-3}$ together with $y_D \sim 0.01$ would point towards a violation of 
quark-hadron duality \index{quark-hadron duality} in $y_D$. 
\par
The observation of $D^0 - \bar D^0$ oscillations thus might or might not signal 
the intervention of New Physics. However as explained in the next Section they can  
create a new stage for CP violation in $D^0$ decays that would unequivocally 
reveal New Physics. Furthermore they can have an important or even crucial impact 
on the CP phenomenology in $B$ decays. One of the most promising ways to 
extract the angle $\phi_3 = \gamma$ of the CKM unitarity triangle is to compare 
$B^{\pm} \to D^{neut}K^{\pm}$ transition rates, where $D^{neut}$ denotes a neutral 
$D$ meson that either reveals its flavour through its decay or it does not. It has been pointed 
out \cite{Silva:1999bd} that even moderate values $x_D, \, y_D \sim 10^{-2}$ -- which 
are not far from the border that SM dynamics can reach -- could have a sizable 
impact on the value of $\phi_3 = \gamma$ thus extracted. Therefore $D^0-\bar D^0$ 
oscillations could either hide the impact New Physics had on 
$B^{\pm} \to D^{neut}K^{\pm}$ -- or fake such an impact!  

\begin{table} 
\caption{Compilation of oscillation  
parameters (95\%\,cl )  
\cite{Hagiwara:fs}. 
\label{TAB:MIX} 
  }  
\footnotesize 
 \begin{center} 
 \begin{tabular}{lcrrr} \hline 
  $P^0$              &  $SD/LD$    & $x$        & $\Delta M(P^0)$  &$y$                \\ 
  \hline 
  $K^0\,(d\bar s)$    & $SD\sim LD$ & $0.946 \pm 0.003$ &  
  $(3.5 \pm 0.003) \cdot 10^{-6}$ eV
  &$0.9963 \pm 0.0036$          
\\ 
  $D^0\,(c\bar u)$    & $SD\ll LD$  & $<0.03$          & 
 $ \leq 4.6 \cdot 10^{-5}$ eV 
  &$-0.002 \pm 0.011$    
\\ 
  $B^0_d\,(d\bar b)$  & $SD\gg LD$  & $0.771\pm 0.012$ & 
  $(3.30 \pm 0.05) \cdot 10^{-4}$ eV
&\?                
\\ 
  $B^0_s\,(s\bar b)$  & $SD\gg LD$  & $>20.0 $      & 
  $\geq 9.3 \cdot 10^{-3}$ eV
  &$<0.15$           
\\ 
  \hline 
 \end{tabular} 
 \vfill 
 \end{center} 
\end{table} 

%
\section{CP violation} 
\label{CPV} 
%
The SM predicts quite small CP asymmetries in charm transitions.  
Therefore searches for CP violation there are mainly motivated  
as probes for New Physics, as it is with $D^0 - \bar D^0$  
oscillations. Yet again the experimental sensitivity has reached the  
very few percent level, and one has to ask how small is small --  
$10^{-2}$, $10^{-3}$, $10^{-4}$? 

Due to CPT invariance (later we add a few comments on CPT violation) 
CP violation can be implemented only through a  
complex phase in some effective couplings.  For it to become  
observable two different, yet coherent amplitudes have to  
contribute to an observable. There are two types of scenarios for 
implementing this requirement: 
\begin{enumerate}
\item 
Two different $\Delta C =1$ amplitudes of fixed ratio --
distinguished by,  say, their isospin content -- exist leading 
{\em coherently} to the same final state. 
\item 
$D^0-\bar D^0$ oscillations driven by $\Delta C=2$ dynamics 
provide the second amplitude, the weight of 
which varies with time. 

\end{enumerate}

\subsection{Direct CP violation} 
\label{DIRECT} 

CP violation appearing in $\Delta C =1$ amplitudes is called  
{\em direct} CP violation. It can occur in the decays of charged and neutral
charm meson and  baryons. 

\subsubsection{Partial widths} 

Consider a final state $f$ that can be reached coherently via two 
different quark level transition amplitudes ${\cal M}_{1}$ and 
${\cal M}_{2}$:  
\beq  
T(D \to f) = \lambda_1 {\cal M}_1  + \lambda_2 {\cal M}_2   
\label{T2M}
\eeq 
We have factored out the weak couplings $\lambda_{1,2}$ while allowing 
the amplitudes ${\cal T}_{1,2}$ to be still complex due to strong or electromagnetic FSI.   
For the CP conjugate reaction one has  
\beq  
T(\bar D \to \bar f) = \lambda_1^* {\cal M}_1 + 
\lambda_2^* {\cal M}_2  
\eeq 
It is important to note that the reduced amplitudes 
${\cal M}_{1,2}$ remain 
unchanged, since strong and electromagnetic forces conserve CP.  
Therefore we find   
\beq  
\Gamma (\bar D \to \bar f) - \Gamma (D \to  f) = 
\frac{2{\rm Im}\lambda_1 \lambda_2^* \cdot {\rm Im}{\cal M}_1{\cal M}_2^*}
{|\lambda_1|^2|{\cal M}_1|^2 + |\lambda_2|^2|{\cal M}_2|^2 
+ 2{\rm Re}\lambda_1 \lambda_2^* \cdot {\rm Re}{\cal M}_1{\cal M}_2^*} 
\label{PARTIAL} 
\eeq 
i.e. for a CP asymmetry to become observable, two  
conditions have to satisfied simultaneously irrespective of the underlying dynamics: 
\begin{itemize}
\item 
Im $\lambda_1 \lambda_2^* \neq 0$, i.e.  there has to be a relative phase between the weak 
coulings $\lambda_{1,2}$. 
\item 
Im${\cal M}_1{\cal M}_2^* \neq 0$, i.e. FSI have to induce a phase shift 
between ${\cal M}_{1,2}$. 
\end{itemize}
For a larger asymmetry one would like to have also 
$|\lambda_1{\cal M}_1| \simeq |\lambda_2{\cal M}_2|$.

The first condition requires {\em within the SM}   
that such an effect can occur in   
singly Cabibbo suppressed, yet neither Cabibbo allowed nor doubly
suppressed channels. There is a subtle exception to this  
general rule in $D^+ \to K_S \pi ^+$,  
as explained later.

The second condition implies there have to be nontrivial FSI
\index{final state interactions}, 
which can happen in particular when the two transition amplitudes 
differ in their isospin content. We know, as discussed in 
Sect.\ref{TWOBODY} that 
FSI are virulent in the charm region leading to in general 
sizeable phase shifts. While we cannot calculate them and therefore 
cannot predict the size of direct CP asymmetries, even when the 
weak phases are know, CPT symmetry implies some relations between 
CP asymmetries in different channels\index{CPT constraints}. 
For CPT invariance enforces 
considerably more than the equality 
of lifetimes for particles and antiparticles; it tells us that the 
widths for {\em sub}classes of transitions for particles and 
antiparticles have to coincide already, either identically or 
at least practically. Just writing down strong phases in an equation 
like Eq.(\ref{T2M}) does {\em not automatically} satisfy CPT constraints 
\index{CPT constraints}.

We will illustrate this feature first with two 
simple examples and then express it in more general terms. 
\begin{itemize}
\item 
CPT invariance already implies $\Gamma (K^- \to \pi ^- \pi ^0) = 
\Gamma (K^+ \to \pi ^+ \pi ^0)$ up to small electromagnetic 
corrections, since in that case there are no other channels it 
can rescatter with. 
\item 
While 
$\Gamma (K^0 \to \pi^+\pi^-) \neq \Gamma (\bar K^0 \to \pi^+\pi^-)$ 
and $\Gamma (K^0 \to \pi^0\pi^0) \neq \Gamma (\bar K^0 \to \pi^0\pi^0)$ 
one has 
$\Gamma (K^0 \to \pi^+\pi^- + \pi^0\pi^0) = 
\Gamma (\bar K^0 \to \pi^+\pi^- + \pi^0\pi^0)$. 
\item 
Let us now consider a scenario where a particle $P$ and its antiparticle 
$\bar P$ can each decay into two final states only, namely $a,b$ and 
$\bar a, \bar b$, respectively 
\cite{WOLFFSI,URIFSI}. Let us further assume that strong (and
electromagnetic) forces drive transitions among $a$ and $b$ -- and 
likewise for $\bar a$ and $\bar b$ -- as described by an S matrix 
${\cal S}$. The latter can then be decomposed into two parts  
\beq 
{\cal S} = {\cal S}^{diag} + {\cal S}^{off-diag} \; , 
\eeq
where ${\cal S}^{diag}$ contains the diagonal transitions 
$a \Rightarrow a$, $b \Rightarrow b$ 
\beq 
{\cal S}^{diag}_{ss} = e^{2i\delta _s} \; , s=a,b 
\eeq
and 
${\cal S}^{off-diag}$ the off-diagonal ones 
$a \Rightarrow b$, $b \Rightarrow a$:
\beq 
{\cal S}^{off-diag}_{ab} = 
2i{\cal T}^{resc}_{ab} e^{i(\delta _a + \delta _b)} 
\eeq 
with 
\beq 
{\cal T}^{resc}_{ab} = {\cal T}^{resc}_{ba} = 
({\cal T}^{resc}_{ab})^* \; , 
\eeq
since the strong and electromagnetic forces driving the 
rescattering \index{rescattering}
conserve CP and T. The resulting S matrix is unitary to first 
order in ${\cal T}^{resc}_{ab}$. 
 CPT invariance implies the following relation between the 
weak decay amplitude of $\bar P$ and $P$: 
\bea 
T(P \to a) &=& e^{i\delta _a}\left[ T_a + T_b i{\cal T}^{resc}_{ab}\right] \\
T(\bar P \to \bar a) &=& e^{i\delta _a}\left[ T^*_a +T^*_bi{\cal T}^{resc}_{ab}\right]
\eea
and thus 
\beq 
\Delta \gamma (a) \equiv |T(\bar P \to \bar a)|^2 - 
|T(P \to  a)|^2 = 4 {\cal T}^{resc}_{ab} {\rm Im}T^*_a T_b \; ; 
\label{DELTAA}
\eeq
likewise
\beq 
\Delta \gamma (b) \equiv |T(\bar P \to \bar b)|^2 - 
|T(P \to  b)|^2 = 4 {\cal T}^{resc}_{ab} {\rm Im}T^*_b T_a 
\eeq
and therefore as expected 
\beq 
\Delta \gamma (b) = - \Delta \gamma (b)
\eeq
Some further features can be read off from Eq.(\ref{DELTAA}):
\begin{enumerate} 
\item 
If the 
two channels that rescatter have comparable widths -- 
$\Gamma (P \to a) \sim \Gamma (P \to b)$ -- one would like the
rescattering 
$b \leftrightarrow a$ to proceed via the usual strong forces; for
otherwise  the asymmetry $\Delta \Gamma $ is suppressed relative to these 
widths by the
electromagnetic coupling. 
\item 
If on the other hand the channels 
command very different widths -- say 
$\Gamma (P \to a) \gg \Gamma (P \to b)$ -- then a large {\em relative} 
asymmetry in $P \to b$ is accompagnied by a tiny one in 
$P \to a$. 
\end{enumerate} 
This simple scenario can easily be extended to two sets $A$ and $B$ of 
final states s.t. for all states $a$ in set $A$ the transition 
amplitudes have the same weak coupling and likewise for states 
$b$ in set $B$. One then finds 
\beq 
\Delta \gamma (a) = 4 \sum _{b\, \in \, B}{\cal T}^{resc}_{ab}{\rm Im}T_a^*T_b 
\eeq
The sum over all CP asymmetries for states $a \, \in \, A$ cancels the 
correponding sum over $b\, \in \, B$: 
\beq 
\sum _{a\, \in \, A} \Delta \gamma (a) 
= 4 \sum _{b\, \in \, B}{\cal T}^{resc}_{ab}{\rm Im}T_a^*T_b = 
- \sum _{b\, \in \, B} \Delta \gamma (b)
\eeq
\end{itemize}
These considerations tell us that the CP asymmetry averaged over certain 
classes of channels defined by their quantum numbers has to 
vanish. Yet these channels can still be very heterogenous, 
namely consisting of two- and quasi-two-body modes, 
three-body channels and other multi-body decays. 
Hence we can conclude: 
\begin{itemize}
\item 
If one finds a direct CP asymmetry in one channel, 
one can infer -- based on rather general grounds -- which other channels
have to exhibit the compensating  asymmetry as required by CPT invariance.
Observing them would enhance  the significance of the measurements very
considerably.  
\item 
Typically there can be several classes of rescattering channels. 
The SM weak dynamics select a subsect of those where the 
compensating asymmetries have to emerge. QCD frameworks like 
generalized factorization can be invoked to estimate the 
relative weight of the asymmetries in the different classes. 
Analyzing them can teach us important lessons about the 
inner workings of QCD. 
\item 
If New Physics generates the required weak phases (or at least 
contributes significantly to them), it can induce 
rescattering with novel classes of channels. The pattern in the 
compensating asymmetries then can tell us something about the 
features of the New Physics 
involved.  
\end{itemize} 

Direct CP violation can affect transitions involving  
$D^0 - \bar D^0$ oscillations in two different ways, as  
described later. 

\subsubsection{Asymmetries in final state distributions} 

For channels with two pseudoscalar mesons or a pseudoscalar and a vector  
meson a CP asymmetry can manifest itself only in a difference between  
conjugate partial widths. If, however, the final state  
is more complex -- being made up by three pseudoscalar or two   
vector mesons etc. -- then it contains more dynamical information than  
expressed by its partial width, and CP violation can emerge also through  
asymmetries in final state distributions. One general comment  
still applies: since also such CP asymmetries require the  
interference of two {\em different} weak amplitudes, within the SM  
they can occur in Cabibbo suppressed modes only.

In the simplest such scenario one compares CP conjugate  
{\em Dalitz plots}\index{Dalitz plots}. It is quite  
possible that different regions of a Dalitz plot exhibit CP  
asymmetries of varying signs that largely cancel each other when  
one integrates over the whole phase space. I.e., subdomains of the  
Dalitz plot could contain considerably larger CP asymmetries  
than the integrated partial width.  
Once a Dalitz plot is fully understood with all its contributions, one 
has a  powerful new probe. This is not an easy goal to  
achieve, though, in particular when looking for effects that  presumably 
are not large. It might be more promising  as a practical matter to start 
out with a more euristic approach.  I.e., one can start a search for  
CP asymmetries by just looking at conjugate Dalitz plots. One simple 
strategy would be to focus on an area  with a resonance band and analyze 
the density in stripes {\em across} the  resonance as to whether there is 
a difference in CP conjugate plots.

For more complex final states containing  
four pseudoscalar mesons etc. other probes have to be  
employed.  Consider, e.g.,   
$  
D^0 \to K^+K^- \pi ^+ \pi ^- \; ,  
$   
where one can form a T-odd correlation with the momenta:  
$   
C_T \equiv \langle \vec p_{K^+}\cdot  
(\vec p_{\pi^+}\times \vec p_{\pi^-})\rangle  
$.  
Under time reversal T one has  
$   
C_T \to - C_T 
$   
hence the name `T-odd'. Yet $C_T \neq 0$ does not necessarily  
establish T violation. Since time reversal is implemented  
by an {\em anti}unitary operator, $C_T \neq 0$ can be induced by  
FSI \index{final state interactions}\cite{CPBOOK}. While in contrast to the situation  
with partial width differences FSI are not required to produce  
an effect, they can act as an `imposter' here, i.e. induce a T-odd  
correlation with T-invariant dynamics. This ambiguity can unequivoally  
be resolved by measuring   
$   
\bar C_T \equiv \langle \vec p_{K^-}\cdot  
(\vec p_{\pi^-}\times \vec p_{\pi^+})\rangle 
$    
in $\bar D^0 \to K^+K^- \pi ^+ \pi ^- $; finding  
$  C_T \neq - \bar C_T  $    
establishes CP violation without further ado. 
\index{T-odd correlations}
Outline of a search carried out at fixed target experiment FOCUS
 \cite{Pedrini02Frontier} was
presented recently in decay $D^0\to K^-K^+\pi^-\pi^+$,
 with a preliminary asymmetry $A_{T}=0.075\pm 0.064$ out of a sample of 400
 decays.
\par
Decays of {\em polarized} charm baryons provide us with a  
similar class of observables; e.g., in  
$\Lambda _c \Uparrow \; \to p \pi ^+\pi ^-$, one can analyse the  
T-odd correlation $\langle \vec \sigma _{\Lambda _c}  
\cdot (\vec p_{\pi ^+} \times \vec p_{\pi ^-})\rangle$ \cite{BENSON}.  
Probing $\Lambda_c^+ \to \Lambda l^+ \nu$ for   
$\langle \vec \sigma_{\Lambda_c}\cdot (\vec p_{\Lambda}\times  
\vec p_l)\rangle$ or  
$\langle \vec \sigma_{\Lambda}\cdot (\vec p_{\Lambda}\times  
\vec p_l)\rangle$ is a particularly intriguing case; for in this  
reaction there are not even electromagnetic FSI that could fake  
T violation. The presence of a net polarization transverse to the decay 
plane depends on the weak phases and Lorentz structures of the 
contributing transition operators. Like in the well-known case 
of the muon transverse polarization in $K^+ \to \mu^+ \pi ^0 \nu$ 
decays the T-odd correlation is controlled by Im$(T_-/T_+)$, where 
$T_-$ and $T_+$ denote the helicity violating and conserving 
amplitudes, respectively. Since the former are basically absent in the 
SM, a transverse polarization requires the intervention of New Physics 
to provide the required helicity violating amplitude. This can happen in 
models with multiple Higgs fields, which can interfere with amplitudes 
due to $W$ exchange. For $b \to u \ell \nu$ the two relevant amplitudes are: 
\bea
{\cal T}_{W-X} & = &
\frac{G_{F}}{ \sqrt{2}}V_{ub}( \bar u \gamma^{ \alpha}(1-
\gamma_{5})b)( \bar \ell \gamma_{ \alpha}(1- \gamma_{5}) \nu_{\ell})\\
{\cal T}_{H-X} & = & \frac{G_{F}}{ \sqrt{2}}V_{ub} \sum_{i} \frac{C_{i}
m_{b} m_\ell}
{\langle v\rangle ^2}( \bar u (1- \gamma_{5})b)( \bar \ell (1- \gamma_{5})
\nu_{\ell}).
\eea 
with $\langle v\rangle$ denoting the average of the vacuum expectation
values for the (neutral) Higgs fields. For an order of magnitude estimate 
one can calculate the transverse polarization on the quark level. 
Unfortunately one finds tiny effects: ${\cal O}(10^{-5})$ for 
$b \to u \tau \nu$ even taking ${\rm Im}(C_i)\approx 1$. For charm 
such effects are even further suppressed since decays into final states 
with $\tau$ leptons are not allowed kinematically. 

\subsection{CP asymmetries involving oscillations} 
\label{CPVOSC} 

In processes involving oscillations there are actually two gateways 
through  which CP violation can enter. In the notation introduced in 
Sect.\ref{DDOSC} one can have      
\beq  
\left| \frac{q}{p} \right| \neq 1 \; ,  
\eeq 
which unequivocally constitutes CP violation in  
$\Delta C =2$ dynamics, and     
\beq  
{\rm Im}\frac{q}{p}\bar \rho _f \neq 0 \; ,  
\eeq 
which reflects the interplay of CP violation in $\Delta C = 1\&2$ dynamics.  

The first effect can be cleanly searched for in semileptonic transitions: 
\bea  
{\rm rate}(D^0(t) \to l^{+}\nu X) &\propto&  
e^{- \Gamma_H t}+e^{-\Gamma_L t} + 2e^{\bar \Gamma t} 
{\rm cos}\Delta M_Dt \\ 
{\rm rate}(D^0(t) \to l^{-}\nu X) &\propto&  
\left| \frac{q}{p}\right| ^2 \left( 
e^{- \Gamma_H t}+e^{-\Gamma_L t} -  
2e^{\bar\Gamma t} {\rm cos}\Delta M_Dt \right)  
\eea 
and therefore  
\beq  
a_{SL}(D^0) \equiv  
\frac{\Gamma (D^0(t) \to l^{-}\nu X) -  
\Gamma(\bar D^0(t) \to l^{+}\nu X)} 
{\Gamma(D^0(t) \to l^{-}\nu X) +  
\Gamma(\bar D^0(t) \to l^{+}\nu X)} =  
\frac{|q|^4 - |p|^4}{|q|^4 + |p|^4} \; .  
\eeq 
While the rates depend on the time of decay in a non-trivial manner, the asymmetry does not. 

The second effect can be looked for in decays to final states that 
are CP eigenstates, like $D \to K^+K^-$, $\pi^+\pi^-$, in close 
qualitative -- though not quantitative -- analogy to 
$B_d \to \psi K_S$ or $\pi^+\pi^-$  
\footnote{In principle one can also use $D \to K_S\phi$; however one has 
to make  sure that the observed final state $K_SK^+K^-$ is really due to 
$K_S\phi$ rather than $K_Sf_0(980)$; for in the latter case it 
would constitute a CP even rather than a CP odd state like the former, 
and therefore the signal would be washed out. }. 
The general expressions 
for the decay rate as a function of (proper) time are lengthy: 
$$ 
{\rm rate}(D^0(t) \to K^+K^-) \propto e^{-\Gamma _1t} 
|T(D^0 \to K^+K^-)|^2 \times 
$$ 
$$
\left[ 
1+ e^{\Delta \Gamma t} + 2 e^{\frac{1}{2}\Delta \Gamma t} 
{\rm cos}\Delta m_Dt + 
\right. 
$$ 
$$
\left. 
\left( 
1+ e^{\Delta \Gamma t} - 2 e^{\frac{1}{2}\Delta \Gamma t} 
{\rm cos}\Delta m_Dt
\right)\left| \frac{q}{p}\right| ^2 |\bar \rho (K^+K^-)|^2 
\right. +
$$
\beq 
\left. 
2\left( 
1- e^{\Delta \Gamma t} 
\right) {\rm Re}\frac{q}{p} \bar \rho (K^+K^-) 
- 4 e^{\frac{1}{2}\Delta \Gamma t} {\rm sin}\Delta m_Dt 
{\rm Im}\frac{q}{p} \bar \rho (K^+K^-)
\right]
\eeq 
$$ 
{\rm rate}(\bar D^0(t) \to K^+K^-) \propto e^{-\Gamma _1t} 
|T(\bar D^0 \to K^+K^-)|^2 \times 
$$ 
$$
\left[ 
1+ e^{\Delta \Gamma t} + 2 e^{\frac{1}{2}\Delta \Gamma t} 
{\rm cos}\Delta m_Dt + 
\right. 
$$ 
$$
\left. 
\left( 
1+ e^{\Delta \Gamma t} - 2 e^{\frac{1}{2}\Delta \Gamma t} 
{\rm cos}\Delta m_Dt
\right)\left| \frac{p}{q}\right| ^2 |\rho (K^+K^-)|^2 
\right. +
$$
\beq 
\left. 
2\left( 
1- e^{\Delta \Gamma t} 
\right) {\rm Re}\frac{p}{q} \rho (K^+K^-) 
- 4 e^{\frac{1}{2}\Delta \Gamma t} {\rm sin}\Delta m_Dt 
{\rm Im}\frac{p}{q} \rho (K^+K^-)
\right]
\eeq 
To enhance the transparency of these expressions we simplify them 
by assuming there is no direct CP violation -- 
$|\bar \rho (K^+K^-)|=1$ -- and no purely superweak CP violation 
-- $|q|=|p|$: 
$$ 
{\rm rate}(D^0(t) \to K^+K^-) \propto 2 e^{-\Gamma _1t} 
|T(D^0 \to K^+K^-)|^2 \times 
$$ 
\beq 
\left[ 
1+ e^{\Delta \Gamma t} +  
\left( 1- e^{\Delta \Gamma t} \right) 
{\rm Re}\frac{q}{p} \bar \rho (K^+K^-) 
- 2 e^{\frac{1}{2}\Delta \Gamma t} 
{\rm sin}\Delta m_Dt {\rm Im}\frac{q}{p} \bar \rho (K^+K^-)
\right]
\nonumber
\eeq 
$$ 
{\rm rate}(\bar D^0(t) \to K^+K^-) \propto 2 e^{-\Gamma _1t} 
|T(D^0 \to K^+K^-)|^2 \times 
$$
\beq 
\left[ 
1+ e^{\Delta \Gamma t} +  
\left( 1- e^{\Delta \Gamma t}\right) 
{\rm Re}\frac{q}{p} \bar \rho (K^+K^-)  
+ 2 e^{\frac{1}{2}\Delta \Gamma t} 
{\rm sin}\Delta m_Dt {\rm Im}\frac{q}{p} \bar \rho (K^+K^-)
\right]
\nonumber
\eeq 
With those simplifications one has for the relative asymmetry as a
function of $t$: 
$$ 
\frac{{\rm rate}(D^0(t) \to K^+K^-) - {\rm rate}(\bar D^0(t) \to K^+K^-)}
{{\rm rate}(D^0(t) \to K^+K^-) + {\rm rate}(\bar D^0(t) \to K^+K^-)} = 
$$
\beq
- \frac{2e^{\frac{1}{2}\Delta \Gamma t} 
{\rm sin}\Delta m_Dt {\rm Im}\frac{q}{p} \bar \rho (K^+K^-)}
{1+ e^{\Delta \Gamma t} +  
\left( 1- e^{\Delta \Gamma t} \right) 
{\rm Re}\frac{q}{p} \bar \rho (K^+K^-)}
\label{DCS} 
\eeq
As mentioned before if CP is conserved, then the mass eigenstates 
have to be CP eigenstates as well. It is interesting to note how this
comes  about in these expressions: CP invariance implies $|q|=|p|$, 
$|\bar \rho (K^+K^-)| =1$ and 
$\frac{q}{p}\bar \rho (K^+K^-) = \pm 1$ and thus 
${\rm rate}(D^0(t) \to K^+K^-) \propto e^{-\Gamma t} 
|T(D^0 \to K^+K^-)|^2$.

Comparing doubly Cabibbo suppressed modes 
$D^0(t) \to K^+\pi^-$, Eq.(\ref{DCS}) with  
$\bar D^0(t) \to K^-\pi^+$ allows 
particularly intriguing CP tests, since SM effects there are highly 
suppressed. Assuming again $|q/p|^2 =1$ and also 
$|T(D^0\to K^+\pi ^-)|^2 = |T(\bar D^0\to K^-\pi ^+)|^2$, 
$|T(D^0\to K^-\pi ^+)|^2 = |T(\bar D^0\to K^+\pi ^-)|^2$ to 
enhance the transparency of the expressions we arrive at:  
$$  
\frac{{\rm rate}(D^0(t) \to K^+ \pi ^-) - 
{\rm rate}(\bar D^0(t) \to K^- \pi ^+)}
{{\rm rate}(D^0(t) \to K^+ \pi ^-) + 
{\rm rate}(\bar D^0(t) \to K^- \pi ^+)} = 
$$
\beq 
\frac{-x_D^{\prime}\frac{t}{\tau _D} {\rm sin}\phi _{K\pi}
|\hat \rho _{K\pi}|}
{{\rm tg}^2\theta_C + \frac{(x_D^2 + y_D^2)\frac{t^2}{\tau _D^2} 
|\hat \rho _{K\pi}|^2}{4{\rm tg}^2\theta_C} + 
y_D^{\prime}\frac{t}{\tau _D} {\rm cos}\phi _{K\pi}
|\hat \rho _{K\pi}|}
\eeq 
This provides such a promising lab since the SM contribution is  
highly suppressed by tg$^2\theta _C$ in {\em amplitude}. 

As mentioned above, {\em direct} CP violation can manifest itself   
in two different ways, when $\d0d0$ oscillations are involved:  
\begin{itemize} 
\item  
there can be a difference in the absolute size of the CP conjugate  
amplitudes   
\beq 
 |T(D^0 \to f)|^2 \neq |T(\bar D^0 \to \bar f )|^2 
\eeq 
This produces a cos$\Delta M_Dt$ term.  
\item  
It can induce a difference in the quantity Im$\frac{q}{p}\bar \rho_f$  
for two different CP eigenstates $f_{1,2}$ 
\beq 
 {\rm Im}\frac{q}{p}\bar \rho_{f_1} \neq 
{\rm Im}\frac{q}{p}\bar \rho_{f_2} 
\eeq  
leading to different coefficients for the sin$\Delta M_Dt$ term. 
\end{itemize} 

These CP asymmetries involving $D^0 - \bar D^0$ oscillations depend 
on the time of decay in an essential manner. Producing neutral $D$ 
mesons in a symmetric $e^+e^-$ machine just above threshold precludes 
measuring such asymmetries. For the time evolution of the difference in, 
say, 
\bea 
&&e^+e^- \to D^0 \bar D^0 \to (l^+X)_{t_1}(K^+K^-)_{t_2} 
\nonumber 
\\
&vs.& \; e^+e^- \to D^0 \bar D^0 \to (l^-X)_{t_1}(K^+K^-)_{t_2}
\eea
is proportional to sin$\Delta m_D(t_1 - t_2)$ since the $D^0-\bar D^0$ 
pair is produced as a C odd state; it vanishes upon integration over 
the times of decay $t_1$ and $t_2$. This quantum mechanical effect is 
of course the very reason why the 
$e^+e^-$ $B$ factories are asymmetric. 

Nevertheless there are two ways to search for CP violation involving 
$D^0-\bar D^0$ oscillations at such a machine, at least in principle: 
\begin{itemize}
\item 
In $e^+e^- \to D^0 \bar D^0 \gamma$, the $D\bar D$ pair is produced in a 
C even state, and the dependance on the times of decay 
$t_{1,2}$ is sin$\Delta m_D(t_1 + t_2)$; one finds for the
asymmetry in 
$(l^+X)_D(K^+K^-)_D$ vs. $(l^-X)_D(K^+K^-)_D$ integrated over all 
times of decay $2x_D{\rm Im}\frac{q}{p}\bar \rho(K^+K^-)$. This result is 
actually easily understood: averaging the asymmetry in a 
coherent C odd and even $D^0 \bar D^0$ pair yields 
$\frac{1}{2} \left[ 0 + 2x_D{\rm Im}\frac{q}{p}\bar \rho(K^+K^-)\right] = 
x_D{\rm Im}\frac{q}{p}\bar \rho(K^+K^-)$, which coincides with what one
finds  for a incoherently produced neutral $D$ meson. 
\item 
The reaction 
\beq 
e^+e^- \to \psi ^{\prime\prime} \to D^0 \bar D^0 \to f_a f_b \; , 
\eeq
where $f_a$ and $f_b$ are CP eigenstates that are either both even or 
both odd, 
can occur {\em only if} CP is violated. For the initial state is CP even,
the  combined final state CP odd since $f_a$ and $f_b$ have to form a P
wave:  
\beq 
CP[\psi ^{\prime\prime}] = +1 \neq 
CP[f_af_b] = (-1)^{l=1}\eta _a \eta _b = -1 
\eeq
It is thus the mere existence of a reaction that establishes CP
violation. It is {\em not} neccessary for the two states $f_{a,b}$ 
to be the same. 

By explicit calculation one obtains 
$$  
BR(D^0 \bar D^0|_{C=-} \to f_a f_b ) \simeq 
BR(D \to f_a)BR(D \to f_b) \cdot 
$$
\beq
\left[ 
2\left| \bar \rho (f_a) - \bar \rho (f_b)\right| ^2 + 
x_D^2
\left| 1- \frac{q}{p}\bar \rho (f_a) \frac{q}{p}\bar \rho (f_b)
\right|^2 
\right]
\label{PSITOFAFB}
\eeq
There are several intriguing subscenarios in this reaction. 
 \begin{itemize}
  \item 
   In the absence of CP violation -- 
   $\frac{q}{p}\bar \rho (f_a) = \pm 1 = \frac{q}{p}\bar \rho (f_b)$ -- 
   the reaction cannot proceed, as already stated. 
  \item 
   In the absence of $D^0 - \bar D^0$ oscillations -- $x_D =0$ -- 
   it can proceed only if $\bar \rho (f_a) \neq \bar \rho (f_b)$, i.e. if 
   $D \to f_a$ and $D\to f_b$ show a different amount of direct CP 
   violation. 
  \item 
   Without such direct CP violation the transition requires $x_D \neq 0$. 
   In contrast to the situation with $D^0(t) \to K^+K^-$ it is 
   actually quadratic in the small quantity $x_D$. 
  \item 
   Eq.\ref{PSITOFAFB} can actually be applied also when $f_a$ 
   and $f_b$ are not CP eigenstates, yet still modes common to 
   $D^0$ and $\bar D^0$. Consider for
   example 
   $f_a = K^+K^-$ and $f_b = K^{\pm}\pi^{\mp}$ or 
   $f_a = f_b =K^+\pi^-$. Measuring the rate will then yield information 
   also on the strong phase shifts. 
  \end{itemize}
\end{itemize}

\subsection{Theory expectations and predictions  } 
\label{THEORYCPASYM}

As outlined above, there are three classes of quantities  
representing CP violation:  
(i) $|T(D^0 \to f)| \neq |T(\bar D^0 \to \bar f)|$;  
(ii) $|q| \neq |p|$;  
(iii) Im$\frac{q}{p}\bar \rho _f \neq 0$. 

The Wolfenstein representation of the CKM matrix reveals 
one of the best pieces of evidence that the SM is incomplete.
\beq
\label{defCKM}
V_{\rm CKM}=\pmatrix{1-\frac{1}{2}\lambda^2 & \lambda & 
A\lambda^3\left(\rho-i\eta+\frac{i}{2}\eta\lambda^2\right)\cr
-\lambda&1-\frac{1}{2}\lambda^2-i\eta A^2 \lambda^4 & 
A\lambda^2\left(1+i\eta\lambda^2\right)\cr 
A\lambda^3\left(1-\rho-i\eta\right) & -A\lambda^2 & 1\cr}
\label{VCKM}
\eeq
up to higher orders in $\lambda = {\rm sin}^2\theta_C$. On general grounds 
$V_{\rm CKM}$ has to be unitary; yet Eq.(\ref{VCKM}) exhibits a peculiar pattern: 
$V_{\rm CKM}$ is `almost' diagonal -- even close to the identity matrix --, almost 
symmetric with elements that get smaller further away from the diagonal; this  
can hardly be accidental.  

On a practical level it shows that up to higher order in $\lambda$ mixing with the 
third family induces only an imaginary part for the charged current couplings of charm 
with light quarks. The most relevant source for CP violation is the  
phase in $V(cs)$, which is $\simeq \eta A^2 \lambda^4 \simeq \eta |V(cb)|^2 \sim 10^{-3}$  
which provides a very rough benchmark number. One can easily draw  
quark box and penguin diagrams where this phase enters. That does not 
mean, though, that one knows how to calculate contributions from  
these diagrams. For since the internal quarks -- the strange  
quarks -- are lighter than charm, these diagrams do {\em not} represent  
local or even short distance contributions. We have no theoretical description reliably  
based on QCD to calculate such quantities. The usual panacea -- lattice  
QCD -- has first to mature to a unquenched state before it has a chance to   
yield reliable results. This is meant as a call for a healthy dose of skepticism rather  than 
negativism. 

There is one contribution to $q/p$ which can be calculated reliably, namely the one from the 
quark box diagram with $b \bar b$ as internal quarks, since it reflects a local  
operator. Yet as already mentioned it makes a tiny contribution to $D^0 - \bar D^0$ 
oscillations. 

Discuss CP violation in the condensate contribution to  
$D^0$ oscillations ... 

Within the SM no direct CP violation can emerge in Cabibbo favoured  
and doubly suppressed modes, since in both cases there is but a single  
weak amplitude. Observing direct CP violation there would establish   
the intervention of New Physics. There is  
one exception to this general statement  \cite{YAMA}: the transition  
$D^{\pm} \to K_S \pi ^{\pm}$ reflects the interference between  
$D^{+} \to \bar K^0 \pi ^+$ and $D^+ \to K^0 \pi ^+$ which  
are Cabibbo favoured and doubly Cabibbo suppressed, respectively.  
Furthermore in all likelihood those two amplitudes will exhibit  
different phase shifts since they differ in their isospin  content.  
The known CP impurity in the $K_S$ state induces a difference without any theory
uncertainty:    
$$  
\frac{\Gamma (D^+ \to K_S \pi ^+) - \Gamma (D^- \to K_S \pi ^-)} 
{\Gamma (D^+ \to K_S \pi ^+) + \Gamma (D^- \to K_S \pi ^-)} =  
-2{\rm Re}\epsilon _K  
$$ 
\beq  
\simeq - 3.3 \cdot 10^{-3}  
\label{DKSSM} 
\eeq   
In that case the same asymmetry both in magnitude as well as sign arises for the 
experimentally much more challenging  final state with a $K_L$.   

If on the other hand New Physics is present in $\Delta C=1$ dynamics,  
most likely in the doubly Cabibbo transition, then both the sign and the  
size of an asymmetry can be different from the number in Eq.(\ref{DKSSM}),  
and by itself it would make a contribution of the opposite sign to the asymmetry in  
$D^+ \to K_L\pi ^+$ vs. $D^- \to K_L\pi ^-$. 

\begin{figure} 
   \centering 
   \includegraphics[width=10.0cm]{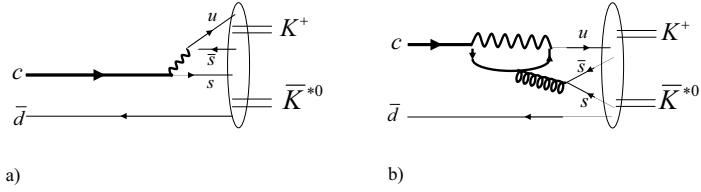} 
\caption{
 Tree and penguin graphs for $D^+\to \bar K^{*0}K^+$.
As explained before the leading term in the penguin diagram here represents a
nonlocal
operator,
since the internal $s$ quark in the loop is lighter than the external charm
quark.
\label{FIG:SCSDCPV} }
\end{figure}

In singly-Cabibbo suppressed modes on the other hand there are  
two amplitudes driven by $c \to d \bar d u$ and $c\to s \bar su$,  
respectively. How they can interfere and generate CP asymmetries can be 
illustrated by the two diagrams in  (Fig.\ref{FIG:SCSDCPV}: the one in a) is a tree diagram, 
 the one in b) of the Penguin type. The latter not only has a different weak phase 
 than the former, but -- being a one-loop diagram with internal quarks $s,d$ being 
 lighter than the $c$ quark -- induces also a strong phase. One cannot rely, though, 
 on this diagram to obtain a quantitative prediction. This penguin diagram 
 with $m_{s,d} < m_c$ does not represent a local operator  and cannot reliably 
 be treated by short-distance dynamics. 
 \par
Searching for {\em direct} CP violation in Cabibbo suppressed $D$ decays as a sign for New 
Physics would however represent a very complex challenge: the CKM benchmark mentioned  
above points to asymmetries of order 0.1 \%. One can perform an analysis like in  
Ref.~\cite{BUCCELLA} based on theoretical engineering of hadronic matrix  
elements and their phase shifts as described above. There one typically  
finds asymmetries $\sim {\cal O}(10^{-4})$, i.e. somewhat smaller  
than the rough benchmark stated above. Yet $10^{-3}$ effects are  
conceivable, and even 1\% effects cannot be ruled out completely. Observing a CP  
asymmetry in charm decays would certainly be a first rate discovery even irrespective of its 
theoretical interpretation. Yet to make a case that a 
signal in a singly  Cabibbo suppressed mode reveals New Physics is quite 
iffy. In all  likelihood one has to analyze at least several channels 
with comparable  sensitivity to acquire a measure of confidence 
in one's interpretation. 

The interpretation is much clearer for a CP asymmetry involving 
oscillations, where one 
\index{$D^0\to K_s\phi$ and $B^0\to K_s\phi$} 
compares the time evolution of transitions like $D^0(t) \to K_S \phi$,  
$K^+ K^-$, $\pi ^+ \pi ^-$ \cite{BSD}  and/or  
$D^0(t) \to K^+ \pi ^-$ \cite{BIGIBERKELEY,Blaylock:1995ay} with their CP conjugate channels. A  difference for a final 
state $f$ would depend on the product  
\beq  
{\rm sin}(\Delta m_D t) \cdot {\rm Im} \frac{q}{p}  
[T(\bar D\to f)/T(D\to \bar f)] \; .  
\eeq    
With  both factors being $\sim {\cal O}(10^{-3})$ in the SM one predicts a practically zero  
asymmetry $\leq 10^{-5}$. Yet New Physics could generate considerably larger values, 
namely $x_D \sim {\cal O}(0.01)$, 
Im$\frac{q}{p} [T(\bar D\to f)/T(D\to \bar f)] \sim  {\cal O}(0.1)$  
leading to an asymmetry of ${\cal O}(10^{-3})$.  One should note that the   
oscillation dependant term is linear in the small quantity $x_D$ (and in $t$) --  
${\rm sin}\Delta m_D t  \simeq x_D t /\tau _D$ --  in contrast to $r_D$ which is  
quadratic:  $ r_D \equiv \frac{D^0 \to l^-X}{D^0 \to l^+X}  
\simeq \frac{x_D^2 + y_D^2}{2}$.  
It would be very hard to see $r_D = 10^{-4}$ in CP insensitive  
rates. It could well happen that $D^0 - \bar D^0$  
oscillations are first discovered in such CP asymmetries! 

Predictions about CP asymmetries in Dalitz
 plot and other final state distributions are at 
present even less reliable beyond the general statement that within the CKM description 
they can arise only in Cabibbo suppressed modes. However it seems conceivable that progress 
could be made there by a more careful evaluation of the available tools rather than having 
to hope for a theoretical breakthrough.

\subsection{Data} 
\label{CPVEXP} 
Experiments so far have basically searched for direct CP asymmetries 
in partial widths and quite recently in Dalitz plot distributions. 
Yet experiments have accumulated data sets sufficiently large for 
meaningful searches for CP asymmetries that can 
occur only in the presence of $\d0d0$ oscillations. As just argued the 
second class of CP asymmetries, if observed, would quite unequivocally 
establish the intervention of New Physics. Recent reviews are in
\cite{Stenson:2002ah}.

\subsubsection{Direct CP asymmetries in partial widths}
\label{DATADIRECTCP}

The data analysis strategy is quite straightforward. The asymmetry between CP 
conjugate partial widths stated in Eq.(\ref{DIRECT}) has the advantage that most 
systematic uncertainties cancel out in the ratio. The particle/antiparticle nature is 
determined by $D^*$-tagging for $D^0$\index{$D^*$ tag} and by the charge 
of the final state for $D^+$, $D_s^+$ and $\Lambda_c^+$. .

As mentioned above, probing for direct CP asymmetries in 
{\em Cabibbo favoured} modes represents 
a search for New Physics, which is required for providing the second weak 
phase. It has, though, some elements of "searching for lost keys in the 
night under the lamp posts": one has not necessarily lost the keys there, but it is 
the first place one can look for them. 
DCSD channels with their highly suppressed SM contributions 
are much more promising in that respect. 
\par
In fixed target experiments (E791, FOCUS) 
\cite{Appel:1993hn,Appel:2000iy,Gottschalk:2002pf}  
one has to correct for the notorious particle/antiparticle {\em production}
 asymmetries 
which occur in hadronization due to associated production, leading particle effects etc. This is 
done by measuring the asymmetry
normalized to a copious mode that is not likely to present a CP 
asymmetry, such as $K^-\pi^+$ or $K^-\pi^+\pi^+$. The main sources of systematic
errors are then the tiny differences in reconstruction efficiency, particle
identification cuts, and absorption of secondaries in the target and
spectrometer.
\par
At $\epem$ colliders  no production 
asymmetry needs to be accounted
for. Standard vertexing techniques are used with  $D^*$-tagging for $D^0$, and
refit of soft pion to the $D^0$ production point. {\it Ad-hoc} analysis is
employed for neutral modes such as $\pi^0\pi^0$ or $K_s\pi^0$ where $D^0$
daughter neutral tracks do not have sufficient precision to reconstruct the
direction of $D^0$. 
\par
The $D^0$ and $D^+$ sections of \cite{Hagiwara:fs} list more than twenty
 measurements of $A_{CP}$ 
(Tab.\ref{TAB:CPVEXP}), all consistent with zero, the best ones at the percent
level. SM direct CP violation is searched for by experiments by investigating
accessable twobody SCS decays channels such as $K^+K^-, \pi^+\pi^-, \pi^0\pi^0$.
Fixed target experiments and $\epem$ experiment CLEO have about the same
sensitivity for charged channels (several thousand events samples), while CLEO
has an obvious advantage in channels with neutral pions. Limits on $A_{CP}$ are
at the 1\% level, with the important recent limit of the mode $K^0_S\pi^+$
discussed in Sect.\ref{THEORYCPASYM}. 
These searches for charm decay rate asymmetries can profitably be 
done at $B$ factories. 
The measurements are relatively simple, since most of systematics cancels in the ratio, and 
do not require any lifetime information. We should expect a factor of two-three 
improvement in $A_{CP}$ limits, once BABAR and BELLE come onto this scene having 
gathered a few $10^6$ charm event sample, which will bring
the present limits below the $1$~\% level. 
A similar level of sensitivity can be expected from CDF. 

A further improvement below the 0.1\% level is expected from BTeV with $10^7$
charm event sample. The only other player in this game could be  hadroproduction
experiment COMPASS\cite{Baum:1996yv}
 at CERN in its eventual phase II, envisioning several $10^6$ 
reconstructed charm meson decays. 
%
\subsubsection{Dalitz plot distributions}
\label{DALITZCPDATA}

The analysis of three-body charged final states  is
complicated by the possibility of intermediate resonant states such as
$K^{*0}K^+$ and $\phi\pi^+$, and thus requires a Dalitz plot
analysis. As a byproduct of their Dalitz plot resonant analysis, 
CLEO has published measurements of $A_{CP}$ {\em integrated} over the 
entire Dalitz plot for several channels\cite{Muramatsu:2002jp,Frolov:2003jf},   
which do not add much additional information to the decay rate asymmetry
measurement. CLEO has also looked for a New Physics CP-violating 
phase in CF mode
$K^-\pi^+\pi^0$, as well as for CP violation in DCS decays under the assumption
that the second phase necessary for CP violating effects to materialize be
provided by $\d0d0$ mixing.
\par
Dalitz plots are promising fields to search for direct CP 
asymmetries, as already mentioned. Since FSI are required it is actually quite 
likely that integrating over the whole Dalitz plot will tend to wash out CP asymmetries.   
Differences in amplitudes and phases of 
resonant structures for
three-body particle and antiparticle final states are regarded as a sensitive
portal for accessing CP violation. No experiment so far has dared
publishing results. FOCUS has only shown \cite{MalvezziICHEP02}
 preliminary results on measured amplitudes and 
phases $\theta \equiv arg(A_i)+\delta_i$
separately for particle and antiparticle, with statistical errors only, finding
zero asymmetry. CLEO \cite{Frolov:2003jf} 
only mention that no difference is observed in
amplitudes and phases without quantitative statements.
 The plethora of arguments
on the interpretation of Dalitz plots in charm decays discussed 
in Sect.\ref{DALITZ}
makes it a complex challenge to apply the Dalitz plot formalism 
to CP studies. 
Yet we would like to re-iterate that it could bear precious 
fruit in charm decays 
while preparing us for similar studies in beauty decays like 
$B\to 3\pi \to \rho \pi, \; \sigma \pi$.

\subsubsection{Indirect CP asymmetries}
\label{DATAINDIRECTCP}

No measurements on time-dependent asymmetries have been 
published so far, although data sets of hefty size exist now for 
some of the interesting channels.

We will limit ourselves to briefly discussing the experimental reach in three case 
studies of indirect CP asymmetries. 
They all require as essential ingredients superb vertexing for
precise lifetime resolution, and large statistics; particle identification and
lepton tag capabilities are also a must. In principle such studies can be 
performed in hadronic collisions, photoproduction and continuum production at 
$\epem$ $B$ factories. At $\tau$-charm factories like BES and CLEO-c one does 
not have access to the lifetime information; yet employing quantum correlations one 
can infer the same information from comparing 
$\epem \to \psi (3770)\to D^0 \bar D^0$ and 
$\epem \to D^{*0}\bar D^0 + h.c. \to D^0\bar D^0 + \gamma/\pi^0$ as explained 
in Sect.\ref{CPVOSC}. 
\begin{enumerate}
\item
$D^0(t) \to l^- X$ vs. $\bar D^0(t) \to l^+ X$ --- 
Although present experiments such as FOCUS have a 20,000 event sample of semileptonic 
decays in hand, it is surprising that no progress has occurred in semileptonic measurements 
concerning oscillations or CP violation beyond the ancient measurement by E791 
\cite{Aitala:1996vz}.
The only new finding is FOCUS' tantalizing (yet unpublished) promise of a 0.2\% ?check 
sensitivity for oscillations in semileptonic $D^0$ decays. 
When determining $A_{CP}(t)$ in a few lifetime bins,
one should expect a 50-10\% statistical error, for each of the FOCUS, BABAR 
and BELLE data sets.
\item $D^0(t) \to K^+K^-$ or $\pi^+\pi^-$ ---
In this case, the observation of {\em any} deviation of the proper time 
distribution from pure single exponential establishes CP violation. 
The distributions from 
FOCUS and BABAR are shown indeed in Fig.\ref{FIG:SIGNALS}.
 The proper time distribution from 
an $\epem$ collider has to be deconvoluted from the  resolution function, thus seemingly 
indicating that fixed target have a distinct advantage in this case. 
However since the CP asymmetry involves oscillations, it needs `time' to build up. 
The maximal effect occurs  around $t/\tau_D \sim x_D^{-1}\pi/2 > 10$; therefore one 
wants to go to as large lifetimes as practically possible.  
The large-lifetime region is critical because a) low-statistics and b) plagued by outliers events 
which constitute a source of systematics for the very measurement of lifetimes.
 \item
   $D^0(t) \to K^+ \pi^-$ vs. $\bar D^0(t) \to K^- \pi^+$ 
   --- Only a few hundred events have been gathered so far of this DCS decay,
   therefore this case looks unfeasible right now below the level of a 50\%
   limit or so.
 \end{enumerate}
 When looking  at experimental possibilities in the medium term the same
 comments presented in the previous section do apply. FOCUS and CLEO  
 are able to provide
 first limits right away. BABAR and BELLE can eventually improve by a factor
  of two-three. CDF could enter the game once they have shown their lifetime
  resolution, and COMPASS when entering their charm phase. After that, a quantum
  leap is expected from BTeV with $10^8$ reconstructed charm decays,
  $5\,10^6$  reconstructed semileptonic decays and great lifetime
  measurement capabilities.

\begin{table}  
 \caption{ CP-violating asymmetries. Data from PDG03 \cite{Hagiwara:fs}
 unless noted.
 Statistical and systematic errors are summed in quadrature.
\label{TAB:CPVEXP} 
 } 
 \footnotesize 
 \begin{center} 
 \begin{tabular}{|llrr|} \hline 
 $D^0 \to$        &    
                  & 
  Events 	  & 
  $A_{CP}$          \\
 $K^+K^-$          &
  PDG03 Avg        &
  7k               &
  $+0.005\pm 0.016$   \\
  $K^+K^-$          &
  BELLE03        &
  36.5k               &
  $-0.002\pm 0.007$   \\
$K^0_SK^0_S$          &
 CLEO        &
 65               &
 $-0.23\pm 0.19$                       \\
$\pi^+\pi^-$          &
 PDG03 Avg        &
 2.5k               &
 $+0.021\pm 0.026$                       \\
$\pi^0\pi^0$          &
 CLEO         &
 810               &
 $+0.001\pm 0048$                       \\
$K^0_S\phi$          &
 CLEO        &
                &
 $-0.028\pm 0.094$                       \\
$K^0_S \pi^0$          &
 CLEO         &
 9.1k               &
 $+0.001\pm 0.013$                       \\
$K^\pm\pi^\mp$          &
 CLEO - assumes no $\d0d0$ mixing        &
 45               &
 $+0.02\pm 0.19$                       \\
$K^\mp\pi^\pm\pi^0$          &
 CLEO - integr. Dalitz pl.        &
    7k            &
 $-0.031\pm 0.086$                       \\
$K^\pm\pi^\mp\pi^0$          &
 CLEO  - assumes no     $\d0d0$ mixing    &
 38               &
 $+0.09^{+0.25}_{-0.22}$                       \\        
$\pi^\mp\pi^\pm\pi^0$          &
 CLEO  \cite{Frolov:2003jf}, integr. Dalitz pl.       &
                &
 $+0.01\pm 0.12$                       \\
\hline
  $D^+ \to$       &
                  & 
            &
             \\  
 $K^0_S\pi^+$   &
 FOCUS            &
 10.6k            &
 $-0.016\pm 0.017$   \\
$K^0_SK^\pm$          &
 FOCUS        &
 949               &
 $+0.071\pm 0.062$                       \\
$K^+K^-\pi^\pm$          &
  PDG03 Avg       &
  14k              &
 $+0.002\pm 0.011$                       \\
$K^\pm K^{*0}$          &
  PDG03 Avg       &
                &
 $-0.02\pm 0.05$                       \\
$\phi\pi^\pm$          &
  PDG03 Avg       &
                &
 $-0.014\pm 0.033$                       \\
$\pi^+\pi^-\pi^\pm$          &
  E791       &
                &
 $-0.017\pm 0.042$                       \\    
\hline 
 \end{tabular} 
  \vfill 
 \end{center}  
\end{table} 
%
\subsection{Searching for CPT violation in charm transitions}
\label{CPTV}
%
CPT symmetry is a very general property of quantum field theory derived by 
invoking little more than locality and Lorentz invariance\index{Lorentz invariance}. 
Despite this impeccable pedigree, it makes sense to ask whether limitations exist. 
Precisely because the CPT theorem rests on such essential pillars of our present paradigm, 
we have to make {\em reasonable} efforts to probe its universal validity. While 
only rather contrived models of CPT violation have been presented, we should keep in mind 
that super-string theories -- suggested as more fundamental than quantum field theory -- 
are intrinsically {\em non}-local and thus do not satisfy one of the basic axioms of 
the CPT theorem; they might thus allow for it, though not demand it. In any case, a veritable 
industry has sprung up \cite{KOSTEL}.

\subsubsection{Experimental limits} 
\label{CPTDATA}

While no evidence for CPT breaking has been found, one should 
note that the purely empirical underpinning of CPT invariance -- in contrast to its 
theoretical one -- is {\em not} overly impressive \cite{CPBOOK}. The `canonical' 
tests concerning the equality in the masses and lifetimes of particles and 
antiparticles yield bounds of typically around $10^{-4}$ for light flavour states. 
Some very fine work has been done on CPT tests in $K_L$ decays. However even 
the often quoted bound $|M_{\bar K^0} - M_{K^0}|/M_K < 9\cdot 10^{-19}$ looks much 
more 
impressive than it actually is. For there is little justification of using $M_K$ as yardstick 
(unless one wants to blame CPT violation on gravity); 
a better -- yet still not well motivated -- calibrator is the $K_S$ width leading to 
$|M_{\bar K^0} - M_{K^0}|/\Gamma_{K_S} < 7\cdot 10^{-5}$. 
Intriguing future tests have been suggested 
for neutral $K$ and $B$ meson transitions \cite{SANDACPT}. Their sensitivity is enhanced 
by involving $K^0-\bar K^0$ and $B^0 - \bar B^0$ oscillations and making use of 
EPR correlations\index{EPR correlations} in $e^+e^-\to \phi(1020)\to K^0\bar K^0$ 
and $e^+e^-\to \Upsilon(4S)\to B_d \bar B_d$, respectively. Their analyses will be 
performed at DA$\Phi$NE and by BELLE and BABAR. 

Masses and lifetimes of $D$ mesons agree to about the $10^{-3}$ level, i.e. an order 
of magnitude worse than for light flavour hadrons. One can also search for CP asymmetries 
that are also CPT asymmetries like $D^+ \to l^+\nu K_S$ vs. $D^- \to l^- \bar \nu K_S$ or 
$D^0 \to l^+\nu K^-$ vs. $\bar D^0 \to l^- \bar \nu K^+$. Oscillation phenomena again 
might 
be the least unlikely place for CPT violation to surface. They present here 
a much less favourable stage, though, than for $K$ and $B$ mesons, since $D^0-\bar D^0$ 
oscillations have not been observed. The {\em phenomenology} is a straightforward 
generalization of the formalism presented in Eq.(\ref{SCHROED}) of Sect. \ref{DDNOT}. 
Without CPT invariance we can have 
$\Lambda_{11} \equiv M_{11} -\frac{i}{2}\Gamma_{11} \neq 
\Lambda_{22} \equiv M_{22} -\frac{i}{2}\Gamma_{22}$. Then one has a complex 
parameter 
for CPT violation: $\xi\equiv \Lambda_{11}- \Lambda_{22}/\Delta \lambda$ with 
$\Delta \lambda$ being the difference in the eigenvalues of the matrix 
$M-\frac{i}{2}\Gamma$. It can be probed by comparing the rates for the Cabibbo 
favoured modes $D\to K\pi$: 
$A_{CPT}(t') \equiv [{\rm rate}(\bar D^0 (t')\to K^+\pi^-) - {\rm rate}(D^0 (t')\to K^-\pi^
+)]/
[{\rm rate}(\bar D^0 (t')\to K^+\pi^-) +{\rm rate}(D^0 (t')\to K^-\pi^+)]$; $t'$ is the 
reduced proper time. 
For slow oscillations 
-- $x_D,y_D \ll 1$ -- one derives 
\beq 
A_{CPT}(t) = [y_D{\rm Re}\xi - x_S {\rm Im}\xi] \cdot \frac{t}{\tau_{D^0}}
\eeq
Measurement of the slope for $A_{CPT}(t')$ thus returns the quantity 
$y_D{\rm Re}\xi- x_D{\rm Im}\xi$.
As a first step a loose limit has been published recently \cite{LINKCPT} based on 
a sample of 17,000 $D^0\to K^-\pi^+$ events. One finds 
\beq 
y_D{\rm Re}\xi- x_D{\rm Im}\xi = 0.0083\pm 0.0065 \pm 0.0041
\eeq
For $x=0, y=1$\% this translates into 
${\rm Re}\xi=0.83\pm 0.65 \pm 0.41$ corresponding to a 50\% limit. 
One order of magnitude increase in statistics expected from B-factory will allow 
to improve the limit to the 10-20\% level.

\subsection{Resume}
\label{CPRESUME}
{\em Within SM dynamics} the breaking of CP invariance manifests itself only in small 
ways: 
\begin{itemize}
\item 
The main stage is in {\em singly Cabibbo suppressed} modes. Direct CP asymmetries in 
partial widths could be `as large as' $10^{-3}$. There is no theorem, though, ruling out 
SM effects of 1\%. 
\item 
In {\em Cabibbo allowed and doubly forbidden} channels {\em no direct} CP violation can 
occur with the exception of modes like $D^{\pm} \to K_S\pi^{\pm}$, where interference 
between 
a Cabibbo allowed and a doubly forbidden amplitude takes place (and where CP violation 
in the $K^0-\bar K^0$ complex induces an asymmetry of 
$2{\rm Re}\epsilon _K \simeq 3.3\cdot 10^{-3}$. 
\item 
CP violation involving $D^0 - \bar D^0$ oscillations is practically absent. 
\end{itemize} 
These expectations should be viewed as carrying an optimistic message: the CP 
phenomenology is a sensitive probe for New Physics. This is further strengthened 
by the availability of the $D^*$ trick to flavour tag the decaying charm meson and by the 
strength of FSI in the charm region, which are a pre-requisite for many asymmetries. 

While so far no CP asymmetry has emerged in charm decays, `the game has only now 
begun', when one has acquired a sensitivity level of down to very few percent. CP 
asymmetries 
of $\sim 10$\% or even more -- in particular for Cabibbo favoured channels --
would have  
been quite a stretch even for New Physics models. 

Analyzing Dalitz plots\index{Dalitz plots} and other final state distributions
is a powerful  
method to probe for CP violation. It is also a challenging one, though, with
potential  
pitfalls, and it is still in its infancy. We hope it will be pushed to maturity. 

Ongoing experiments at $B$ factories and the planned program at BTeV will provide 
data samples that should reveal several effects as small as predicted with the SM. 
%
\section{Charm and the Quark-Gluon Plasma}  
\label{PLASMA}  
%

QCD has been introduced to describe the structure of hadrons and their interactions; 
as such it has to confine quarks and gluons. Yet that is only one of QCD's `faces': based 
on general considerations as well as lattice QCD studies \cite{HEINZ} one 
confidently predicts it to exhibit also a {\em non-}confining phase, where quarks and gluons 
are not arranged into individual hadrons, but roam around freely
\index{deconfinement}. This scenario is referred to 
as quark-gluon plasma. Its simple -- or simplified -- operational characterization is to 
say the {\em non}perturbative dynamics between quarks and gluons can be ignored; 
among other things it leads to the disappearance of the condensate terms\index{condensates} 
for quark and gluon fields mentioned in Sect.\ref{SUMRULES}. 

Great experimental efforts are being made to verify its existence. There is much 
more involved 
than a `merely academic' interest in fulling understanding all aspects of QCD: 
studying the onset of the quark-gluon plasma and its properties  can teach us  
essential lessons on the formation of neutron stars and other exotic celestial 
objects and even on the early universe as a whole. 

Following 
the analogy with QED one undertakes to create a high energy density of roughly  
$1\; {\rm GeV}fm^{-3}$ (compared to $0.15\; {\rm GeV}fm^{-3}$ for ordinary 
cold nuclear matter) by colliding heavy ions against each other. If this energy density 
is maintained for a sufficiently long time, a phase transition to the de-confined quark-gluon 
plasma is expected to occur with ensuing thermalization. 

The crucial question is which 
experimental signatures unequivocally establish this occurrence. 
One of the early suggestions \cite{SATZ} 
has been that a {\em suppression} in the production of heavy quarkonia  
like $J/\psi$, $\psi^{\prime}$, $\chi_c$ etc. can signal the transition to the quark-gluon 
plasma.  This is 
based on a very intuitive picture: the high gluon density prevalent in the plasma 
`Debye'-screens the colour interactions between the initially produced $c$ and 
$\bar c$ quarks and other quarks as well. After the matter has cooled down 
re-establishing the confining phase, hadronization can take place again. Yet the 
$c$ and $\bar c$ quarks have `lost sight' of each other thus enhancing the 
production of charm hadrons at the expense of charmonia even more than 
under normal conditions, i.e. when hadronization is initiated 
{\em without} delay. More 
specific predictions can be made \cite{HEINZ}: the more loosely bound 
$\psi^{\prime}$ is even more suppressed than $J/\psi$, one expects certain 
distributions in $E_{\perp}$ and Feynman $x_F$ \cite{KOPELI} etc. 

Gross features of these predictions have indeed been observed in heavy ion 
collisions at the CERN SPS and at RHIC by the NA38 \cite{NA38}, NA50 \cite{NA50} 
and PHENIX \cite{PHENIX} collaborations, respectively. These findings were indeed 
interpreted as signaling the transition to the quark-gluon plasma. However this conclusion 
has been challenged by authors \cite{CAPELLA}, who offered alternative 
explanations for the 
observed suppression that does not invoke the onset of the quark-gluon plasma. 
A better understanding of charmonia production in ordinary proton-nucleus would 
certainly help to clarify this important issue.  Alternatively one can probe more detailed 
features of charmonium production, like the polarization of $J/\psi$ and 
$\psi^{\prime}$ \cite{IOFFEE}. As described in Sect.\ref{ONIUMPROD} the 
application of 
NRQCD\index{NRQCD} to this problem has so far been  met with less than 
clear success, which could mean that {\em non}perturbative dynamics cannot be 
treated by NRQCD, at least not for charmonia. 
The quark-gluon plasma might offers a scenario 
where NRQCD's {\em perturbative} features can be tested and vindicated. This might 
not only provide a clear signature for the quark-gluon plasma itself, but could also 
serve as a valuable diagnostics of NRQCD and its limitations. 

\section{Summary and Outlook}
\label{OUTLOOK}
\subsection{On Charm's future entries into High Energy Physics' Hall of Fame
aka.   Pantheon
{In "Old Europe's" Romanic parlance} aka. 
Valhalla
{In "Old Europe's" Germanic parlance}}
\label{PANTHEON}
It is a highly popular game in many areas to guess which persons -- athletes, artists, writers, 
actors, politicians -- will be judged by posterity as having acquired 
such relevance due to outstanding achievements that they deserve a special place in history. 
Such an exercise is often applied also to events, even merely conceivable future ones. We  
succumb to this playful urge here, since it can act as a concise summary for how 
future discoveries in charm physics could have a profound impact on our knowledge of 
fundamental dynamics. Such a list is bound to be subjective, of course, yet that is the charm -
- pun intended -- of this exercise. There is {\em no} implication that items left out from the 
list below are unimportant -- merely that we sense no Pantheon potential in them. 

We divide candidate discoveries into three categories, namely those that will certainly make it 
onto the `Valhalla' list, those that are likely to do so and those with no more than an outside
chance; a necessary condition for all three categories of discoveries is of course that they 
happen. 

\subsubsection{Sure bets}
\label{SURE}
Any measurement that unequivocally reveals the intervention of New Physics falls into this 
category. Finding any mode with lepton number violation like $D\to h e^{\pm}\mu^{\mp}$ 
or $D^0 \to e^{\pm}\mu^{\mp}$, Sect.\ref{DMUE}, or with a familon $D^+\to h^+f^0$, 
Sect.\ref{FAMILON}, qualifies-- as would $D\to \mu^+\mu^-, e^+e^-$ in clear excess of the 
SM prediction, Sect.\ref{EVENRARER}.  The last case appears the least unlikely one. 

Maybe the best chance to find a clear manifestation 
of New Physics is to observe an CP 
asymmetry in Cabibbo allowed or doubly suppressed decays 
or one that involves $D^0-\bar D^0$ oscillations, 
Sect.\ref{THEORYCPASYM}.  

\subsubsection{Likely candidates}
\label{LIKELY}
If lattice QCD's predictions on charmonium spectroscopy, on the decay constants $f_D$, 
$f_{D_s}$ and on the form factors in exclusive semileptonic charm decays are fully confirmed 
by future data with uncertainties not exceeding the very few percent level and if its 
simulations can be shown to have their systematics under control to this level -- both 
nontrivial `ifs' -- then 
these results would deserve elite status, since it would demonstrate {\em quantitative} control 
over strong dynamics involving both heavy and light flavours. 

Finding direct CP violation in once Cabibbo suppressed charm decays would be a first-rate 
discovery. To decide whether it requires New Physics or could still be accommodated 
within the SM is, however, a quantitative issue and thus much more iffy. It would be 
overshadowed, though, if CP violation were also found in Cabibbo allowed and doubly 
suppressed modes, as 
listed above.

\subsubsection{On the bubble}
\label{BUBBLE}

Being "on the bubble" means that one hopes to attain a high goal while being 
conscious that one might loose one's footing very quickly. In our context here it 
means that discoveries in this category can hope for elevation to Valhalla only 
if they receive an {\em unequivocal theoretical} interpretation and/or provide important input 
to other measurements of Pantheon standing. This is best illustrated by the first example. 

In Sect.\ref{DDOSC}  we have expressed skepticism about our ability to base a conclusive 
case for New Physics on the observation of $D^0-\bar D^0$ oscillations. This might change, 
though. Yet in addition such a discovery -- whether through $x_D \neq 0$ or $y_D \neq 0$ 
or both -- would certainly be an important one complementing that of 
$K^0-\bar K^0$ and $B^0-\bar B^0$ oscillations, completing the chapter on meson 
oscillations 
and as such represent a text book measurement. Lastly, as explained in
 Sect.\ref{RESDDOSC}, it could have a significant impact on {\em interpreting} the transitions $B\to 
D^{neut}K$ with their anticipated CP asymmetries and how the CKM angle $\phi_3/\gamma$  
is extracted; ignoring  
$D^0-\bar D^0$ oscillations could fake a signal for New Physics or alternatively hide it. 

For other candidates it is even more essential that unambiguous measurements are 
matched with a clear-cut theoretical interpretation. 

Novel insights into light-flavour hadronic spectroscopy inferred from the final states in 
semileptonic and nonleptonic charm decays, Sects.\ref{SPECTSL},\ref{LIGHTFLAVOUR} 
and \ref{BARYONNL} could qualify for elite status, if a {\em comprehensive} picture 
with good theoretical control emerges for a  body of data rather than individual states.

Relying on hidden versus open charm production as signal for the 
quark-gluon plasma, Sect.\ref{PLASMA}, might belong here and maybe the weak 
lifetimes of $C=1$ and $C=2$ hadrons, Sects.\ref{WEAKLIFE}, \ref{CGEQ2BARY}. 
We are not suggesting that all items listed in this third category carry the same intellectual 
weight. 
One should keep in mind that a `Pantheon' is meant to include all gods and goddesses, not 
only those of Olympic standing. 
\subsection{On the Future of Charm Physics}  
\label{FUTURE}  
The discussion in Sects. \ref{THTOOLS} - \ref{CPV} has hopefully 
convinced the reader that  relevant new insights into SM dynamics can be gained 
in charm physics and quite 
possibly even physics beyond the  SM be observed. The pursuit of those 
goals requires  
even larger samples of high quality  data. Fortunately existing machines 
have an ongoing program in that  direction augmented by novel ideas on 
triggering and analysing; even new initiatives have 
been suggested as sketched below. A synopsis of existing and planned
initiatives was shown in Sect.~\ref{KEYCOMP}, Tables \ref{TAB:EXPTS} and
\ref{TAB:FUTURE}.

%
\subsubsection{Photoproduction}  
%
With the end of fixed target programs at the Fermilab Tevatron and the CERN SPS,
no photoproduction experiments with real  photons are planned. 
FOCUS still has significant new analyses in progress and under active 
consideration, and will provide  results 
in channels such as semileptonics, all-charged hadronic decays, and
spectroscopy.
The two experiments H1 and ZEUS at HERA will refine and extend their analysis 
of charm (and even beauty) production and fragmentation 
in the next few years, until their shutdown planned for 2007. Collisions of 
on- and off-shell photons (and off-shell weak bosons) with protons provide a 
dynamical environment more involved than 
$e^+e^-$ annihilation, yet less complex 
than hadronic collisions. As an extra bonus one can even study the transition 
to hadroproduction  by going from the deep inelastic 
regime in momentum transfers 
-- $|q^2| \gg 1$ GeV$^2$ -- to $|q^2|\simeq 0$. 
%
\subsubsection{Hadronic collisions} 
%
After having proved to the scientific community the feasibility of doing
first-class B physics at a hadron collider,
CDF is now opening a new chapter in the study of charm hadrons at 
the TEVATRON by triggering on them.
This new ability
 will give us access 
to huge statistics.
We can expect -- or at least hope -- that, even after severe 
cuts, the available data samples will be of unprecedented size, thus allowing us 
to study single and multiple charm production including their correlations, 
and the 
excitations and decays of single and double charm baryons and their decays. It 
might even extend considerably our sensitivity for $D^0-\bar D^0$ oscillations, 
as well as  CP violation accompanying them and in Dalitz plot distributions, 
where the normalization is not essential. 
The ultimate surprise from CDF would be the ability to decently reconstruct
in their electromagnetic calorimeters photons and neutral pions from charm decays. 
\par 
The COMPASS experiment \cite{Baum:1996yv} at CERN was proposed in 1996 with a "phase
II" devoted to charm hadroproduction and a full spectrum of intended 
measurements, from lifetimes to spectroscopy, measurement of $f_D$ and searches 
for $D^0-\bar D^0$ oscillations and CP violation. The experiment
is currently taking data in its muonproduction mode for
structure functions measurements. The hadroproduction "phase II" foreseen for 2006 
will be devoted to production of exotics and glueballs, and to some charm
physics items \cite{COMPASSWORKSHOP}. They plan to collect $5\,10^6$
fully recostructed charm events in a one-year running of $10^7 s$.
\par
The most emphasized item is $C=2$ baryon spectroscopy. Assuming $C=2$ baryons are 
produced at the SELEX rate with respect to $\Lambda_c$, 
Sect.\ref{C2BAR}, scaling rates up and
considering efficiencies and acceptances via simulation, they end up with an
estimate of $10^4$ reconstructed $C=2$ baryons. A more pessimistic
estimate based on measured hadroproduction cross section for $C=1$ baryons
scaled to $C=2$ production yields about 100 reconstructed $C=2$ baryons.
%
\subsubsection{Beauty Factories}  
%
The $B$ factories at KEK and SLAC have had a spectacularly successful  
start-up with respect to both their technical performance  
\footnote{By the summer of 2002 PEP-II had achieved a luminosity  
of $4.6 \times 10^{33} \, cm^{-2}s^{-1}$ thus exceeding its  
design goal of $3 \times 10^{33} \, cm^{-2}s^{-1}$, while KEK-B  
had established a new world record of  
$7.4 \times 10^{33} \, cm^{-2}s^{-1}$.} and the impact  
of their measurements. For those have already promoted the CKM description of 
heavy flavour dynamics from an ansatz to a tested theory. 
They are also highly efficient factories of charm 
hadrons.  There are three sources for their production:  
(i) Continuum production;  (ii) production of single charm hadrons in $B$ decays driven by  
$b \to c \bar u d,\, c \ell \nu$ and (iii) production of a pair of charm hadrons or of 
charmonium in $B$ decays  due to $b\to c \bar cs$. 

There are further advantages beyond the sample size: the multiplicities  
in the final states are relatively low (though not as low as at a  
tau-charm factory); good vertexing is available and the two $B$ decays  
can in general be separated on an event by event basis; in particular  
for charm emerging from $B$ decays several correlations with quantum  
numbers like beauty, strangeness, lepton number etc. can be exploited. 

\par

Such methods will become especially powerful once one has accumulated  
sample sizes of about 500 $fb^{-1}$ for  
$e^+e^- \to \Upsilon(4S) \to B \bar B$, which might  
be the case by 2005, and has succeeded in fully reconstructing one of 
the beauty mesons in about $10^6$ events 
\cite{SHARMA}. One can then study the decays  
and decay chains of the other beauty mesons with exemplary cleanliness.  
This should allow us to study $D^0 - \bar D^0$ oscillations with  
excellent control over systematics and enable us to measure  
certain quantities for the first time like the absolute values for  
charm {\em baryon} branching ratios \cite{BIGISLBR}. 
\subsubsection{Tau-Charm Factories}  
\label{TAUC}
%
Studying charm decays at threshold in $e^+e^-$ annihilation offers many  
unique advantages:  
\begin{itemize} 
\item  
Threshold production of charm hadrons leads to very clean low  
multiplicity final states with very low backgrounds.   
\item  
One can employ tagged events to obtain the {\em absolute} values  
of charm hadron branching ratios in a model independent way.  
\item  
Likewise one can measure the widths for $D^+ \to \mu ^+ \nu$  
and $D^+_s \to \mu ^+ \nu$ with unrivaled control over  
systematics.  
\item  
With the charm hadrons being produced basically at rest the time  
evolution of $D^0$ decays cannot be measured directly. Yet by comparing  
EPR correlations\index{EPR correlations} in $D$ decays produced in $e^+e^- \to D^0\bar 
D0$,  
$e^+e^- \to D^0\bar D^0 \gamma$ and $e^+e^- \to D^0\bar D^0\pi ^0$  
one can deduce whether oscillations are taking place or not, as explained in 
Sect.\ref{DDOSC}.  
\end{itemize} 
One such machine has been operating since 1990, namely the BEPC  
collider with the BES detector in Beijing. Presently they are  
running with a luminosity of $5\times 10^{30} \, cm^{-2}s^{-1}$  
at the $J/\psi$. There are plans for upgrades leading to  
considerably increased luminosities in the near future. 
\par
To fully exploit the advantages listed above one cannot have  
enough luminosity. Several proposals have been discussed over the  
last 15 years for tau-charm factories with the ambitious goal  
of achieving luminosities of up to the  
$10^{33} - 10^{34} \, cm^{-2}s^{-1}$  range for the c.m. energy interval 
of 3 - 5 GeV. 
\par
One such project is realized at Cornell  
University. They plan to operate CESR-c with luminosities   
$(1.5 - 4.4) \times 10^{32} \, cm^{-2}s^{-1}$ in the range  
$\sqrt{s} =3 - \leq 5$ GeV  for three 
years starting in 2003 using a modified CLEO-III detector.  
The goal is to accumulate  $\sim$  
$1.3 \times 10^9$ $\jp$, $1 \times 10^9$ $\psi ^{\prime}$,  
$3\times 10^7$ $D\bar D$, $1.5\times 10^6$ $D_s^+D_s^-$  
and $4\times 10^5$ $\Lambda_c \bar \Lambda_c$ events \cite{Briere:2001rn},\cite{Cassel:2003vr}.
\par
Data samples of that size and cleanliness would provide ample material  
for many important studies of the SM:  
\begin{itemize} 
 \item  
  They will presumably complete the charmonium spectroscopy, provide  
  authoritative answers concerning charmonium hybrids and clarify  
  the situation with respect to candidates for glueballs and hybrids  
  in charmonium decays like $J/\psi \to \gamma X$.  
 \item  
  CLEO-c's measurements of the widths for $D^+ \to \mu ^+\nu$,  
  $D^+_s \to \mu ^+\nu$ and $D^+_s \to \tau ^+\nu$ will be in a  
  class by themselves: with an integrated luminosity of 3 $fb^{-1}$  
  the uncertainties are expected to be in the 3 - 5 \% regime, about  
  an order of magnitude better than what is achievable at the $B$  
  factories with an integrated luminosity of 400 $fb^{-1}$!  
 \item  
  The absolute branching ratios for nonleptonic decays like  
  $D^0 \to K\pi$ and $D^+ \to K \pi \pi$ [$D^+_s \to \phi \pi$] will be 
  measured with  uncertainties not exceeding the 1\% [2\%] level. The  
  improvement over the present situation would be even greater for  
  charm baryon decays.  
 \item  
  Significant improvements in the direct determination of  
  $|V(cd)|$ and $|V(cs)|$ from $D \to \ell \nu \pi$ and $D \to \ell \nu K$  
  modes could be obtained.  
 \item  
  One could measure the lepton spectra in inclusive semileptonic  
  decays separately for $D^0$, $D^0$ and $D_s^+$ mesons. 
\end{itemize} 
Absolute charm branching ratios and decay sequences represent important  
engineering inputs for $B$ decays, and the present uncertainties  
in them are becoming a bottleneck in the analysis of beauty decays.  

Extracting precise numbers for the decay constants $f_D$ and $f_{D_s}$  
provides important tests for our theoretical control over QCD as it is  
achievable through lattice QCD as described before. The same motivation applies to  
{\em exclusive} semileptonic $D$ decays: to which degree one succeeds  
in extracting the values of $|V(cs)|$ and $|V(cd)|$ -- assumed to be known  
by imposing three-family unitarity on the CKM matrix -- provides a sensitive  
test for the degree to which lattice QCD can provide a quantitative  
description of nonperturbative dynamics and can be trusted when  
applied to {\em exclusive} semileptonic decays of beauty mesons.  

Comparing the measured lepton spectra in {\em inclusive} semileptonic  
decays separately for $D^0$, $D^+$ and $D_s^+$ mesons with expectations  
based on the $1/m_c$ expansion can provide us with novel lessons  
on the on-set of quark-hadron duality. All of this can be summarized  
by stating that a tau-charm factory provides numerous and excellent  
opportunities to probe or even test our mastery over QCD. 
\par
It is not clear, however, whether searches for $D^0 - \bar D^0$  
oscillations and CP violation can be extended in a more than  
marginal way, as long as the luminosity stays below  
$10^{33} \, cm^{-2}s^{-1}$.  

\subsubsection{Gluon-charm Factory at GSI}  
\label{GSI} 
The GSI laboratory in Germany has submitted a proposal for building  
a new complex of accelerators on its site. One of its elements is  
HESR, a storage ring for antiprotons with energies up to 15 GeV.  
It might allow an experimental program contributing to three areas  
of charm physics:  
\begin{itemize} 
 \item  
   In $p\bar p$ annihilation -- implemented with internal targets -- one  
   can exploit the superior energy calibration achievable there --  
   $\Delta E \sim$ 100 KeV vs. $\Delta E \sim$ 10 MeV in  
   $e^+e^-$ annihilation -- to add to our knowledge on charmonium  
   spectroscopy and to search for charm(onium) hybrids through formation  
   as well as production processes. The methodology employed would be  
   an extension of what was pioneered by the ISR experiment R 704 and  
   the FNAL experiments E 760/E 835.  
 \item  
   One can study open charm production in $\bar p$ collisions with heavy  
   nuclei like gold. The goal is to analyze how basic properties of  
   charm quarks like their mass are affected by the nuclear medium,  
   i.e. whether it really lowers their masses, as discussed in  
   Sect.\ref{HVYNUCL}.  
 \item  
   Since $p \bar p$ annihilation is driven by the strong forces, it leads  
   to huge data samples of charm hadrons, which could be employed to  
   look for novel phenomena like CP violation in the charm meson sector.  
   Of course, one needs very clean signatures since there is a huge  
   background of non-charm events. For the same reason this is not an environment 
   for precision measurements. 
\end{itemize} 

\subsubsection{BTeV}  
\label{BTEV}

The BTeV experiment \cite{BTEVPROP}
at the Fermilab Tevatron was proposed to measure oscillations,
 CP violation and rare  decays in both charm and beauty decays. 
The goal is {\it to perform an exhaustive set of measurements of
 beauty and charm particle properties in order  to overdetermine the parameters of the CKM
 matrix, and  to either precisely determine SM parameters or to discover
inconsistencies revealing new physics}.

 \par 

Production of beauty and charm hadrons in $\bar p p$ interactions at 2~TeV center of mass 
energy is peaked in forward-backward regions within a few hundreds milliradians. The 
experiment thus has effectively a fixed-target geometry. Initially conceived as a two-arm 
spectrometer, BTeV, due to funding constraints, is now proposed and awaiting 
approval with a 
one-arm detector.
 \par
 While the BTeV detector is made of the standard set of subdetectors for charm
 and beauty physics (including state-of-the-art em calorimetry),
  the feature that should allow it to be a protagonist in charm
 physics is the first level trigger on detached vertices, which  makes
  BTeV efficient for fully hadronic final states. In one year of data taking
  $(10^7 {\rm s})$ at a luminosity of $2\cdot 10^{32}$\cm2s1 $2\cdot 10^{12}$ $\ccb$ 
pairs   will be produced. With a 1\% trigger efficiency and 10\%
  reconstruction efficiency, BTeV expects $10^9$ {\em reconstructed} charm decays
  in one year (in addition to $4\cdot 10^4$ reconstructed $B^0\to \jp K^0_s$ and 
  1500 $B^0\to \pi^+ \pi^-$). 
\par  
 BTeV will be able to perform all studies done so far by fixed-target
 experiments with the higher hadronic background supposedly
  tamed by superior
 detector solutions both in tracking (pixel detector), and in neutral
 reconstruction (crystal em calorimeter). As far as charm is concerned major emphasis 
 is placed on probing 
 high-impact physics topics such as $\d0d0$  oscillations, CP
 violation and doublecharm spectroscopy.
\par
 With $10^9$ reconstructed charm decays,
  BTeV expects to reach a sensitivity of 
 $(1-2)\cdot 10^{-5}$ on the oscillation parameter $r_D$ in both  semileptonic and hadronic
 final states. With $10^6$ reconstructed, background-free SCS decays, the
 sensitivity to CP violating asymmetries will be of order $10^{-3}$; BTeV will thus enter 
 the range of SM predictions for direct CP violation there. Particular care 
 will be taken to keep the systematics under sufficient control. The definition of the primary 
vertex and secondary interactions in the target, which were the main sources of systematic 
errors at fixed-target experiments, will not be an issue here.
\par
BTeV is anticipated to measure semileptonic formfactors with a percent level precision -- 
similar to what is expected from CLEO-c, yet with very different systematics. Finally BTeV's 
photon reconstruction capability will allow high sensitivity probes of $D\to \gamma X$. 
Within the SM they are -- unlike $B\to \gamma X$ -- dominated by long-distance dynamics 
with predictions for their branching ratios ranging from $10^{-6}$ up to $10^{-4}$. These 
rates 
can be significantly enhanced in SUSY scenarios with the comparison of 
$D\to \gamma \rho/\omega$ versus $D\to \gamma K^*$ providing an important diagnostic. 

\subsubsection{Lattice QCD}  
\label{LATQCD}
%

The tau-charm factory with the precise measurements  
it will allow has been called a "QCD machine". Such a label implies 
that a quantitatively 
precise  theoretical treatment of charm(onium) transitions can be given.  
Lattice QCD represents our best bet in this respect. This  
theoretical technology has made considerable progress through the  
creation of more efficient algorithms together with the availability of  
ever increasing computing power. We have certainly reason to  
expect such progress to continue. One can also benefit from two  
complementary approaches to treating charm physics on the lattice:  
one can approach the charm scale from below largely by `brute force', 
i.e. by using finer and larger lattices;  or one can approach it from 
above by scaling down from the heavy  quark limit and from $b$ quarks. 
In either case it will be essential to perform fully unquenched 
lattice simulations of QCD, which seems to be within reach. In particular 
the recent `manifesto' of Ref. \cite{LATTICEPREC}  expresses 
considerable optimism that a real 
breakthrough in this direction is happening right now. In any case there is 
good hope that the anticipated progress in our experimental knowledge 
of charm dynamics will be matched by progress in our theoretical understanding 
with {\em defendable} theoretical uncertainties in the decay constants and semileptonic 
formfactors on the percent level.  

\section{`Fabula docet' }  
\label{FABULA}  

The discovery of charm states achieved much more than `merely' establish the existence 
of a second complete quark family -- it marked a true paradigm shift in how the community 
viewed quarks: before the observation of Bjorken scaling \index{Bjorken scaling} 
in deep inelastic lepton-nucleon scattering quarks were regarded by many as objects 
of mathematical convenience; certainly after the discoveries of $J/\psi$ and 
$\psi ^{\prime}$ they were seen as real physical objects albeit confined ones. 
Many important lessons of philosophical, historical and sociological relevance on 
progress in general and in the sciences in particular can be drawn from  this transition

It also gave nature an opportunity to show its kindness, which is much more than its 
customary lack of malice: the discovery of charm mesons with lifetimes 
$\sim 4\cdot 10^{-13}$ sec provided the impetus for developing a new 
{\em electronic} technology to resolve such lifetimes, namely silicon 
microvertex detectors \index{silicon microvertex detectors}. Those were ready 
`just in time' to take on another challenge, namely measuring lifetimes of 
beauty hadrons. Resolving track lengths of a particle with a lifetime $\sim 4\cdot 10^{-13}$ 
sec is to first order equivalent to that of a particle with three times the lifetime and mass. 
The silicon technology has been and is still experiencing spectacular success in 
tracking $B^0 - \bar B^0$ oscillations, CP asymmetries in $B_d \to J/\psi K_S$ and 
tagging top quark decays through beauty hadrons in their final state.  
$B$ tagging is also an essential tool in searching for Higgs bosons. 
More generally the search for and analysis of charm hadrons gave rise to new 
detector components, trigger devices and experimental setups and strategies that are now 
firmly established in the mainstream of HEP. They have met with spectacular successes in 
beauty physics. 

We have also described how 
new combinations of previously existing theoretical technologies as well as novel ones 
contributed to our progress in understanding charm and subsequently beauty dynamics. 

Every truly good story  actually points to the future as well and in more than one way, and 
this is certainly the case here as well. The thesis of this review is that a strong case 
can be made for continuing dedicated studies of charm dynamics  to be pursued 
at existing facilities like CLEO-c, the $B$ factories at KEK and SLAC and FNAL's 
TEVATRON collider. This case rests on three pillars, one experimental and two more 
of a theoretical nature:
\begin{itemize}
\item
More precise data on charm spectroscopy and decays are needed as 
{\em engineering input} for refining our analysis of beauty decays. To cite but one 
non-trivial example concerning the determination of $V(cb)$: to extract 
$\Gamma (B \to \ell \nu X_c)$ including its uncertainty accurately from real data, 
one has to know the masses, quantum numbers and decays of the various 
charm resonances and combinations occurring there. This applies also when 
measuring $B\to \ell \nu D^*$, $l \nu D$ at zero recoil. 
\item 
Likewise the theoretical tools for treating beauty decays rely on input from the 
charm sector. Consider the just mentioned example: sum rules derived from 
QCD proper relate the contributions of different charm resonances to semileptonic 
$B$ decays to the basic heavy quark parameters that in turn control the 
theoretical expressions for $B$ decay rates. 

Furthermore the tools for describing beauty decays have not only been developed 
first for charm decays, but can still be calibrated and refined with more precise and 
comprehensive charm data. This will enhance considerably their credibility in 
treating beauty decays. This is true for considerations based on quark models, 
light cone sum rules, HQE and in particular for lattice QCD as described in the 
previous section. In particular for the validation of the latter charm can act 
as an important bridge, since lattice QCD should be able to approach charm 
dynamics from higher as well as lower scales.  

\item 
The existence of charm hadrons and their basic properties has provided essential 
confirmation for the SM. Yet at the same time it offers a unique angle to searches 
for physics beyond the SM, which is `orthogonal' to other approaches. In contrast 
to $s$ and $b$ quarks charm is an up-type quark; unlike top it hadronizes and can 
thus exhibit coherent phenomena that enhance CP asymmetries 
\footnote{Hadronization is typically decried as an evil feature curtailing our ability 
to treat CP violation in strange decays quantitatively. This is however a short-sighted 
view: indeed it makes it harder to extract the microscopic quantities describing 
CP violation; yet without it there would be neither $K^0 - \bar K^0$ nor 
$B^0 - \bar B^0$ oscillations, and CP violation could not manifest itself through 
$K_L \to \pi \pi$ or in $B_d(t) \to J/\psi K_S$.} 
This will however not happen `automatically' -- dedicated efforts will be required. 
The fact that no New Physics has shown up yet in charm transitions should not at all 
deter us from continuing our searches there. On the contrary it has been only very recently 
that we have entered the experimental sensitivity level in 
$D^0 - \bar D^0$ oscillations and CP violation, where a signal for New Physics 
would be believable. 
Finally,  experience has shown time and again that when one 
gives HEP groups data and time, they will find novel ways to formulate questions 
to nature and understand its replies. 
\end{itemize} 
An artist once declared that true art is due to 10\% inspiration and 90\% perspiration, 
i.e. committed long term efforts. This aphorism certainly applies to progress in 
fundamental physics as demonstrated by charm's tale. 
%
\vspace*{.2cm}

{\bf Acknowledgements:}~
We have benefitted from exchanges with Profs. 
 M. Artuso,
 E. Barberio, 
 W. Bardeen, 
 R. Cester, 
 B. Guberina, 
 P. Migliozzi,
 J. Miranda, 
 P. Nason, 
 D. Pedrini, 
 J. Russ, 
 K. Stenson, 
 N. Uraltsev.
Technical help by R. Baldini and P. Biosa (Frascati) is
gratefully acknowledged.
This work was supported by the
National Science Foundation under grant number PHY00-87419, and by 
 the Italian Istituto
 Nazionale di Fisica Nucleare and
 Ministero dell'Istruzione, dell'Universit\`a e della Ricerca.

%
%
\end{document}